\documentclass[conf]{new-aiaa}

\usepackage{graphicx}
\usepackage{float}
\usepackage{placeins}

\usepackage{amsmath}

\usepackage{amssymb}
\usepackage{booktabs}
\usepackage{longtable}
\graphicspath{{figs/}}

\newcommand{\con}{c_{\mathrm{on}}}

\newcommand{\tocentry}[2]{\phantomsection\addcontentsline{toc}{#1}{#2}}
\makeatletter
\renewcommand*\l@section[2]{%
  \addpenalty\@secpenalty\addvspace{1.0em \@plus\p@}%
  \setlength\@tempdima{2.1em}%
  \begingroup
    \parindent\z@ \rightskip\@pnumwidth \parfillskip-\@pnumwidth
    \leavevmode\bfseries \advance\leftskip\@tempdima \hskip-\leftskip
    #1\nobreak\hfil\nobreak\hb@xt@\@pnumwidth{\hss #2}\par
  \endgroup}
\renewcommand*\l@subsection{\@dottedtocline{2}{2.1em}{3.0em}}
\makeatother

\newcommand{\figcont}[1]{\addtocounter{figure}{-1}\caption{\emph{Continued.} #1}}

\title{Two Regimes, One Equation: The Spalart--Allmaras with Autogenous
Inception (SA-AI) Transition-Aware RANS Model}

\author[1,2]{Qiqi Wang\footnote{Founding Technologist and Chief Taoist Officer, Flexcompute, Inc.;
Associate Professor, Department of Aeronautics and Astronautics, Massachusetts
Institute of Technology. AIAA Associate Fellow.}}
\affil[1]{Flexcompute, Inc.}
\affil[2]{Department of Aeronautics and Astronautics, Massachusetts Institute of Technology}

\begin{document}

\thispagestyle{empty}\setcounter{page}{0}
\vspace*{\fill}
\begin{center}\includegraphics{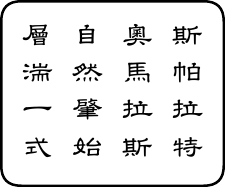}\end{center}
\vspace*{\fill}\vspace*{0.12\textheight}
\clearpage

\maketitle

\begin{abstract}
The envelope $e^N$ method predicts natural transition from three inputs: the
shape of the boundary-layer profile, the momentum-thickness Reynolds number,
and the freestream turbulence intensity. The first two are integrals across
the boundary layer, and the amplification factor $N$ is an integral along it.
An unstructured Reynolds-averaged Navier--Stokes (RANS) solver evaluates only
local quantities and cannot apply the method directly. The model described
here predicts natural transition with local quantities, inside the one-equation
Spalart--Allmaras (SA) turbulence model. Where the SA variable $\tilde\nu$ is
below the molecular viscosity $\nu$, the eddy viscosity of SA is negligible,
and $\ln\tilde\nu$ is used as $N$. Its growth rate is a function of two
dimensionless indicators of the profile shape, a vorticity fraction and an
accumulated inflection. Both are formed from the velocity, its first and
second derivatives, and the wall distance at a point. Its growth begins where
a vorticity Reynolds number exceeds a threshold. Its freestream value is a
function of the turbulence intensity. Near
$\tilde\nu=\nu$ the production changes from the amplification source to the
production of SA. One equation therefore has two regimes: the laminar growth
of a disturbance, and the turbulence that follows it. The constants of the
amplification are determined by three results of stability theory: the growth
rate of a free shear layer, the linear-stability onset of the Falkner--Skan
family, and the Blasius $e^N$ envelope. The model is computed on a
zero-pressure-gradient flat plate, the NLF(1)-0416 and Eppler 387 airfoils, a
two-element section, a circular cylinder through the drag crisis, the Daedalus
human-powered-aircraft wing, and a 6:1 prolate spheroid, and compared with
experimental data and with other transition models.
\end{abstract}

\par\vspace{9.6pt}\begin{center}\includegraphics{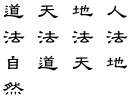}\end{center}\vspace{-9.6pt}

\clearpage
\setcounter{tocdepth}{2}
{\hypersetup{hidelinks}\tableofcontents}

\section*{\tocentry{section}{Nomenclature}Nomenclature}
{\renewcommand\arraystretch{1.0}
\noindent\begin{longtable*}{@{}l @{\quad=\quad} p{4.5in}@{}}
$A,\;B,\;C$ & constants of the onset threshold ($175$, $2$, $2600$), Eq.~\eqref{w:onset} \\
$a,\;a_{\max}$ & amplification rate and its ceiling, the rate of a free shear layer ($0.19$), Eq.~\eqref{w:a} \\
$C_L,\;C_D$ & lift and drag coefficients of a wing \\
$C_l,\;C_d,\;C_m$ & lift, drag, and pitching-moment coefficients of a section \\
$C_p,\;C_f$ & pressure and skin-friction coefficients \\
$c,\;x/c$ & chord and chordwise coordinate \\
$c_{b1},\,c_{b2},\,c_{v1},\,c_{w1},\,c_{w2},\,c_{w3}$ & SA constants, Eq.~\eqref{w:closure} and Table~\ref{t:saconstants} \\
$c_{\nu,\mathrm{ai}}$ & factor on the molecular viscosity in the diffusion term of $\tilde\nu$ ($1/6$), Eq.~\eqref{w:sa} and Section~II.D \\
$\con$ & onset factor ($0.712$), Eq.~\eqref{w:onset} and Section~II.D \\
$D,\;D_\mathrm{SA}$ & destruction of $\tilde\nu$, Eq.~\eqref{w:blend}, and SA's destruction, Eq.~\eqref{w:closure} \\
$d,\;\mathbf d$ & distance to the nearest wall; wall-distance vector $d\,\nabla d$, Eq.~\eqref{w:ratio} \\
$f_{v1}$ & SA's damping function of the eddy viscosity, Eq.~\eqref{w:closure} and Section~II.F \\
$H$ & shape factor, the displacement thickness divided by the momentum thickness, Section~I \\
$I,\;\hat I$ & accumulated inflection $Y-X-Z$ and its nondimensionalized form $I/U_\mathrm{loc}$, Eqs.~\eqref{w:ihat} and~\eqref{w:integral} \\
L0, L1, L2 & the three grid-refinement levels, coarse to fine, Section~III \\
$\boldsymbol\ell$ & curvature vector, Eq.~\eqref{w:ratio} \\
$M$ & Mach number \\
$N,\;N_\mathrm{crit}$ & amplification factor of the $e^N$ method and its critical value, Section~I and Eq.~\eqref{w:mack} \\
$P,\;P_\mathrm{SA},\;P_\mathrm{AI}$ & production of $\tilde\nu$, Eq.~\eqref{w:blend}; SA's production, Eq.~\eqref{w:closure}; amplification source $a\,\sigma_{Re}\,\omega\,\tilde\nu$, Eq.~\eqref{w:blend} \\
$p$ & pressure, Eq.~\eqref{w:ns} \\
$q$ & ratio of $|\mathbf u|\,d/\nu$ to its value in a wall layer with the same $\chi$, Eq.~\eqref{w:fv1q} \\
$Re$ & chord Reynolds number \\
$Re_D,\;Re_L$ & cylinder-diameter and spheroid-length Reynolds numbers, Sections~III.F and~III.H \\
$Re_x,\;Re_\theta$ & streamwise and momentum-thickness Reynolds numbers \\
$Re_\Omega,\;Re_\Omega^{\mathrm c}$ & vorticity Reynolds number $d^2\omega/\nu$, Section~II.C, and its onset threshold, Eq.~\eqref{w:onset} \\
$\tilde S$ & SA modified vorticity, Eq.~\eqref{w:closure} \\
$Tu$ & freestream turbulence intensity, Eq.~\eqref{w:mack} \\
$U_\infty,\;U_e$ & freestream speed; velocity at the edge of the boundary layer \\
$U_\mathrm{loc}$ & local velocity scale, $\sqrt{X^2+Y^2+Z^2}$, Eq.~\eqref{w:ind} \\
$U^+$ & velocity divided by $u_\tau$, Eq.~\eqref{w:fv1q} \\
$u;\;u',\,u''$ & velocity parallel to the wall; its first and second wall-normal derivatives, Section~II.B \\
$\mathbf u$ & velocity vector, Eqs.~\eqref{w:ns} and~\eqref{w:ratio} \\
$u_\tau$ & friction velocity, the square root of the wall shear stress divided by $\rho$, Section~II.D \\
$w$ & relative width of the onset gate ($0.35$), Eq.~\eqref{w:rate} \\
$X,\;Y,\;Z$ & velocity, shear, and curvature indicators, each with the dimension of velocity, Eq.~\eqref{w:ind}; in three dimensions, Eq.~\eqref{w:ratio} \\
$\hat X,\;\hat Y,\;\hat Z$ & the same, divided by $U_\mathrm{loc}$: direction cosines on the unit sphere, Eq.~\eqref{w:ind} \\
$x_\mathrm{tr};\;x_{LS},\,x_R$ & transition location; laminar-separation and reattachment stations, Section~III \\
$y^+$ & wall coordinate, $d\,u_\tau/\nu$, Section~II.E \\
$\alpha$ & angle of attack \\
$\beta$ & Falkner--Skan wedge parameter, Section~II.B \\
$\delta_\nu$ & boundary-layer length scale, $\sqrt{\nu x/U_e}$, Section~IV.A \\
$\eta$ & fraction of semi-span, Section~III.G \\
$\kappa$ & von K\'arm\'an constant ($0.41$) \\
$\mu,\;\mu_t$ & molecular viscosity and eddy viscosity (dynamic), Eq.~\eqref{w:ns} \\
$\nu$ & kinematic molecular viscosity, $\mu/\rho$ \\
$\tilde\nu$ & SA working variable, Eq.~\eqref{w:sa} \\
$\rho$ & density \\
$\sigma$ & SA diffusion constant ($2/3$), Eq.~\eqref{w:sa} \\
$\sigma_{Re}$ & onset gate on $Re_\Omega$, Eq.~\eqref{w:rate} \\
$\sigma_P,\;\sigma_D$ & handover weights of the production and the destruction, Eq.~\eqref{w:sigma} and Section~II.E \\
$\sigma_{v1};\;\sigma_\chi,\;\sigma_q$ & bypass weight in the eddy viscosity and its two gates, Eq.~\eqref{w:fv1bypass} and Section~II.F \\
$\tau$ & width of the handover in $\chi$ ($4$), Eq.~\eqref{w:sigma} \\
$\boldsymbol\tau$ & stress tensor, Eq.~\eqref{w:ns} \\
$\chi,\;\chi_\infty$ & $\tilde\nu/\nu$ and its freestream value, Eq.~\eqref{w:mack} \\
$\hat\Omega$ & vorticity fraction, $Y/\sqrt{X^2+Y^2}$, Eq.~\eqref{w:omhat} \\
$\omega,\;\boldsymbol\omega$ & vorticity magnitude and vorticity vector \\
\end{longtable*}}

\section{Introduction}

\par\vspace{7.5pt}\begin{center}\includegraphics{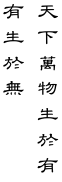}\end{center}\vspace{-7.5pt}

A laminar boundary layer in a stream of low turbulence intensity does not
turn turbulent all at once. It first becomes unstable to small-amplitude
two-dimensional waves, the
Tollmien--Schlichting (TS) waves~\cite{mack_1977,schmid_henningson_2001}. Waves
of different frequencies grow at different rates. The envelope, the largest
amplitude over all frequencies, grows nearly exponentially along the surface,
by a factor of several thousand. The distance that this growth takes determines
where transition occurs. Transition itself takes a much shorter distance: three-dimensional instabilities become
dominant, and turbulent spots form, spread, and
merge~\cite{schubauer_klebanoff_1956}. The location of transition changes the
forces on a body. On the
natural-laminar-flow airfoil of Section~III.B, a computation that assumes
transition at the leading edge gives $1.6$ to $2.5$ times the drag of a
computation in which the model determines the transition location.

The envelope $e^N$ method represents the growth of the envelope by one number, the amplification factor $N$ \cite{smith_gamberoni_1956,vaningen_2008}. It is the logarithm of
the factor by which the envelope has grown along the boundary layer. Transition is declared
where $N$ reaches $N_\mathrm{crit}$, and $N_\mathrm{crit}$ is a function of the freestream turbulence intensity~\cite{mack_1977}. Drela and Giles reduced the stability
calculation behind $N$ to two functions of the shape factor $H$ \cite{drela_giles_1987}. The shape factor is the ratio of the
displacement thickness to the momentum thickness, one dimensionless number for the shape of the
boundary-layer profile. The envelope does not grow until the momentum-thickness Reynolds number $Re_\theta$ passes a critical
value $Re_{\theta0}(H)$. Past it, $N$ grows at a rate $dN/dRe_\theta$ that depends on $H$ alone. The method therefore does three jobs,
and each job has its own input:
\begin{enumerate}\itemsep2pt
\item How fast the disturbance grows. The input is the shape of the boundary-layer profile, through $H$.
\item When the growth starts. The input is $Re_\theta$, compared with a threshold that also depends on $H$.
\item How far the disturbance must grow. The input is the freestream turbulence intensity.
\end{enumerate}

A general RANS solver cannot apply this method directly. Both $H$ and $Re_\theta$ are integrals across the boundary layer, and $N$ is an integral along it. An unstructured solver can evaluate only
local quantities at each grid point: the flow variables, their gradients, and
the distance to the nearest wall.

The model described here does the three jobs with local quantities, inside
the Spalart--Allmaras (SA) equation~\cite{spalart_allmaras_1992}. SA
transports one field, $\tilde\nu$. Where $\chi=\tilde\nu/\nu$ is below one,
the resulting eddy viscosity is negligible. The baseline model leaves this
range ``empty''. Here $\ln\tilde\nu$ fills it, in the role of $N$. Its starting level, the
freestream value, is a function of the freestream turbulence intensity. Its growth
begins where a local Reynolds number, formed from the vorticity and the wall
distance, exceeds a threshold. Its growth rate depends only on the local shape of
the boundary-layer profile. Near $\chi=1$ the production switches to that of baseline SA. One
equation therefore has two regimes: the laminar growth of a disturbance, and
the turbulence that follows it. The model is called SA-AI\footnote{``AI'' here is not the other kind making headlines in the 2020s.}, Spalart--Allmaras with
Autogenous Inception: the turbulence variable carries the inception of its
own turbulence.

SA-AI is a transition-aware RANS model for natural, Tollmien--Schlichting transition at low Mach number.
Bypass transition ($Tu\gtrsim1\%$), crossflow instability, and
compressibility effects are not modeled.

Section~\ref{sec:model} builds the model one piece at a time and ends with
its constants. Each piece is the simplest form found to do its job, and no
piece has been refined beyond that. Section~\ref{sec:results} reports what it computes on a flat
plate, two airfoils, a two-element section, a circular cylinder through the
drag crisis, the Daedalus wing, and a 6:1 prolate spheroid.
Section~\ref{sec:envelope} describes what the model computes in five
situations beyond those cases: a rotating frame, a curved wall, a swept
attachment line, a wall with suction, and the wake of another surface.

\section{Model}\label{sec:model}

\subsection{Transport Equation and Blended Sources}

\par\vspace{7.6pt}\begin{center}\includegraphics{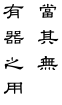}\end{center}\vspace{-7.6pt}

SA's eddy viscosity is $\rho\tilde\nu f_{v1}(\chi)$, with
$f_{v1}=\chi^3/(\chi^3+c_{v1}^3)$ and $c_{v1}=7.1$. At $\chi=1$ that is about
$0.3\%$ of $\rho\tilde\nu$. Below $\chi\approx1$ the eddy viscosity is negligible whatever the value of $\tilde\nu$. This is the empty range.

In that empty range, a source proportional to $\tilde\nu$ makes $\tilde\nu$
grow exponentially along a particle path. When $\tilde\nu\ll\nu$,
\begin{equation}
  \frac{D\ln\tilde\nu}{Dt} = a\,\sigma_{Re}\,\omega ,
  \label{w:lngrowth}
\end{equation}
and $\ln\tilde\nu$ accumulates as $N$ does, at the rate $a\,\sigma_{Re}$ on
the time scale $1/\omega$, the inverse of the vorticity magnitude. Three quantities each do one of the three jobs of the $e^N$ method listed in the Introduction. The rate $a$ determines how fast the
disturbance grows. It is a dimensionless number that depends only on the shape of the boundary-layer profile. The gate $\sigma_{Re}$ determines when the growth starts. A gate is a factor
between zero and one that multiplies a term: at zero the gate is shut and the
term vanishes, at one it is open and the term acts in full. This gate,
$0\le\sigma_{Re}\le1$, opens as a local Reynolds number exceeds a threshold.
Every $\sigma$ with a subscript in this paper is such a factor; the $\sigma$
without a subscript is the SA diffusion constant. The freestream value $\chi_\infty$ determines how far the disturbance must grow: $\ln\tilde\nu$ increases from that value
until $\chi$ reaches one.

The full model is SA with that source and four adjustments: a reduced
molecular diffusion, a handover in the production, a matching weight on the
destruction, and a bypass in the eddy viscosity. It is stated in full in this
subsection, for reference. The subsections that follow reach it from
Eq.~\eqref{w:lngrowth} one step at a time. Sections~II.B and~II.C define $a$
and $\sigma_{Re}$, which completes Eq.~\eqref{w:lngrowth}. Section~II.D adds
the molecular diffusion of $\tilde\nu$. Section~II.E adds the handover to SA.
Section~II.G writes the indicators for a three-dimensional flow.
Section~II.F turns to the momentum equation, which the model affects only
through the eddy viscosity:
\begin{gather}
  \frac{\partial\rho}{\partial t} + \nabla\!\cdot\!(\rho\mathbf u) = 0,
  \qquad
  \frac{\partial(\rho\mathbf u)}{\partial t}
    + \nabla\!\cdot\!(\rho\,\mathbf u\!\otimes\!\mathbf u)
    = -\nabla p + \nabla\!\cdot\!\boldsymbol\tau, \notag\\[3pt]
  \boldsymbol\tau = (\mu+\mu_t)\big[\nabla\mathbf u
    + (\nabla\mathbf u)^{\!\top}
    - \tfrac23(\nabla\!\cdot\!\mathbf u)\,\mathbf I\big],
  \qquad
  \mu_t = \big[(1-\sigma_{v1})\,f_{v1}(\chi)+\sigma_{v1}\big]\,\rho\tilde\nu,
  \label{w:ns}
\end{gather}
with $\sigma_{v1}$ the lifted-layer bypass weight of Eq.~\eqref{w:fv1bypass}. The
working variable is transported by
\begin{equation}
  \frac{D\tilde\nu}{Dt} = P - D
  + \frac{1}{\sigma}\nabla\!\cdot\!\big[(c_{\nu,\mathrm{ai}}\,\nu+\tilde\nu)\,\nabla\tilde\nu\big]
  + \frac{c_{b2}}{\sigma}\,|\nabla\tilde\nu|^2,
  \label{w:sa}
\end{equation}
\begin{equation}
  P = \max\!\big[\,(1-\sigma_P)\,a\,\sigma_{Re}\,\omega\,\tilde\nu,\;\;\sigma_P\,P_\mathrm{SA}\,\big],
  \qquad
  D = \sigma_D\,D_\mathrm{SA} .
  \label{w:blend}
\end{equation}
The rate $a$ is defined by Eq.~\eqref{w:a}, the gate $\sigma_{Re}$ by Eq.~\eqref{w:rate}, and the
weights $\sigma_P$ and $\sigma_D$ by Eq.~\eqref{w:sigma}.
The production $P$ is the amplification source,
$P_\mathrm{AI}=a\,\sigma_{Re}\,\omega\,\tilde\nu$, while $\chi<1$, the laminar
regime, and SA's own production once $\sigma_P$ has reached one, the
turbulent regime. The destruction is SA's destruction multiplied by $\sigma_D$. The molecular viscosity in the diffusion term of
$\tilde\nu$ is multiplied by $c_{\nu,\mathrm{ai}}$. In the eddy viscosity, $f_{v1}$ is
replaced by $(1-\sigma_{v1})f_{v1}+\sigma_{v1}$, with $\sigma_{v1}$ the bypass weight.
Note that baseline SA is recovered exactly by setting $\sigma_P\!\equiv\!1$
(and with it $\sigma_D\!=\!1$), $c_{\nu,\mathrm{ai}}\!=\!1$, and
$\sigma_{v1}\!\equiv\!0$. Note also that Eq.~\eqref{w:lngrowth} is recovered
by setting $\sigma_P\!=\!0$ and $c_{\nu,\mathrm{ai}}\!=\!0$ where
$\chi\ll1$: the destruction and the remaining diffusion terms are then of
second order in $\tilde\nu$. The SA closure functions are unchanged:
\begin{gather}
  P_\mathrm{SA} = c_{b1}\tilde S\,\tilde\nu, \qquad
  D_\mathrm{SA} = c_{w1}f_w\!\left(\frac{\tilde\nu}{d}\right)^{\!2}, \qquad
  \tilde S = \omega + \frac{\tilde\nu}{\kappa^2 d^2}\,f_{v2}, \qquad
  f_{v2} = 1 - \frac{\chi}{1+\chi f_{v1}}, \notag\\[3pt]
  f_{v1} = \frac{\chi^3}{\chi^3+c_{v1}^3}, \qquad
  f_w = g\!\left(\frac{1+c_{w3}^6}{g^6+c_{w3}^6}\right)^{\!1/6}, \qquad
  g = r + c_{w2}(r^6-r), \qquad
  r = \min\!\left(\frac{\tilde\nu}{\tilde S\,\kappa^2 d^2},\,10\right).
  \label{w:closure}
\end{gather}
The
non-negative $\tilde S$ modification and the negative-$\tilde\nu$
safeguards of Allmaras et al.~\cite{allmaras_2012} are retained, with one
consistency requirement: the negative-branch diffusion factor $f_n$ is
formed with the scaled coefficient $c_{\nu,\mathrm{ai}}\,\nu$, which
keeps the total diffusivity positive,
$c_{\nu,\mathrm{ai}}\nu+f_n\tilde\nu>0$.
The boundary conditions are also those of SA. At a wall $\tilde\nu=0$. In the
free stream $\tilde\nu=\chi_\infty\nu$, with $\chi_\infty$ given in
Section~III.A. The handover weights are
\begin{equation}
  \sigma_P = \sigma_t(\chi;\tau), \qquad
  \sigma_t(\chi;\tau) = \max\!\left[\,1-e^{-(\chi-1)/\tau},\;0\right], \qquad
  \sigma_D = 1-\frac{c_{b1}}{\kappa^2c_{w1}}\,(1-\sigma_P).
  \label{w:sigma}
\end{equation}

\subsection{Local Indicators of the Boundary-Layer Profile Shape}

\par\vspace{7.6pt}\begin{center}\includegraphics{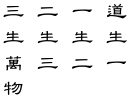}\end{center}\vspace{-7.6pt}

The rate $a$ of Eq.~\eqref{w:lngrowth}, and the threshold at which the gate
$\sigma_{Re}$ opens, are functions of the shape of the boundary-layer profile.
The model must therefore determine that shape from local quantities.

SA-AI assumes that the shapes of laminar boundary-layer profiles form a
\textbf{one}-parameter family. The shape of a profile is the velocity divided
by the edge velocity, as a function of the wall distance divided by the
momentum thickness. It is a whole curve, and two boundary layers could in
principle differ in any part of it. The assumption is that in practice they
do not: a single number selects the curve, and once that number is known the
rest of the shape is known. The Falkner--Skan profiles, selected by $\beta$,
are such a family. The assumption is an approximation, and it holds best where
the pressure gradient changes slowly along the surface. It is the assumption
behind one-parameter integral methods such as that of Thwaites
\cite{thwaites_1949}, and behind the envelope $e^N$ method
\cite{drela_giles_1987}, where the number is the shape factor $H$ and
$dN/dRe_\theta$ is a function of $H$ alone. However, SA-AI cannot use $H$,
which is an integral across the boundary layer and cannot be determined from the values and
derivatives of the velocity at a point. A local quantity must replace $H$.

But one local quantity is not enough, because any local quantity changes across the boundary layer. Consider the ratio of $d\,u'$ to $u$, with $u$ the velocity, $u'$ its
wall-normal derivative, and $d$ the distance to the wall. It is one at the
wall and zero at the edge of every attached boundary layer. Every value in between occurs for every profile shape, at some wall distance. Identifying the shape from local quantities requires \textbf{two} independent ratios, because two numbers are unknown at a point. One is the
number that parameterizes the shape. The other is the wall-normal coordinate of the point: its wall distance
divided by the momentum thickness.

Two independent ratios need \textbf{three} local quantities of the same dimension, because two quantities form only one ratio. At a point, a solver
can evaluate $u$, $u'$, and the second derivative $u''$. Multiplied by powers
of $d$, they become the terms of the Taylor series over the distance to the
wall: $u$, $d\,u'$, and $\tfrac12 d^2u''$, all with the dimension of velocity.

\begin{figure}[H]\centering
\begin{minipage}[b]{0.495\textwidth}\centering
\includegraphics[width=\textwidth]{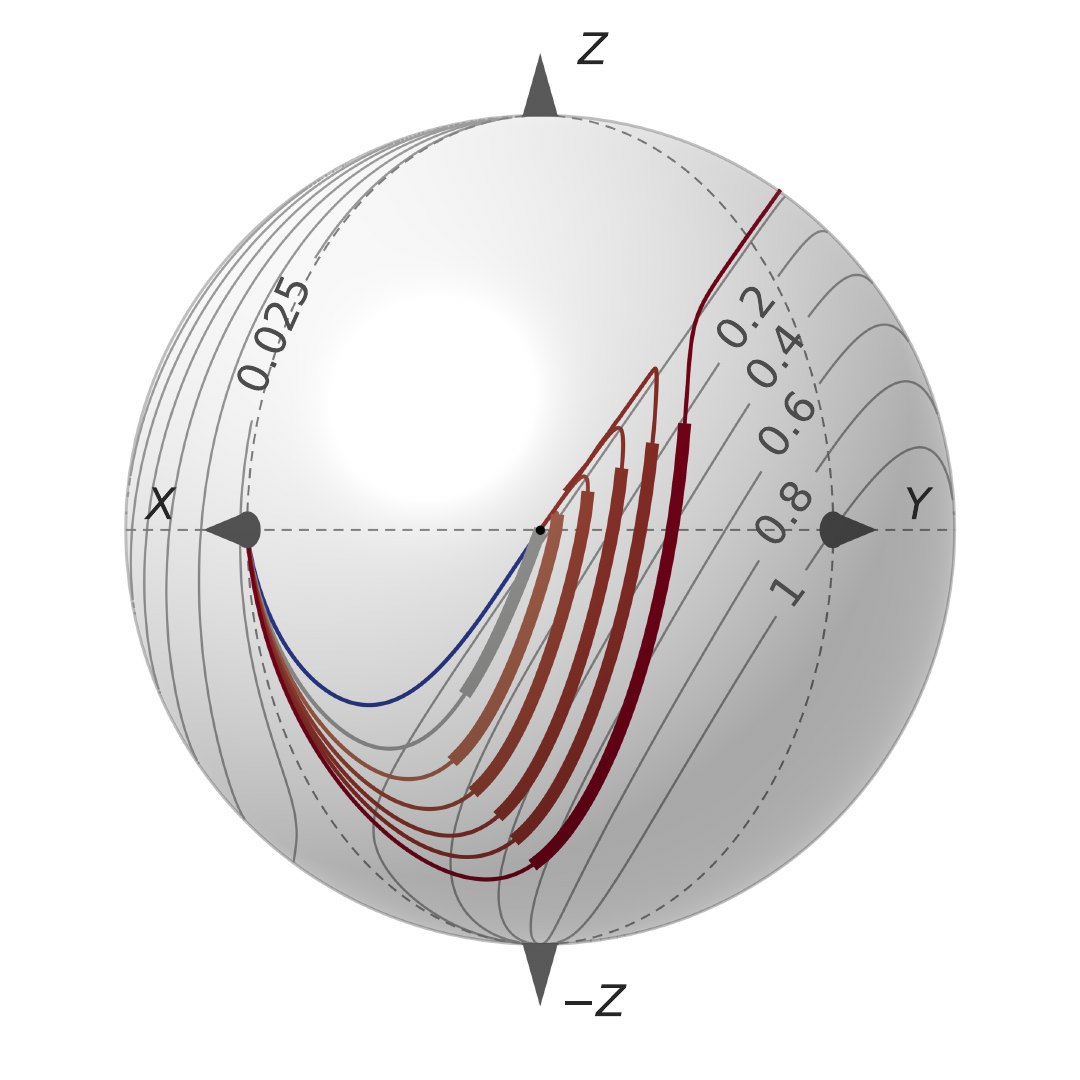}\\ a)
\end{minipage}\hfill
\begin{minipage}[b]{0.495\textwidth}\centering
\includegraphics[width=\textwidth]{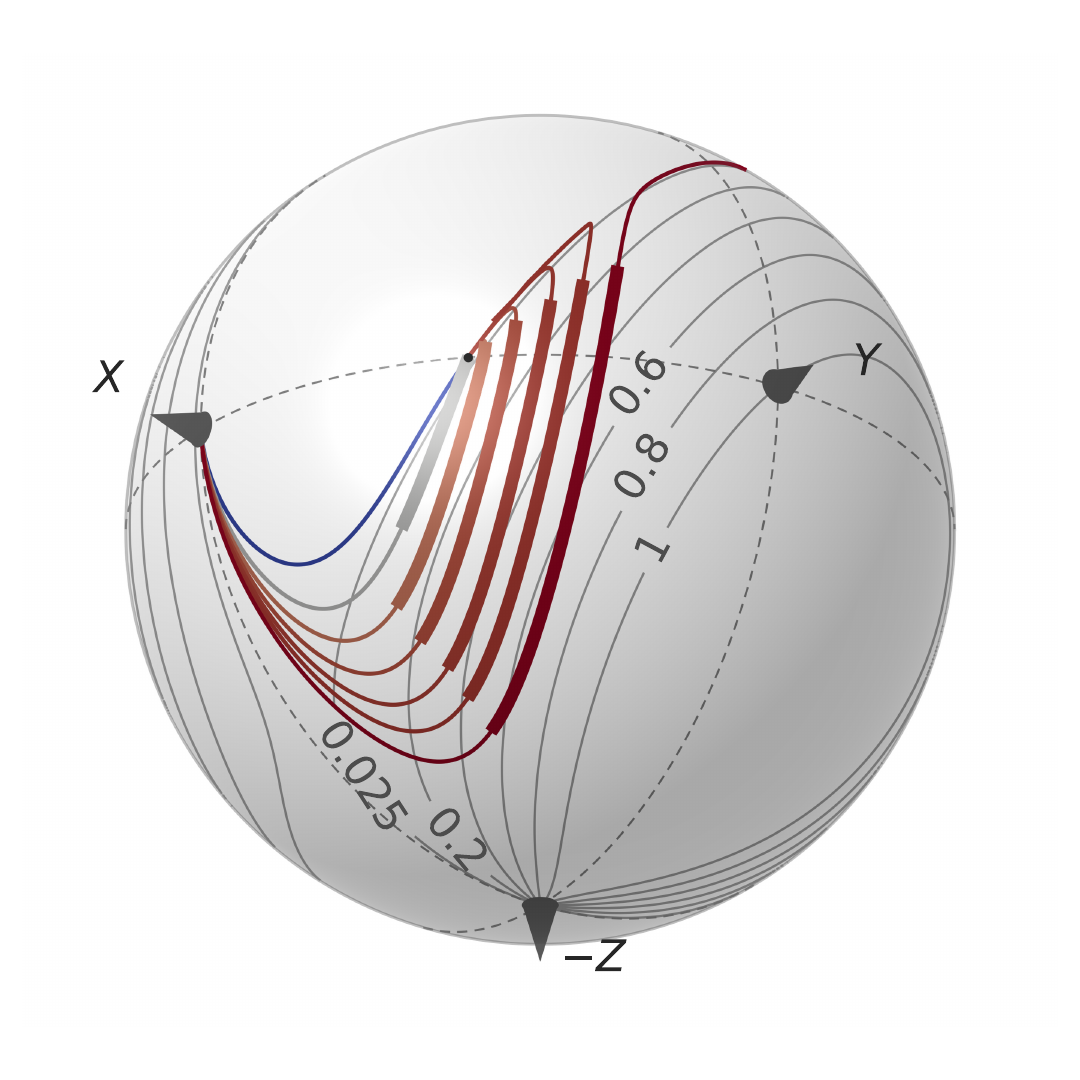}\\ b)
\end{minipage}
\caption{Indicator sphere with contours of $\hat\Omega\hat I$, in two views:
a) facing the wall point, b) turned $10^\circ$ toward the $Y$ pole and seen
from $25^\circ$ below, so that $-Z$ faces the reader and $Z$ is hidden. Dashed lines: the great circles through the poles. Curves: Falkner--Skan
profiles $\beta=+0.30$, $0$, $-0.10$, $-0.16$, $-0.19$, $-0.1988$ and the
separated Stewartson profile ($\beta=-0.19$, $H\approx4.9$), colored from blue
(favorable) through gray (Blasius) to dark red (separated). Only the near side
of the sphere is drawn; the separated profile starts on the far side, at the
antipodal wall point. Thick: the part of each curve where the Reynolds-stress
production of the most amplified Orr--Sommerfeld wave, scaled with the local
length $u/\omega$, exceeds half of its peak value on that curve, at
$Re_\theta=500$ ($200$ for the separated profile). The $\beta=+0.30$ profile
is stable at that Reynolds number and has no band. At the peak, the spatial
growth rate of the wave scaled with $u/\omega$ is $0.007$, $0.030$, $0.051$,
$0.078$, $0.104$ and $0.128$ for $\beta=0$, $-0.10$, $-0.16$, $-0.19$,
$-0.1988$ and the separated profile.}\label{f:sphere}
\end{figure}

And three is sufficient. In a thin quasi-two-dimensional boundary layer these three
velocity scales determine the Taylor expansion of the profile to second order
about any wall-normal point. Normalized by their
common magnitude, the local velocity scale
$U_\mathrm{loc}=\sqrt{u^2+(du')^2+(\tfrac12 d^2u'')^2}$, they define a point on the unit sphere of directions
\begin{equation}
  (X,Y,Z) = \big(u,\;d\,u',\;\tfrac12 d^2u''\big),
  \qquad
  (\hat X,\hat Y,\hat Z) = (X,Y,Z)\big/U_\mathrm{loc},
  \label{w:ind}
\end{equation}
so unhatted symbols have the dimension of velocity and hatted ones are the
direction cosines on the unit sphere. The normalization removes the
magnitude and keeps the direction of $(X,Y,Z)$. Any ratio of the three terms,
such as $Y/X$ or $Z/X$, can be computed from that direction. The direction is
used in place of two chosen ratios because each of the three terms vanishes
somewhere in a flow, so none can be the denominator everywhere.
Figure~\ref{f:sphere} plots boundary-layer profiles of the Falkner--Skan family
\cite{falkner_skan_1931} on the sphere. Each profile maps to
a curve, parameterized by the wall distance from the wall to the edge of the boundary layer. Every profile with nonzero wall shear starts at a wall point,
$\hat X=\hat Y$, $\hat Z=0$, because $u\approx u'd\gg u''d^2$ next to the
wall. Every profile ends at a velocity pole, $\hat X=\pm1$: toward the edge of the boundary layer $u'$ and $u''$ decay at least
exponentially with $d$, so $Y$ and $Z$ vanish.
Between their end points no two curves cross. A point on the sphere therefore
corresponds to exactly one profile shape and one wall-normal coordinate, the two unknowns.

The model is built on one feature of Fig.~\ref{f:sphere}. Close to the wall
every curve lies on the same great circle, whatever the pressure gradient:
favorable, zero, adverse, or strong enough to reverse the flow at the wall.
Each curve leaves its wall point along that circle, which is the image of every parabola through the wall: any $u=c_1d+c_2d^2$ satisfies
$\hat Y=\hat X+\hat Z$ at every wall distance, shear equals velocity plus
curvature. What distinguishes one profile from another is its departure from
that circle. One indicator measures the departure:
\begin{equation}
  \hat I = \hat Y - \hat X - \hat Z
  \qquad\text{(accumulated inflection)}.
  \label{w:ihat}
\end{equation}
It is zero on the circle. Its name comes from the unnormalized form,
\begin{equation}
  I(d) \equiv U_\mathrm{loc}\,\hat I = d\,u'-u-\tfrac12 d^2u''
  = -\int_0^d \tfrac12\,s^2\,u'''(s)\,\mathrm{d}s .
  \label{w:integral}
\end{equation}
The accumulated inflection is the integral of $-u'''$ from the wall to the
point, weighted by $\tfrac12 s^2$. It measures how much the curvature $u''$
has decreased between the wall and the point. It is zero for a parabola,
whose curvature is constant. The same three terms have a second interpretation. The second-order Taylor expansion of the profile at the point, extrapolated
to the wall, gives a wall velocity of $X-Y+Z=-I$. The no-slip condition
requires zero. The accumulated inflection is the true wall velocity minus the
extrapolated one. The circle is the line of zero accumulated inflection: no
AI, in a second sense of the abbreviation.

Away from the wall the profiles of Fig.~\ref{f:sphere} differ in the maximum of
$\hat I$. The strongly favorable profile, $\beta=+0.30$, has a maximum $\hat I$ of only $0.007$ and is
negative farther from the wall. For the Blasius profile the maximum is $0.14$.
In an adverse pressure gradient the maximum is larger. The separated profile, the reversed-flow solution
of Stewartson~\cite{stewartson_1954}, starts at the antipodal wall point, with velocity and shear both negative, and has the largest maximum of all.

\subsection{Amplification Rate and Onset Gate Built from the Indicators}

\par\vspace{7.6pt}\begin{center}\includegraphics{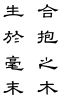}\end{center}\vspace{-7.6pt}

The amplification rate $a$ of Eq.~\eqref{w:lngrowth} is defined first. It is a
single dimensionless number at every point of the sphere, and it measures how
unstable a profile of that shape is at that wall distance.
Figure~\ref{f:sphere} shows where linear stability theory places the
instability. The amplification of an Orr--Sommerfeld wave is a spatial rate,
with the dimension of inverse length. It is usually made dimensionless with
the boundary-layer thickness, which is not available at a point.
Equation~\eqref{w:lngrowth} supplies a local length in its place: a steady
disturbance grows in space at the rate $a\,\omega/u$, so the length is
$u/\omega$. The thick part of each curve is the band of wall distances where
the Reynolds-stress production of the most amplified wave, scaled with
$u/\omega$, exceeds half its peak. The objective is
a simple function on the sphere that is large where those bands lie.

The accumulated inflection $\hat I$ is the starting point for modeling the
amplification rate. A profile with no inflection point is stable to inviscid
disturbances, by Rayleigh's theorem~\cite{schmid_henningson_2001}, and a
parabola has none. The great circle $\hat I=0$, the image of every parabola
through the wall, is the neutral line of the model. On the side where
inflection has accumulated, $\hat I$ is positive, and every band lies there.
The accumulated inflection cannot be the rate by itself: a flow and its
reverse, $u\to-u$, are equally unstable, but $\hat I$ changes sign under the
reversal. The simplest way to respect this symmetry is to multiply $\hat I$ by a second indicator that also changes sign under $u\to-u$. An obvious candidate is the shear cosine $\hat Y$, but it fails for a shear layer far from the wall. The curvature scale $Z$ is proportional to $d^2$. In
such a shear layer $Z$ dominates $U_\mathrm{loc}$, and $\hat Y$ tends to zero
however strong the shear.

The vorticity fraction changes sign under $u\to-u$ and does not contain $Z$:
\begin{equation}
  \hat\Omega = \frac{\hat Y}{\sqrt{\hat X^2+\hat Y^2}}
  = \frac{Y}{\sqrt{X^2+Y^2}}
  \qquad\text{(vorticity fraction)}.
  \label{w:omhat}
\end{equation}
It is the fraction of shear in the combined magnitude of velocity and shear.
It is zero in a uniform stream and at the center of a jet or a wake. It
approaches one in a detached shear layer, at any distance from the wall. The
product $\hat\Omega\hat I$ is unchanged under $u\to-u$.

The contours of $\hat\Omega\hat I$ in Fig.~\ref{f:sphere} follow the thick
bands. The product is positive in every band, and across the Falkner--Skan
family it grows with the adverse pressure gradient. For $\beta\ge\tfrac12$ it
is never positive, because $\hat I$ is never positive there while
$\hat\Omega$ is. Its maximum is below $0.005$ on the stable profile,
$\beta=+0.30$, is $0.08$ on the Blasius profile, and reaches $0.66$ on the
separated one. On every unstable profile the maximum lies inside the band. Within every profile the product is zero where either indicator is
zero: on the great circle $\hat I=0$, and in the free stream, where the shear
vanishes and $\hat\Omega$ with it.

The amplification rate of SA-AI, $a$ in Eqs.~\eqref{w:lngrowth}
and~\eqref{w:blend}, is this product, clipped to the interval from zero to
one and multiplied by $a_{\max}$:
\begin{equation}
  a = a_{\max}\,\operatorname{clip}\!\big\langle\,\hat\Omega\,\hat I\,\big\rangle_{0}^{1} .
  \label{w:a}
\end{equation}
The clip
sets the rate to zero wherever the product is not positive, so $\tilde\nu$
does not grow at the wall or in the free stream. The ceiling $a_{\max}$ is the rate of the free shear
layer, the most unstable profile of all. The fastest wave on a
hyperbolic-tangent shear layer grows at $0.1897$ of the peak vorticity
\cite{michalke_1964}, which gives $a_{\max}=0.19$. The rate of any other
profile is $a_{\max}$ multiplied by $\hat\Omega\hat I$.
Section~\ref{sec:results} reports what the model computes with this rate.

The rate $a$ does not depend on the viscosity, just as $dN/dRe_\theta$ does
not in the $e^N$ method. Viscosity appears only in the onset gate,
$\sigma_{Re}$ in Eqs.~\eqref{w:lngrowth} and~\eqref{w:blend}. The local
Reynolds number is the vorticity Reynolds number $Re_\Omega=d^2\omega/\nu$.
The gate is a standard sigmoid in the relative excess of $Re_\Omega$ over a
threshold $Re_\Omega^{\mathrm c}$:
\begin{equation}
  \sigma_{Re} = \tfrac12\big[1+\tanh(\zeta/w)\big],
  \qquad
  \zeta = \frac{Re_\Omega-Re_\Omega^{\mathrm c}}{Re_\Omega^{\mathrm c}} .
  \label{w:rate}
\end{equation}
It is one half at the threshold. The constant $w$ is the relative width of
the rise. With $w=0.35$ the gate is $0.003$ at $Re_\Omega=0$ and $0.997$ at
twice the threshold. The threshold depends on the same shape number
$\hat\Omega\hat I$, as $Re_{\theta0}$ depends on $H$.

\begin{figure}[H]\centering
\includegraphics[width=0.62\textwidth]{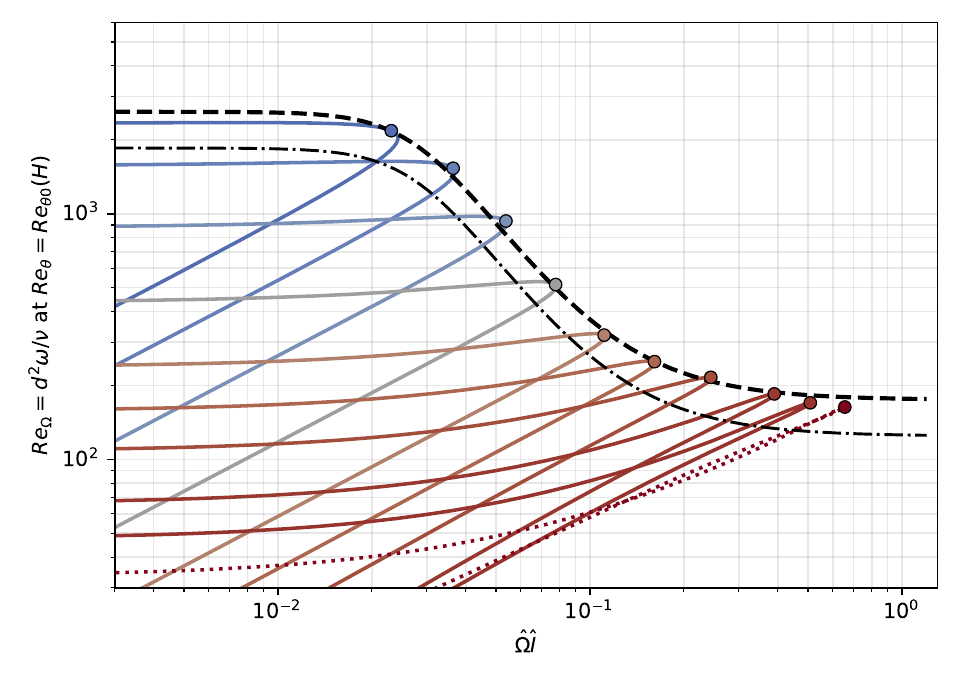}
\caption{Onset threshold as the upper bound of the Falkner--Skan curves.
Solid curves: attached profiles $\beta=+0.15$, $+0.10$, $+0.05$, $0$,
$-0.05$, $-0.10$, $-0.15$, $-0.19$, $-0.1988$, in the colors of
Fig.~\ref{f:sphere}, each at its Drela--Giles $Re_{\theta 0}(H)$. Dotted: separated Stewartson profile,
$\beta=-0.19$ ($H\approx4.9$, $Re_{\theta 0}\approx26$). Circles: the point of each curve closest to the dashed line. Dashed: Eq.~\eqref{w:onset} with $\con=1$, the upper bound. Dash-dot: Eq.~\eqref{w:onset} with $\con=0.712$, the model
threshold.}\label{f:graze}
\end{figure}

The threshold has one job: to start the growth where the $e^N$ method starts
it, at the critical Reynolds number $Re_{\theta0}(H)$. Figure~\ref{f:graze} shows
how. A boundary layer at one station traces a curve in the plane of
$\hat\Omega\hat I$ and $Re_\Omega$, one point for each wall distance. Each
colored curve in the figure is one Falkner--Skan profile. Increasing
$Re_\theta$ without changing the profile shape multiplies $Re_\Omega$ by the
same factor at every wall distance. On the logarithmic scale of the figure the
whole curve moves up without changing its form. The gate
opens at the first wall distance where the curve reaches the threshold. For
the growth to start at $Re_{\theta0}$, the curve at that Reynolds number must
touch the threshold at one point and lie below it everywhere else. The same must
hold for every profile shape, each taken at its own critical Reynolds number,
where the boundary layer has just become thick enough for a disturbance to
grow. The threshold is therefore the upper bound of all those curves, one that touches each of them.

The curves in Fig.~\ref{f:graze} are drawn at those critical Reynolds
numbers, for nine attached Falkner--Skan profiles from $\beta=+0.15$ to
incipient separation. Each curve starts at the
lower left, next to the wall, where $\hat\Omega\hat I$ and $Re_\Omega$ are both
small. Away from the wall it rises to the right, turns back at the wall
distance where $\hat\Omega\hat I$ is largest, and returns to the left as
$\hat\Omega\hat I$ decreases toward the edge of the boundary layer. The turning points mark this upper bound. They run from the upper left to the lower right. A
favorable profile becomes critical at a large $Re_\Omega$, about $2000$ at
$\beta=+0.15$, and has a small $\hat\Omega\hat I$. A profile near separation
becomes critical at $Re_\Omega\approx170$ and has a large $\hat\Omega\hat I$. On the logarithmic scales of the figure the upper bound is level on the left, level again on the
right at about one fifteenth of that value, and falls from one level to the other between
$\hat\Omega\hat I\approx0.03$ and $0.2$.

The following function, with three constants, is level at both ends and
decreases in between:
\begin{equation}
  Re_\Omega^{\mathrm c}(\hat\Omega\hat I) = \con\,\operatorname{softmin}_2\!\Big(C,\;
      A + B\,\langle\hat\Omega\hat I\rangle_+^{-2}\Big),
  \qquad
  \operatorname{softmin}_2(x_1,x_2)=\frac{x_1x_2}{\sqrt{x_1^2+x_2^2}},
  \qquad \langle x\rangle_+=\max(x,0).
  \label{w:onset}
\end{equation}
The constant $C$ is the level on the left, where $\hat\Omega\hat I$ is small
and the second argument is large. The sum in the second argument is a smooth
maximum of its two terms, and it tends to $A$, the level on the right, where
$\hat\Omega\hat I$ is large. The constant $B$ determines the range of $\hat\Omega\hat I$ over which the
threshold decreases from $C$ to $A$: the sum equals $C$ near $\hat\Omega\hat I=\sqrt{B/C}$ and is twice $A$ at
$\hat\Omega\hat I=\sqrt{B/A}$. With $\con=1$ and $(C,A,B)=(2600,\,175,\,2)$ the
function is the dashed line of Fig.~\ref{f:graze}.

The constants $(C,A,B)$ were fitted for Eq.~\eqref{w:lngrowth}, in which the
growth begins as soon as the gate opens. Section~II.A noted that the full
equation reduces to Eq.~\eqref{w:lngrowth} only when
$c_{\nu,\mathrm{ai}}=0$. With $c_{\nu,\mathrm{ai}}>0$, diffusion carries
$\tilde\nu$ away from the band where it is produced and removes it at the
wall, where $\tilde\nu$ is zero, so the net growth begins later than the gate
opens. The threshold is lowered to compensate, through the onset factor $\con$ in Eq.~\eqref{w:onset}.

\subsection{Laminar Diffusion and the Onset Factor}

\par\vspace{7.6pt}\begin{center}\includegraphics{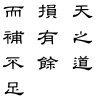}\end{center}\vspace{-7.6pt}

With $a$ and $\sigma_{Re}$ defined, Eq.~\eqref{w:lngrowth} is complete. Where
$\chi\ll1$ the transport equation, Eq.~\eqref{w:sa}, contains one more term:
the molecular diffusion of $\tilde\nu$. This subsection determines the two constants connected with it: the factor
$c_{\nu,\mathrm{ai}}$, which reduces the diffusion, and the onset factor
$\con$ of Eq.~\eqref{w:onset}.

The factor $c_{\nu,\mathrm{ai}}<1$ multiplies the molecular viscosity in the
diffusion term of Eq.~\eqref{w:sa} only. The mean-flow viscosity, the SA
production and destruction, and $\chi=\tilde\nu/\nu$, formed with the true
$\nu$, are unchanged. The effect of $c_{\nu,\mathrm{ai}}$ is measured on the Blasius boundary layer.
Where $\chi\ll1$, Eq.~\eqref{w:sa} in the boundary-layer approximation is
\begin{equation}
  u\,\frac{\partial\tilde\nu}{\partial x}+v\,\frac{\partial\tilde\nu}{\partial y}
  = a\,\sigma_{Re}\,\omega\,\tilde\nu
  + \frac{c_{\nu,\mathrm{ai}}\,\nu}{\sigma}\,
    \frac{\partial^2\tilde\nu}{\partial y^2},
  \label{w:lamdiff}
\end{equation}
with $x$ along the plate and $y$ normal to it. This is
Eq.~\eqref{w:lngrowth} with the laminar diffusion added. The table below lists
its solution on the Blasius velocity field $(u,v)$, marched downstream from the
leading edge, with $\tilde\nu=0$ at the wall and $\tilde\nu=\chi_\infty\nu$ at
the leading edge and in the free stream. The equation is linear in
$\tilde\nu$, so the amplification factor
\begin{equation*}
  N=\max_y\,\ln\frac{\chi}{\chi_\infty}
\end{equation*}
does not depend on $\chi_\infty$. The table gives the growth rate
$\Delta N/\Delta Re_\theta$ for a range of $c_{\nu,\mathrm{ai}}$, where
$\Delta Re_\theta$ is the increase of $Re_\theta$ from $N=1$ to $N=5$ and
$\Delta N=4$, as a fraction of the Drela--Giles slope
$dN/dRe_\theta=1.039\times10^{-2}$.
\begin{center}\small
\begin{tabular}{lcccccccc}
  \toprule
  $c_{\nu,\mathrm{ai}}$ & $1$ & $1/2$ & $1/4$ & $1/6$ & $1/12$ & $1/24$ & $1/48$ & $0$ \\
  \midrule
  $\Delta N/\Delta Re_\theta$, fraction of Drela--Giles
  & $61\%$ & $70\%$ & $79\%$ & $83\%$ & $90\%$ & $96\%$ & $100\%$ & $107\%$ \\
  \bottomrule
\end{tabular}
\end{center}
The growth rate is nearly independent of the onset factor $\con$. The table is
for $\con=0.712$, and between $\con=1$ and $\con=\tfrac12$ no entry changes by
more than $2\%$ of the Drela--Giles slope. Choosing $c_{\nu,\mathrm{ai}}$ is a trade-off: a smaller value brings the growth rate closer to the
Drela--Giles slope, but it sharpens the wall-normal profile of $\tilde\nu$,
which the grid must resolve. The wall-normal length scale of that profile is proportional to
$\sqrt{c_{\nu,\mathrm{ai}}}$, by dimensional analysis: the diffusion
coefficient $c_{\nu,\mathrm{ai}}\nu/\sigma$ and the growth rate
$a\,\sigma_{Re}\,\omega$ of Eq.~\eqref{w:lamdiff} form only one length,
$\sqrt{c_{\nu,\mathrm{ai}}\,\nu/(\sigma\,a\,\sigma_{Re}\,\omega)}$. The model uses
\begin{equation*}
  c_{\nu,\mathrm{ai}}=\tfrac16 .
\end{equation*}

The factor $c_{\nu,\mathrm{ai}}$ leaves SA's solution of a turbulent wall layer
unchanged. That solution, $\tilde\nu=\kappa u_\tau d$ with $u_\tau$ the friction velocity, is linear in the wall
distance, and the molecular diffusion of a linear profile is zero whatever its
coefficient. The factor is also absent from the free shear flows on which
Spalart and Allmaras~\cite{spalart_allmaras_1994} calibrated $c_{b1}$,
$\sigma$ and $c_{b2}$: their equation for those flows does not contain the
molecular viscosity. The flat-plate boundary layer at $Re_\theta=10^4$, on
which they calibrated $c_{w2}$, does change, because $\chi$ decreases through
one at its outer edge. With $c_{\nu,\mathrm{ai}}=\tfrac16$ in SA, its
skin-friction coefficient is $0.07\%$ higher, and its eddy viscosity differs by
at most $0.25\%$ of its peak.

Diffusion also delays the growth, and the onset factor $\con$ compensates for
the delay. It lowers the whole threshold of Eq.~\eqref{w:onset}, so the growth
starts farther upstream. On the Drela--Giles envelope $N=1$ at
$Re_\theta=338$. With $c_{\nu,\mathrm{ai}}=\tfrac16$ and $\con=1$ the marched
Blasius boundary layer reaches $N=1$ at $Re_\theta=398$, $18\%$ late. With
$\con=\tfrac12$ it reaches $N=1$ at $Re_\theta=281$. The model uses
\begin{equation*}
  \con=0.712 ,
\end{equation*}
with which $N=1$ at $Re_\theta=331$, $2\%$ before the Drela--Giles station.
The march uses $3200$ streamwise steps and $9600$ wall-normal cells. Halving
either spacing again lowers the $N=1$ stations by less than $1\%$ and changes
the growth rates by less than $0.2\%$. Halving the width $w$ of the onset gate,
Eq.~\eqref{w:rate}, moves the $N=1$ stations of the attached Falkner--Skan
family by at most $1.6\%$, and doubling it by at most $3.3\%$.

\begin{figure}[H]\centering
\includegraphics[width=0.99\textwidth]{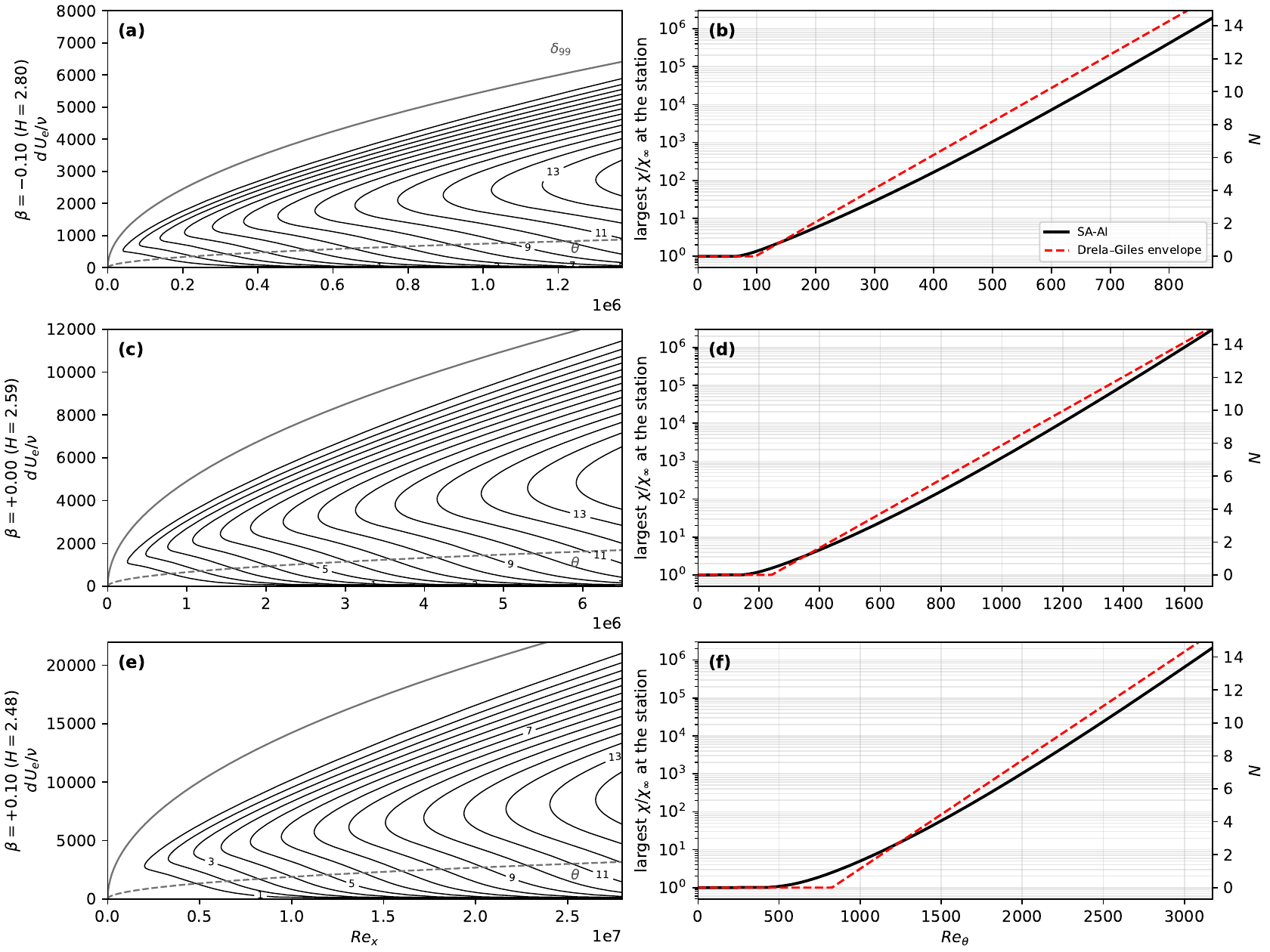}
\caption{Solution of Eq.~\eqref{w:lamdiff} on three Falkner--Skan profiles
(rows): $\beta=-0.10$, $0$ and $+0.10$, on which the largest
$\hat\Omega\hat I$ is $0.162$, $0.078$ and $0.037$. Left: contours of
$\ln(\chi/\chi_\infty)$ against $Re_x$ and the wall distance $d\,U_e/\nu$, with
the $99\%$ thickness (solid gray) and the momentum thickness (dashed gray).
Right: the largest $\chi/\chi_\infty$ at each station against $Re_\theta$, with
$N$ on the right axis, and the Drela--Giles envelope (dashed).}\label{f:nuhat}
\end{figure}

\begin{figure}[H]\centering
\includegraphics[width=\textwidth]{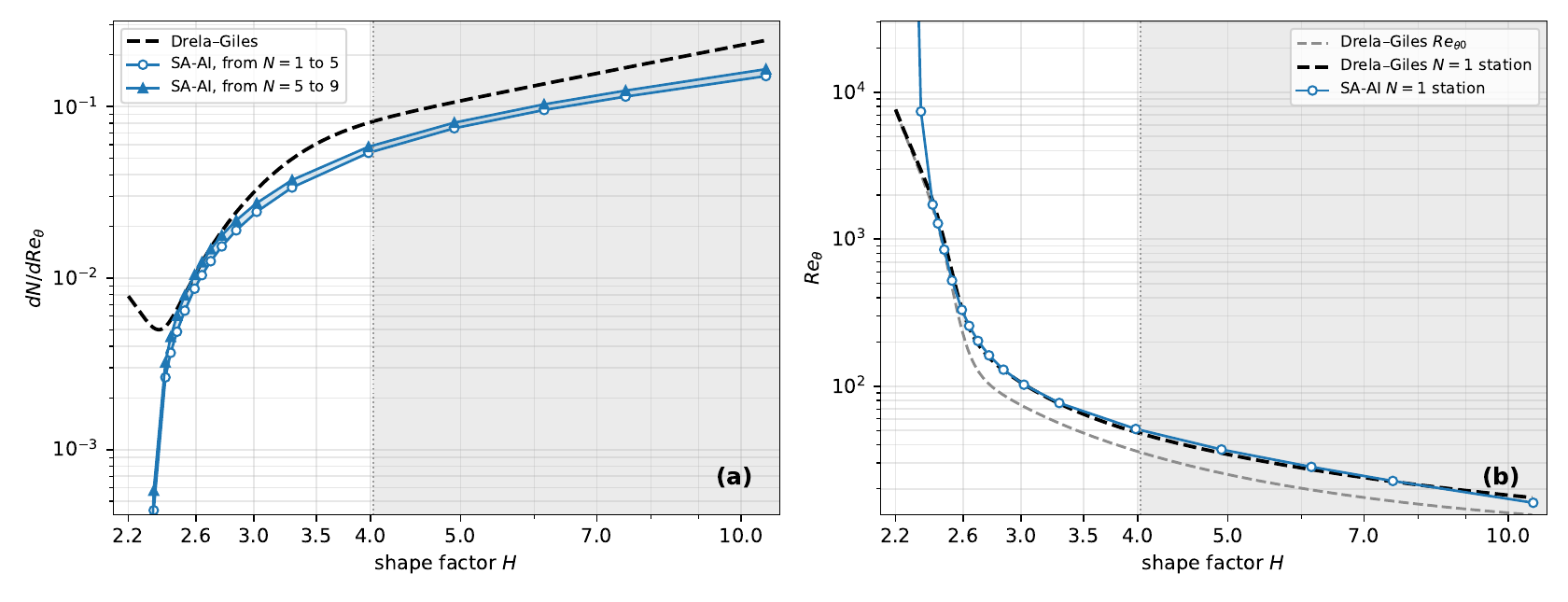}
\caption{The model and the Drela--Giles envelope over the Falkner--Skan
family, against the shape factor, from $\beta=+0.35$ ($H=2.34$) through the
Blasius profile ($H=2.59$) and incipient separation ($H=3.98$) to the
reversed-flow profiles (shaded, to $H=10.6$). (a)~Growth rate
$dN/dRe_\theta$ from $N=1$ to $5$ and from $N=5$ to $9$. (b)~Station where
$N=1$, with the critical $Re_{\theta0}(H)$.}\label{f:calibrate}
\end{figure}

Figure~\ref{f:nuhat} shows the solution of Eq.~\eqref{w:lamdiff} on three
Falkner--Skan profiles: $\beta=-0.10$, the Blasius profile, and $\beta=+0.10$.
The left panels are contours of $\ln(\chi/\chi_\infty)$ in the boundary layer.
The right panels compare its largest value at each station, which is $N$, with
the Drela--Giles envelope. Figure~\ref{f:calibrate} makes the same comparison
over the whole Falkner--Skan family, against the shape factor: the growth rate
in panel (a), and the station where $N=1$ in panel (b).

The station where $N=1$ is close to the Drela--Giles station wherever the
pressure gradient is zero or adverse. It is $2\%$ early on the Blasius
profile. On the attached profiles from $\beta=-0.05$ to incipient separation
it is between $0.5\%$ early and $5\%$ late. In a favorable pressure gradient
it is $9\%$ to $13\%$ early, from $\beta=+0.05$ to $+0.15$. The
root-mean-square difference over these nine profiles is $7\%$. On the
reversed-flow profiles, up to $H=10.6$, the difference is at most $8\%$.

The growth rate of the model increases along the boundary layer, and over
most of the family it is lower than the Drela--Giles rate. From $N=1$ to $N=5$
it is $83\%$ of that rate on the Blasius profile, the value in the table
above, $67\%$ at incipient separation, and $62\%$ to $70\%$ on the
reversed-flow profiles. From $N=5$ to $N=9$ it is $101\%$, $72\%$, and $68\%$
to $76\%$. The station where $N=9$ is therefore later than the Drela--Giles
station. On the Blasius profile it is at $Re_\theta=1175$ against $1108$,
which is $6\%$ later. It is $12\%$ later at $\beta=-0.10$ and $31\%$ later at
incipient separation.

In a strongly favorable pressure gradient the model departs from the envelope
method. At $\beta=+0.35$ the growth rate is about one tenth of the
Drela--Giles rate, and $N=1$ is reached at $2.7$ times the critical
$Re_\theta$. For $\beta\ge\tfrac12$ the product $\hat\Omega\hat I$ is never
positive, so $a=0$ and $\tilde\nu$ does not grow.

\subsection{Handover of the Production to SA}

\par\vspace{7.6pt}\begin{center}\includegraphics{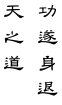}\end{center}\vspace{-7.6pt}

Equation~\eqref{w:lamdiff} describes $\tilde\nu$ while $\chi\ll1$, where the eddy
viscosity is negligible. Above $\chi\approx1$ the eddy viscosity affects the
mean flow significantly, and the equation of $\tilde\nu$ must become SA's. Near $\chi=1$ the
production must therefore change from the amplification source to SA's own. In Eq.~\eqref{w:blend} the handover weight $\sigma_P$ of
Eq.~\eqref{w:sigma} multiplies SA's production, and $1-\sigma_P$ multiplies
the amplification source. The weight $\sigma_P$ is zero for $\chi\le1$ and
rises toward one over the width $\tau$. With $\tau=4$, $\sigma_P=0.78$ at
$\chi=c_{v1}=7.1$ and $\sigma_P=0.96$ at $\chi=2c_{v1}=14.2$.

The production of Eq.~\eqref{w:blend} is the larger of the two weighted
sources. Below $\chi=1$ it is the amplification source. Above $\chi=1$ it
remains the amplification source, multiplied by $1-\sigma_P$, up to the value
of $\chi$ at which SA's weighted production exceeds it. Beyond that value the
production is SA's alone, multiplied by $\sigma_P$, and the amplification
source does not affect the equation. The maximum, in place of the sum of the
two weighted sources, is an empirical choice. With the sum, many of the
airfoil cases of Section~\ref{sec:results} do not converge when the
$\tilde\nu$ equation and the mean-flow equations are solved in turn, without a
joint Jacobian. A feedback loop is a possible cause. The displacement
thickness changes rapidly across transition. That change alters the pressure
gradient upstream, and with it the shape of the boundary-layer profile there.
The amplification source is a function of that shape. With the sum it remains
in the equation wherever $\sigma_P<1$, and so the altered shape changes
transition in turn. With the maximum it does not.

Like the factor $c_{\nu,\mathrm{ai}}$ of Section~II.D, the handover weights
must leave SA's solution of a turbulent wall layer unchanged. SA's solution in
the constant-stress wall layer is the linear profile
$\tilde\nu=\kappa u_\tau d$, so $\chi=\kappa y^+$. Every value of $\sigma_P$
between zero and one therefore occurs in that layer: $\sigma_P$ is zero for
$y^+<2.4$, $0.78$ at $y^+=17$ and $0.96$ at $y^+=35$. On that profile SA's
production, destruction and diffusion are $c_{b1}u_\tau^2$,
$c_{w1}\kappa^2u_\tau^2$ and $(1+c_{b2})\kappa^2u_\tau^2/\sigma$. Each is
independent of the wall distance, and they balance by the definition of
$c_{w1}$. Multiplying the production by $\sigma_P$ breaks the balance wherever
$\sigma_P<1$, unless the destruction changes with it.
Requiring the balance for every value of $\sigma_P$,
\begin{equation*}
  \sigma_P\,c_{b1}-\sigma_D\,c_{w1}\kappa^2+\frac{(1+c_{b2})\,\kappa^2}{\sigma}=0,
\end{equation*}
gives the destruction weight of Eq.~\eqref{w:sigma}. The weight $\sigma_D$ is a
linear function of $\sigma_P$, with slope $c_{b1}/(\kappa^2c_{w1})\approx0.249$
and value $(1+c_{b2})/(\sigma c_{w1})\approx0.751$ at $\sigma_P=0$.

With this tie the linear profile remains a solution of the full equation, for
any $\tau$, to within $8\times10^{-6}$ of SA's production. The only term left
over is the amplification source, which is the whole production where
$\sigma_P=0$, that is for $y^+<2.4$. It is that small because there the onset
gate $\sigma_{Re}$ is $0.003$ and $\hat\Omega\hat I$ is at most $0.002$.
Wherever $\sigma_P$ exceeds about $10^{-5}$, SA's weighted production is the
larger of the two, and the balance holds to rounding error.

In the laminar regime $\sigma_P=0$ and $\sigma_D=0.751$, so three quarters of
SA's destruction remains. It does not change Eq.~\eqref{w:lamdiff}: SA's
destruction is of second order in $\tilde\nu$, and Eq.~\eqref{w:lamdiff}
contains the terms of first order. The destruction is
below $10^{-3}$ of the amplification production up to the $\chi=1$ crossing,
and at most $1.4\%$ of the $\sigma_P$-weighted SA production at $\chi=c_{v1}$.

\subsection{Eddy Viscosity without Wall Damping in Separated Shear Layers}

\par\vspace{7.6pt}\begin{center}\includegraphics{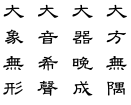}\end{center}\vspace{-7.6pt}

The transport equation of $\tilde\nu$ is complete with the handover of
Section~II.E. The mean flow depends on $\tilde\nu$ only through the eddy
viscosity $\mu_t$ of Eq.~\eqref{w:ns}. SA's eddy viscosity is
$\rho\tilde\nu f_{v1}(\chi)$, and the model alters it in one situation: a
shear layer that has separated from the wall. The reason is an observation.
With SA's eddy viscosity unaltered, the laminar separation bubbles computed
with the model reattach too late, or not at all. With the alteration they
reattach earlier, although still later than measured (Sections~III.C
and~III.D).

SA's damping function $f_{v1}$ is designed for the wall layer: the inner
layer of a fully developed turbulent boundary layer, where the velocity
obeys the law of the wall.
Spalart and Allmaras~\cite{spalart_allmaras_1994} introduce it for the
near-wall region only: in their calibration on free shear flows the
transported quantity is the eddy viscosity itself, without damping. SA's
solution in the wall layer, $\tilde\nu=\kappa u_\tau d$, is linear all the way to the wall,
while the eddy viscosity of a turbulent wall layer decreases toward the wall
as a higher power of the wall distance. Multiplying by $f_{v1}$ produces that
decrease: $\tilde\nu f_{v1}$ is proportional to $d^4$ where $\chi\ll c_{v1}$.
The function $f_{v1}$ can be a function of $\chi$ alone because
$\chi=\kappa y^+$ in that layer: $f_{v1}$ is one half at $\chi=c_{v1}$, which
is $y^+\approx17$.

A separated shear layer passes through the same values of $\chi$ with no wall
layer present. Downstream of a laminar separation the disturbance grows in a
shear layer that has lifted off the wall, and $\chi$ increases through one
inside it. The damping then removes most of the eddy viscosity while $\chi$ is
of the order of $c_{v1}$: $f_{v1}=0.07$ at $\chi=3$. That eddy viscosity is
what reattaches the flow.

The model must therefore distinguish a wall layer from a lifted
shear layer, with local quantities. The law of the wall provides the test. In
a wall layer $\chi$ determines the wall coordinate,
$y^+=\chi/\kappa$, and the law of the wall then determines the velocity,
$U^+(y^+)$. Their product, $y^+U^+$, is the Reynolds number $|\mathbf u|\,d/\nu$ of a
wall-layer point with that $\chi$. The indicator $q$ is the ratio of the
actual $|\mathbf u|\,d/\nu$ to that product,
\begin{equation}
  q = \frac{|\mathbf u|\,d/\nu}{\,(\chi/\kappa)\;U^+(\chi/\kappa)\,},
  \qquad
  U^+(y^+)=\tfrac1\kappa\ln(1+\kappa y^+)
  +7.8\big(1-e^{-y^+/11}-\tfrac{y^+}{11}e^{-y^+/3}\big),
  \label{w:fv1q}
\end{equation}
with Reichardt's formula~\cite{reichardt_1951} for $U^+$. In the wall layer
$q$ is close to one. With the velocity of SA's own solution in that layer,
$q$ is between $0.96$ and $1.07$ at every $\chi$. In a lifted shear layer $q$ is well above one, because the point
is farther from the wall and moves faster than a point of a wall layer with
the same $\chi$.

The bypass weight $\sigma_{v1}$ of Eq.~\eqref{w:ns} is the product of two
gates,
\begin{equation}
  \sigma_{v1} = \sigma_\chi\,\sigma_q, \qquad
  \sigma_\chi=\operatorname{clip}\!\big\langle\chi-1\big\rangle_0^1, \qquad
  \sigma_q=\operatorname{clip}\!\big\langle(q-2)/2\big\rangle_0^1 .
  \label{w:fv1bypass}
\end{equation}
Where $\sigma_{v1}=1$ the eddy viscosity is $\rho\tilde\nu$, without damping.
The gate $\sigma_\chi$ is zero for $\chi\le1$ and one for $\chi\ge2$. It keeps
the eddy viscosity out of the laminar regime, where $q$ is far above four.
Without it the eddy viscosity at $\chi=0.5$ would be half the molecular
viscosity. The range $\chi<1$ would then not be ``empty'': $\tilde\nu$ would
affect the mean flow there. Above $\chi=30$, $f_{v1}>0.98$, making the eddy viscosity in
Eq.~\eqref{w:ns} insensitive to $\sigma_{v1}$, so $\sigma_{v1}$ affects it only for $\chi$ between $1$ and
$30$.

The gate $\sigma_q$ opens between $q=2$ and $q=4$. It starts at two because the
outer part of an attached turbulent boundary layer is faster than the law of
the wall. On Coles' composite profile~\cite{coles_1956} this excess increases
with the adverse pressure gradient and decreases with the Reynolds number. In
zero pressure gradient at $Re_\theta\approx780$ ($H=1.54$), the largest $q$
for $\chi$ between $1$ and $30$ is $1.02$. In an adverse pressure gradient
about $7$ times the wall shear stress divided by the displacement
thickness~\cite{das_1987}, it is $1.15$ at $Re_\theta\approx1700$ ($H=1.95$)
and $1.88$ at $Re_\theta\approx520$ ($H=2.27$).

\subsection{Three-Dimensional Form of the Indicators}

In three dimensions the triple $(X,Y,Z)$ of Eq.~\eqref{w:ind} is formed from
four local vectors: the
velocity $\mathbf u$, the vorticity $\boldsymbol\omega$, the wall-distance
vector $\mathbf d=d\,\nabla d$, and the curvature vector
$\boldsymbol\ell=\tfrac12\langle\mathbf d,\mathbf d\rangle\,\nabla^2\mathbf u$,
with $\nabla^2\mathbf u=-\nabla\times\boldsymbol\omega$ giving the second
derivative:
\begin{equation}
  (X,\,Y,\,Z) \;=\;
  \big(\,|\mathbf u|,\;\; d\,|\boldsymbol\omega|,\;\;
  \langle\boldsymbol\ell,\hat{\mathbf u}\rangle\,\big),
  \qquad \hat{\mathbf u}=\mathbf u/|\mathbf u| .
  \label{w:ratio}
\end{equation}
Each term has the dimension of velocity. In a thin two-dimensional boundary
layer with positive $u$ and $u'$, the three reduce to the terms of
Eq.~\eqref{w:ind}. The solver assembles the same triple multiplied by
$|\mathbf u|$,
\begin{equation}
  |\mathbf u|\,(X,\,Y,\,Z) \;=\;
  \Big(\,\langle\mathbf u,\mathbf u\rangle,\;
  \sqrt{\langle\mathbf u,\mathbf u\rangle\,
        \langle\mathbf d,\mathbf d\rangle\,
        \langle\boldsymbol\omega,\boldsymbol\omega\rangle},\;
  \langle\boldsymbol\ell,\mathbf u\rangle\,\Big),
  \label{w:gram}
\end{equation}
which requires no division by $|\mathbf u|$. The common factor leaves
$\hat\Omega$ and $\hat I$ unchanged, because both depend only on ratios of the
three terms. Every term of Eqs.~\eqref{w:ratio} and~\eqref{w:gram} is a
function of inner products of vectors alone. Both forms are therefore
invariant under any rotation or reflection: a rotated or mirrored flow gives
the same $(X,Y,Z)$ at corresponding points, in any Cartesian coordinate
system.

\subsection{Constants}

The constants of baseline SA are unchanged. Table~\ref{t:saconstants} lists
them with the sections and equations of Spalart and
Allmaras~\cite{spalart_allmaras_1994} that determine them.
Table~\ref{t:constants} lists the constants of SA-AI in the order in which
Section~\ref{sec:model} determines them. Constants that are determined together
share a column.

\begin{table}[H]\centering\small
\caption{Baseline SA constants. Sections and equations are those of Spalart
and Allmaras~\cite{spalart_allmaras_1994}.}\label{t:saconstants}
\begin{tabular}{lccccc}
\toprule
constant & $(c_{b1},\,\sigma,\,c_{b2})$ & $\kappa$ & $c_{w1}$ & $(c_{w2},\,c_{w3})$ & $c_{v1}$ \\
\midrule
value & $(0.1355,\,\tfrac23,\,0.622)$ & $0.41$
  & $\dfrac{c_{b1}}{\kappa^2}+\dfrac{1+c_{b2}}{\sigma}$ & $(0.3,\,2)$ & $7.1$ \\[6pt]
section & II.2 & II.3, Appendix & II.3 & II.3 & II.4 \\
equation & (1), (2) & (5) & (4) & (6) & (7) \\
\bottomrule
\end{tabular}
\end{table}

\begin{table}[H]\centering\small
\caption{SA-AI constants. For $\sigma_\chi$ and $\sigma_q$ the values are the limits
between which each gate opens.}\label{t:constants}
\begin{tabular}{lccccccc}
\toprule
constant & $a_{\max}$ & $w$ & $(C,\,A,\,B)$ & $c_{\nu,\mathrm{ai}}$ & $\con$
  & $\tau$ & $\sigma_\chi;\;\sigma_q$ \\
\midrule
value & $0.19$ & $0.35$ & $(2600,\,175,\,2)$ & $1/6$ & $0.712$
  & $4$ & $(1,\,2);\;(2,\,4)$ \\
section & II.C & II.C & II.C & II.D & II.D & II.E & II.F \\
equation & \eqref{w:a} & \eqref{w:rate} & \eqref{w:onset} & \eqref{w:lamdiff}
  & \eqref{w:onset}, \eqref{w:lamdiff} & \eqref{w:sigma} & \eqref{w:fv1bypass} \\
\bottomrule
\end{tabular}
\end{table}

With these constants the sources of Eq.~\eqref{w:blend} are continuous
functions of the solution. Their derivatives are discontinuous in four
places: at $\hat\Omega\hat I=0$ and at $\hat\Omega\hat I=1$, where the
clip of Eq.~\eqref{w:a} begins and ends; at $\chi=1$, where the handover
weight $\sigma_P$ of Eq.~\eqref{w:sigma} begins to increase; and where the
maximum in Eq.~\eqref{w:blend} changes from one weighted source to the other.

\section{Results}\label{sec:results}

\par\vspace{7.6pt}\begin{center}\includegraphics{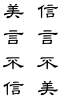}\end{center}\vspace{-7.6pt}

Each case in this section is chosen for one thing that it shows about the
model. The flat plate shows the role of the freestream value $\chi_\infty$,
which is the model's $N_\mathrm{crit}$. The NLF(1)-0416 airfoil shows
transition in an attached boundary layer, at a location that changes with the
pressure gradient. The Eppler 387 airfoil shows transition in a laminar
separation bubble. The same airfoil at lower Reynolds numbers shows the
condition at which the computed bubble no longer reattaches while the measured
bubble does. A two-element section shows a boundary layer that receives
$\tilde\nu$ from the wake of another element. The circular cylinder shows
transition moving from the separated wake onto the wall as the Reynolds number increases
through ten decades. The Daedalus wing shows the separation bubble on a wing whose chord Reynolds
number changes along the span.
The 6:1 spheroid is a three-dimensional body on which crossflow instability,
which the model does not represent, causes transition at the higher
incidences.

The results are computed with Flexcompute Flow (formerly known as Flow360), an unstructured, node-centered,
second-order finite-volume solver with the Roe flux, at $M=0.1$ unless stated
otherwise. The airfoil
cases use the freestream value $\chi_\infty=c_{v1}e^{-9}\approx8.76\times10^{-4}$,
which is $N_\mathrm{crit}=9$ and, by Mack's correlation, $Tu\approx0.07\%$
(Eq.~\eqref{w:mack} in Section~III.A).

The airfoil cases are computed on two independently
generated grid families. The unstructured grids are generated with
cavity-based operations~\cite{loseille_alauzet_menier_2017} and consist of
triangles everywhere, including in the boundary layers. The structured grids
are all-quadrilateral O-grids. Each family is refined at three levels, from
$1.5\times10^4$ to $2.6\times10^5$ nodes. The levels are called L0, L1 and L2,
from the coarsest to the finest, and every length scale of the grid is halved
from one level to the next.

A transition location is
the station where the largest $\chi$ near the wall crosses one. On the L2
grids the $\chi=c_{v1}$ crossing, where there is one, is $0.002$ to
$0.07\,c$ farther aft on the
NLF(1)-0416, with a median of $0.02\,c$, and $0.004$ to $0.30\,c$ farther aft
on the Eppler 387, with a median of $0.07\,c$. Bubble stations are
the zero crossings of the signed $C_f$.

The $e^9$ reference in the airfoil
figures is a coupled viscous--inviscid panel computation with the $e^N$ method
at $N_\mathrm{crit}=9$. It is computed with mfoil~\cite{fidkowski_mfoil_2022},
or with XFOIL~\cite{drela_xfoil_1989} or Flexcompute
Foil~\cite{smith_flexfoil_2026}, a reimplementation of XFOIL in Rust, where
mfoil does not converge. The model is also implemented in a second solver,
OpenFOAM v2412, the OpenCFD release. This replication is computed with simpleFoam, a cell-centered,
pressure-based solver for steady incompressible flow, on a flat-plate
grid of the same dimensions and on the same structured airfoil grids. It omits the $f_{v1}$ bypass of Section~II.F. It covers thirty
airfoil cases and the flat-plate sweep.

Every computation is converged with a reduced CFL number for the
$\tilde\nu$ equation in the laminar regime, so that the mean flow settles
before $\tilde\nu$ responds to the startup transient. The pseudo-time step of
the $\tilde\nu$ equation is multiplied by $10^{-2}$ where $\chi\le c_{v1}$.
Above $c_{v1}$ the factor is $10^{-2\exp[-(\chi-c_{v1})/4]}$, which
increases to one.

\FloatBarrier
\subsection{The Blasius Flat Plate}

The flat plate is the case in which the freestream turbulence intensity is
varied. The freestream value $\chi_\infty$ is the input that determines how far
the disturbance must grow, as $N_\mathrm{crit}$ is in the $e^N$ method. The two are
related by
\begin{equation}
  \chi_\infty = c_{v1}\,e^{-N_\mathrm{crit}},
  \qquad
  N_\mathrm{crit}(Tu) = -8.43 - 2.4\,\ln Tu ,
  \label{w:mack}
\end{equation}
where $N_\mathrm{crit}(Tu)$ is Mack's correlation~\cite{mack_1977}, with the
turbulence intensity $Tu$ as a fraction. With this definition the amplification
factor $N=\max_y\ln(\chi/\chi_\infty)$ of Section~II.D equals
$N_\mathrm{crit}$ where the largest $\chi$ in the boundary layer is
$c_{v1}=7.1$. It is $N_\mathrm{crit}-2.0$ where the largest $\chi$ is one. A
change of $\chi_\infty$ by the factor $e$ changes $N_\mathrm{crit}$ by one. On
the Blasius envelope, where $dN/dRe_\theta\approx10^{-2}$, that moves the
onset by $\Delta Re_\theta\approx100$. At $Tu=1\%$ the correlation gives
$N_\mathrm{crit}\approx2.6$, and at $Tu\approx3\%$ it gives $\chi_\infty=c_{v1}$.
These are conditions of bypass transition, which the model does not represent.

The plate is computed at five turbulence intensities from $Tu=0.04\%$ to
$0.6\%$, for which Eq.~\eqref{w:mack} gives $\chi_\infty\approx2.3\times10^{-4}$,
$1.2\times10^{-3}$, $6.3\times10^{-3}$, $2.9\times10^{-2}$ and
$1.5\times10^{-1}$. The grid is structured, $320\times80$, with
$y^+\lesssim0.3$, and extends to $Re_x=6\times10^6$.

\begin{figure}[!htbp]\centering
\includegraphics[width=0.90\textwidth]{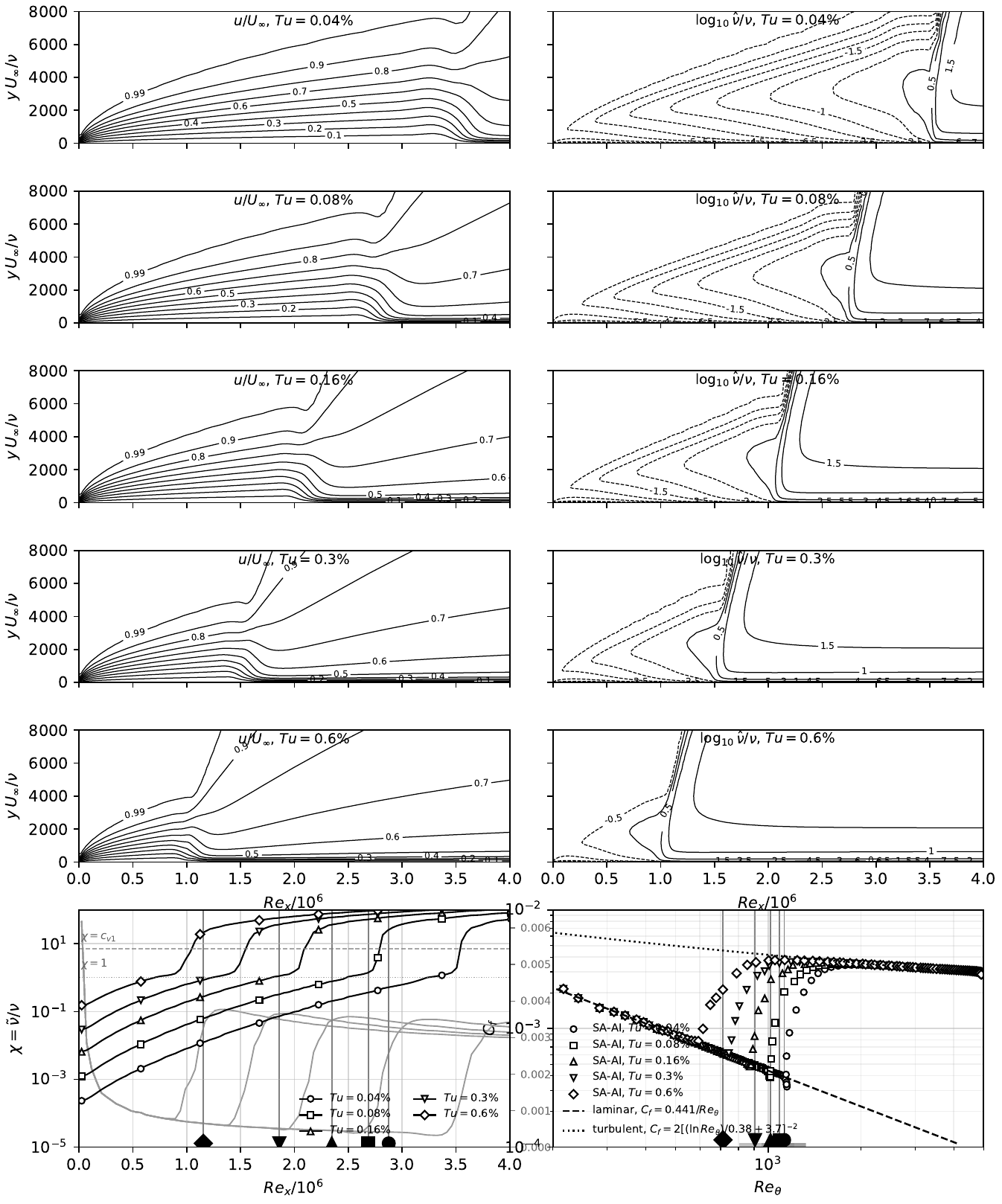}
\caption{Flat plate, Flexcompute Flow, at the five turbulence intensities. Upper rows: contours of $u/U_\infty$ (left) and of $\log_{10}\chi$ (right). Bottom left: the largest $\chi$ in the boundary layer against $Re_x$, with $C_f$ in gray on the right axis. Bottom right: $C_f$ against $Re_\theta$. Filled symbols and vertical lines: the onset correlation of Abu-Ghannam and Shaw.}\label{f:flatplate}
\end{figure}

Figure~\ref{f:flatplate} shows the five computations. Along the plate $\chi$
grows nearly exponentially from $\chi_\infty$, while the skin friction remains
that of the Blasius boundary layer. Where the largest $\chi$ crosses one the
skin friction begins to rise. It reaches its turbulent level shortly after the
largest $\chi$ crosses $c_{v1}$. A higher turbulence intensity gives a larger
$\chi_\infty$, and both crossings occur at a lower $Re_x$.

\begin{figure}[!htbp]\centering
\includegraphics[width=0.90\textwidth]{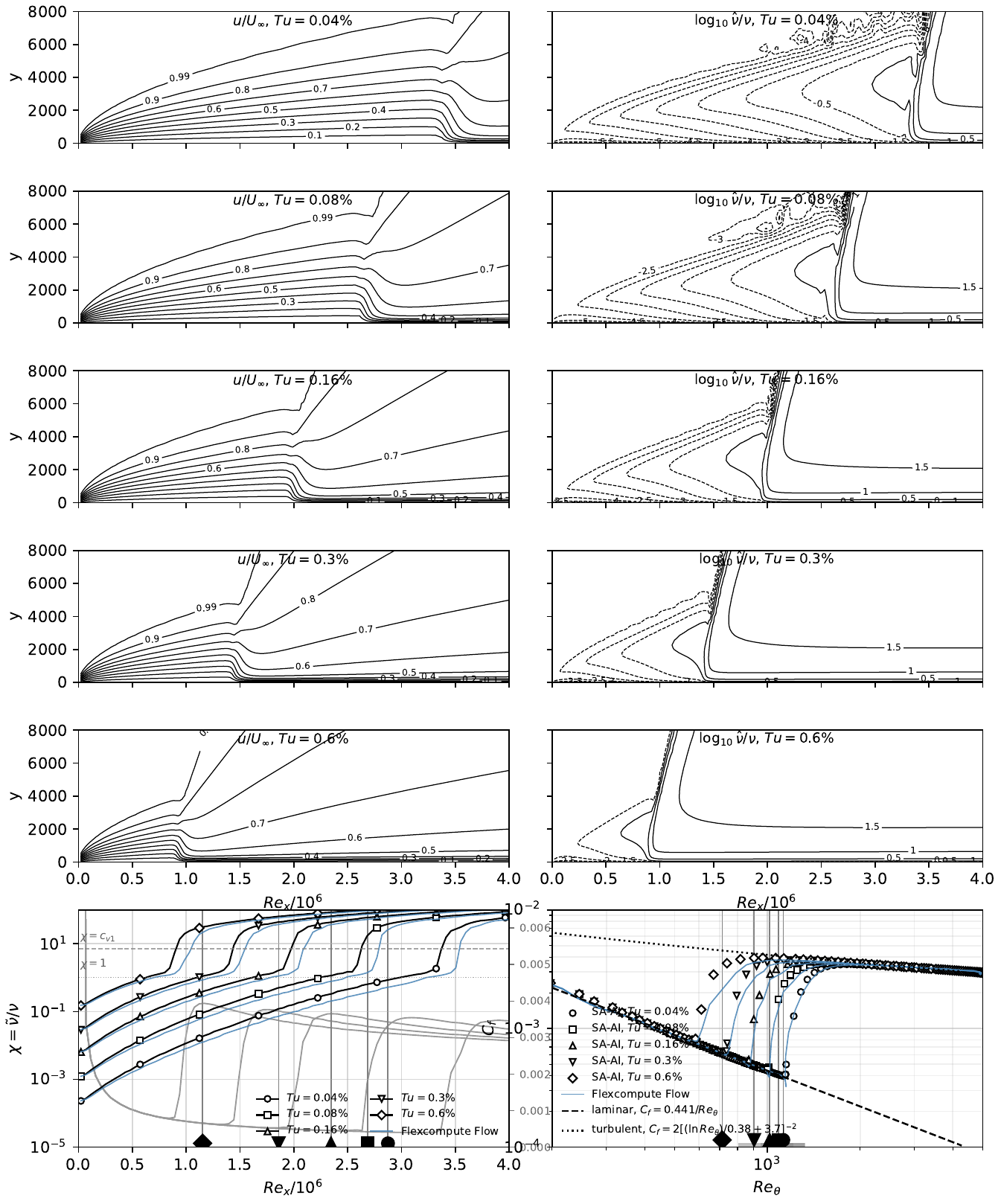}
\caption{Flat plate, OpenFOAM replication, in the layout of Fig.~\ref{f:flatplate}. The thin blue curves in the two bottom panels are the Flexcompute Flow results.}\label{f:flatplateof}
\end{figure}

Table~\ref{t:flatplate}
lists the $Re_\theta$ at which the largest $\chi$ crosses one and $c_{v1}$,
for Flexcompute Flow and for the OpenFOAM replication, next to the correlation of
Abu-Ghannam and Shaw~\cite{abu_ghannam_shaw_1980} for the onset of transition.
The $\chi=c_{v1}$ crossing of Flexcompute Flow is within $10\%$ of the correlation at
every turbulence intensity. The $\chi=1$ crossing is $2\%$ above the
correlation at $Tu=0.04\%$ and $27\%$ below it at $Tu=0.6\%$. The two
crossings are $54$ apart in $Re_\theta$ at $Tu=0.04\%$ and $161$ apart at
$Tu=0.6\%$. Between adjacent turbulence intensities the Flexcompute Flow $\chi=1$
crossing moves by $88$ to $106$ in $Re_\theta$ per unit of $N_\mathrm{crit}$,
which is the estimate $\Delta Re_\theta\approx100$ made above. The OpenFOAM
$\chi=1$ crossings are at an $Re_\theta$ $4$--$7\%$ lower than those of
Flexcompute Flow (Fig.~\ref{f:flatplateof}).

\begin{table}[!htbp]
  \centering\small
  \caption{Flat plate: $Re_\theta$ at which the largest $\chi$ crosses one and $c_{v1}$, for Flexcompute Flow and for the OpenFOAM replication (OF), and the onset $Re_\theta$ of the correlation of Abu-Ghannam and Shaw (AGS). The momentum thickness is integrated from the computed profiles.}
  \label{t:flatplate}
  \begin{tabular}{cccccc}
    \toprule
    $Tu$ [\%] & AGS $Re_\theta$ & $\chi\!=\!1$ & $\chi\!=\!c_{v1}$ & OF $\chi\!=\!1$ & OF $\chi\!=\!c_{v1}$ \\
    \midrule
    0.04 & 1126 & 1149 & 1203 & 1098 & 1154 \\
    0.08 & 1088 & 1002 & 1069 & 949 & 1004 \\
    0.16 & 1017 & 845 & 934 & 798 & 887 \\
    0.3 & 905 & 700 & 815 & 656 & 739 \\
    0.6 & 713 & 524 & 685 & 487 & 589 \\
    \bottomrule
  \end{tabular}
\end{table}

\FloatBarrier
\subsection{The NLF(1)-0416 Airfoil}

The NLF(1)-0416 is an airfoil designed for natural laminar flow. At
$Re=4\times10^6$ its boundary layers are attached, or nearly attached, where
they become turbulent, and the transition location on each surface changes with the
incidence as the pressure distribution changes. The case therefore shows the rate $a$ and the
onset gate $\sigma_{Re}$ of Section~II.C responding to the shape of the
boundary-layer profile. Somers~\cite{somers_1981} measured the forces and the
transition locations of this airfoil in the NASA Langley Low-Turbulence
Pressure Tunnel (LTPT). The case is also one of the test cases of the first AIAA Transition
Modeling and Prediction Workshop~\cite{coder_tmpw_summary}.

The computations cover thirteen incidences from $\alpha=-12^\circ$ to
$15^\circ$ (Table~\ref{t:nlf}). Six of them, $-8^\circ$, $-4^\circ$, $0^\circ$, $4^\circ$,
$9^\circ$ and $15^\circ$, are computed on all six grids. The other seven are
computed on the two L2 grids.

\begin{figure}[!htbp]\centering
\includegraphics[width=0.99\textwidth]{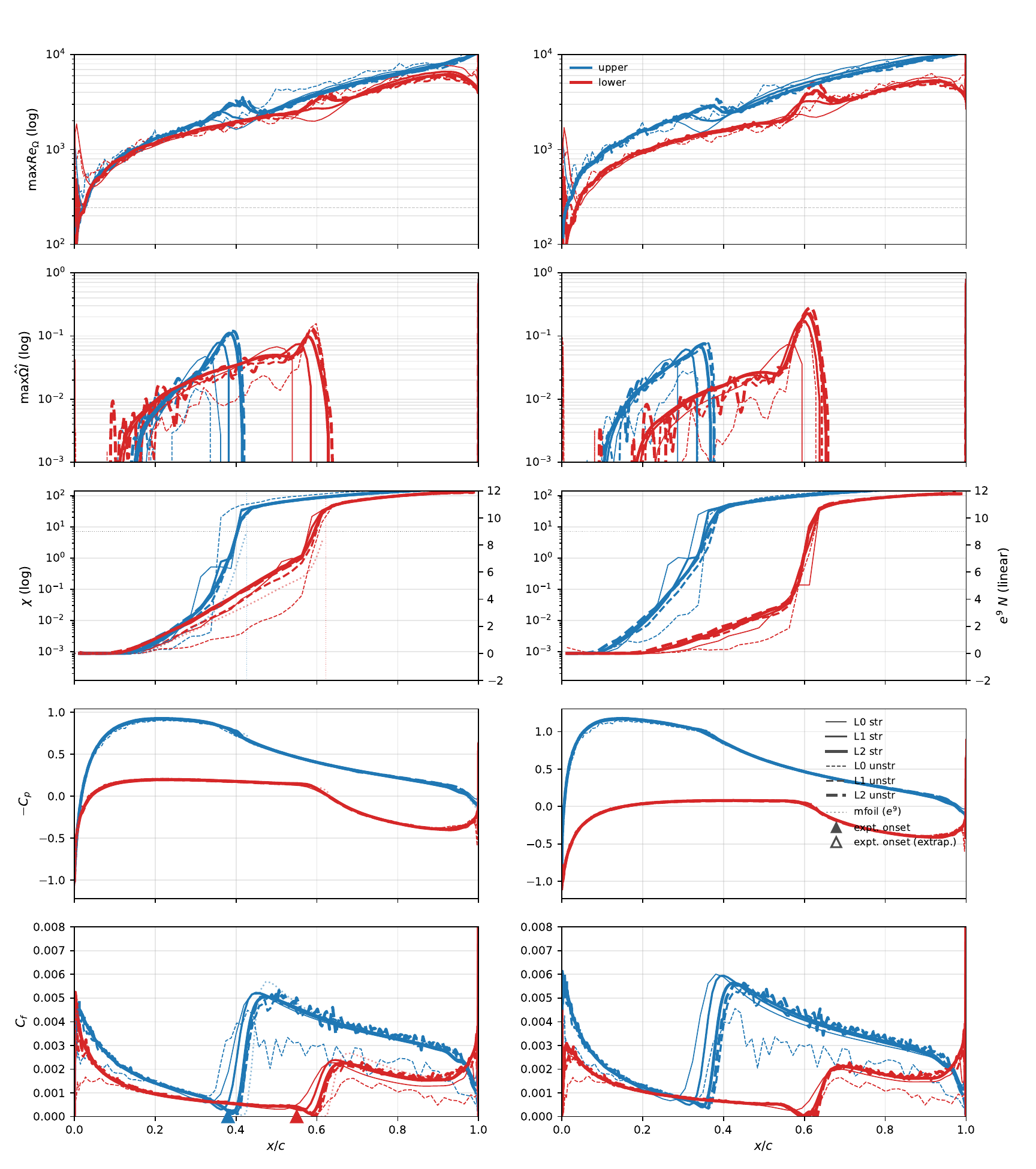}
\caption{NLF(1)-0416 at $\alpha=0^\circ$ (left) and $2^\circ$ (right), $Re=4\times10^6$, all six grids. Rows: largest $Re_\Omega$ on the wall normal; largest $\hat\Omega\hat I$; largest $\chi$, with the amplification factor $N$ of the $e^9$ reference on the right axis; $-C_p$; $C_f$.}\label{f:nlfcf0}
\end{figure}

Figure~\ref{f:nlfcf0} shows the quantities of the model along the chord at
$\alpha=0^\circ$ and $2^\circ$. Near the leading edge, where the pressure
gradient is favorable, the product $\hat\Omega\hat I$ is below $10^{-3}$ and
$\chi$ remains at $\chi_\infty$. Farther aft $\hat\Omega\hat I$ is positive
and $\chi$ grows. The right axis of the $\chi$ panel is the amplification
factor $N$ of the $e^9$ reference. One unit of $N$ on the right axis is one
factor of $e$ in $\chi$ on the left axis, so the computed $\chi$ and the
reference $N$ can be compared directly. On each surface $\hat\Omega\hat I$ is
largest, between $0.07$ and $0.3$, where the pressure begins to rise toward
the trailing edge. The largest $\chi$ crosses one there, and the skin friction
rises to its turbulent level.

\begin{figure}[!htbp]\centering
\includegraphics[width=\textwidth]{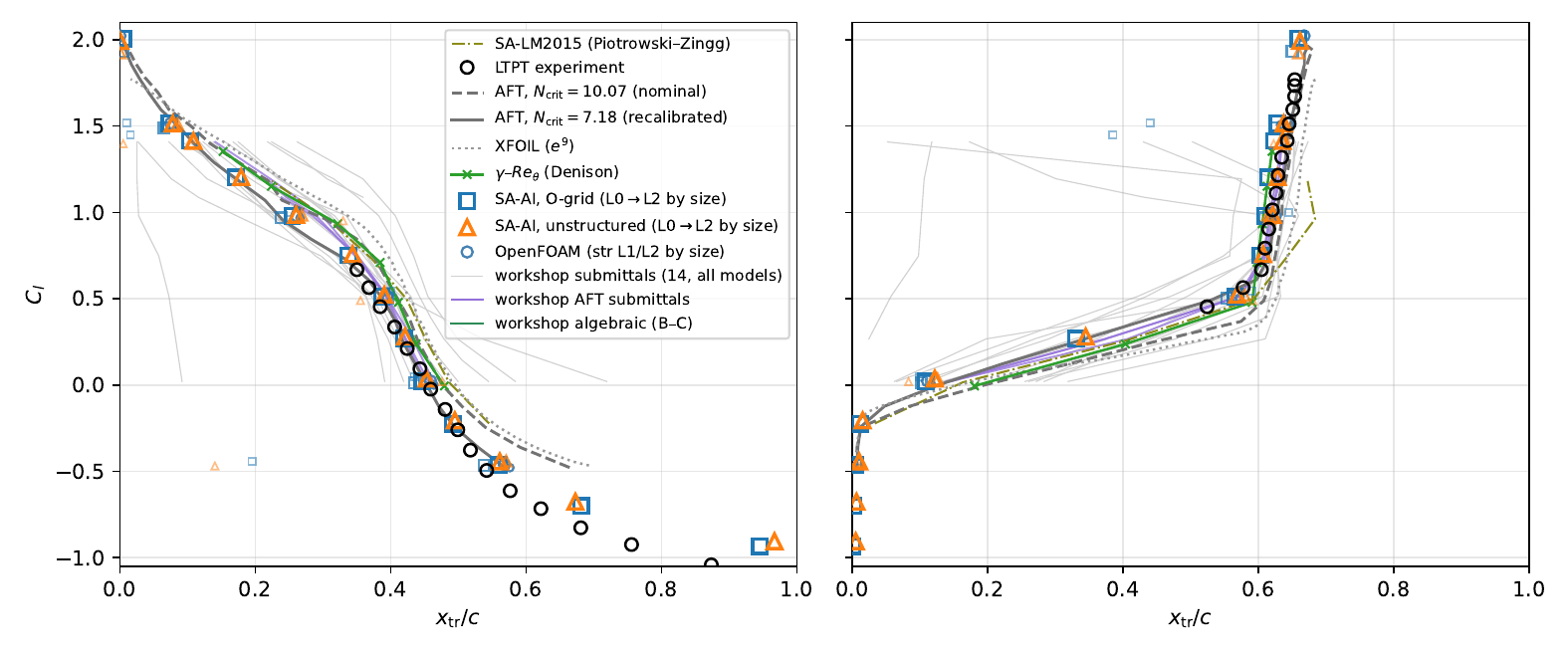}
\caption{NLF(1)-0416, $Re=4\times10^6$: transition location on the upper surface (left) and on the lower surface (right) against the lift coefficient. The SA-AI location is the station where the largest $\chi$ near the wall crosses one. The measurement is that of Somers~\cite{somers_1981}. The AFT locations are those computed by Coder~\cite{coder_thesis_2014}, the SA-LM2015 locations those of Piotrowski and Zingg~\cite{piotrowski_zingg_2021}, and the $\gamma$--$Re_\theta$ locations those of Denison~\cite{denison_2021_overflow}. The workshop submittals are those summarized by Coder~\cite{coder_tmpw_summary}.}\label{f:nlfaft}
\end{figure}

Figure~\ref{f:nlfaft} shows the transition locations on both surfaces against the lift
coefficient. On the structured L2 grid the location on the upper surface moves from
$0.945\,c$ at $\alpha=-12^\circ$ to $0.005\,c$ at $15^\circ$, and the
location on the lower surface from $0.001\,c$ to $0.659\,c$. Between the two grid
families at L2 the locations differ by at most $0.022\,c$. The OpenFOAM replication on the
structured L2 grid is within $0.016\,c$ of Flexcompute Flow in every transition location and within
$0.017$ in lift coefficient. The figure also contains the measurement, the $e^9$
reference, the amplification-factor-transport (AFT) model as computed by
Coder~\cite{coder_thesis_2014} at two values of $N_\mathrm{crit}$, and the
fourteen submittals to the workshop. At the workshop incidences $-2^\circ$,
$2^\circ$, $6^\circ$ and $8^\circ$, the transition locations on both L2 grids are inside the range of
the submittals on both surfaces. At
$2^\circ$, for example, the locations on the upper surface are
at $0.338\,c$ and $0.344\,c$ on the two grids, and the submittals range from $0.040\,c$ to
$0.453\,c$. On the lower surface they are at $0.604\,c$ and $0.607\,c$, and the
submittals range from $0.080\,c$ to $0.654\,c$.

Somers reports the last laminar and the first turbulent orifice, so each
measured transition location is a bracket $0.05\,c$ wide, not a point.
Table~\ref{t:extnlfsomers} compares the computed transition locations with these brackets at
the same lift, at $\alpha=0^\circ$ and $4^\circ$ and four Reynolds numbers
from $10^6$ to $4\times10^6$. Of the $27$ comparisons, $20$ are inside the
bracket. The others are on the upper surface, $0.012$--$0.029\,c$ aft of the
bracket at $\alpha=4^\circ$, $Re=2\times10^6$ and $3\times10^6$, and
on the lower surface, $0.020\,c$ ahead of it at $\alpha=4^\circ$, $Re=10^6$.
As the Reynolds number increases from $10^6$ to $4\times10^6$ the computed
transition location on the upper surface moves forward, from $0.446\,c$ to $0.387\,c$ at
$\alpha=0^\circ$ and from $0.379\,c$ to $0.254\,c$ at $4^\circ$
(Table~\ref{t:extnlfresweep}). The brackets are digitized from all four panels of
Somers' Fig.~9. The same data as replotted in
Coder's dissertation~\cite{coder_thesis_2014} consist of $28$ points, and all
of them are inside the digitized brackets.

\begin{figure}[!htbp]\centering
\includegraphics[width=0.75\textwidth]{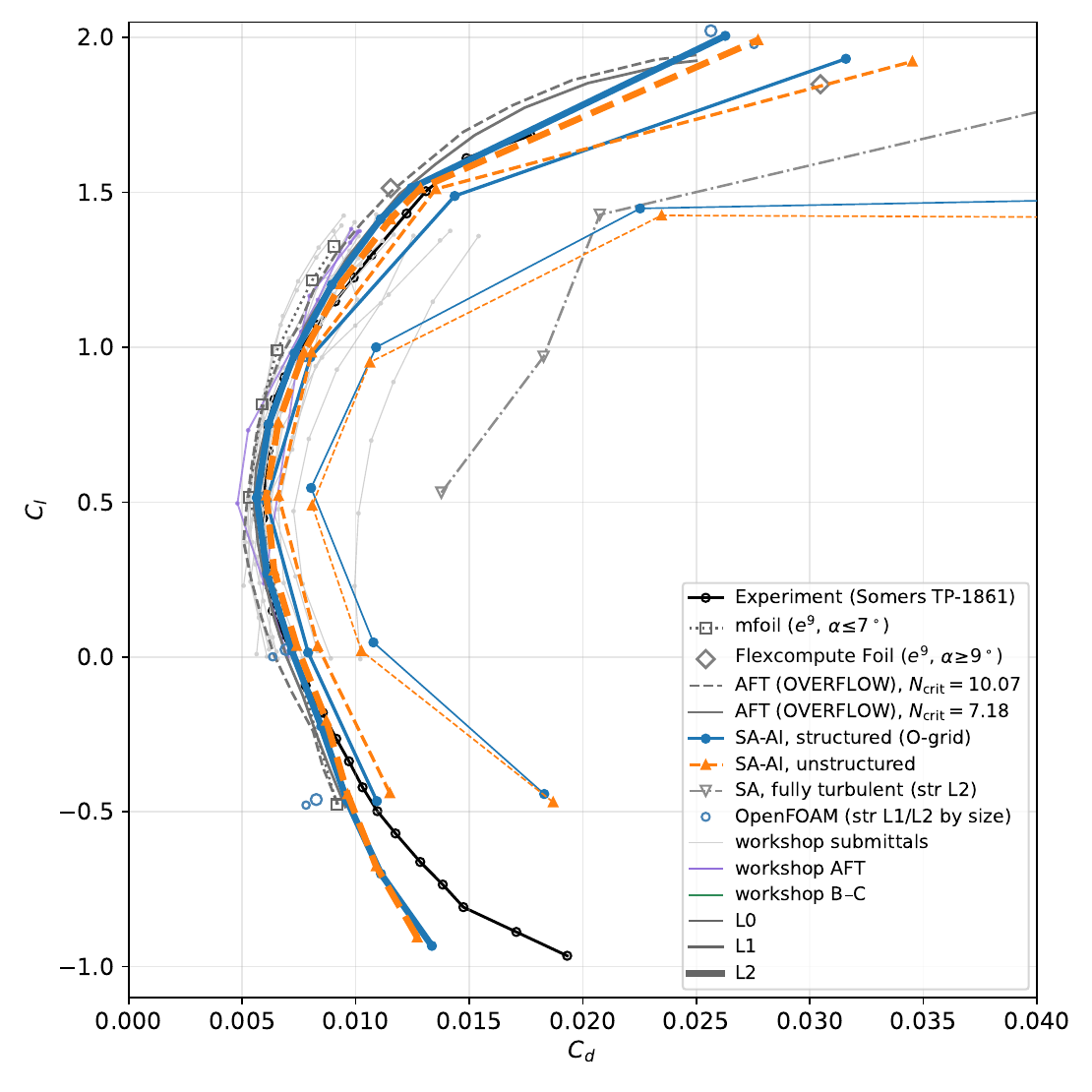}
\caption{NLF(1)-0416 drag polar, $Re=4\times10^6$. Fully turbulent SA is computed with $\chi_\infty=3$.}\label{f:nlfpolar}
\end{figure}

Figure~\ref{f:nlfpolar} shows the drag polar. From $\alpha=-6^\circ$ to $6^\circ$ the drag on the L2 grids is between
$2\%$ below and $10\%$ above the measured drag at the same lift. On the L0 grids the drag is $1.3$ to $4.2$
times the drag on the L2 grids. Fully turbulent SA on the structured L2 grid,
computed with $\chi_\infty=3$, gives $1.6$ to $2.5$ times the SA-AI drag: $2.4$
to $2.5$ times at $\alpha=0^\circ$ and $4^\circ$, and $1.7$ times at $9^\circ$
and $15^\circ$. At
the same incidence the computed lift is higher than the measured lift, by
$0.01$ at $\alpha=-8^\circ$ and by $0.12$ at $8^\circ$. At
$\alpha=15^\circ$ the computed lift coefficient is $2.0$, and the measured
polar ends near $1.7$. Table~\ref{t:nlf} lists the forces and transition locations of every
computation. The Supplementary Figures contain the surface distributions at
every incidence (Fig.~\ref{f:nlfcflow}), the grids (Figs.~\ref{f:nlfmeshstr}
and~\ref{f:nlfmeshcav}), and contours of the velocity and of $\chi$ near the
wall (Fig.~\ref{f:nlfsheets}).

\begin{table}[!htbp]
  \centering\small
  \caption{NLF(1)-0416: computed transition locations on the structured (str) and unstructured (unstr) L2 grids at four Reynolds numbers, and the brackets of Somers~\cite{somers_1981}, Fig.~9, at the same lift. A bracket extends from the last laminar orifice to the first turbulent orifice. $^\dagger$Also in Table~\ref{t:nlf}.}
  \label{t:extnlfsomers}
  \begin{tabular}{cc c ccc ccc}
    \toprule
    & & & \multicolumn{3}{c}{upper surface $x_\mathrm{tr}$} & \multicolumn{3}{c}{lower surface $x_\mathrm{tr}$} \\
    \cmidrule(lr){4-6}\cmidrule(lr){7-9}
    $\alpha$ & $Re/10^6$ & $C_l$ & str & unstr & measured & str & unstr & measured \\
    \midrule
    $0$ & $1$ & 0.4898 & 0.446 & 0.456 & $0.421$--$0.472$ & 0.613 & 0.616 & $0.601$--$0.651$ \\
    $0$ & $2$ & 0.5053 & 0.418 & 0.414 & $0.401$--$0.451$ & 0.604 & 0.611 & $0.602$--$0.652$ \\
    $0$ & $3$ & 0.5109 & 0.400 & 0.400 & $0.367$--$0.417$ & 0.586 & 0.586 & $0.585$--$0.634$ \\
    $0$ & $4$\rlap{$^\dagger$} & 0.5139 & 0.387 & -- & $0.352$--$0.400$ & 0.566 & -- & $0.532$--$0.582$ \\
    $4$ & $1$ & 0.9584 & 0.379 & 0.379 & $0.350$--$0.400$ & 0.631 & 0.630 & $0.651$--$0.701$ \\
    $4$ & $2$ & 0.9744 & 0.320 & 0.323 & $0.258$--$0.308$ & 0.619 & 0.628 & $0.601$--$0.651$ \\
    $4$ & $3$ & 0.9786 & 0.279 & 0.287 & $0.211$--$0.260$ & 0.612 & 0.626 & $0.602$--$0.650$ \\
    $4$ & $4$\rlap{$^\dagger$} & 0.9805 & 0.254 & -- & -- & 0.610 & -- & $0.601$--$0.650$ \\
    \bottomrule
  \end{tabular}
\end{table}

\begin{table}[!htbp]
  \centering\small
  \caption{NLF(1)-0416 at $Re=1$, $2$ and $3\times10^6$, on the L2 grids: forces and transition locations.}
  \label{t:extnlfresweep}
  \begin{tabular}{cc cccc cccc}
    \toprule
    & & \multicolumn{4}{c}{structured} & \multicolumn{4}{c}{unstructured} \\
    \cmidrule(lr){3-6}\cmidrule(lr){7-10}
    $\alpha$ & $Re/10^6$ & $C_l$ & $C_d$ & $x_\mathrm{tr}^\mathrm{up}$ & $x_\mathrm{tr}^\mathrm{lo}$
             & $C_l$ & $C_d$ & $x_\mathrm{tr}^\mathrm{up}$ & $x_\mathrm{tr}^\mathrm{lo}$ \\
    \midrule
    $0$ & $1$ & 0.4898 & 0.00777 & 0.446 & 0.613 & 0.4937 & 0.00825 & 0.456 & 0.616 \\
    $0$ & $2$ & 0.5053 & 0.00636 & 0.418 & 0.604 & 0.5112 & 0.00685 & 0.414 & 0.611 \\
    $0$ & $3$ & 0.5109 & 0.00588 & 0.400 & 0.586 & 0.5181 & 0.00633 & 0.400 & 0.586 \\
    $4$ & $1$ & 0.9584 & 0.00890 & 0.379 & 0.631 & 0.9574 & 0.00942 & 0.379 & 0.630 \\
    $4$ & $2$ & 0.9744 & 0.00766 & 0.320 & 0.619 & 0.9764 & 0.00814 & 0.323 & 0.628 \\
    $4$ & $3$ & 0.9786 & 0.00740 & 0.279 & 0.612 & 0.9819 & 0.00784 & 0.287 & 0.626 \\
    \bottomrule
  \end{tabular}
\end{table}

\begin{table}[!htbp]
  \centering\scriptsize\setlength{\tabcolsep}{2pt}
  \caption{NLF(1)-0416, $Re\!=\!4\!\times\!10^6$: forces, and transition locations where the largest $\chi$ near the wall crosses one. The OpenFOAM replication omits the $f_{v1}$ bypass of Section~II.F. Seven incidences are computed on the L2 grids only. The measurements are those of the LTPT, digitized from Figs.~5-22 to 5-25 of Coder~\cite{coder_thesis_2014}. Measured drag and transition are read at the measured $C_l$. Blank entries are outside the digitized range or beyond stall.}
  \label{t:nlf}
  \begin{tabular}{ll cccc cccc cccc cccc}
    \toprule
    & & \multicolumn{4}{c}{Flexcompute Flow, structured O-grid} & \multicolumn{4}{c}{Flexcompute Flow, unstructured} & \multicolumn{4}{c}{OpenFOAM, structured} & \multicolumn{4}{c}{Experiment (LTPT)} \\
    \cmidrule(lr){3-6}\cmidrule(lr){7-10}\cmidrule(lr){11-14}\cmidrule(lr){15-18}
    $\alpha$ & grid & $C_l$ & $C_d$ & $x_\mathrm{tr}^\mathrm{up}$ & $x_\mathrm{tr}^\mathrm{lo}$ & $C_l$ & $C_d$ & $x_\mathrm{tr}^\mathrm{up}$ & $x_\mathrm{tr}^\mathrm{lo}$ & $C_l$ & $C_d$ & $x_\mathrm{tr}^\mathrm{up}$ & $x_\mathrm{tr}^\mathrm{lo}$ & $C_l$ & $C_d$ & $x_\mathrm{tr}^\mathrm{up}$ & $x_\mathrm{tr}^\mathrm{lo}$ \\
    \midrule
    $-12$ & L2 & -0.9329 & 0.01335 & 0.945 & 0.001 & -0.9050 & 0.01271 & 0.967 & 0.005 & -- & -- & -- & -- & -- & -- & -- & -- \\
    \addlinespace
    $-10$ & L2 & -0.7004 & 0.01111 & 0.681 & 0.002 & -0.6756 & 0.01090 & 0.673 & 0.006 & -- & -- & -- & -- & -- & -- & -- & -- \\
    \addlinespace
    $-8$ & L0 & -0.4426 & 0.01830 & 0.195 & 0.011 & -0.4689 & 0.01869 & 0.140 & 0.012 & -0.4460 & 0.01241 & 0.556 & 0.007 & -0.4745 & 0.01052 & 0.538 & -- \\
    $-8$ & L1 & -0.4656 & 0.01093 & 0.538 & 0.006 & -0.4397 & 0.01150 & 0.571 & 0.011 & -0.4785 & 0.00781 & 0.576 & 0.000 &  &  &  &  \\
    $-8$ & L2 & -0.4646 & 0.00962 & 0.559 & 0.004 & -0.4426 & 0.00964 & 0.561 & 0.010 & -0.4606 & 0.00826 & 0.568 & 0.000 &  &  &  &  \\
    \addlinespace
    $-6$ & L2 & -0.2240 & 0.00848 & 0.492 & 0.012 & -0.2052 & 0.00870 & 0.495 & 0.016 & -- & -- & -- & -- & -0.2410 & 0.00881 & 0.496 & -- \\
    \addlinespace
    $-4$ & L0 & 0.0469 & 0.01078 & 0.435 & 0.100 & 0.0197 & 0.01025 & 0.476 & 0.083 & 0.0388 & 0.00748 & 0.424 & 0.018 & -0.0122 & 0.00695 & 0.457 & -- \\
    $-4$ & L1 & 0.0142 & 0.00790 & 0.435 & 0.102 & 0.0357 & 0.00832 & 0.452 & 0.126 & 0.0005 & 0.00634 & 0.464 & 0.117 &  &  &  &  \\
    $-4$ & L2 & 0.0208 & 0.00718 & 0.447 & 0.109 & 0.0363 & 0.00739 & 0.453 & 0.122 & 0.0239 & 0.00693 & 0.459 & 0.111 &  &  &  &  \\
    \addlinespace
    $-2$ & L2 & 0.2683 & 0.00609 & 0.419 & 0.330 & 0.2800 & 0.00640 & 0.421 & 0.345 & -- & -- & -- & -- & 0.2140 & 0.00595 & 0.424 & -- \\
    \addlinespace
    $0$ & L0 & 0.5458 & 0.00803 & 0.400 & 0.590 & 0.4896 & 0.00808 & 0.355 & 0.592 & 0.5473 & 0.00669 & 0.387 & 0.564 & 0.4553 & 0.00567 & 0.384 & 0.525 \\
    $0$ & L1 & 0.5048 & 0.00607 & 0.395 & 0.553 & 0.5209 & 0.00660 & 0.392 & 0.579 & 0.5034 & 0.00603 & 0.393 & 0.572 &  &  &  &  \\
    $0$ & L2 & 0.5139 & 0.00564 & 0.387 & 0.566 & 0.5219 & 0.00606 & 0.391 & 0.569 & 0.5153 & 0.00573 & 0.395 & 0.568 &  &  &  &  \\
    \addlinespace
    $2$ & L2 & 0.7510 & 0.00616 & 0.338 & 0.604 & 0.7573 & 0.00659 & 0.344 & 0.607 & -- & -- & -- & -- & 0.6694 & 0.00587 & -- & 0.605 \\
    \addlinespace
    $4$ & L0 & 1.0001 & 0.01090 & 0.250 & 0.645 & 0.9513 & 0.01062 & 0.330 & 0.621 & 1.0114 & 0.00955 & 0.271 & 0.618 & 0.8903 & 0.00639 & -- & 0.615 \\
    $4$ & L1 & 0.9678 & 0.00800 & 0.238 & 0.610 & 0.9839 & 0.00805 & 0.269 & 0.614 & 0.9670 & 0.00778 & 0.266 & 0.616 &  &  &  &  \\
    $4$ & L2 & 0.9805 & 0.00730 & 0.255 & 0.610 & 0.9848 & 0.00772 & 0.260 & 0.621 & 0.9793 & 0.00749 & 0.261 & 0.625 &  &  &  &  \\
    \addlinespace
    $6$ & L2 & 1.2019 & 0.00895 & 0.171 & 0.614 & 1.2042 & 0.00933 & 0.179 & 0.629 & -- & -- & -- & -- & 1.1049 & 0.00830 & -- & 0.626 \\
    \addlinespace
    $8$ & L2 & 1.4139 & 0.01109 & 0.104 & 0.624 & 1.4137 & 0.01153 & 0.109 & 0.636 & -- & -- & -- & -- & 1.2941 & 0.01001 & -- & 0.633 \\
    \addlinespace
    $9$ & L0 & 1.4483 & 0.02252 & 0.015 & 0.385 & 1.4251 & 0.02345 & 0.105 & 0.641 & 1.5277 & 0.01496 & 0.063 & 0.618 & 1.4026 & 0.01104 & -- & 0.642 \\
    $9$ & L1 & 1.4886 & 0.01436 & 0.065 & 0.625 & 1.5112 & 0.01353 & 0.085 & 0.635 & 1.5066 & 0.01273 & 0.076 & 0.633 &  &  &  &  \\
    $9$ & L2 & 1.5146 & 0.01245 & 0.073 & 0.628 & 1.5139 & 0.01286 & 0.078 & 0.638 & 1.5151 & 0.01270 & 0.075 & 0.644 &  &  &  &  \\
    \addlinespace
    $15$ & L0 & 1.5172 & 0.07109 & 0.010 & 0.440 & 1.3976 & 0.11602 & 0.005 & 0.623 & 1.7210 & 0.05409 & 0.000 & 0.634 & -- & -- & -- & -- \\
    $15$ & L1 & 1.9317 & 0.03159 & 0.004 & 0.650 & 1.9232 & 0.03454 & 0.001 & 0.659 & 1.9780 & 0.02754 & 0.000 & 0.665 &  &  &  &  \\
    $15$ & L2 & 2.0057 & 0.02628 & 0.005 & 0.659 & 1.9924 & 0.02772 & 0.000 & 0.662 & 2.0221 & 0.02564 & 0.000 & 0.667 &  &  &  &  \\
    \bottomrule
  \end{tabular}
\end{table}

\FloatBarrier
\subsection{The Eppler 387 Airfoil}

The Eppler 387 at $Re=2\times10^5$ is the standard case for transition in a
laminar separation bubble. The laminar boundary layer on the upper surface
separates before $\chi$ reaches one. The inflection point of the separated
shear layer's profile is far from the wall, so $\hat\Omega\hat I$ is close to
one and $\chi$ grows at nearly the largest rate, $a_{\max}\omega$. The eddy
viscosity that follows returns the flow to the wall as a turbulent boundary
layer. The region of reversed flow between separation and reattachment is the
bubble. McGhee, Walker and Millard~\cite{mcghee_1988} measured the forces of
this airfoil in the LTPT, and the separation and reattachment stations by oil
flow. The Eppler 387 is also the airfoil on which Drela demonstrated XFOIL at
low Reynolds numbers~\cite{drela_xfoil_1989}, against Langley measurements by
McGhee and coworkers.

\begin{figure}[!htbp]\centering
\includegraphics[width=0.99\textwidth]{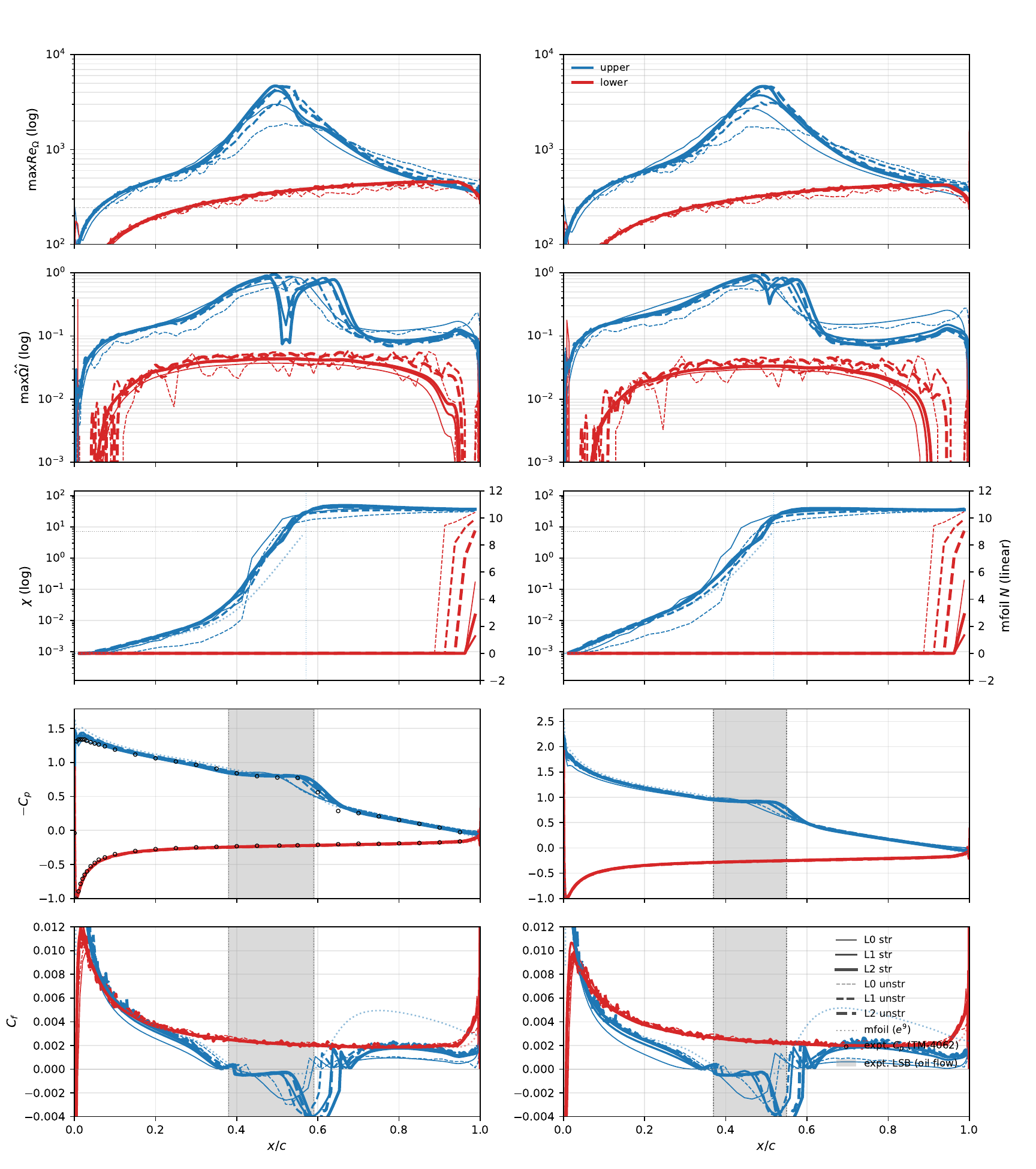}
\caption{Eppler~387 at $\alpha=5^\circ$ (left) and $6^\circ$ (right), $Re=2\times10^5$, all six grids; rows as in Fig.~\ref{f:nlfcf0}. Circles: measured pressure. Gray band: the bubble measured by oil flow.}\label{f:eppcf5}
\end{figure}

The computations cover incidences from $\alpha=-4^\circ$ to $9^\circ$.
Figure~\ref{f:eppcf5} shows the quantities of the model along the chord at
$\alpha=5^\circ$ and $6^\circ$. On the upper surface $\hat\Omega\hat I$
increases from about $0.1$ in the attached boundary layer to nearly one over
the bubble. The largest $\chi$ grows from $\chi_\infty$ and crosses one
inside the bubble. The pressure is nearly constant over the forward part of
the bubble and rises steeply after $\chi$ exceeds $c_{v1}$. The skin friction
is negative between separation and reattachment. The gray band is the bubble
measured by oil flow. On the lower surface $\hat\Omega\hat I$ is below $0.06$,
and $Re_\Omega$ remains below the onset threshold of Eq.~\eqref{w:onset}. The
rate $a\,\sigma_{Re}\,\omega$ is therefore small, and $\chi$ remains near
$\chi_\infty$ over most of the chord.

\begin{figure}[!htbp]\centering
\includegraphics[width=0.99\textwidth,height=0.90\textheight,keepaspectratio]{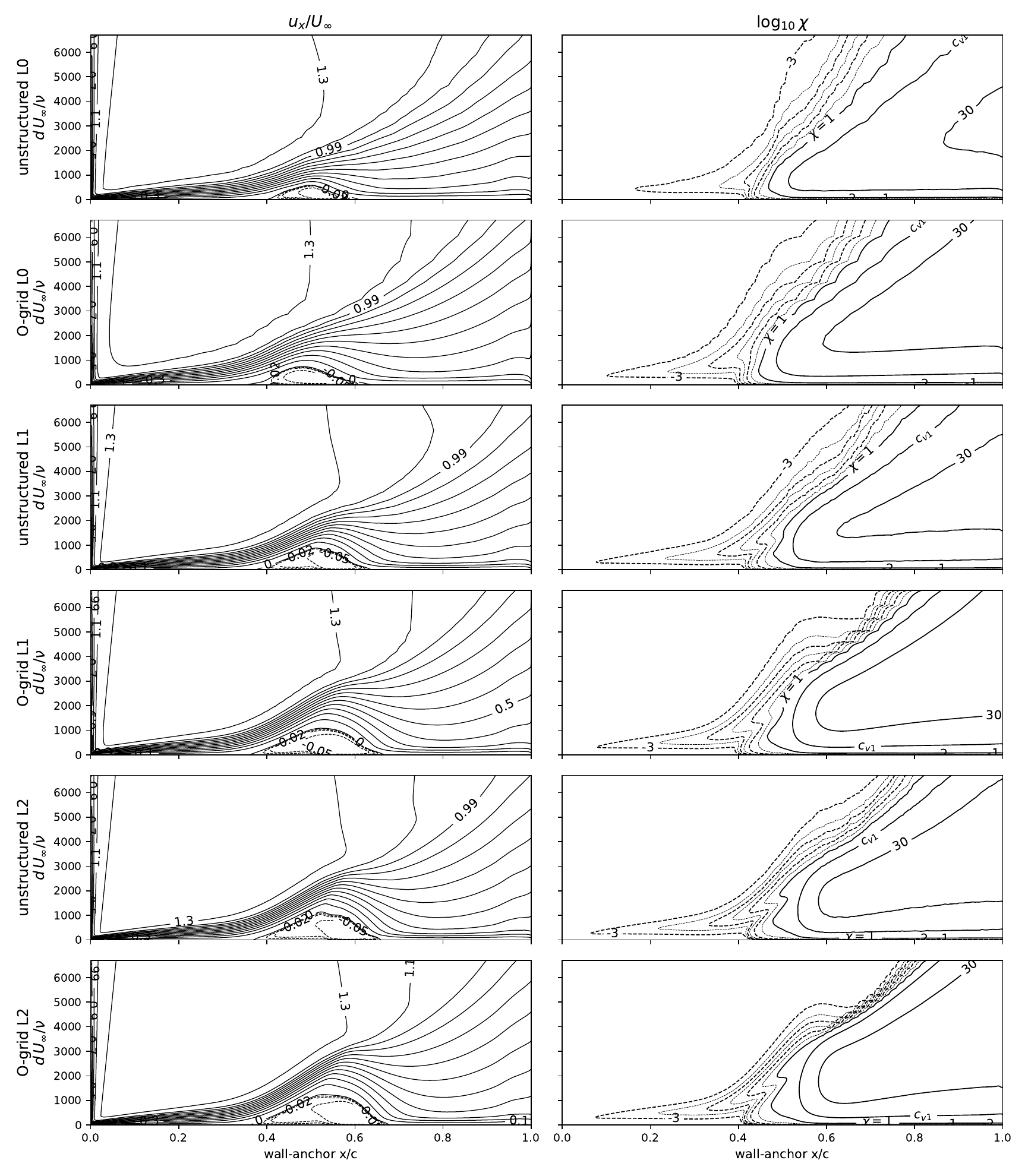}
\caption{Eppler~387 at $\alpha=5^\circ$, $Re=2\times10^5$, upper surface, all six grids: $u_x/U_\infty$ (left) and $\log_{10}\chi$ (right) against the chordwise station and the wall distance.}\label{f:eppsheet5}
\end{figure}

\begin{figure}[!htbp]\centering
\includegraphics[width=\textwidth]{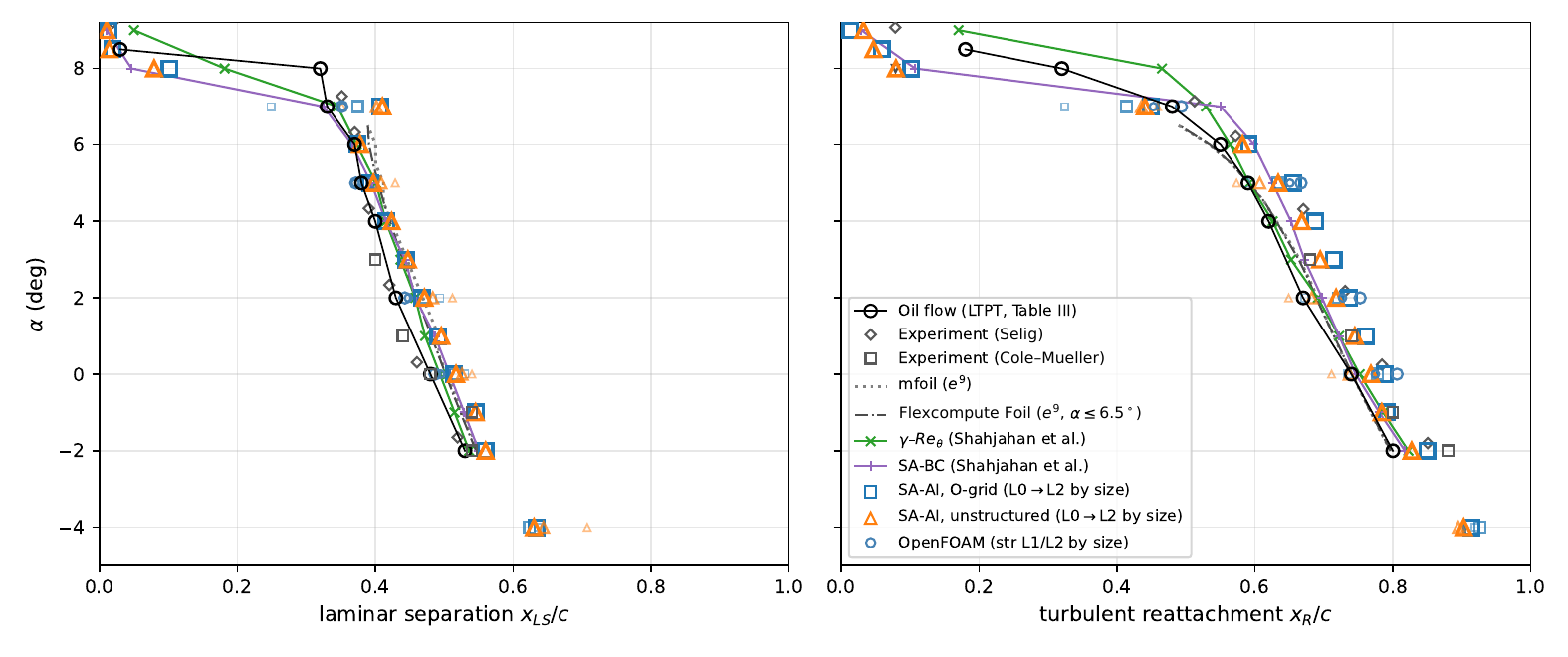}
\caption{Eppler 387, $Re=2\times10^5$: stations of laminar separation (left) and turbulent reattachment (right) against the incidence. The computed stations are the zero crossings of the signed $C_f$. The oil flow is Table~III of McGhee et al.~\cite{mcghee_1988}. The other measurements are those of Selig~\cite{selig_2003}, as replotted by Shahjahan et al.~\cite{shahjahan_2024}, and of Cole and Mueller~\cite{cole_mueller_1990}. The $\gamma$--$Re_\theta$ and SA-BC stations are those of Shahjahan et al.}\label{f:eppbubble}
\end{figure}

Figure~\ref{f:eppsheet5} shows the velocity and $\chi$ near the upper surface
at $\alpha=5^\circ$, on all six grids. In the separated shear
layer, away from the wall, $\chi$ grows and first reaches one.
Figure~\ref{f:eppbubble} compares the computed separation and reattachment
stations with the oil flow, with the measurements of Selig~\cite{selig_2003} and of Cole and
Mueller~\cite{cole_mueller_1990}, with the $e^9$
reference, and with the $\gamma$--$Re_\theta$ and SA-BC models as computed by
Shahjahan et al.~\cite{shahjahan_2024}. From $\alpha=-2^\circ$ to $6^\circ$
the separation station on the L2 grids is $0.00$ to $0.04\,c$ aft of the
oil-flow station, and the reattachment station is $0.03$ to $0.07\,c$ aft. The
computed bubble is therefore longer than the measured bubble: the flow
reattaches late, as stated in Section~II.F. At $\alpha=7^\circ$ the computed
bubble is $0.03$--$0.04\,c$ long, and the measured bubble is $0.15\,c$ long.
At $8^\circ$ and $9^\circ$ no bubble forms on the computed upper surface, and
the transition location on the structured L2 grid is at $0.119\,c$ and $0.016\,c$. SA-BC gives
$0.099\,c$ and $0.022\,c$, and $\gamma$--$Re_\theta$ gives $0.451\,c$ and
$0.150\,c$~\cite{shahjahan_2024}. The measurement at $8^\circ$ records transition at $0.32\,c$
without a bubble.

\begin{figure}[!htbp]\centering
\includegraphics[width=0.75\textwidth]{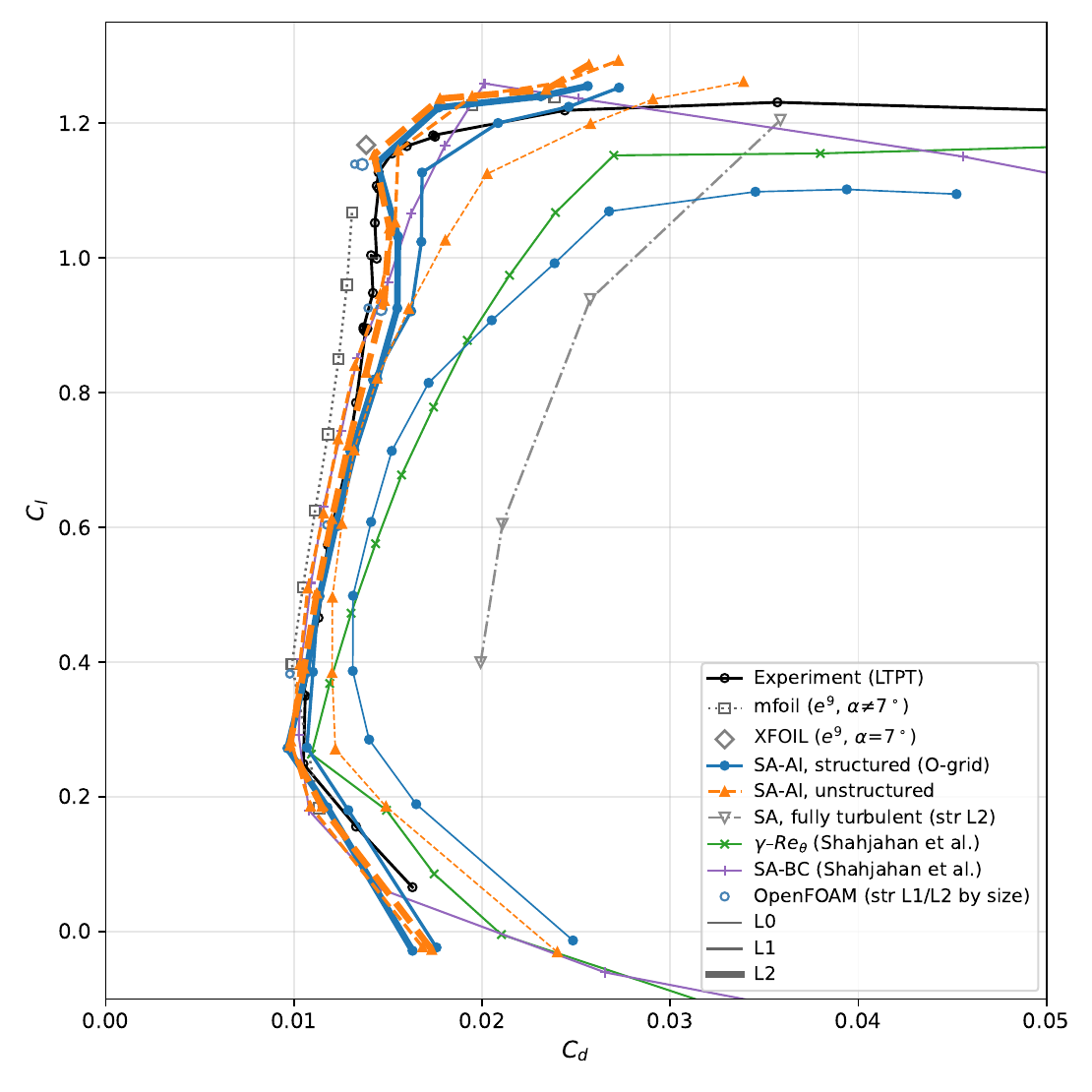}
\caption{Eppler 387 drag polar, $Re=2\times10^5$. Fully turbulent SA is computed with $\chi_\infty=3$.}\label{f:epppolar}
\end{figure}

Figure~\ref{f:epppolar} shows the drag polar. From $\alpha=0^\circ$ to
$3^\circ$ the drag on the L2 grids is within $3\%$ of the measured drag at the
same lift. From $4^\circ$ to $6^\circ$ it is $2\%$ to $12\%$ higher. Fully
turbulent SA on the structured L2 grid gives a drag $66\%$ to $87\%$ above the
SA-AI drag where the drag is lowest, and $149\%$ above it at
$\alpha=7^\circ$. The drag coefficients of the two grid families at L2 differ by at most
$7\times10^{-4}$, or $7$ counts, from $\alpha=-2^\circ$ to $7^\circ$. A computation without the
bypass of Section~II.F, on the structured L2 grid, gives a drag $6$ to $10$
counts higher at $\alpha=0^\circ$ to $5^\circ$. The OpenFOAM replication on the
structured L2 grid is within $0.6\%$ of Flexcompute Flow in lift. It separates
$0.02$--$0.06\,c$ ahead of Flexcompute Flow and reattaches $0.01$--$0.04\,c$ behind it.
Table~\ref{t:epp} lists the forces and bubble stations of every computation.
The Supplementary Figures contain the surface distributions and the
near-wall contours at every incidence (Figs.~\ref{f:eppcflow}
and~\ref{f:eppsheets}).

\begin{table}[!htbp]
  \centering\scriptsize\setlength{\tabcolsep}{2pt}
  \caption{Eppler 387, $Re\!=\!2\!\times\!10^5$: forces, and stations of laminar separation $x_{LS}$ and turbulent reattachment $x_R$ on the upper surface. The stations are the zero crossings of the signed $C_f$. Where no bubble forms, both stations are the transition location and are equal. A dash in $x_R$ alone means that the flow does not reattach ahead of the trailing edge. Nine incidences are computed on the L2 grids only. Measured forces are from Table~B1 of McGhee et al.~\cite{mcghee_1988}, and measured stations from the oil flow of their Table~III. At $8^\circ$ the measurement records transition at $0.32\,c$ and no bubble. The OpenFOAM replication omits the $f_{v1}$ bypass of Section~II.F. Its column $x_\mathrm{tr}^\mathrm{up}$ is the station where the largest $\chi$ near the wall crosses one.}
  \label{t:epp}
  \begin{tabular}{ll cccc cccc ccccc cccc}
    \toprule
    & & \multicolumn{4}{c}{Flexcompute Flow, structured O-grid} & \multicolumn{4}{c}{Flexcompute Flow, unstructured} & \multicolumn{5}{c}{OpenFOAM, structured} & \multicolumn{4}{c}{Experiment (LTPT)} \\
    \cmidrule(lr){3-6}\cmidrule(lr){7-10}\cmidrule(lr){11-15}\cmidrule(lr){16-19}
    $\alpha$ & grid & $C_l$ & $C_d$ & $x_{LS}$ & $x_R$ & $C_l$ & $C_d$ & $x_{LS}$ & $x_R$ & $C_l$ & $C_d$ & $x_\mathrm{tr}^\mathrm{up}$ & $x_{LS}$ & $x_R$ & $C_l$ & $C_d$ & $x_{LS}$ & $x_R$ \\
    \midrule
    $-4$ & L0 & $-0.0130$ & 0.02483 & 0.636 & 0.919 & $-0.0303$ & 0.02400 & 0.707 & 0.900 & -- & -- & -- & -- & -- & -- & -- & -- & -- \\
    $-4$ & L1 & $-0.0234$ & 0.01758 & 0.623 & 0.927 & $-0.0229$ & 0.01684 & 0.644 & 0.895 & -- & -- & -- & -- & -- &  &  &  &  \\
    $-4$ & L2 & $-0.0282$ & 0.01630 & 0.634 & 0.915 & $-0.0269$ & 0.01733 & 0.630 & 0.903 & -- & -- & -- & -- & -- &  &  &  &  \\
    \addlinespace
    $-2$ & L2 & 0.1845 & 0.01178 & 0.560 & 0.850 & 0.1857 & 0.01148 & 0.560 & 0.827 & -- & -- & -- & -- & -- & 0.156 & 0.0133 & 0.53 & 0.80 \\
    \addlinespace
    $-1$ & L2 & 0.2723 & 0.00966 & 0.546 & 0.791 & 0.2763 & 0.00978 & 0.545 & 0.784 & -- & -- & -- & -- & -- & 0.249 & 0.0105 & -- & -- \\
    \addlinespace
    $0$ & L0 & 0.3869 & 0.01313 & 0.530 & 0.735 & 0.3841 & 0.01203 & 0.540 & 0.711 & 0.3887 & 0.01167 & 0.638 & 0.493 & 0.784 & 0.350 & 0.0106 & 0.48 & 0.74 \\
    $0$ & L1 & 0.3855 & 0.01101 & 0.516 & 0.778 & 0.3972 & 0.01032 & 0.526 & 0.737 & 0.3826 & 0.00978 & 0.605 & 0.495 & 0.775 &  &  &  &  \\
    $0$ & L2 & 0.3880 & 0.01068 & 0.514 & 0.788 & 0.3908 & 0.01053 & 0.517 & 0.768 & 0.3857 & 0.01058 & 0.621 & 0.486 & 0.806 &  &  &  &  \\
    \addlinespace
    $1$ & L2 & 0.4979 & 0.01136 & 0.491 & 0.761 & 0.5022 & 0.01121 & 0.496 & 0.745 & -- & -- & -- & -- & -- & 0.466 & 0.0113 & -- & -- \\
    \addlinespace
    $2$ & L0 & 0.6082 & 0.01410 & 0.493 & 0.672 & 0.6055 & 0.01255 & 0.512 & 0.649 & 0.6133 & 0.01406 & 0.568 & 0.447 & 0.739 & 0.574 & 0.0118 & 0.43 & 0.67 \\
    $2$ & L1 & 0.6054 & 0.01228 & 0.470 & 0.721 & 0.6207 & 0.01157 & 0.483 & 0.682 & 0.6037 & 0.01174 & 0.543 & 0.448 & 0.727 &  &  &  &  \\
    $2$ & L2 & 0.6070 & 0.01225 & 0.468 & 0.737 & 0.6126 & 0.01201 & 0.471 & 0.718 & 0.6039 & 0.01229 & 0.564 & 0.443 & 0.752 &  &  &  &  \\
    \addlinespace
    $3$ & L2 & 0.7147 & 0.01321 & 0.444 & 0.714 & 0.7219 & 0.01289 & 0.447 & 0.695 & -- & -- & -- & -- & -- & 0.680 & 0.0127 & -- & -- \\
    \addlinespace
    $4$ & L2 & 0.8206 & 0.01443 & 0.416 & 0.687 & 0.8298 & 0.01384 & 0.424 & 0.668 & -- & -- & -- & -- & -- & 0.785 & 0.0133 & 0.40 & 0.62 \\
    \addlinespace
    $5$ & L0 & 0.9075 & 0.02051 & 0.398 & 0.587 & 0.9246 & 0.01610 & 0.429 & 0.573 & 0.9292 & 0.01665 & 0.457 & 0.344 & 0.654 & 0.891 & 0.0138 & 0.38 & 0.59 \\
    $5$ & L1 & 0.9209 & 0.01623 & 0.394 & 0.633 & 0.9462 & 0.01459 & 0.408 & 0.607 & 0.9256 & 0.01394 & 0.468 & 0.371 & 0.651 &  &  &  &  \\
    $5$ & L2 & 0.9254 & 0.01549 & 0.393 & 0.655 & 0.9365 & 0.01481 & 0.397 & 0.634 & 0.9234 & 0.01463 & 0.487 & 0.372 & 0.667 &  &  &  &  \\
    \addlinespace
    $6$ & L2 & 1.0318 & 0.01553 & 0.374 & 0.591 & 1.0440 & 0.01508 & 0.378 & 0.582 & -- & -- & -- & -- & -- & 1.004 & 0.0141 & 0.37 & 0.55 \\
    \addlinespace
    $7$ & L0 & 1.0691 & 0.02675 & 0.249 & 0.324 & 1.1248 & 0.02027 & 0.351 & 0.459 & 1.1173 & 0.01817 & 0.139 & -- & 0.201 & 1.103 & 0.0145 & 0.33 & 0.48 \\
    $7$ & L1 & 1.1268 & 0.01681 & 0.374 & 0.414 & 1.1596 & 0.01554 & 0.402 & 0.435 & 1.1391 & 0.01323 & 0.329 & 0.353 & 0.453 &  &  &  &  \\
    $7$ & L2 & 1.1411 & 0.01442 & 0.407 & 0.449 & 1.1531 & 0.01426 & 0.410 & 0.440 & 1.1388 & 0.01363 & 0.379 & 0.351 & 0.493 &  &  &  &  \\
    \addlinespace
    $8$ & L2 & 1.2232 & 0.01768 & 0.101 & 0.101 & 1.2364 & 0.01776 & 0.079 & 0.079 & -- & -- & -- & -- & -- & 1.180 & 0.0175 & 0.32 & 0.32 \\
    \addlinespace
    $8.5$ & L2 & 1.2399 & 0.02312 & 0.019 & 0.059 & 1.2496 & 0.02339 & 0.014 & 0.047 & -- & -- & -- & -- & -- & -- & -- & 0.03 & 0.18 \\
    \addlinespace
    $9$ & L2 & 1.2552 & 0.02562 & 0.014 & 0.014 & 1.2869 & 0.02568 & 0.010 & 0.032 & -- & -- & -- & -- & -- & 1.219 & 0.0244 & -- & -- \\
    \bottomrule
  \end{tabular}
\end{table}

\FloatBarrier
\subsection{Reynolds-Number Study of the Eppler 387 Airfoil}

The length of the bubble increases as the Reynolds number decreases. Below
some Reynolds number the separated shear layer does not reattach ahead of the
trailing edge, the lift decreases and the drag increases severalfold. This is
the bursting of the bubble. McGhee's measurements cover $Re=6\times10^4$ to
$4.6\times10^5$. This study shows the Reynolds number at which the computed
bubble bursts, and compares it with the measured one.

\begin{figure}[!htbp]\centering
\includegraphics[width=\textwidth]{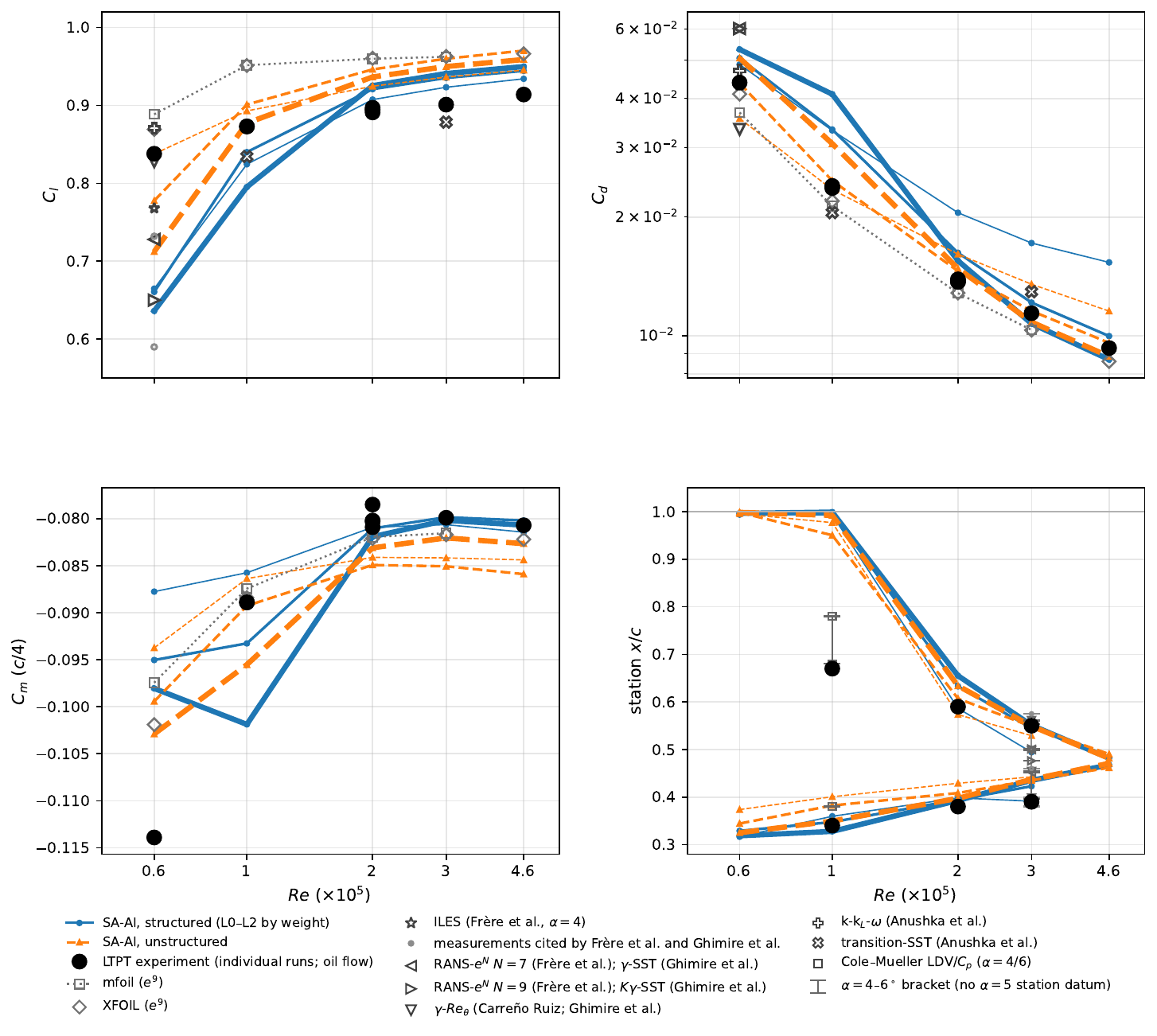}
\caption{Eppler 387 at $\alpha=5^\circ$ against the Reynolds number: lift, drag, quarter-chord moment, and stations of separation and reattachment. Each LTPT run within $0.1^\circ$ of $5^\circ$ is one filled circle: one, two, four, one and one runs at $Re=0.6$, $1$, $2$, $3$ and $4.6\times10^5$. A computed reattachment station at $x/c=1$ means that the flow does not reattach. The other computations are the large-eddy simulation and the coupled RANS--$e^N$ computations of Fr\`ere et al.~\cite{frere_2016} at $Re=6\times10^4$, and the transition models computed by Carre\~no Ruiz and D'Ambrosio~\cite{carreno_ruiz_2022}, Anushka et al.~\cite{anushka_2019} and Ghimire et al.~\cite{ghimire_2025}. The other measurements are those of Cole and Mueller~\cite{cole_mueller_1990} and those cited by Fr\`ere et al.\ and by Ghimire et al.}\label{f:eppresweepforces}
\end{figure}

The airfoil is computed at $\alpha=5^\circ$ and $Re=6\times10^4$, $10^5$,
$3\times10^5$ and $4.6\times10^5$, on the six grids of Section~III.C, with the
same $\chi_\infty$. Figure~\ref{f:eppresweepforces} shows the lift, drag,
moment and bubble stations against the Reynolds number, with the results at
$2\times10^5$ from Section~III.C. Table~\ref{t:eppresweep} lists the
coefficients on the L2 grids. At $2\times10^5$ and above, the three grid
levels converge toward the measurement, and the computed bubble shortens as
the Reynolds number increases. The L2 drag is $0.0155$ and $0.0148$ against a
measured $0.0138$ at $2\times10^5$, and $0.0087$ and $0.0089$ against
$0.0093$ at $4.6\times10^5$.

\begin{figure}[!htbp]\centering
\includegraphics[width=\textwidth]{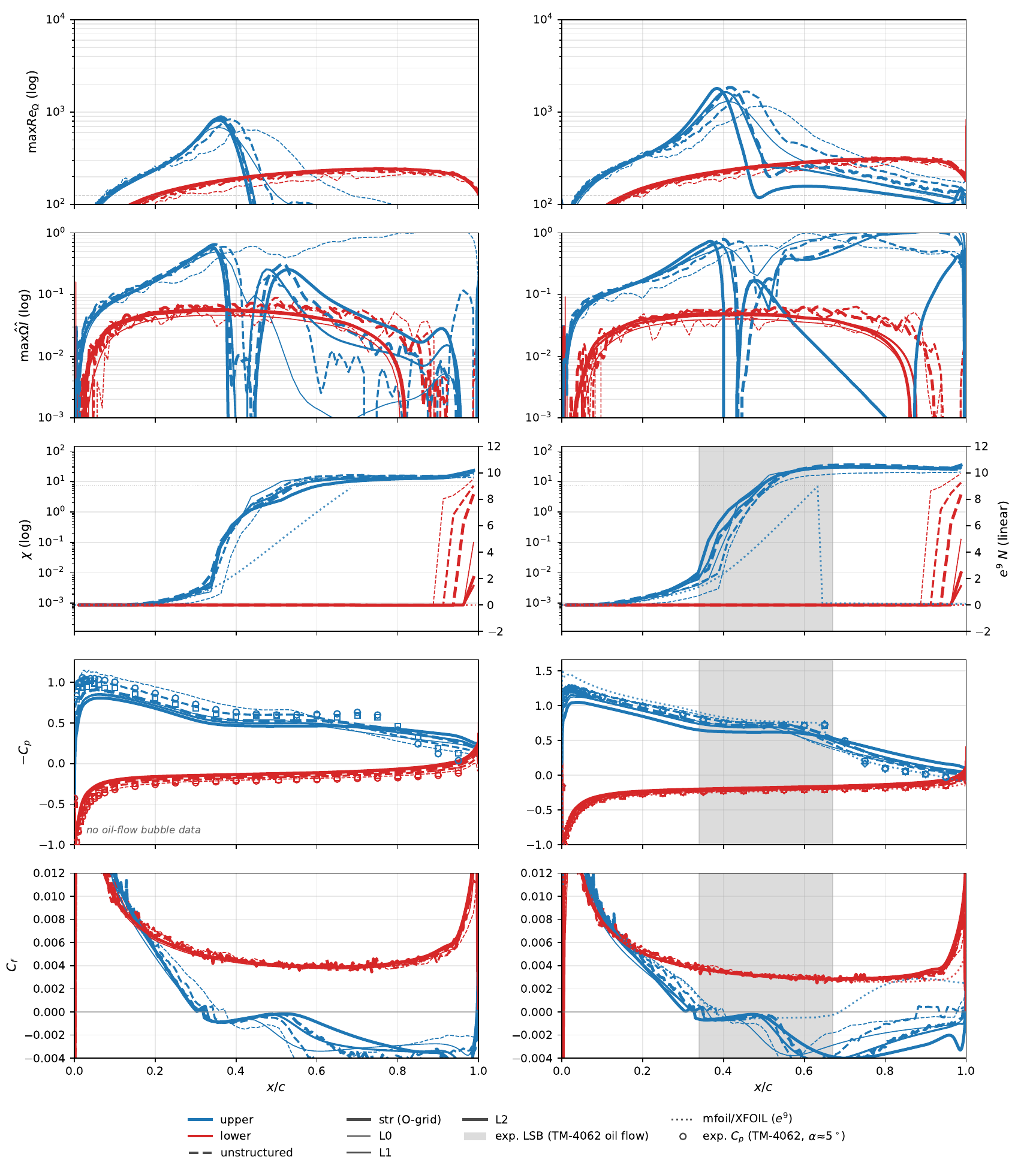}
\caption{Eppler~387 at $\alpha=5^\circ$: surface distributions at $Re=6\times10^4$ (left) and $10^5$ (right); rows as in Fig.~\ref{f:nlfcf0}.}\label{f:eppresweeplow}
\end{figure}

At $Re=10^5$ the measured flow reattaches at $0.67\,c$. The computed skin
friction returns to zero only within $0.05\,c$ of the trailing edge
(Fig.~\ref{f:eppresweeplow}). The L2 drag is $0.0410$ on the structured grid
and $0.0307$ on the unstructured grid, $73\%$ and $30\%$ above the measured
$0.0237$. Refinement increases the difference. At $Re=6\times10^4$ the
computed flow does not reattach. The measured flow at this Reynolds number is
found both with and without reattachment at the same
incidence~\cite{mcghee_1988}. The computed bubble therefore bursts
between $Re=10^5$ and $2\times10^5$ at this incidence. The measured bubble
reattaches at $10^5$ and is at its limit at $6\times10^4$.

\begin{table}[!htbp]
  \centering\footnotesize
  \setlength{\tabcolsep}{3.5pt}
  \caption{Eppler 387 at $\alpha\!=\!5^\circ$: section coefficients on the L2 grids, the $e^9$ reference, and the LTPT measurement at the tabulated incidence nearest $5^\circ$. The computed moment is about the quarter chord.}
  \label{t:eppresweep}
  \begin{tabular}{l ccc ccc ccc ccc}
    \toprule
    & \multicolumn{3}{c}{SA-AI, structured L2} & \multicolumn{3}{c}{SA-AI, unstructured L2}
    & \multicolumn{3}{c}{$e^9$ panel} & \multicolumn{3}{c}{Exp (LTPT)}\\
    \cmidrule(lr){2-4}\cmidrule(lr){5-7}\cmidrule(lr){8-10}\cmidrule(lr){11-13}
    $Re$ & $C_l$ & $C_d$ & $C_m$ & $C_l$ & $C_d$ & $C_m$ & $C_l$ & $C_d$ & $C_m$ & $C_l$ & $C_d$ & $C_m$ \\
    \midrule
    $6\!\times\!10^4$   & 0.636 & 0.0533 & $-$0.098 & 0.712 & 0.0506 & $-$0.103 & 0.869$^\ddagger$ & 0.0411$^\ddagger$ & $-$0.102$^\ddagger$ & 0.838 & 0.0439 & $-$0.114 \\
    $1\!\times\!10^5$   & 0.795 & 0.0410 & $-$0.102 & 0.877 & 0.0307 & $-$0.096 & 0.952 & 0.0213 & $-$0.087 & 0.873 & 0.0237 & $-$0.089 \\
    $2\!\times\!10^5$   & 0.925 & 0.0155 & $-$0.082 & 0.937 & 0.0148 & $-$0.083 & 0.960 & 0.0128 & $-$0.082 & 0.891 & 0.0138 & $-$0.081 \\
    $3\!\times\!10^5$   & 0.941 & 0.0107 & $-$0.080 & 0.950 & 0.0108 & $-$0.082 & 0.962 & 0.0103 & $-$0.082 & 0.901 & 0.0114 & $-$0.080 \\
    $4.6\!\times\!10^5$ & 0.950 & 0.0087 & $-$0.081 & 0.959 & 0.0089 & $-$0.083 & 0.966 & 0.0086 & $-$0.082 & 0.914 & 0.0093 & $-$0.081 \\
    \bottomrule
  \end{tabular}
  \\[2pt]{\footnotesize $^\ddagger$XFOIL. At this Reynolds number mfoil
  gives $C_l\!=\!0.889$, $C_d\!=\!0.0368$, $C_m\!=\!-0.097$, and
  Flexcompute Foil gives $C_l\!=\!0.868$, $C_d\!=\!0.0406$,
  $C_m\!=\!-0.102$.}
\end{table}

\begin{figure}[!htbp]\centering
\includegraphics[width=\textwidth]{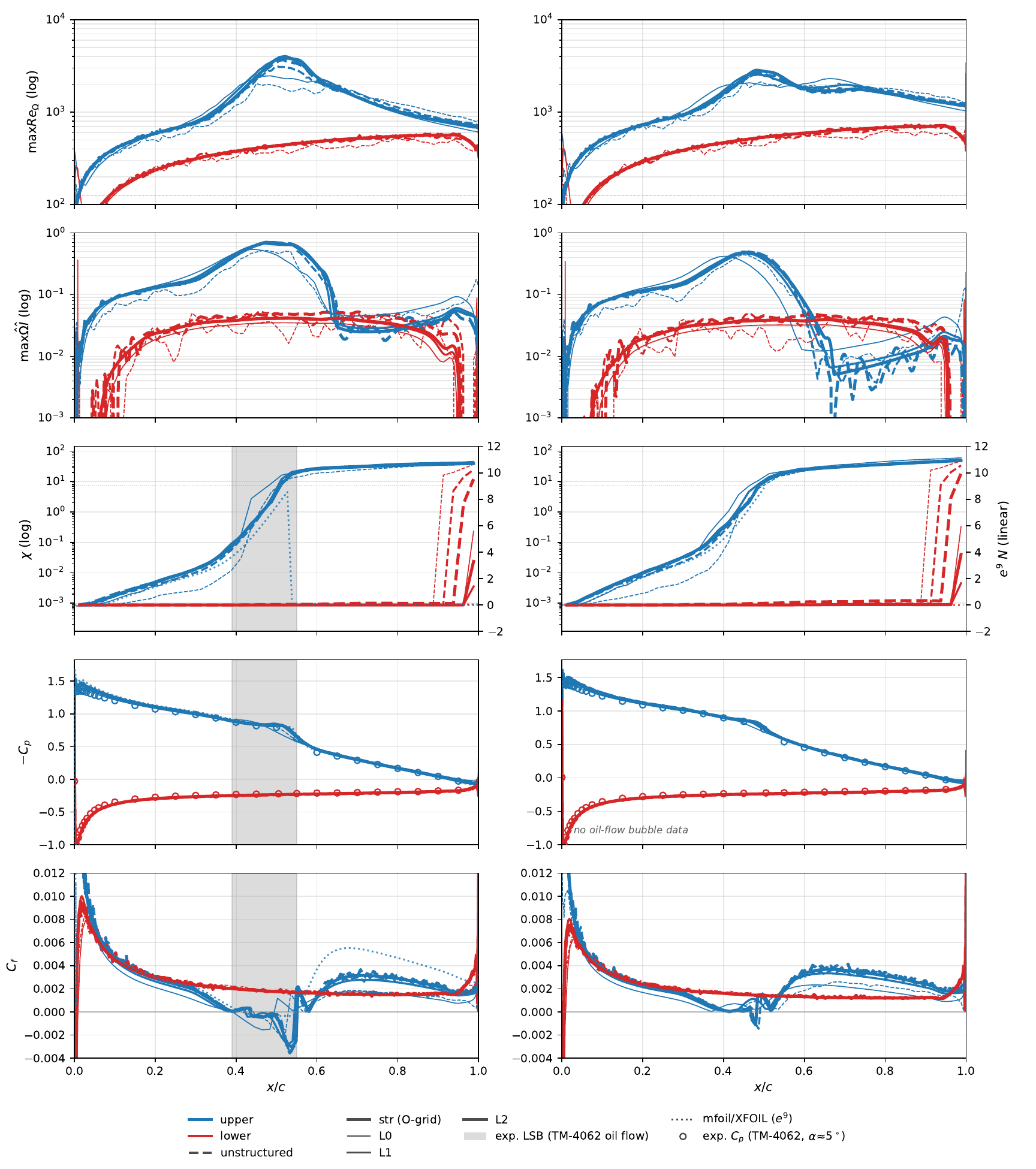}
\figcont{$Re=3\times10^5$ (left) and $4.6\times10^5$ (right).}
\end{figure}

Incidence sweeps at $Re=10^5$ and $3\times10^5$, on the L2 grids, show which
part of the computation is late. They are compared with the bubble stations measured by Cole and
Mueller~\cite{cole_mueller_1990}, with the transition locations of the $\gamma$--$Re_\theta$ and SA-BC
models computed by Shahjahan et al.~\cite{shahjahan_2024} at $10^5$, and with the three
transition models computed in OpenFOAM by Ghimire et al.~\cite{ghimire_2025}
at $3\times10^5$ (Tables~\ref{t:extepp300k} and~\ref{t:extepp100k}). At
$3\times10^5$ the computed transition location is at $0.579\,c$, $0.532\,c$ and $0.443\,c$
at $\alpha=2^\circ$, $4^\circ$ and $6^\circ$. The measured transition is at
$0.66\,c$, $0.55\,c$ and $0.46\,c$, and the three models of Ghimire et al.\
range over $0.592$--$0.646\,c$, $0.490$--$0.542\,c$ and $0.470$--$0.479\,c$.
The flow separates within $0.06\,c$ and reattaches within $0.07\,c$ of the
stations of Cole and Mueller. From $\alpha=-2^\circ$ to $6^\circ$ it
reattaches $0.03$ to $0.07\,c$ downstream of the transition location.

At $10^5$ the computed transition location is $0.07$--$0.12\,c$ ahead of the measured
transition at every incidence. The growth of $\chi$ is therefore not late. The
flow nevertheless reattaches late or not at all. At $\alpha=0^\circ$ it
reattaches at $0.978\,c$ on the structured grid, against a measured
$0.90\,c$, and at $6^\circ$ it does not reattach on either grid, against a
measured $0.68\,c$. Where it does reattach, the reattachment station is
$0.17$ to $0.39\,c$ downstream of the transition location. The drag coefficient reaches
$0.0499$ at $6^\circ$. The distance from the transition location to reattachment is
thus the quantity that is too long at low Reynolds number. Downstream of the
transition location the production is in the handover of Section~II.E, and the mean flow
responds through the eddy viscosity of Section~II.F.

\begin{figure}[!htbp]\centering
\includegraphics[width=0.82\textwidth]{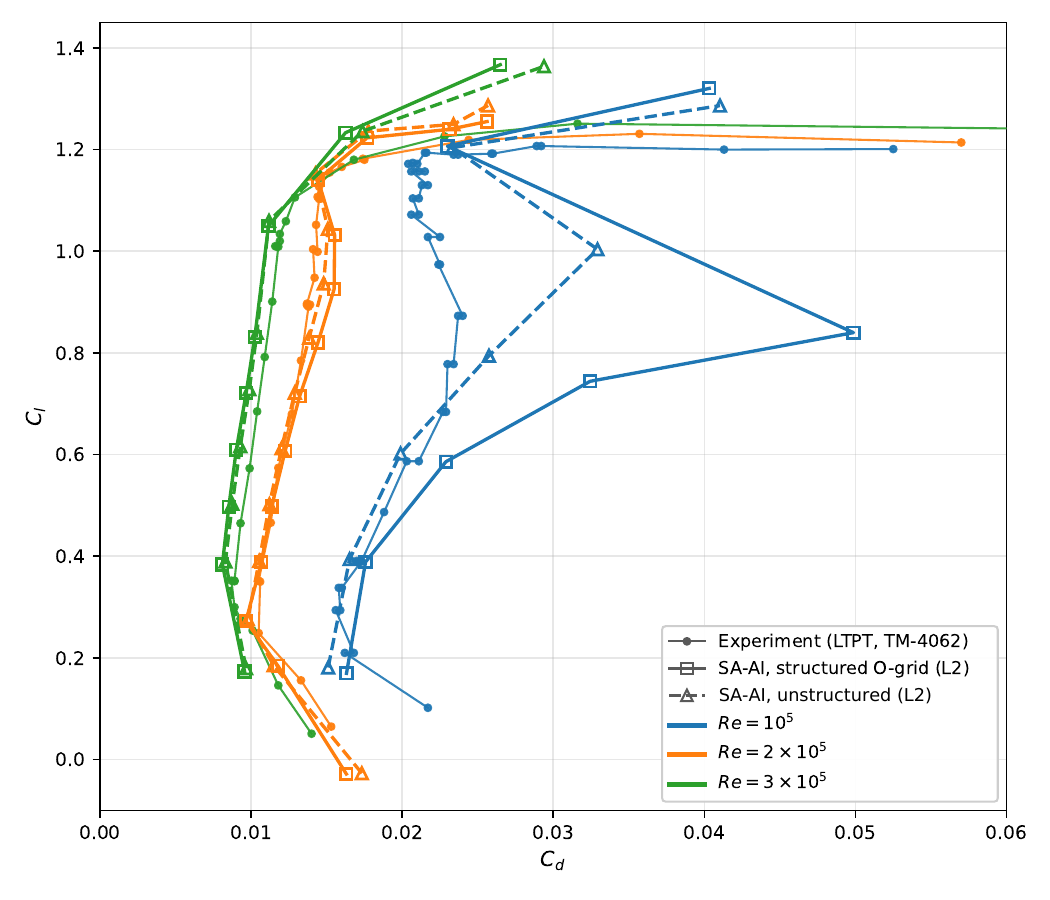}
\caption{Eppler 387 drag polar at $Re=10^5$, $2\times10^5$ and $3\times10^5$, on the L2 grids, with the measurement.}\label{f:epprepolar}
\end{figure}

\begin{figure}[!htbp]\centering
\includegraphics[width=\textwidth]{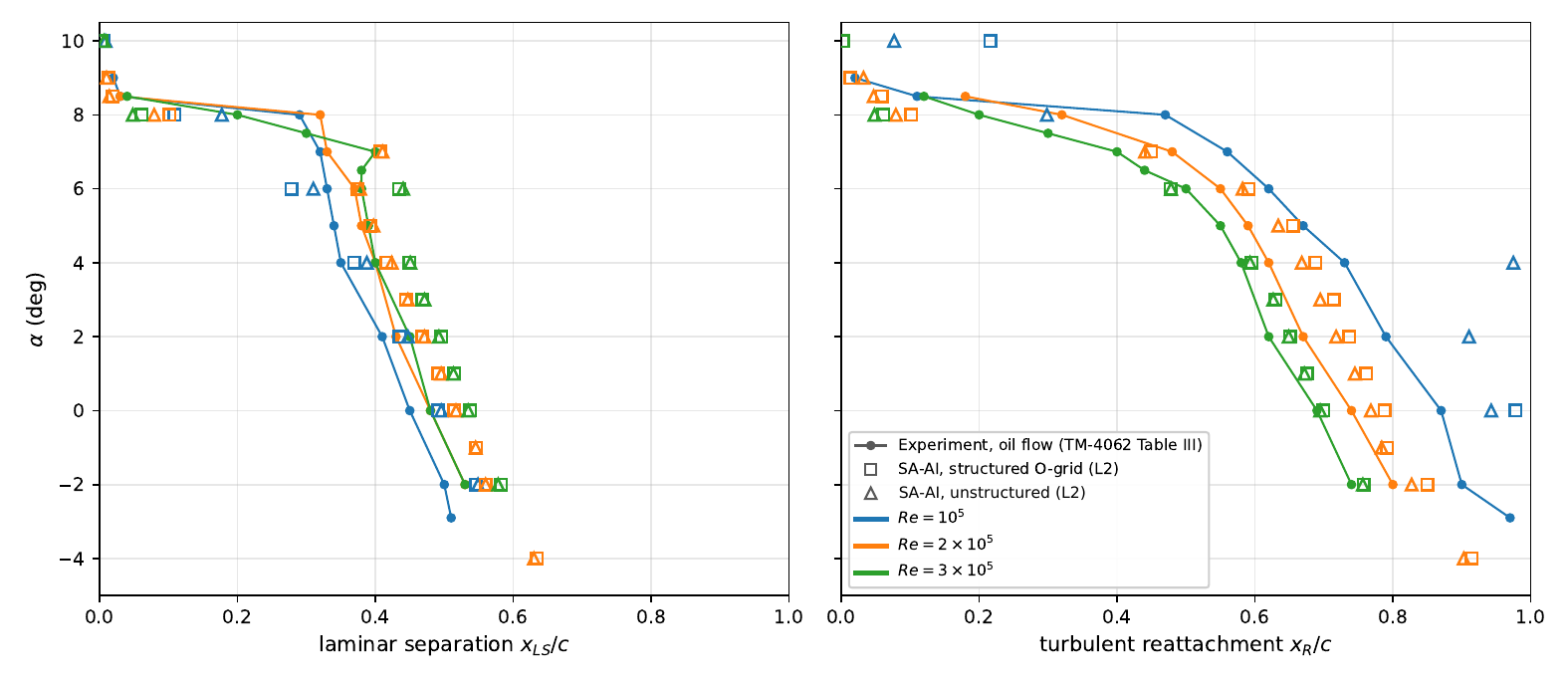}
\caption{Eppler 387 stations of separation and reattachment at $Re=10^5$, $2\times10^5$ and $3\times10^5$, on the L2 grids, with the oil flow. Where no bubble forms, both stations are the transition location, and the point is at the same $x/c$ in both panels.}\label{f:epprebubble}
\end{figure}

Figures~\ref{f:epprepolar} and~\ref{f:epprebubble} show the polars and the
bubble stations at $10^5$, $2\times10^5$ and $3\times10^5$ against the
measurement. The Supplementary Figures contain the surface distributions and
the near-wall contours of both sweeps
(Figs.~\ref{f:eppresweepsheets}--\ref{f:eppsheetRe300k}).

\begin{table}[!htbp]
  \centering\small
  \caption{Eppler 387 at $Re=3\times10^5$, on the L2 grids. $x_\mathrm{tr}$ is the station where the largest $\chi$ near the wall crosses one. $x_{LS}$ and $x_R$ are zero crossings of the signed $C_f$. A dash means that no bubble closes ahead of the trailing edge.}
  \label{t:extepp300k}
  \begin{tabular}{c ccccc ccccc}
    \toprule
    & \multicolumn{5}{c}{structured} & \multicolumn{5}{c}{unstructured} \\
    \cmidrule(lr){2-6}\cmidrule(lr){7-11}
    $\alpha$ & $C_l$ & $C_d$ & $x_\mathrm{tr}$ & $x_{LS}$ & $x_R$
             & $C_l$ & $C_d$ & $x_\mathrm{tr}$ & $x_{LS}$ & $x_R$ \\
    \midrule
    $-2$ & 0.1733 & 0.00957 & 0.698 & 0.582 & 0.758 & 0.1797 & 0.00966 & 0.699 & 0.578 & 0.756 \\
    $0$ & 0.3840 & 0.00810 & 0.632 & 0.537 & 0.699 & 0.3896 & 0.00830 & 0.629 & 0.535 & 0.695 \\
    $1$ & 0.4974 & 0.00852 & 0.607 & 0.514 & 0.675 & 0.5035 & 0.00876 & 0.602 & 0.513 & 0.672 \\
    $2$ & 0.6098 & 0.00902 & 0.579 & 0.496 & 0.651 & 0.6164 & 0.00929 & 0.579 & 0.492 & 0.649 \\
    $3$ & 0.7212 & 0.00967 & 0.561 & 0.468 & 0.630 & 0.7284 & 0.00988 & 0.558 & 0.471 & 0.627 \\
    $4$ & 0.8315 & 0.01024 & 0.532 & 0.449 & 0.596 & 0.8394 & 0.01039 & 0.531 & 0.451 & 0.593 \\
    $6$ & 1.0496 & 0.01115 & 0.443 & 0.435 & 0.477 & 1.0596 & 0.01118 & 0.444 & 0.440 & 0.478 \\
    $8$ & 1.2333 & 0.01625 & 0.070 & -- & -- & 1.2359 & 0.01737 & 0.055 & -- & -- \\
    $10$ & 1.3670 & 0.02649 & 0.005 & -- & -- & 1.3638 & 0.02938 & 0.003 & 0.006 & 0.987 \\
    \bottomrule
  \end{tabular}
\end{table}

\begin{table}[!htbp]
  \centering\small
  \caption{Eppler 387 at $Re=10^5$, on the L2 grids; columns as in Table~\ref{t:extepp300k}. The structured solution at $\alpha=8^\circ$ is unsteady: its transition location varies between $0.24\,c$ and $0.57\,c$ over $8\times10^4$ iterations, and the last value is tabulated.}
  \label{t:extepp100k}
  \begin{tabular}{c ccccc ccccc}
    \toprule
    & \multicolumn{5}{c}{structured} & \multicolumn{5}{c}{unstructured} \\
    \cmidrule(lr){2-6}\cmidrule(lr){7-11}
    $\alpha$ & $C_l$ & $C_d$ & $x_\mathrm{tr}$ & $x_{LS}$ & $x_R$
             & $C_l$ & $C_d$ & $x_\mathrm{tr}$ & $x_{LS}$ & $x_R$ \\
    \midrule
    $-2$ & 0.1698 & 0.01631 & 0.830 & 0.546 & -- & 0.1813 & 0.01510 & 0.821 & 0.549 & 0.987 \\
    $0$ & 0.3888 & 0.01758 & 0.750 & 0.491 & 0.978 & 0.3944 & 0.01652 & 0.731 & 0.496 & 0.943 \\
    $2$ & 0.5860 & 0.02289 & 0.677 & 0.435 & -- & 0.6021 & 0.01989 & 0.661 & 0.447 & 0.910 \\
    $4$ & 0.7442 & 0.03244 & 0.590 & 0.369 & -- & 0.7944 & 0.02574 & 0.587 & 0.388 & 0.975 \\
    $6$ & 0.8399 & 0.04986 & 0.482 & 0.278 & -- & 1.0039 & 0.03293 & 0.523 & 0.310 & -- \\
    $8$ & 1.2067 & 0.02299 & 0.571 & 0.109 & -- & 1.2039 & 0.02332 & 0.279 & 0.177 & 0.990 \\
    $10$ & 1.3204 & 0.04032 & 0.018 & 0.006 & 0.216 & 1.2865 & 0.04103 & 0.025 & 0.009 & 0.077 \\
    \bottomrule
  \end{tabular}
\end{table}

\FloatBarrier
\subsection{A Two-Element Section}

In every case so far, a single boundary layer develops on each surface, and
$\chi$ in it grows from the freestream value. On a multi-element section the wake of one
element passes over the next element. This case shows what the model computes
then. The section was designed for the purpose, and there is no measurement
of it. Its main element is laminar on both surfaces over a narrow range of
incidence. A laminar separation bubble forms on the upper surface of its
flap, a NACA~9416. Changing the incidence changes which surface of the main element is
turbulent, and with it the $\chi$ of the fluid that passes over the flap.
Table~\ref{t:twoelementgeom} defines the geometry. The Reynolds number is
$10^6$ on the total chord, and $\chi_\infty=c_{v1}e^{-9}$. The incidence is
swept from $\alpha=-9^\circ$ to $+13^\circ$ on the L0 and L1 grids and to
$+11^\circ$ on the L2 grid. The solutions are converged to a momentum residual
of $10^{-9}$, except the L2 solution at $-9^\circ$.

\begin{table}[!htbp]
  \centering\footnotesize
  \setlength{\tabcolsep}{4pt}
  \caption{Two-element section, in units of the total chord. Main element: mean line $z_c=x_e[\hat x(1-\hat x)\sum_{i=0}^{10}A^{c}_iB^{10}_i(\hat x)]+\hat x\,\Delta z_{te}$ and half-thickness $t=x_e[\hat x^{1/2}(1-\hat x)\sum_{i=0}^{9}A^{t}_iB^{9}_i(\hat x)+\hat x\,t_{te}/(2x_e)]$, with $\hat x=x/x_e$ and $B^{n}_i$ the Bernstein polynomials. The surfaces are offset by $\pm t$ normal to the mean line, and the leading edge is at the origin.}
  \label{t:twoelementgeom}
  \begin{tabular}{ll}
    \toprule
    \multicolumn{2}{l}{\emph{Main element}}\\
    $x_e$ & 0.7039 \\
    $A^{c}_0\ldots A^{c}_{10}$ &
      $-0.05320$, $-0.16218$, $+0.17632$, $-0.65937$, $+0.84104$, $-1.32895$, \\
    & $+1.13471$, $-1.25791$, $+0.67079$, $-0.83361$, $-0.25106$ \\
    $\Delta z_{te}$ & $+0.08272$ \\
    $A^{t}_0\ldots A^{t}_{9}$ &
      $+0.10668$, $+0.07402$, $+0.29533$, $-0.36313$, $+0.99921$, $-1.10675$, \\
    & $+1.44742$, $-0.91005$, $+0.83270$, $+0.08962$ \\
    \midrule
    \multicolumn{2}{l}{\emph{Flap}}\\
    section & NACA 9416 \\
    chord & 0.30 \\
    incidence & $-8.0^\circ$ \\
    leading edge & $(0.70,\,0.04)$ \\
    \midrule
    \multicolumn{2}{l}{\emph{Both}}\\
    trailing-edge base $t_{te}$ & 0.003 \\
    \bottomrule
  \end{tabular}
\end{table}

\begin{figure}[!htbp]\centering
\includegraphics[width=\textwidth,height=0.90\textheight,keepaspectratio]{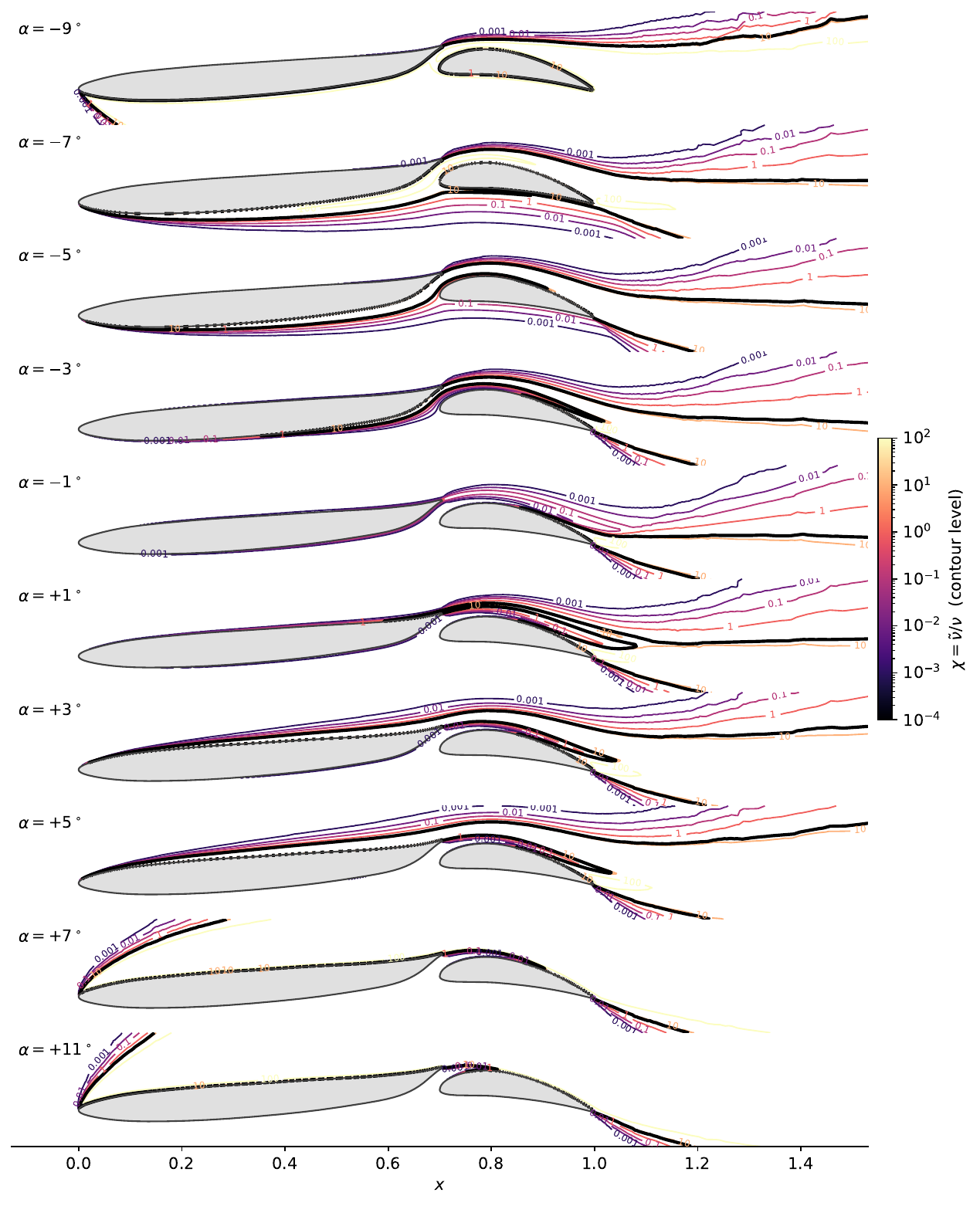}
\caption{Two-element section, L2 grid: contours of $\chi$ at each incidence. The levels are decades from $10^{-3}$ to $10^{2}$; heavy black is $\chi=c_{v1}$.}\label{f:twoelemchiL2}
\end{figure}

\begin{figure}[!htbp]\centering
\includegraphics[width=0.58\textwidth]{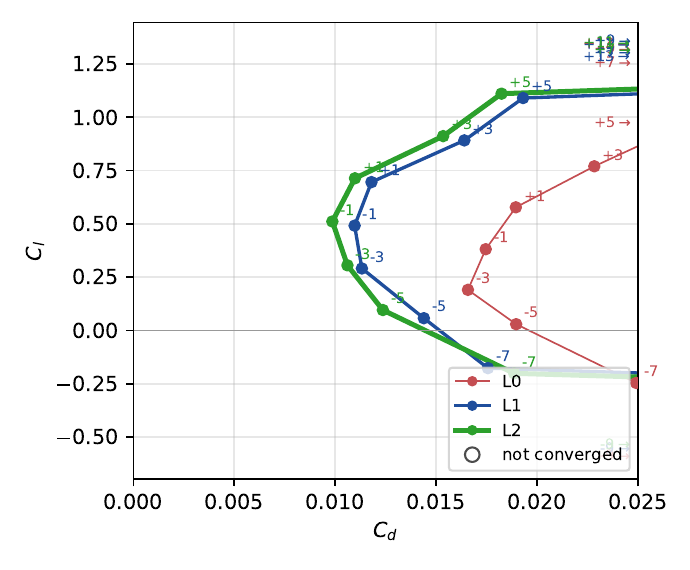}
\caption{Two-element section: polar on the three grid levels, labeled by incidence. Points with $C_d>0.025$ are marked at the right edge. Open marker: the L2 solution at $\alpha=-9^\circ$, which is not converged to a momentum residual of $10^{-9}$.}
\label{f:twoelempolar}
\end{figure}

Figure~\ref{f:twoelemchiL2} shows contours of $\chi$ at each incidence on the
L2 grid. At $\alpha=-1^\circ$ both surfaces of the main element are laminar:
the heavy contour $\chi=c_{v1}$ does not touch the main element. At
$+3^\circ$ and $+5^\circ$ the upper surface of the main element is turbulent,
and its wake passes above the flap. At $-3^\circ$ and below, the lower
surface of the main element is turbulent, and its fluid passes through the
slot onto the upper surface of the flap. On the L1 and L2 grids the drag is
lowest at $\alpha=-1^\circ$ (Fig.~\ref{f:twoelempolar}). At $-9^\circ$ and at
$+7^\circ$ and above, the section is stalled.

\begin{figure}[!htbp]\centering
\includegraphics[width=\textwidth]{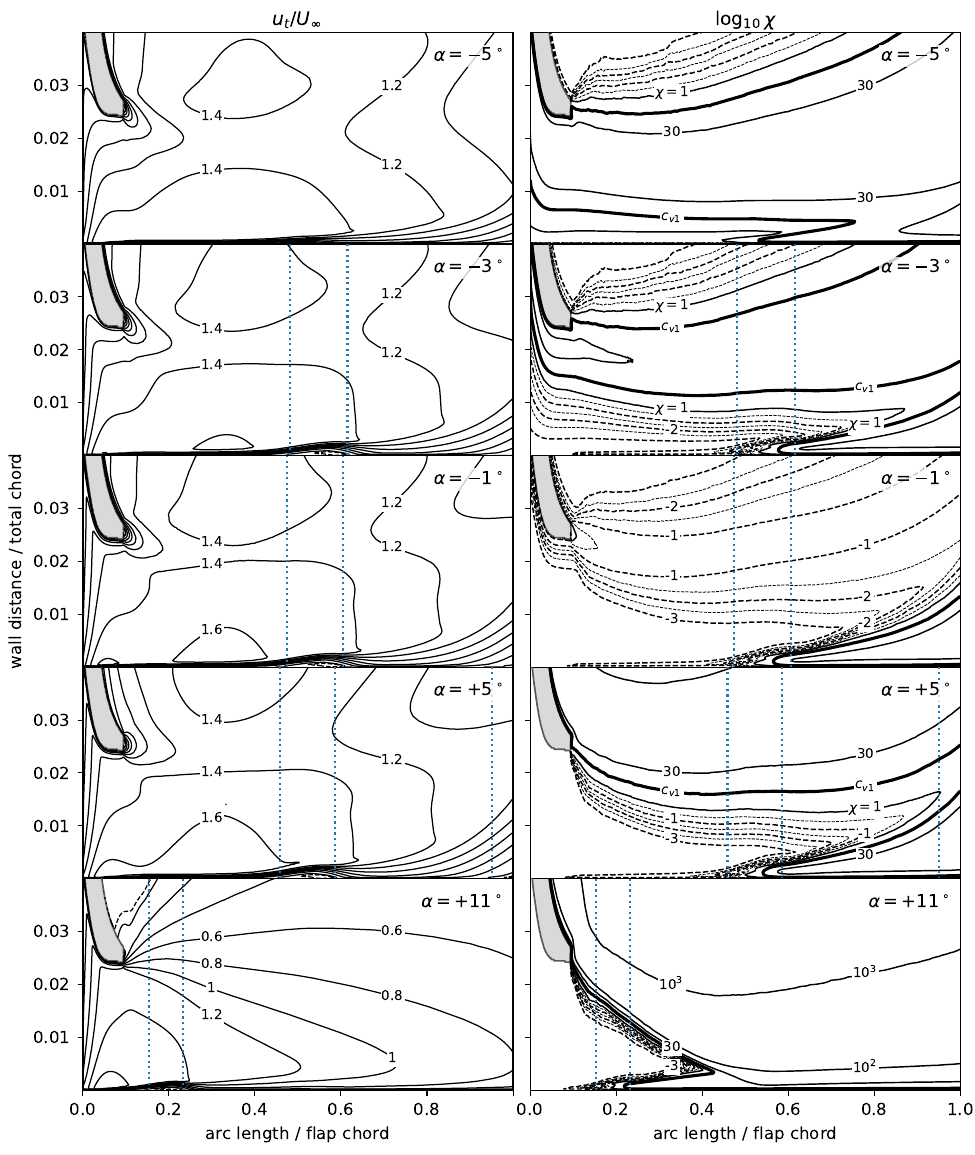}
\caption{Upper surface of the flap, L2 grid, unrolled: arc length from the leading edge of the flap against wall distance, one row per incidence. Left: tangential velocity $u_t/U_\infty$. Right: $\log_{10}\chi$, dashed below $\chi=1$, with $\chi=c_{v1}$ heavy. Gray: the trailing edge of the main element, which the wall normals of the flap meet over the first $0.09$ of the arc length. Dotted vertical lines: separation and reattachment.}
\label{f:twoelemflapchi}
\end{figure}

Figure~\ref{f:twoelemflapchi} shows the flow next to the upper surface of the
flap, with the surface unrolled. The dotted vertical lines mark separation
and reattachment. At $\alpha=-3^\circ$, $-1^\circ$ and $+5^\circ$, $\chi$ in
the boundary layer of the flap ahead of the bubble is between the freestream
value, $9\times10^{-4}$, and $0.004$. At $-3^\circ$ and $+5^\circ$ the main
element sheds fluid with $\chi>c_{v1}$, and that fluid passes above the
boundary layer of the flap, at a wall distance of $0.012$ of the total chord or
more. The flow separates at $0.46$ to $0.48$ of the flap chord and reattaches
at $0.59$ to $0.62$. At $-5^\circ$ the turbulent boundary layer of the lower
surface of the main element passes through the slot, next to the flap. There
$\chi$ exceeds $c_{v1}$ at a wall distance of $0.005$ to $0.006$, and inside the boundary
layer of the flap it is between $0.5$ and $0.8$, more than five hundred times
the freestream value. From that level $\chi$ reaches one in less than one
e-fold of growth, where the other incidences require seven. The smallest
skin-friction coefficient is $2\times10^{-6}$, at $0.51$ of the flap chord: the
boundary layer reaches the point of separation, and the flow at the wall does
not reverse.

Section~IV.E explains why the $\tilde\nu$ of a wake reaches the flap, and
what the same mechanism implies for any surface in the wake of another.
The Supplementary Figures contain the $\chi$ contours on all three grid
levels and the surface distributions (Figs.~\ref{f:twoelemchiL1}
and~\ref{f:twoelemsuite1}).

\FloatBarrier
\subsection{The Cylinder Drag Crisis}

Near $Re_D=3\times10^5$ the drag coefficient of a circular cylinder decreases
to a fraction of its value: the drag crisis. Below the crisis the boundary
layer separates while it is laminar, ahead of the widest point of the
cylinder, and the wake is wide. Above it the boundary layer is turbulent
before its final separation, which is farther aft, and the wake is narrow.
The case shows the model moving transition from the separated shear layer
onto the wall as the Reynolds number increases, with the constants and the
equations of the airfoil cases.

The computations are steady and quasi-two-dimensional. They cover twenty-five
Reynolds numbers from $Re_D=1$ to $10^{10}$ with one freestream value,
$\chi_\infty=1.1\times10^{-2}$, which is $Tu=0.2\%$ in Eq.~\eqref{w:mack}.
Each computation starts from the converged solution at the next lower
Reynolds number. Four O-grids with $y^+\lesssim1$ cover the range. They are
identified by the thickness of their first layer, $y_1/R$: $2\times10^{-3}$
($600\times121$ points) for $Re_D\le10^3$, $1.6\times10^{-5}$
($1200\times148$) up to $1.78\times10^6$, $2\times10^{-6}$ ($1600\times170$)
up to $10^7$, and $2\times10^{-9}$ ($1600\times170$) above. At a change of
grid the solution is interpolated onto the new grid. A computation is
accepted as steady when $C_d$ changes by less than $10^{-3}$ over $1000$
iterations. Five of the twenty-five do not meet this criterion, and
Table~\ref{t:dragcrisis} gives the range of their $C_d$ in brackets.

\begin{figure}[!htbp]\centering
\includegraphics[width=\textwidth,height=0.94\textheight,keepaspectratio]{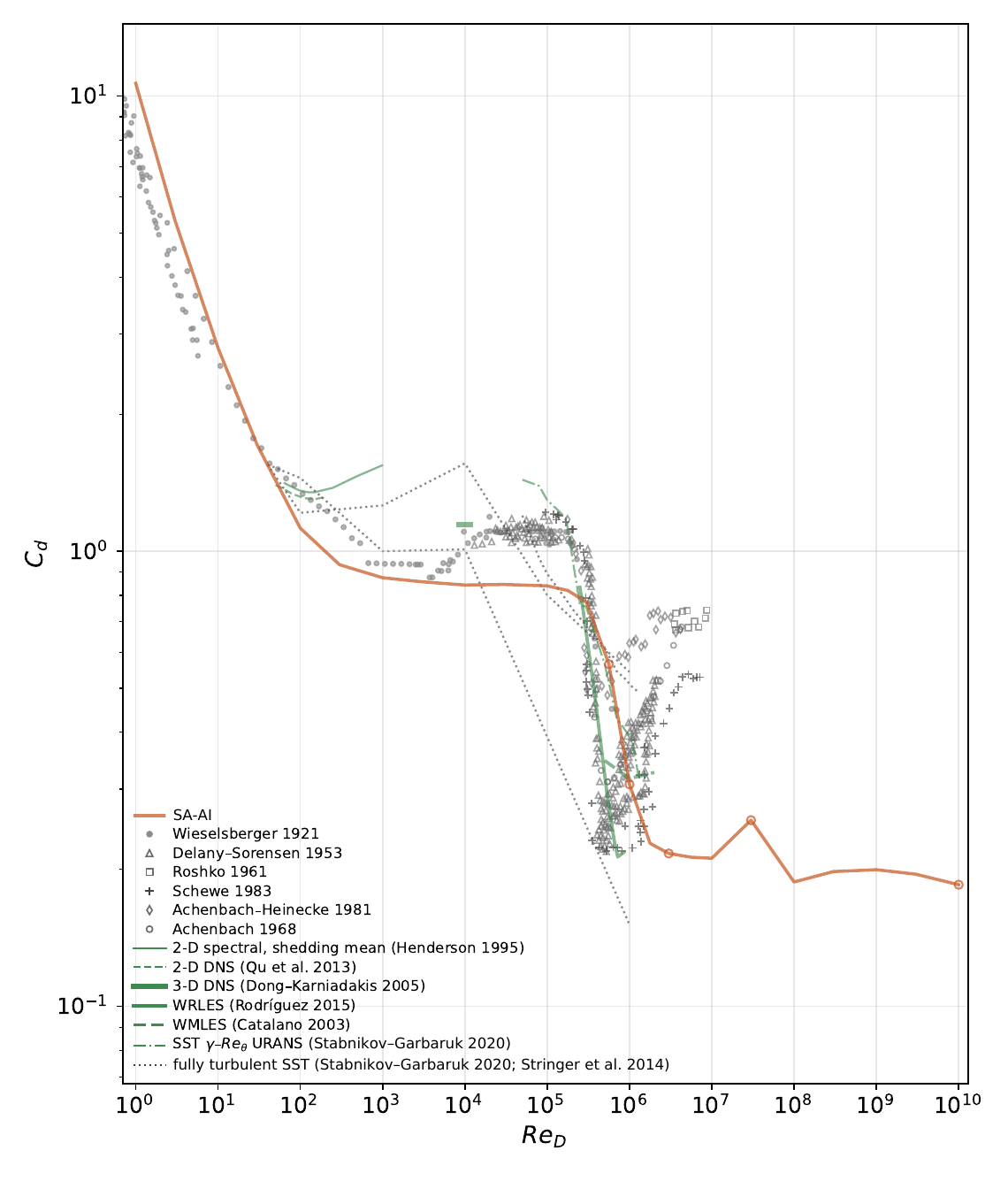}
\caption{Circular cylinder: drag coefficient against $Re_D$, with $\chi_\infty=1.1\times10^{-2}$ ($Tu=0.2\%$). Orange: SA-AI, with open circles where the computation does not meet the criterion for a steady solution. Symbols: the measurements of Wieselsberger~\cite{wieselsberger_1921}, Delany and Sorensen~\cite{delany_sorensen_1953}, Roshko~\cite{roshko_1961}, Achenbach~\cite{achenbach_1968}, Achenbach and Heinecke~\cite{achenbach_heinecke_1981} and Schewe~\cite{schewe_1983}. Green lines: unsteady computations~\cite{henderson_1995,qu_2013,dong_karniadakis_2005,rodriguez_2015,catalano_2003,stabnikov_garbaruk_2020}. Dotted lines: fully turbulent computations with the SST model~\cite{stabnikov_garbaruk_2020,stringer_2014}.}\label{f:dragcrisiscd}
\end{figure}

\begin{figure}[!htbp]\centering
\includegraphics[width=\textwidth]{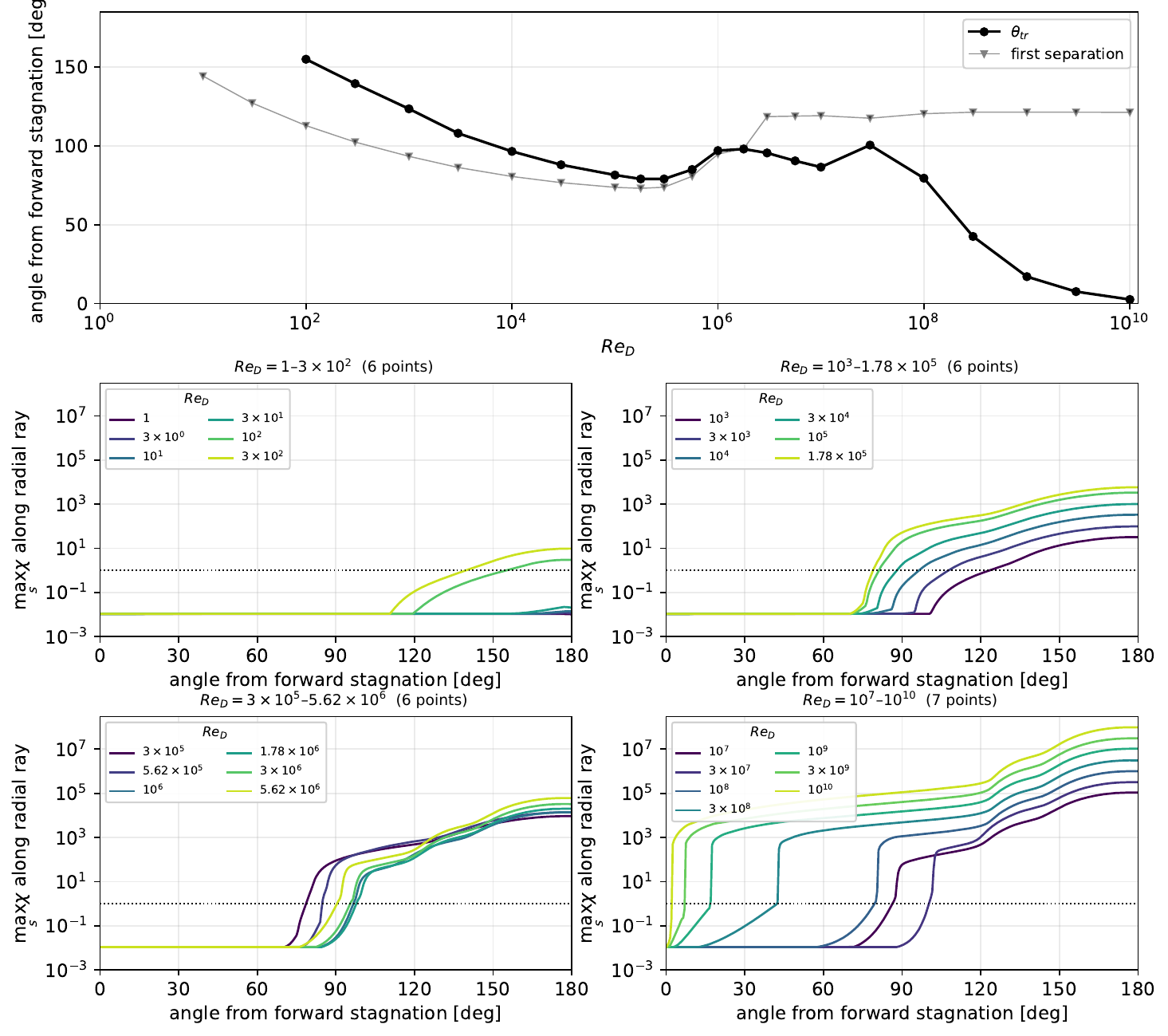}
\caption{Circular cylinder. Top: transition angle $\theta_{tr}$, the smallest angle at which $\chi$ reaches one on a radial ray, and angle of the first separation, against $Re_D$. Lower panels: the largest $\chi$ on the radial ray against the angle from the forward stagnation point, at every Reynolds number.}\label{f:dragcrisisangles}
\end{figure}

Every solution is symmetric, with $|C_l|<2\times10^{-6}$. A steady symmetric
solution contains no vortex shedding. Between the onset of shedding at
$Re_D\approx46$ and the crisis, the computed drag is lower than the measured
drag: the computed $C_d$ is $0.84$ between $Re_D=10^4$ and $10^5$, and the
measured $C_d$ is between $1.0$ and $1.2$ (Fig.~\ref{f:dragcrisiscd}). Below
the onset of shedding the steady solution is the physical flow. The computed
$C_d$ is $10.67$ at $Re_D=1$ and $2.797$ at $Re_D=10$, which is $1.7\%$ below
the value of Dennis and Chang~\cite{dennis_chang_1970}. The largest $\chi$
remains near $\chi_\infty$ up to $Re_D=30$ (Fig.~\ref{f:dragcrisisangles}).

\begin{figure}[!htbp]\centering
\includegraphics[width=\textwidth,height=0.92\textheight,keepaspectratio]{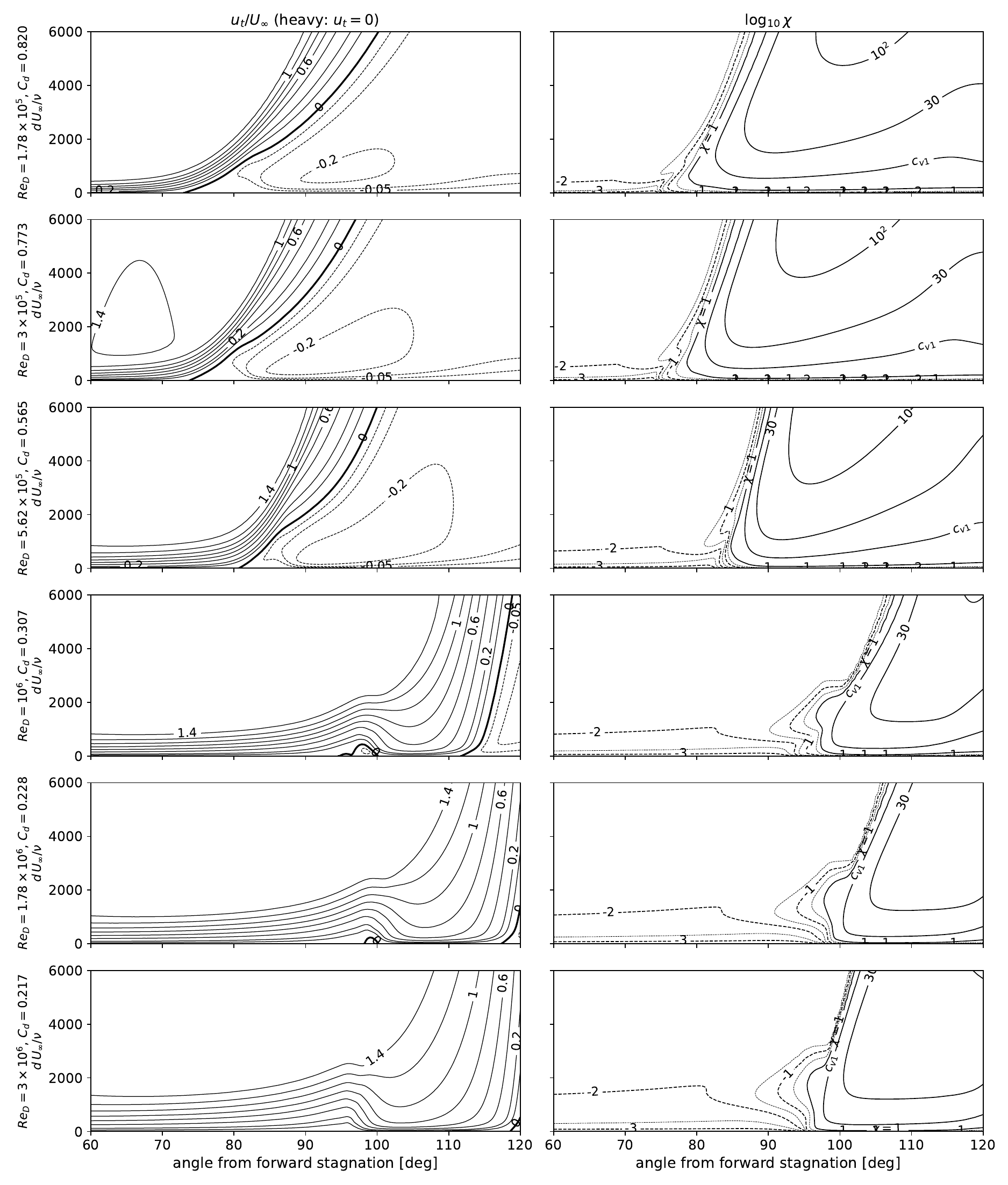}
\caption{Circular cylinder near the wall, $Re_D=1.78\times10^5$ to $3\times10^6$ (rows): tangential velocity $u_t/U_\infty$ (left, $u_t=0$ heavy) and $\log_{10}\chi$ (right), against the angle from the forward stagnation point and the wall distance $d\,U_\infty/\nu$.}\label{f:dragcrisisblsheet}
\end{figure}

The computed crisis is between $Re_D=3\times10^5$, where $C_d=0.773$, and
$1.78\times10^6$, where $C_d=0.228$. Table~\ref{t:dragcrisis} lists, for each
Reynolds number, the angle of the first separation and the transition angle
$\theta_{tr}$, both measured from the forward stagnation point. The
transition angle is the smallest angle at which $\chi$ reaches one on a
radial ray from the wall. From $Re_D=10^5$ to $3\times10^5$ the boundary layer
separates at $73^\circ$--$74^\circ$ while laminar, and $\chi$ reaches one
$5^\circ$ to $8^\circ$ farther, in the separated shear layer. The flow does not
reattach. At $10^6$ and $1.78\times10^6$ the first separation is at
$95^\circ$--$98^\circ$, and it is the start of a separation bubble: the flow
reattaches as a turbulent boundary layer and separates again near
$120^\circ$. From $3\times10^6$ the tabulated first separation is the
turbulent one, at $118^\circ$--$121^\circ$. Figure~\ref{f:dragcrisisblsheet}
shows this sequence in the velocity and $\chi$ near the wall.

Above the crisis the computed $C_d$ is $0.21$--$0.23$ from $1.78\times10^6$
to $10^7$ and $0.19$--$0.26$ from $3\times10^7$ to $10^{10}$. The measured
$C_d$ increases again, to about $0.7$ near $Re_D=10^7$~\cite{roshko_1961}. The
computed $C_d$ does not. The computed
transition angle is $98^\circ$ at $1.78\times10^6$ and $2.5^\circ$ at
$10^{10}$, while the
turbulent separation remains at $118^\circ$--$121^\circ$ and the drag remains
at its level. In this computation the drag above the crisis is therefore a
function of the turbulent separation, which SA determines, and not of the
transition location. The Supplementary Figures contain the velocity and
$\chi$ fields at every Reynolds number (Fig.~\ref{f:dragcrisisfields}).

\begin{table}[!htbp]
  \centering\small
  \caption{Circular cylinder, $\chi_\infty=1.1\times10^{-2}$: drag coefficient, transition angle $\theta_{tr}$, and angle of the first separation $\theta_{sep}$. $C_d$ is the median over the last iterations. Brackets: the range of $C_d$ where the criterion for a steady solution is not met. $|C_l|<2\times10^{-6}$ in every computation.}
  \label{t:dragcrisis}
  \begin{tabular}{l l c cc}
    \toprule
    $Re_D$ & $y_1/R$ & $C_d$ & $\theta_{tr}\,(^\circ)$ & $\theta_{sep}\,(^\circ)$ \\
    \midrule
    $1$ & $2\times10^{-3}$ & 10.665 & -- & -- \\
    $3$ & $2\times10^{-3}$ & 5.330 & -- & -- \\
    $10$ & $2\times10^{-3}$ & 2.797 & -- & 144.3 \\
    $30$ & $2\times10^{-3}$ & 1.710 & -- & 127.2 \\
    $100$ & $2\times10^{-3}$ & 1.123 & 155.0 & 112.9 \\
    $300$ & $2\times10^{-3}$ & 0.934 & 139.5 & 102.4 \\
    $1000$ & $2\times10^{-3}$ & 0.874 & 123.5 & 93.4 \\
    $3000$ & $1.6\times10^{-5}$ & 0.857 & 108.0 & 86.3 \\
    $10^{4}$ & $1.6\times10^{-5}$ & 0.843 & 96.5 & 80.6 \\
    $3\!\times\!10^{4}$ & $1.6\times10^{-5}$ & 0.845 & 88.0 & 76.7 \\
    $10^{5}$ & $1.6\times10^{-5}$ & 0.839 & 81.5 & 73.7 \\
    $1.78\!\times\!10^{5}$ & $1.6\times10^{-5}$ & 0.820 & 79.0 & 73.0 \\
    $3\!\times\!10^{5}$ & $1.6\times10^{-5}$ & 0.773 & 79.0 & 73.8 \\
    $5.62\!\times\!10^{5}$ & $1.6\times10^{-5}$ & 0.565\,[0.075] & 85.0 & 80.7 \\
    $10^{6}$ & $1.6\times10^{-5}$ & 0.307\,[0.044] & 97.0 & 94.8 \\
    $1.78\!\times\!10^{6}$ & $1.6\times10^{-5}$ & 0.228 & 98.0 & 98.3 \\
    $3\!\times\!10^{6}$ & $2\times10^{-6}$ & 0.217\,[0.010] & 95.5 & 118.6 \\
    $5.62\!\times\!10^{6}$ & $2\times10^{-6}$ & 0.213 & 90.5 & 118.9 \\
    $10^{7}$ & $2\times10^{-6}$ & 0.212 & 86.5 & 119.2 \\
    $3\!\times\!10^{7}$ & $2\times10^{-9}$ & 0.256\,[0.171] & 100.5 & 117.6 \\
    $10^{8}$ & $2\times10^{-9}$ & 0.188 & 79.5 & 120.4 \\
    $3\!\times\!10^{8}$ & $2\times10^{-9}$ & 0.198 & 42.5 & 121.4 \\
    $10^{9}$ & $2\times10^{-9}$ & 0.200 & 17.0 & 121.4 \\
    $3\!\times\!10^{9}$ & $2\times10^{-9}$ & 0.195 & 7.5 & 121.3 \\
    $10^{10}$ & $2\times10^{-9}$ & 0.185\,[0.009] & 2.5 & 121.3 \\
    \bottomrule
  \end{tabular}
\end{table}

\FloatBarrier
\subsection{The Daedalus Wing}

The wing of the Daedalus human-powered aircraft~\cite{drela_daedalus_1988} is
the first three-dimensional case. A laminar separation bubble forms on the upper
surface of each section of the wing, as on the Eppler 387. The wing adds the
variation along the span: the chord Reynolds number decreases over the
tapered outer part, and the flow at the tip is three-dimensional.

\begin{figure}[!htbp]\centering
\includegraphics[width=0.99\textwidth]{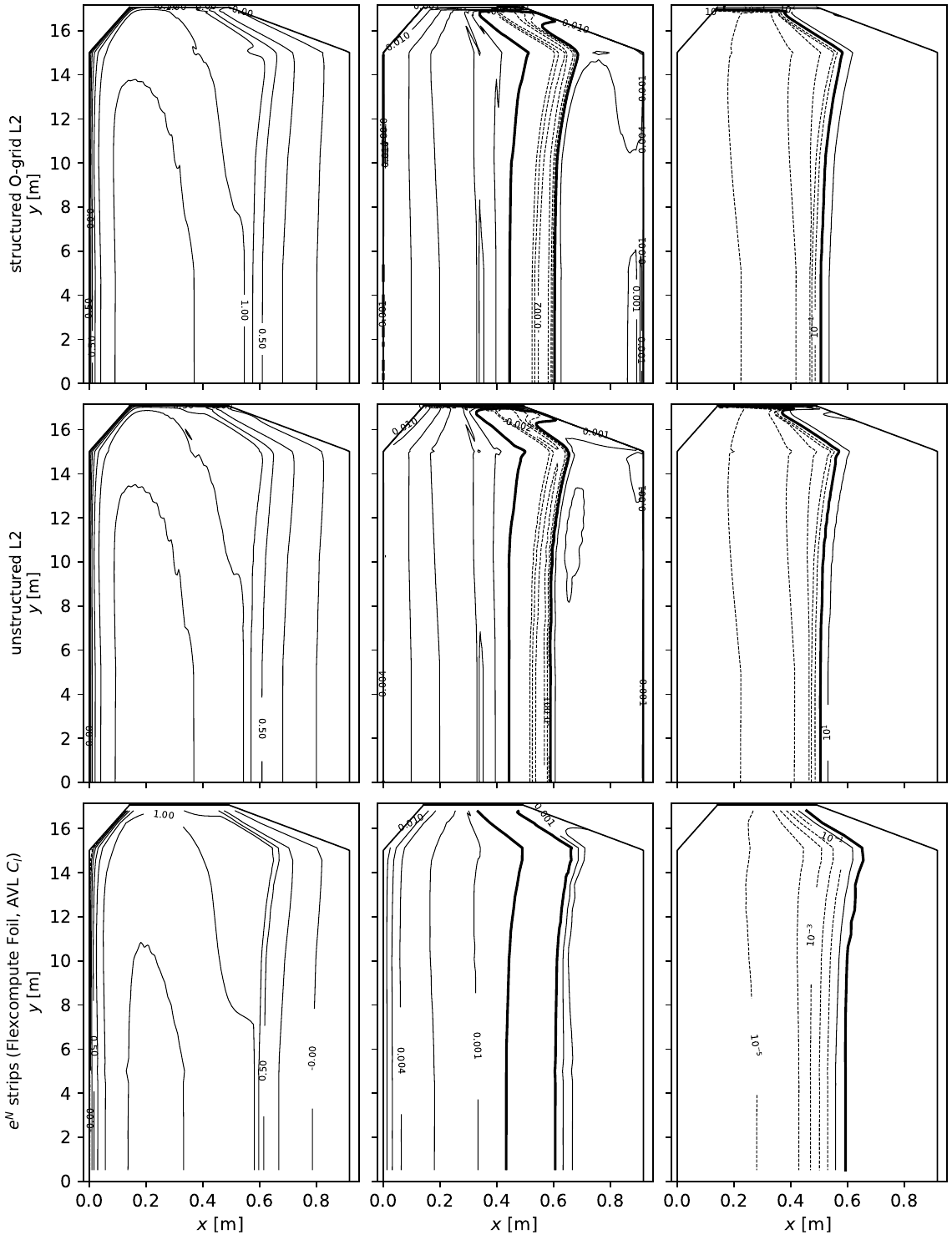}
\caption{Daedalus wing, upper surface, $\alpha=4^\circ$: $-C_p$ (left), streamwise skin friction (middle, zero contour heavy), and the largest $\chi$ near the wall in decades (right, $\chi=1$ heavy). Rows: structured L2 grid, unstructured L2 grid, and the $e^N$ reference section by section. In the reference row the right panel is $\chi_\infty e^N$, and its heavy contour is $N=N_\mathrm{crit}$.}\label{f:daesurf}
\end{figure}

\begin{figure}[!htbp]\centering
\includegraphics[width=0.99\textwidth]{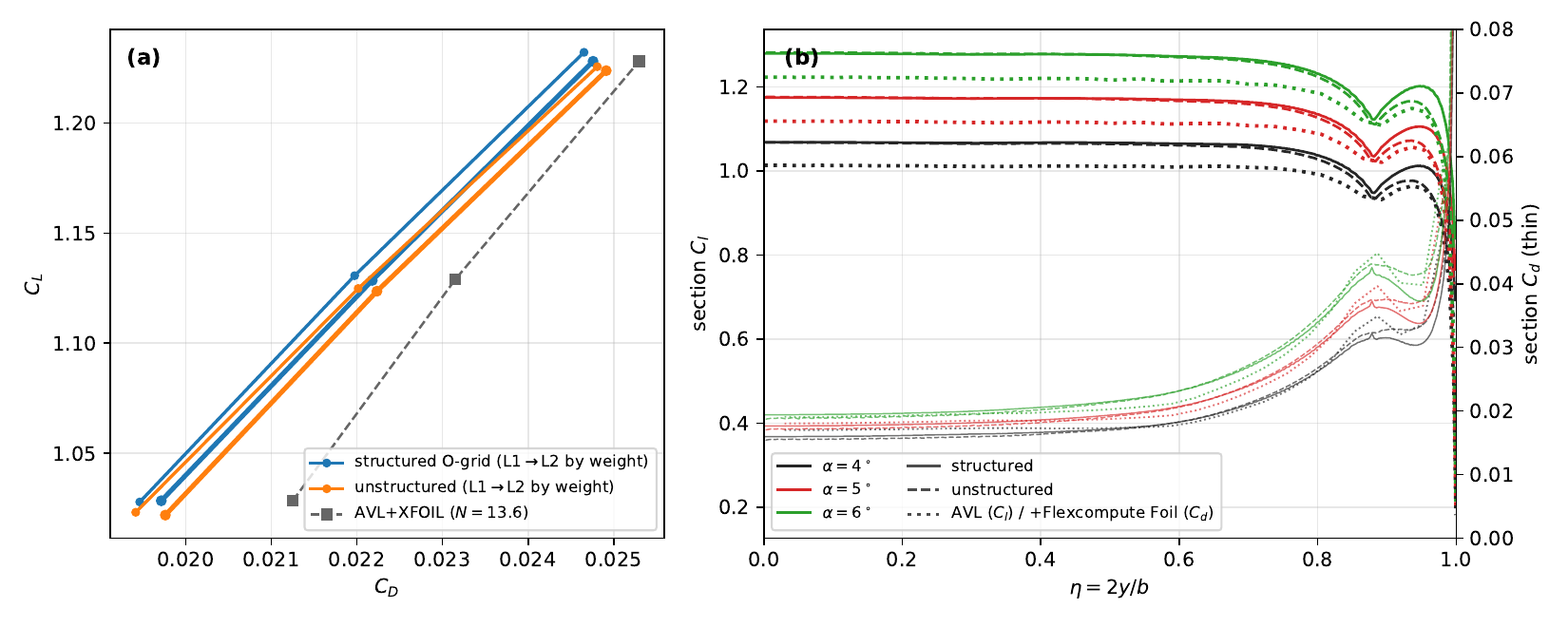}
\caption{Daedalus wing. (a) Polar on the L1 and L2 grids of both grid families, with the reference. (b) Section lift coefficient (thick, left axis) and section drag coefficient (thin, right axis) along the span.}\label{f:daepolar}
\end{figure}

The half-span is $17.07\,$m, the root chord $0.917\,$m, and the aspect ratio
$37.8$. The chord is constant to the spanwise station $\eta=0.88$ and tapers
from there to the tip. The sections are the DAE-11, DAE-21 and DAE-31
airfoils. The root-chord Reynolds number is $5\times10^5$, and the incidences
are $\alpha=4^\circ$, $5^\circ$ and $6^\circ$, for which
$C_L\approx1.02$--$1.23$. The freestream value is
$\chi_\infty=8.76\times10^{-6}$, which is $N_\mathrm{crit}\approx13.6$, a
value representative of flight in still air. There are two grid families, each at three
levels: a structured O-grid stacked along the span, with $0.72$ to $36.7$
million nodes, and an unstructured grid of prisms, tetrahedra and hexahedra,
with $1.70$ to $54.9$ million nodes. There is no measurement of the
transition on this wing. The reference is the vortex-lattice code
AVL~\cite{drela_youngren_avl} with section polars from
XFOIL~\cite{drela_xfoil_1989} at the local lift coefficient and the same
$N_\mathrm{crit}$, at the total lift of the present computation.\footnote{The distributions of the reference along the chord in
Figs.~\ref{f:daesurf}, \ref{f:daesurfall} and~\ref{f:daesections}, and its section drag in
Fig.~\ref{f:daepolar}b, are computed with Flexcompute Foil~\cite{smith_flexfoil_2026}
at the same section lift, Reynolds number and $N_\mathrm{crit}$.}

Figure~\ref{f:daesurf} shows the upper surface at $\alpha=4^\circ$: the
pressure, the streamwise skin friction, and the largest $\chi$ near the wall.
The upper two rows are the L2 grids of the two grid families, and the bottom
row is the reference, section by section. The bubble is the band between the two
heavy contours of zero skin friction. It is at the same chordwise position
along the span up to $\eta=0.88$, and it moves aft over the taper, where the
chord Reynolds number decreases. At $\eta=0.31$ the computed separation is
within $0.017\,c$ of the reference and the reattachment within $0.009\,c$, at
every incidence. The transition location, where $\chi=1$, is about $0.09\,c$ ahead of the
$N=N_\mathrm{crit}$ contour of the reference. The two contours are not at
the same level of amplification. With $N=\ln(\chi/\chi_\infty)$, the level
$\chi=1$ is $N=N_\mathrm{crit}-2.0$, and the level $N=N_\mathrm{crit}$ is
$\chi=c_{v1}$ (Section~III.A). The transition location is therefore upstream of the
reference contour. Over the
last $0.8\%$ of the span, $\eta\gtrsim0.992$, the largest $\chi$ exceeds one
from the leading edge. The tip vortex is there, and the two-dimensional
reference contains no such region.

Figure~\ref{f:daepolar} shows the polar of the wing and the lift and drag of
its sections. At L2 the two grid families agree within $0.35$--$0.63\%$ in
$C_L$ and within $0.7$--$1.7$ counts in $C_D$. At L1 they agree within
$0.46$--$0.53\%$ and $0.4$--$1.6$ counts, and at L0 within $0.74$--$0.97\%$ and
$18$--$21$ counts. The drag on the structured L2
grid is $15.6$, $10.3$ and $5.6$ counts ($7.3\%$, $4.4\%$ and $2.2\%$) below
the reference at $\alpha=4^\circ$, $5^\circ$ and $6^\circ$.
Table~\ref{t:daetotals} lists the forces at each level. The Supplementary
Figures contain the surface maps at $5^\circ$ and $6^\circ$, the section
distributions at three stations, and the near-wall contours
(Figs.~\ref{f:daesurfall}, \ref{f:daesections} and~\ref{f:daesheets}).

\begin{table}[!htbp]
  \centering\small
  \caption{Daedalus wing: forces at each grid level for both grid families, and the reference at the lift of the structured L2 solution.}
  \label{t:daetotals}
  \begin{tabular}{cl cc cc}
    \toprule
    & & \multicolumn{2}{c}{structured O-grid} & \multicolumn{2}{c}{unstructured}\\
    \cmidrule(lr){3-4}\cmidrule(lr){5-6}
    $\alpha$ & level & $C_L$ & $C_D$ & $C_L$ & $C_D$\\
    \midrule
    $4^\circ$ & L0 & 1.0244 & 0.01958 & 1.0156 & 0.02134\\
              & L1 & 1.0279 & 0.01946 & 1.0232 & 0.01941\\
              & L2 & 1.0284 & 0.01969 & 1.0219 & 0.01976\\
              & AVL$+$XFOIL & 1.0284 & 0.02125 & &\\
    \midrule
    $5^\circ$ & L0 & 1.1254 & 0.02219 & 1.1145 & 0.02427\\
              & L1 & 1.1307 & 0.02197 & 1.1248 & 0.02201\\
              & L2 & 1.1290 & 0.02212 & 1.1236 & 0.02223\\
              & AVL$+$XFOIL & 1.1290 & 0.02315 & &\\
    \midrule
    $6^\circ$ & L0 & 1.2262 & 0.02505 & 1.2171 & 0.02713\\
              & L1 & 1.2321 & 0.02465 & 1.2256 & 0.02481\\
              & L2 & 1.2282 & 0.02474 & 1.2238 & 0.02491\\
              & AVL$+$XFOIL & 1.2282 & 0.02530 & &\\
    \bottomrule
  \end{tabular}
\end{table}

\FloatBarrier
\subsection{The 6:1 Spheroid}

The 6:1 prolate spheroid at incidence is the standard three-dimensional case
for transition models. Kreplin, Vollmers and Meier~\cite{kreplin_1985}
measured the wall shear stress over the body with surface hot films in the
DFVLR tunnel at G\"ottingen, and there is one more measurement from the ONERA
F1 tunnel. The stability analysis of Stock~\cite{stock_2006} identifies the
cause of transition at each condition. At $Re_L=1.5\times10^6$ the boundary
layer is laminar up to its separation, and the measured line is the
separation line. At $Re_L\approx7\times10^6$ and zero or small incidence, the
TS waves cause transition. At $Re_L\gtrsim6.5\times10^6$ and
$\alpha\ge5^\circ$, the crossflow instability causes it, alone or together
with the TS waves. The model represents the TS waves only. The case shows
what it computes in each of the three situations.

The geometry is the analytic spheroid of length $L$ and largest diameter
$L/6$, as a half-model on a structured O-grid of $2.2$ million nodes with
a first wall spacing of $1.5\times10^{-6}L$. The constants of the model are
those of every other case. The Mach number is that of the tunnel. The
G\"ottingen tunnel is atmospheric, so its Mach number is proportional to
$Re_L$: $M=1.994\times10^{-8}\,Re_L$, which is $0.030$ at
$Re_L=1.52\times10^6$ and $0.144$ at $7.2\times10^6$. This relation is fitted
to the two published pairs of $Re_L$ and velocity, $7.70\times10^6$ at
$55.0$~m/s and $6.54\times10^6$ at $45.0$~m/s, and reproduces them within
$2\%$. The ONERA F1 tunnel is pressurized, and its condition is computed at
the Mach number of the G\"ottingen condition with the same $Re_L$. Each
computation is iterated $45\,000$ times from a uniform initial condition.

Two values of the turbulence intensity are available for each tunnel, and
they differ by a factor of up to three. Each condition is therefore computed
with two values of $\chi_\infty$. The fitted value is the one that
Eq.~\eqref{w:mack} gives for the $N_\mathrm{crit}$ that Stock fitted to the
measured transition: $8.0$ for G\"ottingen and $7.0$ for ONERA F1. The
measured value is the one for the turbulence intensity measured with a hot
wire: $0.33\%$ for G\"ottingen and less than $0.1\%$ for
ONERA F1. Table~\ref{tab:sphtunnel} lists the computed transition location minus the
measured one, in units of $L$, at every azimuth with a measurement.

\begin{figure}[!htbp]\centering
\includegraphics[width=0.80\textwidth]{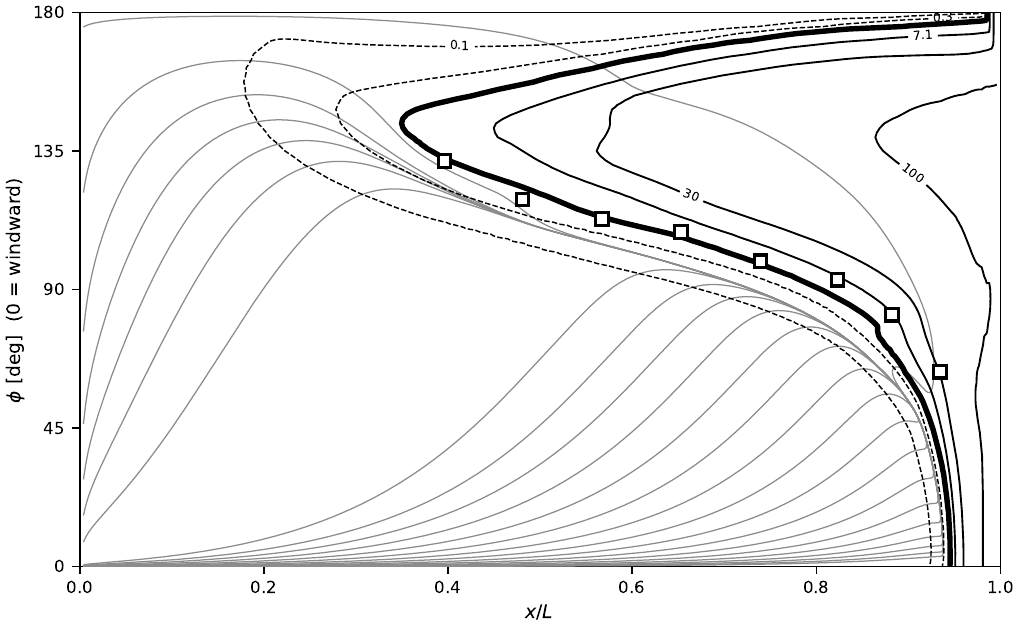}\\ a)\\[4pt]
\includegraphics[width=0.80\textwidth]{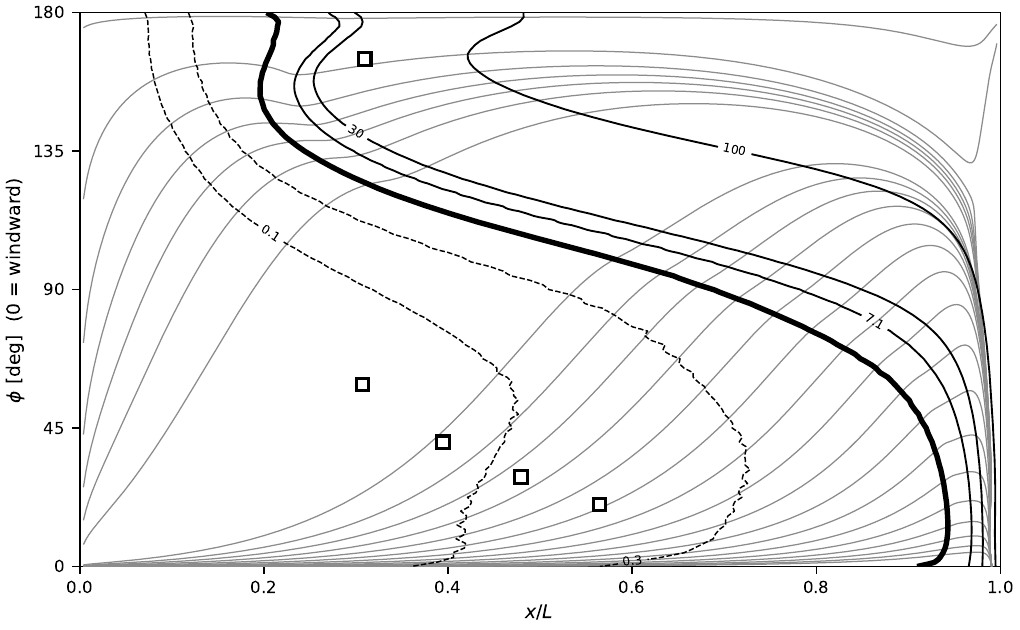}\\ b)
\caption{6:1 spheroid at $\alpha=10^\circ$, with the measured turbulence intensity of the G\"ottingen tunnel: a) $Re_L=1.52\times10^6$, b) $Re_L=6.56\times10^6$. The surface is unrolled into $(x/L,\phi)$, with $\phi=0$ windward. Gray: skin-friction lines. Black: contours of the largest $\chi$ on the wall normal, $\chi=1$ heavy. Squares: the measured line, which is laminar separation in a) and transition in b).}\label{f:ov_a10hi}
\end{figure}

At $Re_L=1.5\times10^6$ the computed transition location is within $0.03\,L$ to
$0.10\,L$ of the measured separation line, in the root mean square over the
measured azimuths. Figure~\ref{f:ov_a10hi}a shows $\alpha=10^\circ$ with the
measured turbulence intensity, where the root-mean-square difference is
$0.027\,L$ over all eight azimuths. With the fitted $\chi_\infty$ it is
$0.082\,L$. At $\alpha=5^\circ$ it is $0.068\,L$ with the measured turbulence
intensity, where the largest $\chi$ reaches one ahead of $x/L=0.97$ at seven
of the eight azimuths, and $0.028\,L$ with the fitted $\chi_\infty$, at four
of the eight. At $\alpha=29.7^\circ$ and $Re_L=1.53\times10^6$ it is
$0.098\,L$ and $0.095\,L$, over twelve azimuths.

At $Re_L=7.2\times10^6$ and zero incidence, where the TS waves cause
transition, the measured transition location is between the locations computed with the two
values of $\chi_\infty$. The computed location is $0.077\,L$ aft of the measured
one with the fitted $\chi_\infty$, and $0.097\,L$ ahead of it with the
measured turbulence intensity. It is uniform within $\pm0.001\,L$ over the
seventeen azimuths.

At $Re_L\approx6.5\times10^6$ and $\alpha=10^\circ$, where the crossflow
instability causes transition, the computed transition location on the windward side is
$0.375\,L$ to $0.549\,L$ aft of the measured one
(Fig.~\ref{f:ov_a10hi}b). It is near the tail, and the measured location is
between $0.3\,L$ and $0.6\,L$. Stock~\cite{stock_2006} attributes the measured
transition there to the crossflow instability, alone or together with the TS
waves. The difference increases with the incidence. At $\alpha=2.5^\circ$ and
the fitted $\chi_\infty$, the computed transition location is $0.016\,L$ aft of the
measurement on the leeward side and $0.132\,L$ aft on the windward side, and
the measured location varies by $0.018\,L$ around the body. At $\alpha=5^\circ$
and $Re_L=6.49\times10^6$ the difference ranges from $-0.088\,L$ on the
leeward side to $+0.246\,L$ on the windward side.

The response of the computed transition location to $\chi_\infty$ changes in the same way.
Between the two values of $\chi_\infty$ of the G\"ottingen tunnel, the
computed location on the windward side moves by $0.174\,L$ at $\alpha=0^\circ$,
$0.181\,L$ at $2.5^\circ$ and $0.239\,L$ at $5^\circ$. At $10^\circ$ it moves
by $0.012\,L$, and between the two values of the ONERA F1 tunnel it does not
move. The Supplementary Figures contain the unrolled surface at every
condition (Fig.~\ref{f:ov_a5lo_gmeas}).

\begin{table}[!htbp]
  \centering\small
  \caption{6:1 spheroid: computed minus measured transition location, in units of $L$, at the azimuths with a measurement. The computed location is the station where the largest $\chi$ near the wall crosses one. The fitted $\chi_\infty$ is that of the $N_\mathrm{crit}$ that Stock~\cite{stock_2006} fitted for the tunnel, and the measured $\chi_\infty$ is that of the measured turbulence intensity $Tu$. The column $n$ is the number of azimuths at which $\chi$ reaches one ahead of $x/L=0.97$, over the number of measured azimuths. The columns lee and wind are the differences at the most leeward and the most windward measured azimuth. The last two columns are the same differences for the $e^N$ computation of Stock, digitized from his figures. Upper block: $Re_L\approx1.5\times10^6$, where the measured line is the laminar separation line. Lower block: $Re_L\ge6.49\times10^6$.}
  \label{tab:sphtunnel}
  \begin{tabular}{cc l l c c cccc cc}
    \toprule
    $\alpha$ & $Re_L$ & tunnel & $\chi_\infty$ & $Tu$ & $n$ & \multicolumn{4}{c}{SA-AI $-$ measured} & \multicolumn{2}{c}{Stock $-$ measured} \\
    & $[10^6]$ & & & [\%] & & mean & rms & lee & wind & mean & rms \\
    \midrule
    $5^\circ$ & $1.52$ & G\"ottingen & measured & 0.330 & 7/8 & -0.020 & 0.068 & +0.074 & -0.037 & +0.020 & 0.058 \\
    $5^\circ$ & $1.52$ & G\"ottingen & fitted & 0.106 & 4/8 & -0.006 & 0.028 & +0.026 & -0.040 & -0.064 & 0.064 \\
    $10^\circ$ & $1.52$ & G\"ottingen & measured & 0.330 & 8/8 & -0.013 & 0.027 & +0.002 & -0.041 & -0.091 & 0.100 \\
    $10^\circ$ & $1.52$ & G\"ottingen & fitted & 0.106 & 8/8 & +0.019 & 0.082 & +0.160 & -0.053 & -0.091 & 0.100 \\
    $29.7^\circ$ & $1.53$ & G\"ottingen & measured & 0.330 & 12/12 & -0.048 & 0.098 & +0.185 & -0.045 & -- & -- \\
    $29.7^\circ$ & $1.53$ & G\"ottingen & fitted & 0.106 & 12/12 & +0.041 & 0.095 & +0.264 & -0.024 & -- & -- \\
    \midrule
    $0^\circ$ & $7.2$ & G\"ottingen & measured & 0.330 & 17/17 & -0.097 & 0.097 & -0.097 & -0.098 & -0.013 & 0.013 \\
    $0^\circ$ & $7.2$ & G\"ottingen & fitted & 0.106 & 17/17 & +0.077 & 0.077 & +0.078 & +0.076 & -0.013 & 0.013 \\
    $2.5^\circ$ & $7.2$ & G\"ottingen & measured & 0.330 & 3/3 & -0.093 & 0.102 & -0.150 & -0.049 & +0.006 & 0.021 \\
    $2.5^\circ$ & $7.2$ & G\"ottingen & fitted & 0.106 & 3/3 & +0.082 & 0.096 & +0.016 & +0.132 & +0.006 & 0.021 \\
    $5^\circ$ & $6.49$ & G\"ottingen & measured & 0.330 & 4/4 & -0.098 & 0.177 & -0.292 & +0.007 & -0.024 & 0.051 \\
    $5^\circ$ & $6.49$ & G\"ottingen & fitted & 0.106 & 4/4 & +0.107 & 0.201 & -0.088 & +0.246 & -0.024 & 0.051 \\
    $10^\circ$ & $6.56$ & G\"ottingen & measured & 0.330 & 5/5 & +0.365 & 0.440 & -0.105 & +0.375 & +0.073 & 0.098 \\
    $10^\circ$ & $6.56$ & G\"ottingen & fitted & 0.106 & 5/5 & +0.421 & 0.463 & +0.068 & +0.387 & +0.073 & 0.098 \\
    $10^\circ$ & $6.56$ & ONERA F1 & fitted & 0.161 & 5/5 & +0.399 & 0.475 & +0.002 & +0.549 & +0.088 & 0.134 \\
    $10^\circ$ & $6.56$ & ONERA F1 & measured & 0.100 & 5/5 & +0.420 & 0.486 & +0.051 & +0.549 & +0.088 & 0.134 \\
    \bottomrule
  \end{tabular}
\end{table}

\FloatBarrier

\section{Remarks on the Operating Envelope}\label{sec:envelope}

\par\vspace{7.6pt}\begin{center}\includegraphics{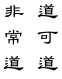}\end{center}\vspace{-7.6pt}

\begin{figure}[!htbp]\centering
\includegraphics[width=0.5\textwidth]{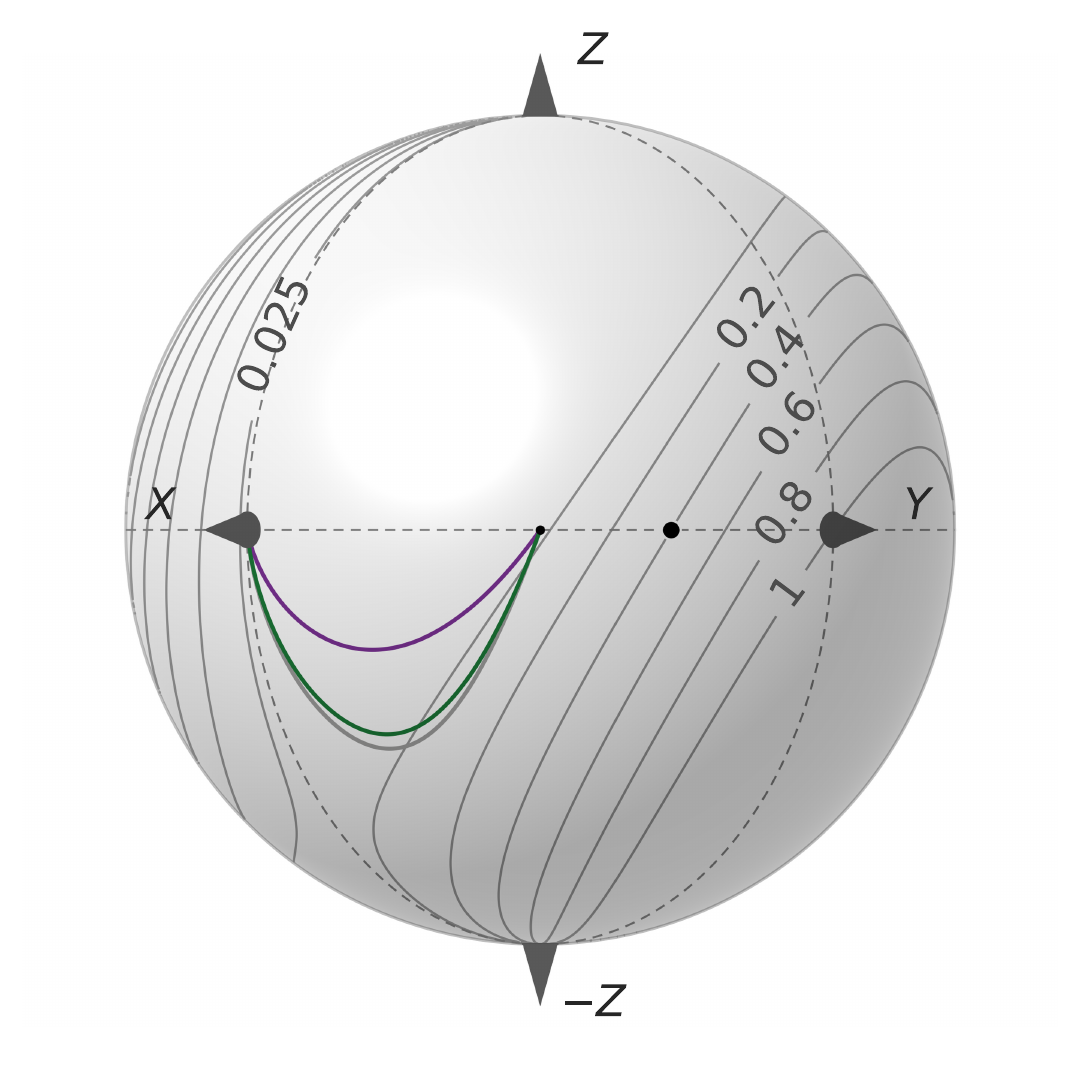}
\caption{Indicator sphere with contours of $\hat\Omega\hat I$, in the view of
Fig.~\ref{f:sphere}a. Gray: Blasius profile. Green: spanwise profile on a
swept attachment line. Purple: asymptotic suction profile. Large dot: a fluid
in solid-body rotation with its wall, with the absolute vorticity. Small dot:
the wall point.}\label{f:envsphere}
\end{figure}

\begin{figure}[H]\centering
\includegraphics[width=0.99\textwidth]{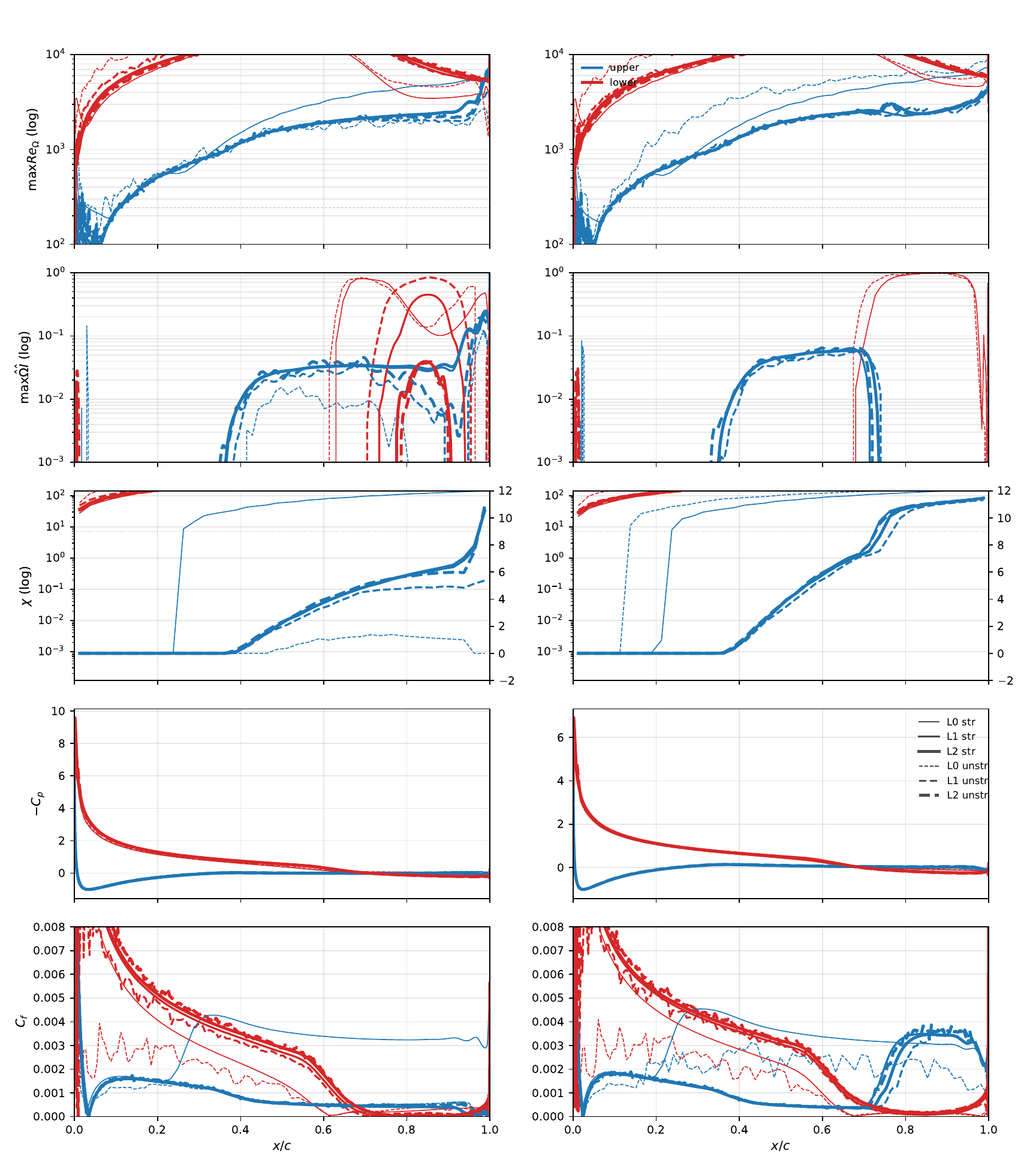}
\caption{NLF(1)-0416 surface distributions, $Re=4\times10^6$, all six grids, in ascending incidence; rows as in Fig.~\ref{f:nlfcf0}. $\alpha=-12^\circ$, $\alpha=-10^\circ$.}\label{f:nlfcflow}
\end{figure}

Section~III computes the model in boundary layers on fixed walls of small
curvature. This section evaluates it in five situations outside that range: a
rotating frame, a curved wall, a swept attachment line, a wall with suction,
and the wake of another surface. No flow is computed here. The first four are
exact solutions, or exact perturbations of the Blasius profile, on which the
three terms of Eq.~\eqref{w:ind} can be written down. The fifth is discussed
with the computation of Section~III.E. Each remark states what stability
theory or experiment finds in the flow, and what the model computes.
Figure~\ref{f:envsphere} shows three of the flows on the
indicator sphere of Fig.~\ref{f:sphere}.

\subsection{Rotation}

Consider a fluid in solid-body rotation together with its container, at the
angular velocity $\Omega_\mathrm{f}$. In the frame of the container the fluid
is at rest. It contains no shear, and it is
stable~\cite{drazin_reid_1981}. In an inertial frame the vorticity of the same
fluid is $2\Omega_\mathrm{f}$, and at the distance $d$ from the wall it moves
at the speed $\Omega_\mathrm{f}d$ relative to the nearest point of the wall.
With these two quantities the triple of Eq.~\eqref{w:ratio} is
$(X,Y,Z)=(\Omega_\mathrm{f}d,\;2\Omega_\mathrm{f}d,\;0)$. Its direction is
$(1,2,0)/\sqrt5$ at every point of the fluid, the large dot in
Fig.~\ref{f:envsphere}. There $\hat\Omega=2/\sqrt5$, $\hat I=1/\sqrt5$ and
$\hat\Omega\hat I=2/5$. The rate of Eq.~\eqref{w:a} is $0.4\,a_{\max}$, and
the gate of Eq.~\eqref{w:rate} is open where
$Re_\Omega=2\Omega_\mathrm{f}d^2/\nu$ exceeds the threshold of
Eq.~\eqref{w:onset}, which is $133$ at this value of $\hat\Omega\hat I$. The
model therefore amplifies $\tilde\nu$ in a fluid that is at rest relative to
its walls.

A boundary layer in a frame that rotates about a spanwise axis is found on
the blade of a vertical-axis turbine and in the passage of a radial impeller.
Stability theory finds two effects of the rotation. A two-dimensional TS wave
is unaffected by it~\cite{schmid_henningson_2001}. Streamwise roll cells,
driven by the Coriolis force, grow where the frame rotates in the sense
opposite to the vorticity of the boundary layer. Where the frame rotates in
the sense of that vorticity the flow is stabilized: in a rotating channel the
transition to turbulence is suppressed on that side~\cite{johnston_1972}.

With the absolute vorticity the model's rate depends on the rotation. It
increases where the rotation stabilizes the flow, and it decreases where the
roll cells grow. On the Blasius profile the largest $\hat\Omega\hat I$ changes
by the fraction $18.3\,(\delta_\nu/L)(\Omega_\mathrm{f}L/U_\infty)$, with
$\delta_\nu=\sqrt{\nu x/U_e}$. For $\Omega_\mathrm{f}=U_\infty/L$, with $L$ the chord or the diameter,
that is $0.6\%$ at the transition location of the NLF(1)-0416 at
$\alpha=0^\circ$, $2.9\%$ on the Eppler 387 at $\alpha=0^\circ$, and $4.9\%$
on the cylinder at $Re_D=10^5$. It is larger at a lower Reynolds number,
because $\delta_\nu/L$ is proportional to $(U_\infty L/\nu)^{-1/2}$. With the vorticity and the velocity relative to the
rotating frame, the model is independent of the rotation, as the TS waves
are, and it represents no roll cells.

\subsection{Wall Curvature}

The curvature of the wall changes the terms of Eq.~\eqref{w:ind} by a small
fraction. That fraction matters because $\hat I$ is a difference: $Y-X-Z$ can
be small compared with the three terms. At the wall distance of the largest
$\hat\Omega\hat I$, the ratio $|Y-X-Z|/(|X|+|Y|+|Z|)$ is $0.04$ on the
Falkner--Skan profile $\beta=+0.10$, $0.08$ on the Blasius profile, $0.16$ at
$\beta=-0.10$ and $0.39$ at incipient separation. A given small change of one
term is therefore a larger fraction of $\hat I$ in a favorable pressure
gradient than near separation.

A wall of radius $R$ that curves in the streamwise direction turns the flow
with it. On a convex wall the centrifugal force is
stabilizing~\cite{drazin_reid_1981}. The model contains the curvature in the
Laplacian of the velocity. For a velocity $u(r)$ along circles about the
center of curvature,
$\nabla^2\mathbf u\cdot\hat{\mathbf u}=u''+u'/r-u/r^2$, with the derivatives
taken along $r$, and $r=R+d$ on a convex wall. To first order in $d/R$ the
curvature term of Eq.~\eqref{w:ratio} increases by $(d/2R)\,Y$, and $\hat I$
decreases. On the Blasius profile the largest $\hat\Omega\hat I$ changes by
the fraction $-4.8\,\delta_\nu/R$. The rate of the model is therefore lower on
a convex wall, the direction of the stabilizing effect. This change is part
of every computation of Section~III. On the upper surfaces at
$\alpha=0^\circ$, which are convex, it is $-0.15\%$ at the transition
location of the NLF(1)-0416 and $-0.32\%$ at the separation of the
Eppler 387.

On a concave wall the centrifugal force is destabilizing, and the boundary
layer forms streamwise vortices, the G\"ortler
vortices~\cite{saric_1994}. Their growth is a function of the G\"ortler
number $G_\theta=Re_\theta\sqrt{\theta/R}$. In the model $r=R-d$ on a
concave wall, the curvature term decreases by $(d/2R)\,Y$, and the largest
$\hat\Omega\hat I$ of the Blasius profile changes by the fraction
$+4.8\,\delta_\nu/R$. The rate of the model is therefore higher on a concave
wall, the direction of the destabilizing effect. The increase does not
contain the Reynolds number. At $\delta_\nu/R=10^{-3}$ it is $0.48\%$ at
every Reynolds number, while $G_\theta$ is $17$ at $Re_x=10^6$ and $171$ at
$Re_x=10^8$. The model represents no streamwise vortices.

On a slender body of revolution of radius $r_0$ in axial flow, the wall
curves across the stream. Stability theory finds this boundary layer more
stable than that of a flat plate: on a long thin cylinder it is stable below
a critical Reynolds number, and above it the flow is unstable only over a
finite distance~\cite{tutty_2002}. In the model, an axial velocity $u(r)$
with $r=r_0+d$ gives $\nabla^2\mathbf u\cdot\hat{\mathbf u}=u''+u'/r$, and
the curvature term increases by $(d/2r)\,Y$. To first order this is the
change on a convex wall with $r_0$ in place of $R$, and the rate of the model
is lower. At the eight measured stations of the spheroid at
$\alpha=10^\circ$ and $Re_L=1.5\times10^6$, the largest $\hat\Omega\hat I$ of
a Blasius profile is lower by $3.0\%$ to $8.8\%$. On a body of fineness ratio
$40$ at the same $Re_L$ and stations it is lower by $18.3\%$ to $46.1\%$, and
$\hat\Omega\hat I$ is nowhere positive where $\delta_\nu$ equals $r_0$.

The term of transverse curvature depends on the form of the second
derivative. Equation~\eqref{w:ratio} uses $\nabla^2\mathbf u$. The wall-normal
derivative of the magnitude of the vorticity, which is the same quantity in a
plane boundary layer, contains the term of streamwise curvature and not that
of transverse curvature.

\subsection{Sweep}

On a swept wing the flow divides along the attachment line of the leading
edge. Outside the boundary layer the chordwise velocity increases linearly
with the distance $x$ from that line, $U_e=(dU_e/dx)\,x$, and the spanwise
velocity is $W_\infty$, the component of the freestream along the leading
edge. The swept Hiemenz flow is the exact solution of the Navier--Stokes
equations with this outer flow~\cite{hall_malik_poll_1984}. On the attachment
line its chordwise velocity is zero, and the boundary layer flows along the
line, in the spanwise direction. The thickness of that boundary layer is the
same at every spanwise station, because the flow toward the wall balances
the diffusion of vorticity away from it. The spanwise velocity is $W_\infty$
multiplied by a function of $d/\delta$ alone, with
$\delta=\sqrt{\nu/(dU_e/dx)}$, and the Reynolds number of the boundary layer
is $\bar R=W_\infty\delta/\nu$.

Linear theory finds waves that travel along the attachment line unstable
above $\bar R=583$~\cite{hall_malik_poll_1984}. With large disturbances,
turbulent flow along the attachment line is observed above
$\bar R\approx245$~\cite{poll_1979}. The model evaluates the spanwise
profile, the green curve in Fig.~\ref{f:envsphere}. Its largest
$\hat\Omega\hat I$ is $0.064$, which is $82\%$ of the $0.078$ of the Blasius
profile. The onset gate $\sigma_{Re}$ of Eq.~\eqref{w:rate} opens
gradually as $\bar R$ increases. The largest $a\,\sigma_{Re}$ across the
profile is $10\%$ of the largest $a$ at $\bar R=371$, one half at
$\bar R=603$ and $90\%$ at $\bar R=839$. The linear critical value, $583$, is
inside that range. At $\bar R=245$ the largest $a\,\sigma_{Re}$ is $3\%$ of
the largest $a$: the value $245$ is a property of large disturbances, and the
amplification of Section~II describes the growth of small ones.

Downstream of the attachment line the boundary layer of a swept wing contains
a crossflow. Its instability is not represented in the model, as
Section~III.H shows on the spheroid.

\subsection{Suction}

Uniform suction through the wall, at the velocity $v_w$, makes the boundary
layer thinner and lowers its shape factor $H$. Far downstream the profile is
the asymptotic suction profile, $u=U_\infty\big(1-e^{-d|v_w|/\nu}\big)$, an
exact solution of the Navier--Stokes equations. Its
shape factor is $2$, against $2.59$ for the Blasius profile. Its
displacement thickness is $\nu/|v_w|$, so $Re_{\delta^*}=U_\infty/|v_w|$. Its
critical Reynolds number is
$Re_{\delta^*}\approx5.4\times10^4$~\cite{hocking_1975}, about one hundred
times the $519$ of the Blasius profile~\cite{schmid_henningson_2001}. On this
profile the three terms of Eq.~\eqref{w:ind}, divided by $U_\infty$, and the
accumulated inflection that they form are
\begin{gather}
  X=1-e^{-d|v_w|/\nu}, \qquad
  Y=\frac{d\,|v_w|}{\nu}\,e^{-d|v_w|/\nu}, \qquad
  Z=-\frac12\Big(\frac{d\,|v_w|}{\nu}\Big)^{\!2}e^{-d|v_w|/\nu}, \notag\\[3pt]
  I = Y-X-Z = e^{-d|v_w|/\nu}\,\bigg[1+\frac{d\,|v_w|}{\nu}
      +\frac12\Big(\frac{d\,|v_w|}{\nu}\Big)^{\!2}\,\bigg]-1 .
  \label{w:suction}
\end{gather}
The polynomial in the square brackets of Eq.~\eqref{w:suction} consists of the
first three terms of the Taylor series of $e^{d|v_w|/\nu}$ in powers of
$d\,|v_w|/\nu$. It is smaller than $e^{d|v_w|/\nu}$ wherever $d>0$, so $I$ is
negative at every wall distance: the purple curve in
Fig.~\ref{f:envsphere} lies on the side of the sphere where
$\hat\Omega\hat I$ is negative. The rate of Eq.~\eqref{w:a} is zero at every
Reynolds number. The model therefore gives the direction of the effect of
suction and overstates its size: a boundary layer with
$|v_w|/U_\infty<1.8\times10^{-5}$ is unstable, and it remains laminar in the
model.

\subsection{The Wake of Another Surface}

The two-element section of Section~III.E is an example of the wake of an
upstream surface changing the transition location on a downstream surface.
In experiments the turbulence of such a wake causes the boundary layer of the
downstream surface to become turbulent farther upstream. In turbomachines
this wake-induced transition occurs on a blade row behind
another~\cite{mayle_1991}. In the model the wake acts through its
$\tilde\nu$. The destruction term of SA is proportional
to $(\tilde\nu/d)^2$ and is negligible far from a wall, so the $\tilde\nu$ of
a wake is convected and diffused, and it is not destroyed. A boundary layer
next to the wake receives fluid with a $\chi$ far above the freestream value,
also where the $\chi$ of that fluid is still below one.
On the flap of Section~III.E, $\chi$ in the boundary layer is $0.5$ to $0.8$
at the incidence where the wake of the main element passes next to it, and
$9\times10^{-4}$ to $0.004$ at the other incidences. It reaches one in less
than one e-fold of growth in place of seven. Where fluid with $\chi>c_{v1}$
enters a boundary layer, the production is already that of SA, and the
amplification of Section~II does not contribute. The transition in the
model is therefore farther upstream next to a wake, the direction of the
measured effect.

\FloatBarrier

\section*{\tocentry{section}{Acknowledgments}Acknowledgments}

\par\vspace{7.6pt}\begin{center}\includegraphics{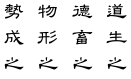}\end{center}\vspace{-7.6pt}

The motivation for this work came from implementing the
amplification-factor-transport (AFT) model of James G.\ Coder
\cite{coder_maughmer_2014} in Flexcompute Flow.
The idea for this approach was sparked by conversations with Mark Drela
and Philippe R.\ Spalart, whose insights on the amplification-factor envelope
and on the structure of the SA equation were formative.
The \emph{Daodejing} epigraphs are set in the clerical-script font of the
Ministry of Education, Republic of China (Taiwan), used under the Creative
Commons Attribution-NoDerivatives 3.0 Taiwan license.

\clearpage
\section*{\tocentry{section}{Supplementary Figures}Supplementary Figures}\label{app:atlas}

These figures complete the sets from which Section~III shows one page each. They are in the order of the
cases of Section~III.

\subsection*{\tocentry{subsection}{The NLF(1)-0416 Airfoil}The NLF(1)-0416 Airfoil}

\begin{figure}[H]\centering
\includegraphics[width=0.99\textwidth]{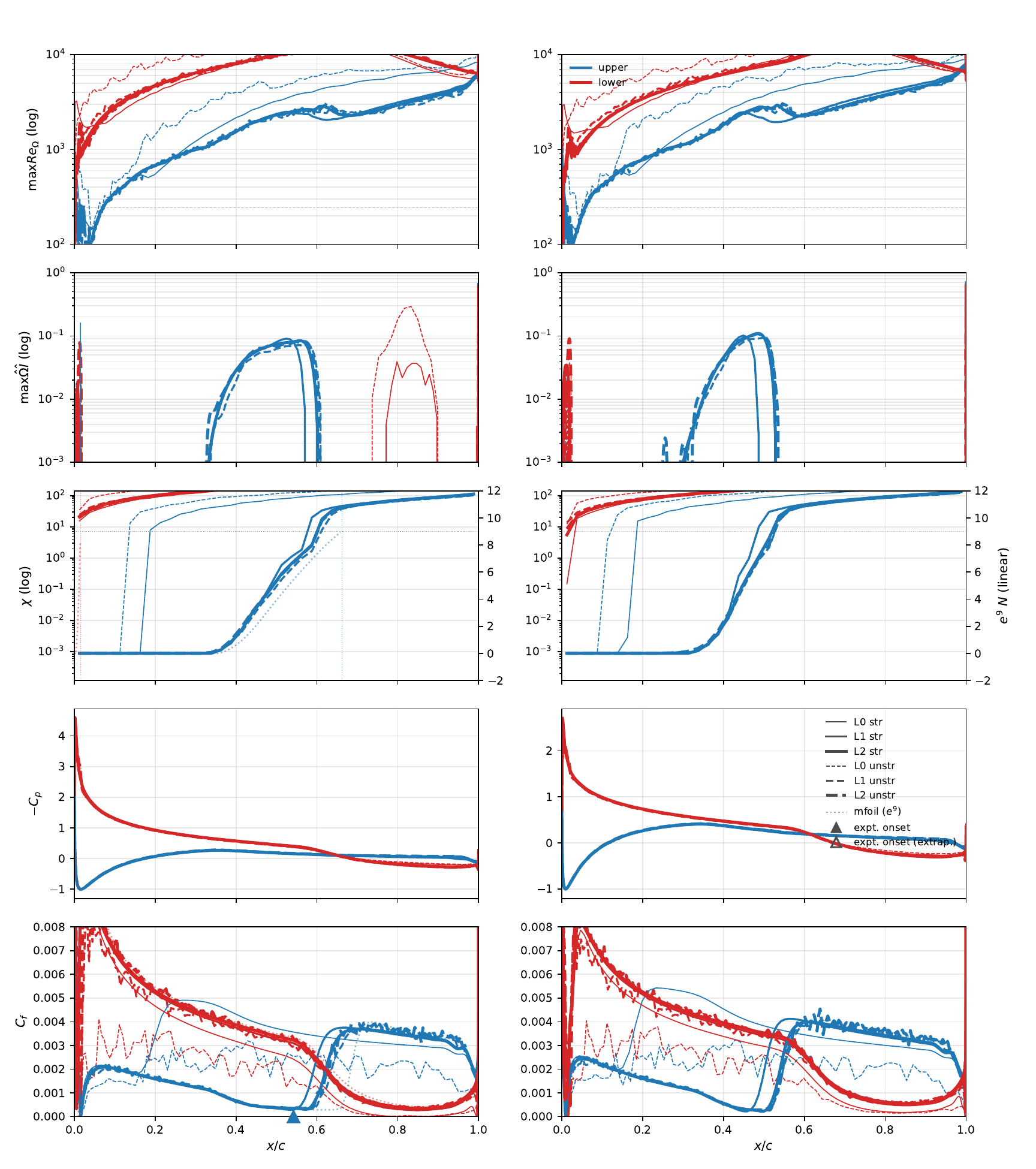}
\figcont{$\alpha=-8^\circ$, $\alpha=-6^\circ$.}
\end{figure}

\begin{figure}[H]\centering
\includegraphics[width=0.99\textwidth]{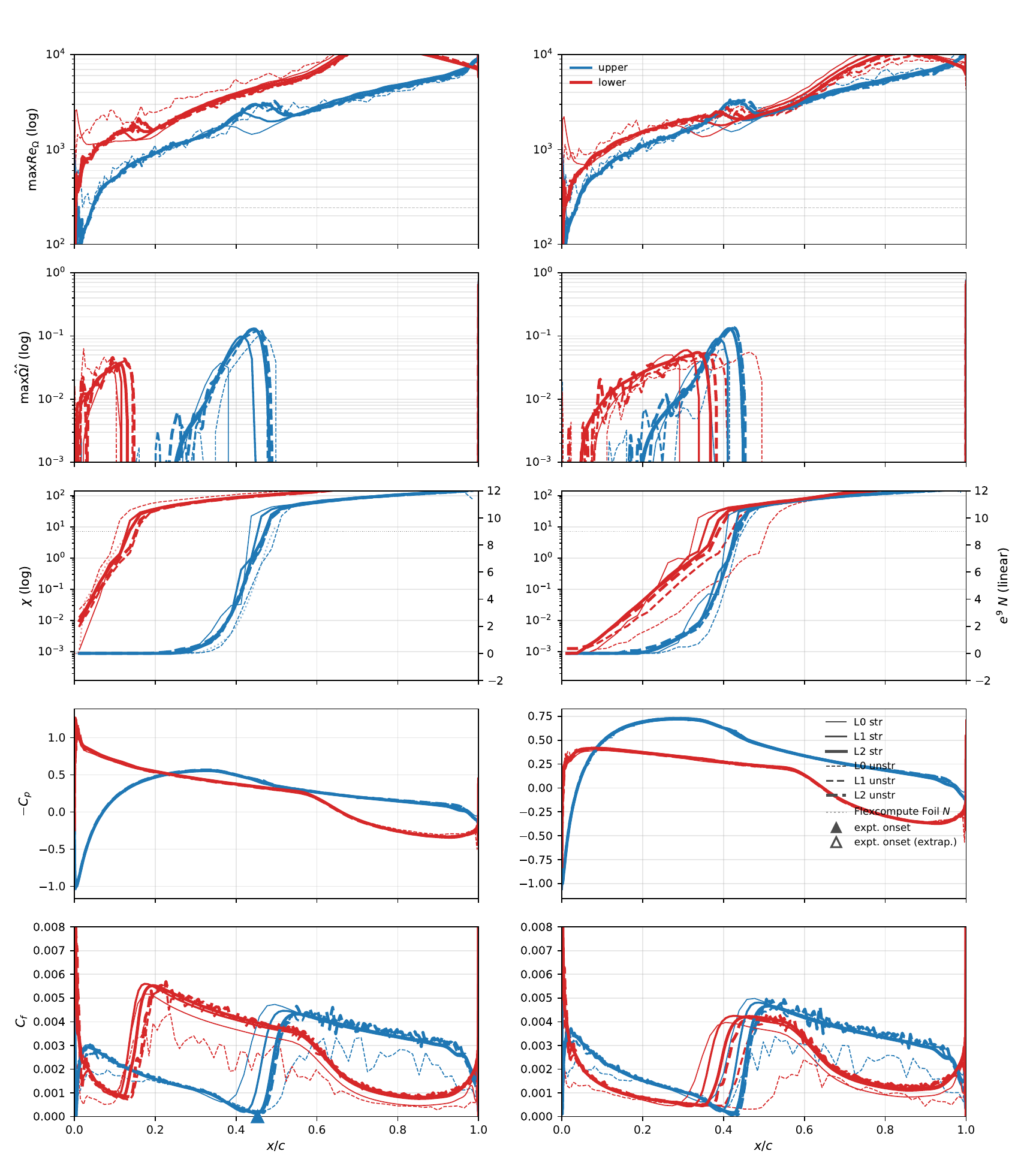}
\figcont{$\alpha=-4^\circ$, $\alpha=-2^\circ$.}
\end{figure}

\begin{figure}[H]\centering
\includegraphics[width=0.99\textwidth]{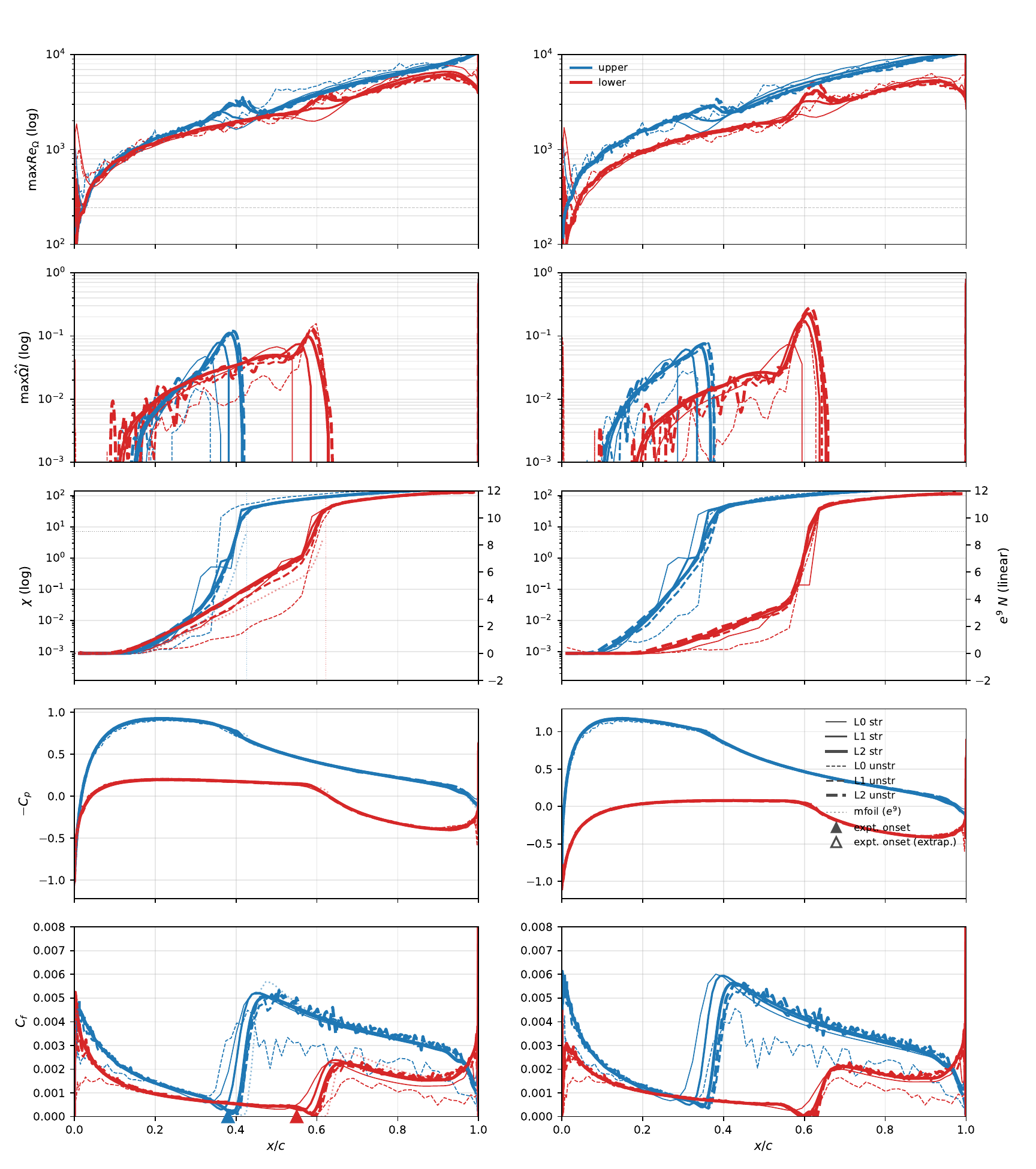}
\figcont{$\alpha=0^\circ$, $\alpha=2^\circ$.}
\end{figure}

\begin{figure}[H]\centering
\includegraphics[width=0.99\textwidth]{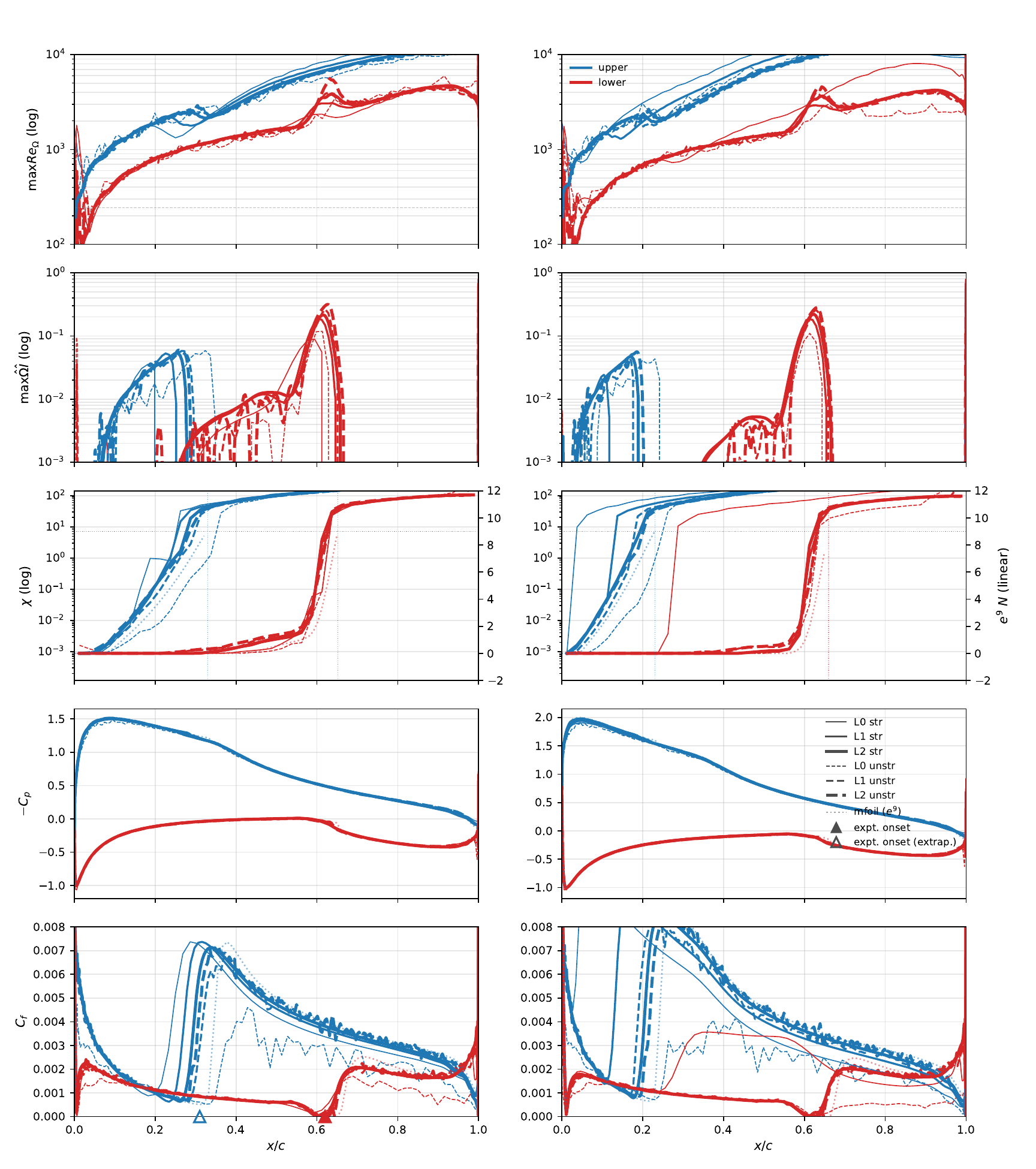}
\figcont{$\alpha=4^\circ$, $\alpha=6^\circ$.}
\end{figure}

\begin{figure}[H]\centering
\includegraphics[width=0.99\textwidth]{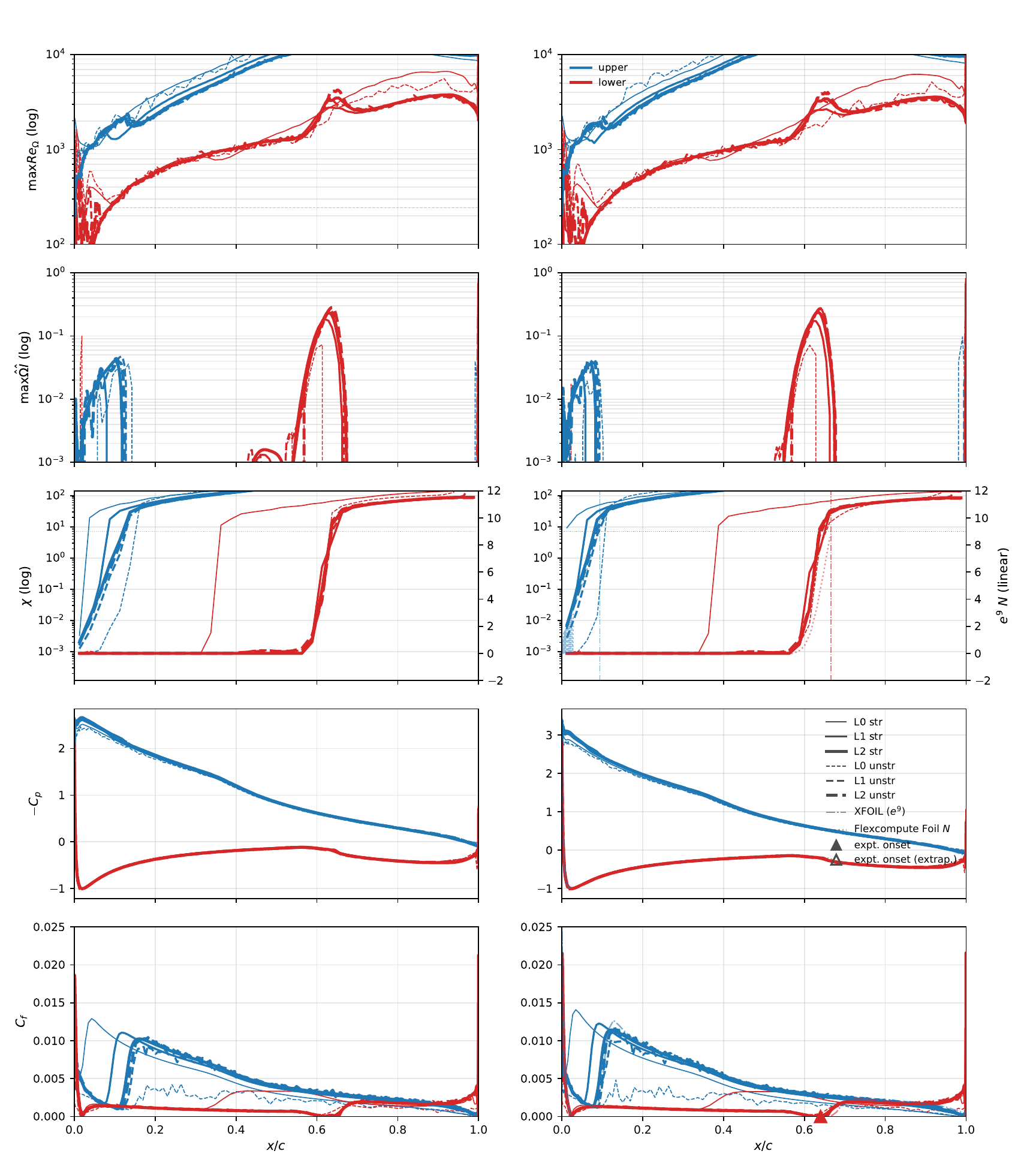}
\figcont{$\alpha=8^\circ$, $\alpha=9^\circ$.}
\end{figure}

\begin{figure}[H]\centering
\includegraphics[width=0.55\textwidth]{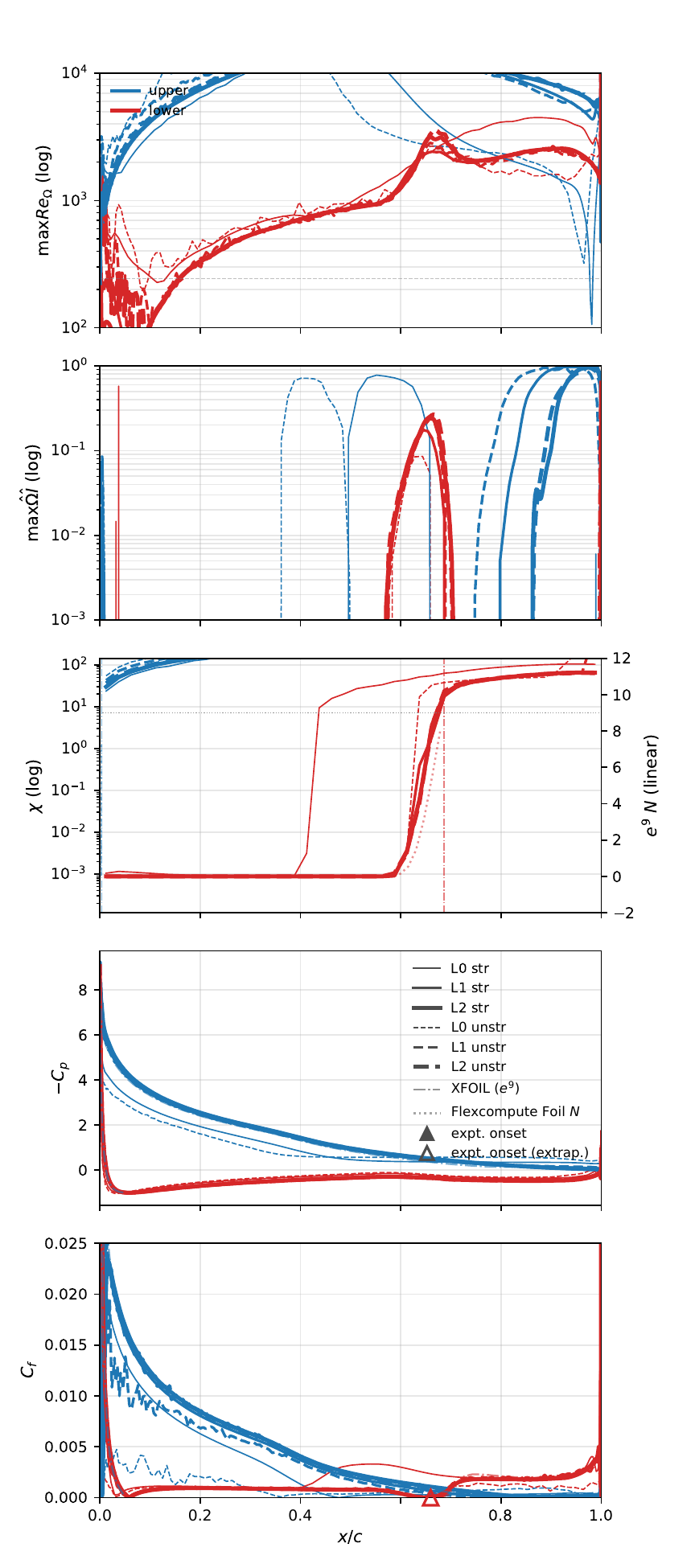}
\figcont{$\alpha=15^\circ$.}
\end{figure}

\begin{figure}[H]\centering
\includegraphics[width=\textwidth,height=0.94\textheight,keepaspectratio]{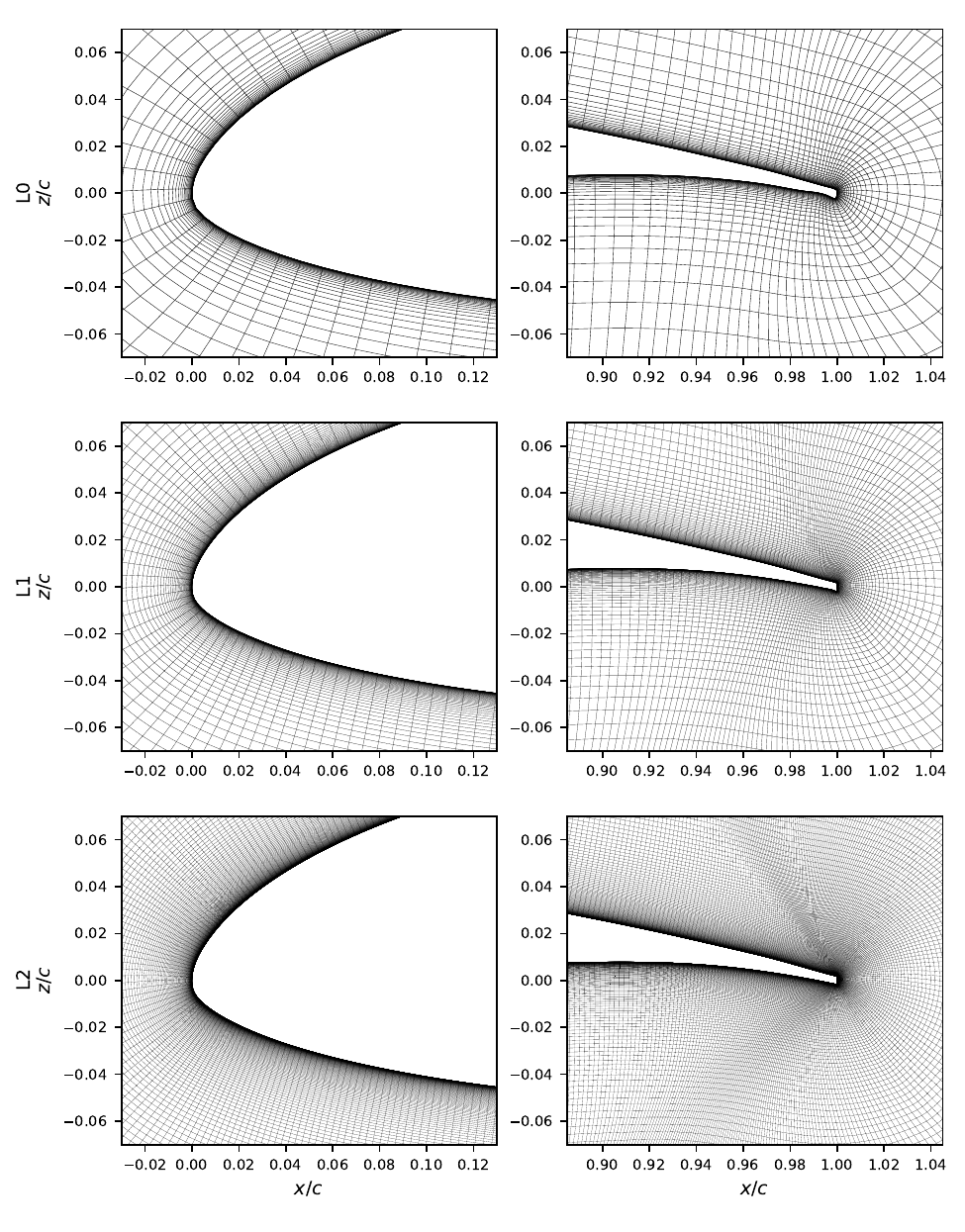}
\caption{NLF(1)-0416, structured O-grid: leading edge and trailing edge (columns) at levels L0 to L2 (rows).}\label{f:nlfmeshstr}
\end{figure}

\begin{figure}[H]\centering
\includegraphics[width=\textwidth,height=0.94\textheight,keepaspectratio]{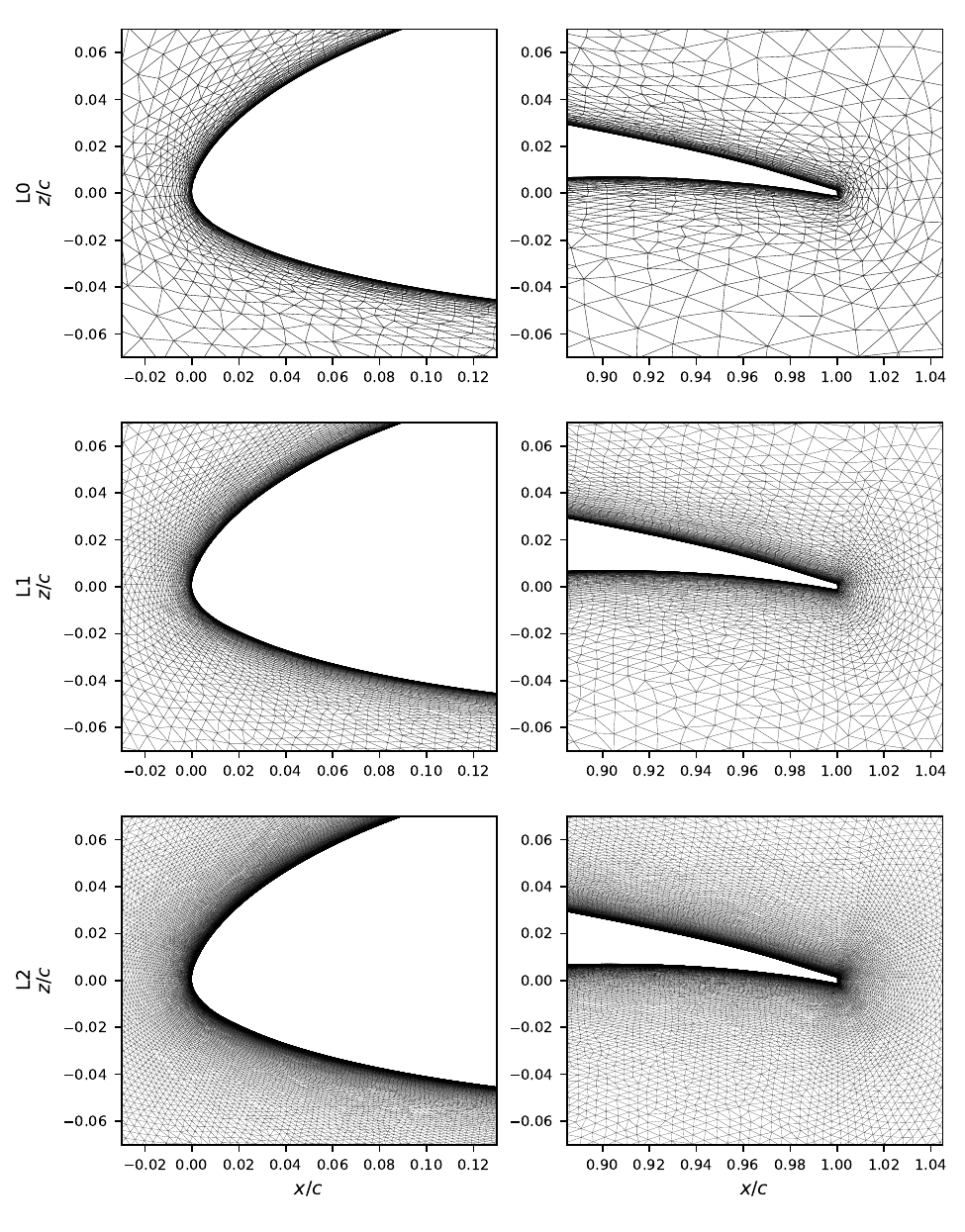}
\caption{NLF(1)-0416, unstructured grid: leading edge and trailing edge (columns) at levels L0 to L2 (rows).}\label{f:nlfmeshcav}
\end{figure}

\begin{figure}[H]\centering
\includegraphics[width=\textwidth,height=0.94\textheight,keepaspectratio]{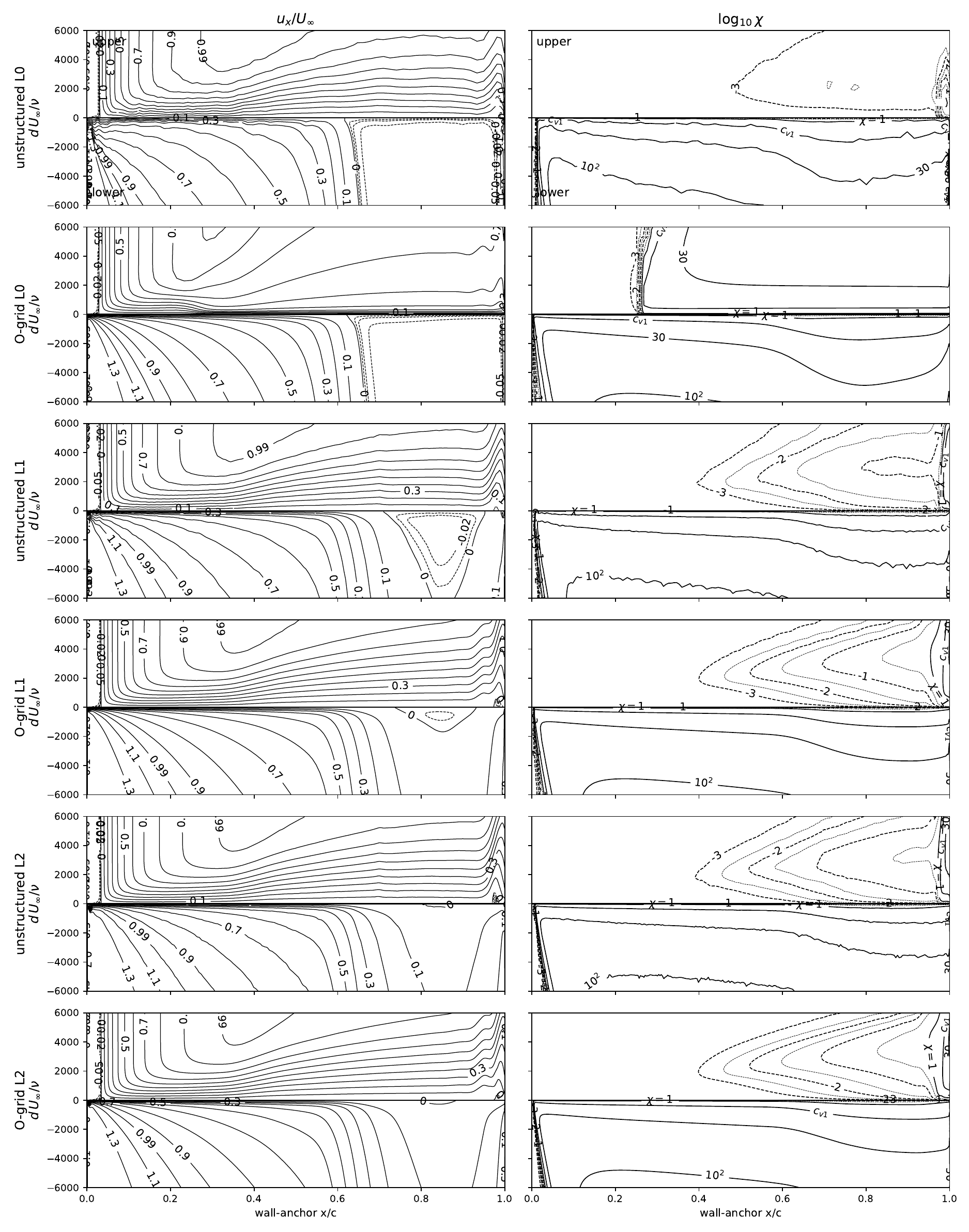}
\caption{NLF(1)-0416 near the wall, $Re=4\times10^6$, all six grids: $u_x/U_\infty$ (left) and $\log_{10}\chi$ (right) against the chordwise station and the wall distance. The upper surface is above $y=0$, and the lower surface is reflected below it. $\alpha=-12^\circ$.}\label{f:nlfsheets}
\end{figure}

\begin{figure}[H]\centering
\includegraphics[width=\textwidth,height=0.94\textheight,keepaspectratio]{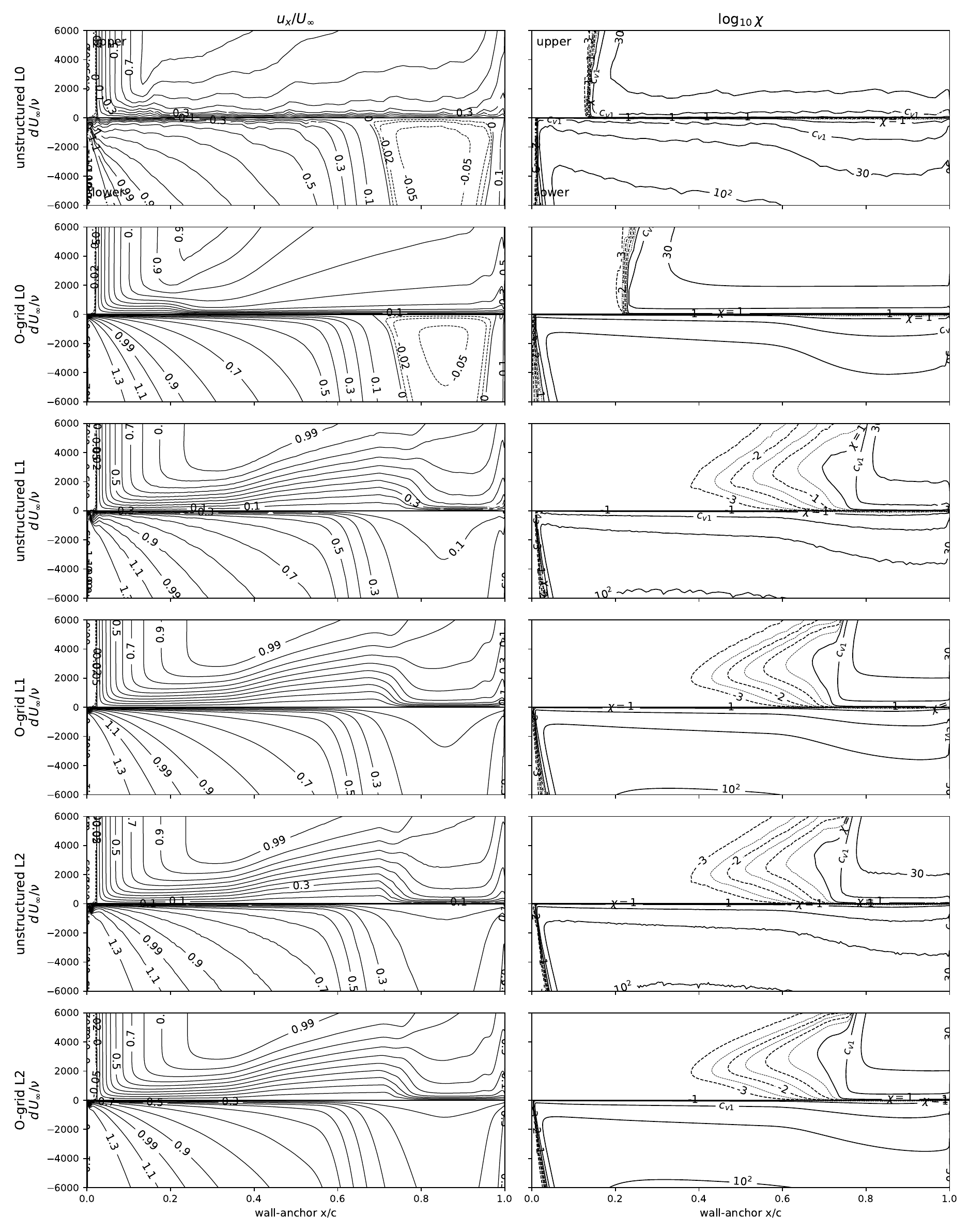}
\figcont{$\alpha=-10^\circ$.}
\end{figure}

\begin{figure}[H]\centering
\includegraphics[width=\textwidth,height=0.94\textheight,keepaspectratio]{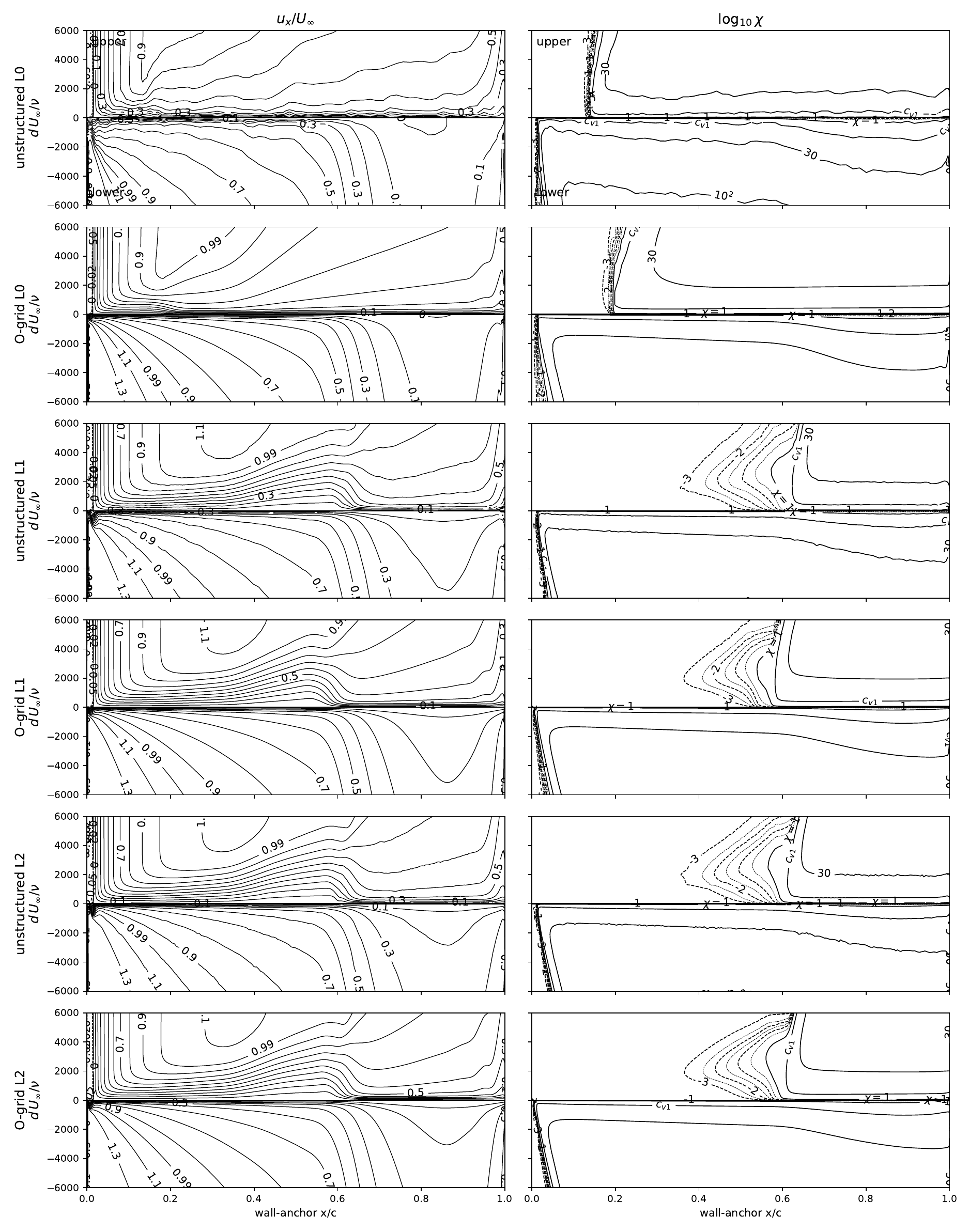}
\figcont{$\alpha=-8^\circ$.}
\end{figure}

\begin{figure}[H]\centering
\includegraphics[width=\textwidth,height=0.94\textheight,keepaspectratio]{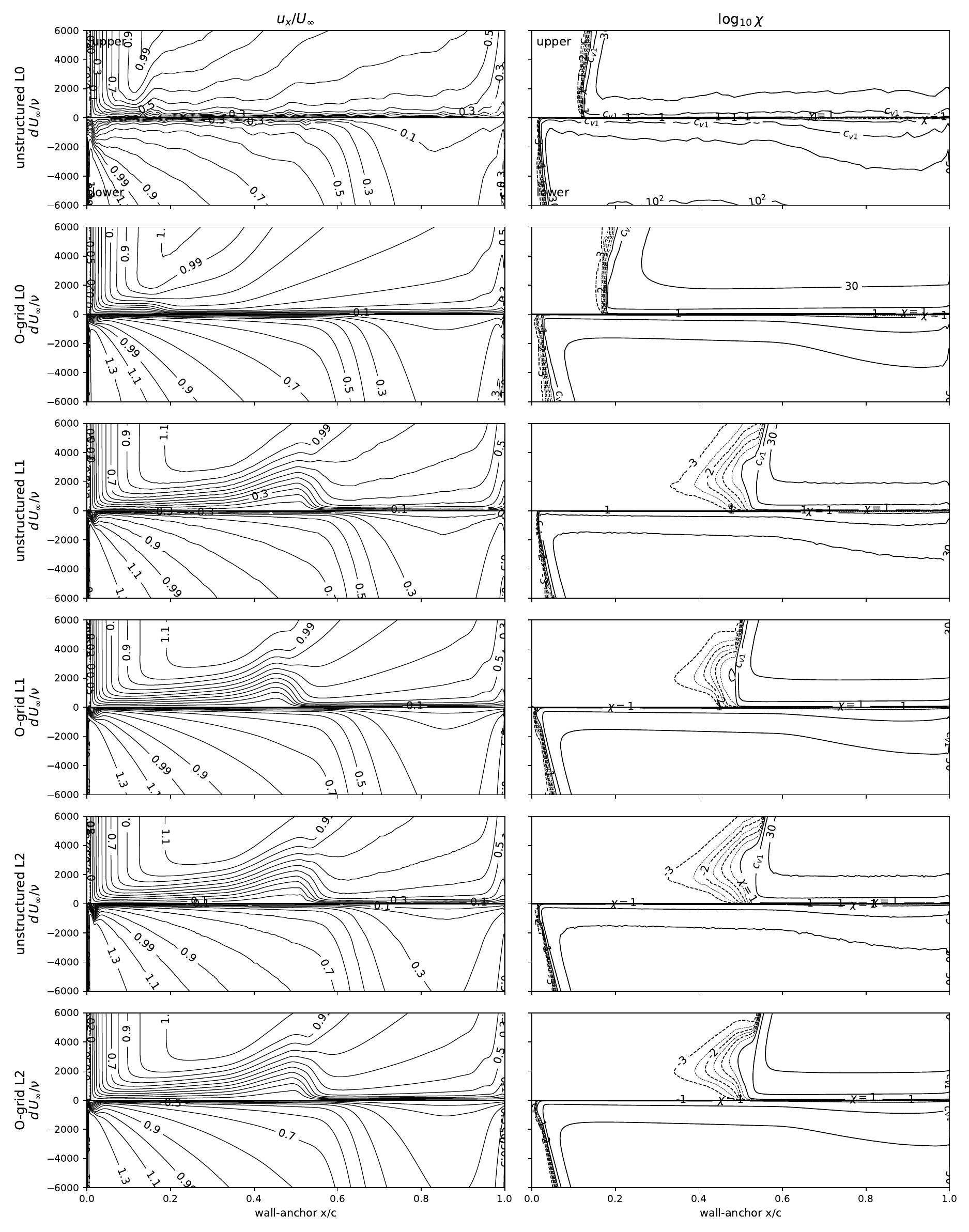}
\figcont{$\alpha=-6^\circ$.}
\end{figure}

\begin{figure}[H]\centering
\includegraphics[width=\textwidth,height=0.94\textheight,keepaspectratio]{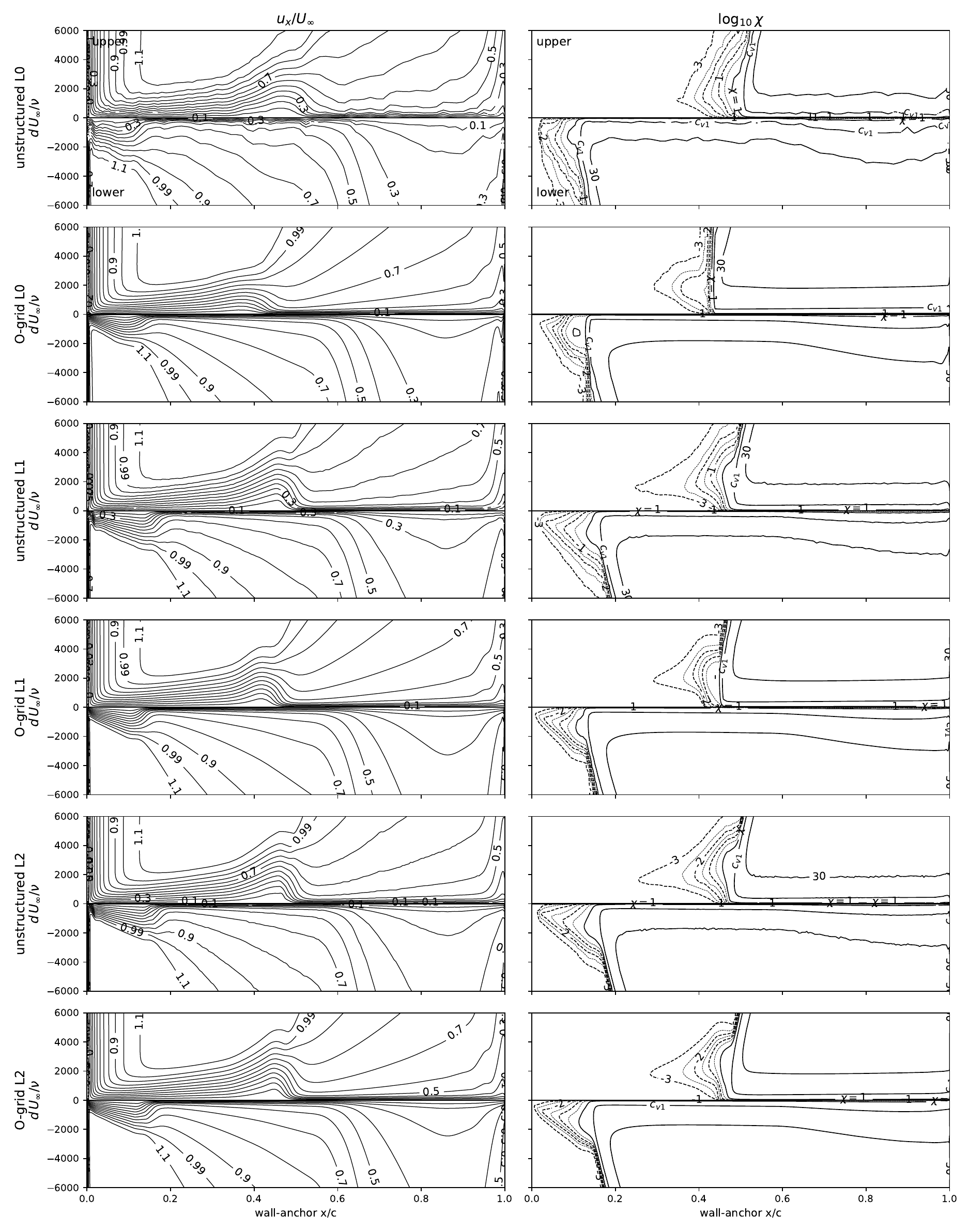}
\figcont{$\alpha=-4^\circ$.}
\end{figure}

\begin{figure}[H]\centering
\includegraphics[width=\textwidth,height=0.94\textheight,keepaspectratio]{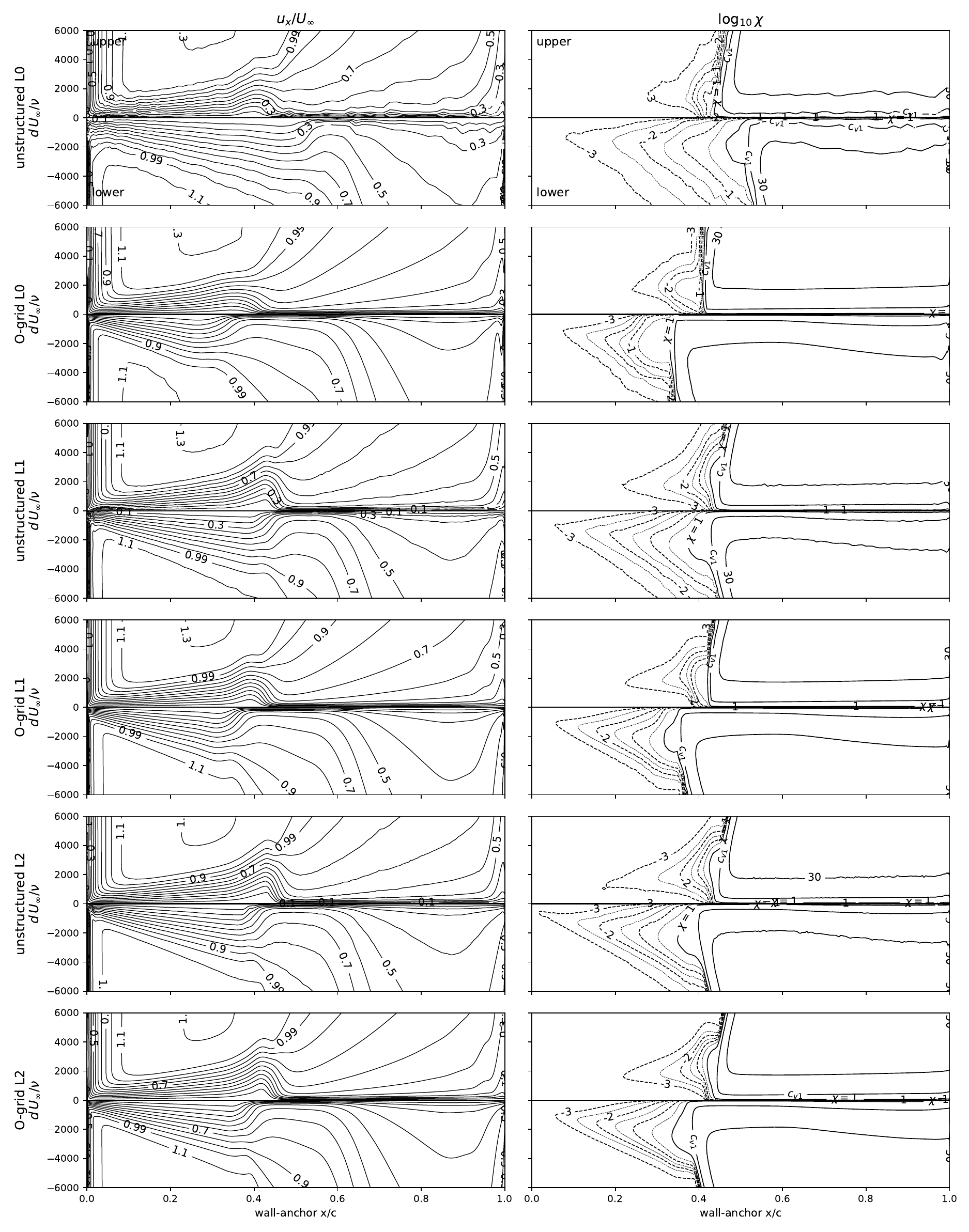}
\figcont{$\alpha=-2^\circ$.}
\end{figure}

\begin{figure}[H]\centering
\includegraphics[width=\textwidth,height=0.94\textheight,keepaspectratio]{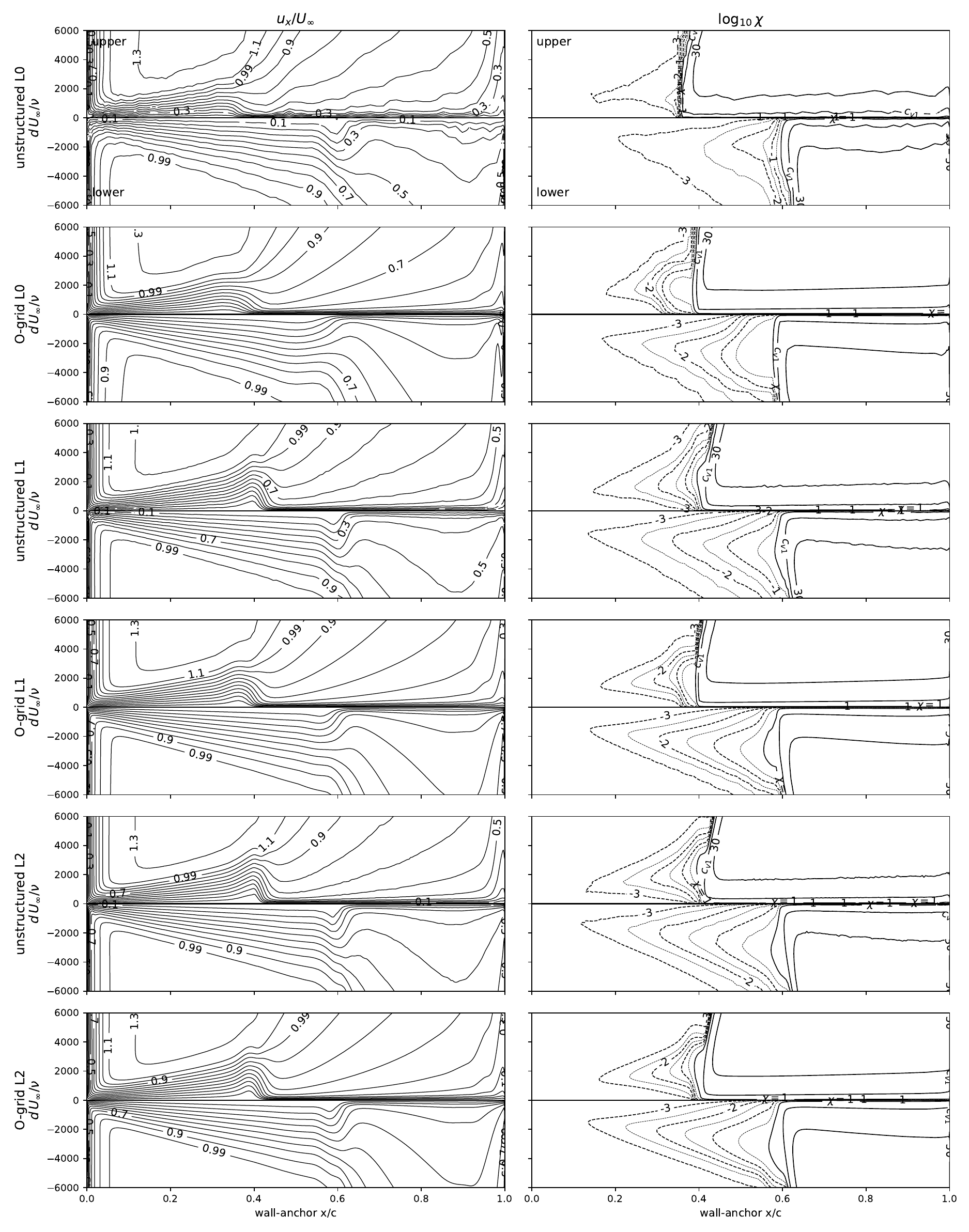}
\figcont{$\alpha=0^\circ$.}
\end{figure}

\begin{figure}[H]\centering
\includegraphics[width=\textwidth,height=0.94\textheight,keepaspectratio]{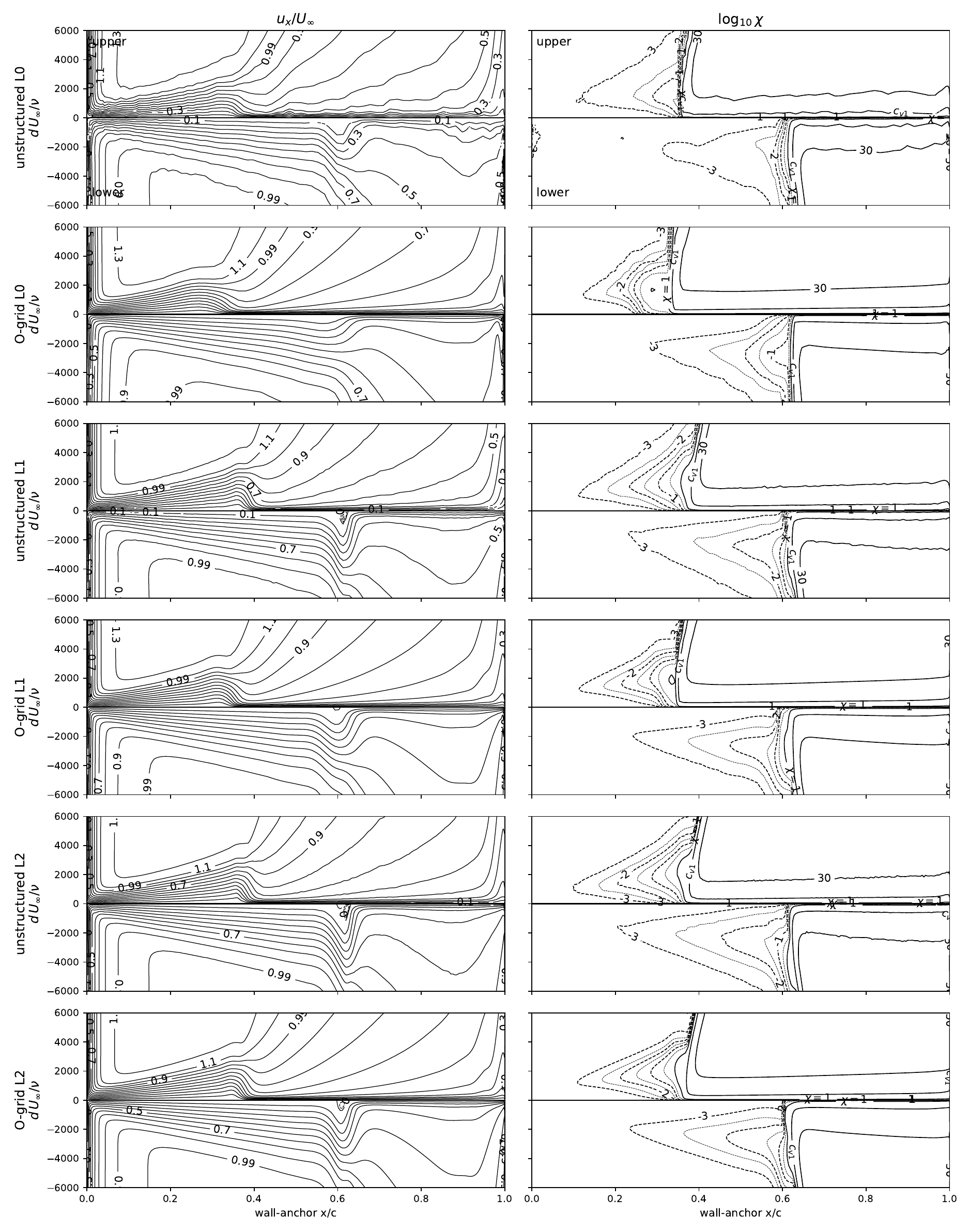}
\figcont{$\alpha=2^\circ$.}
\end{figure}

\begin{figure}[H]\centering
\includegraphics[width=\textwidth,height=0.94\textheight,keepaspectratio]{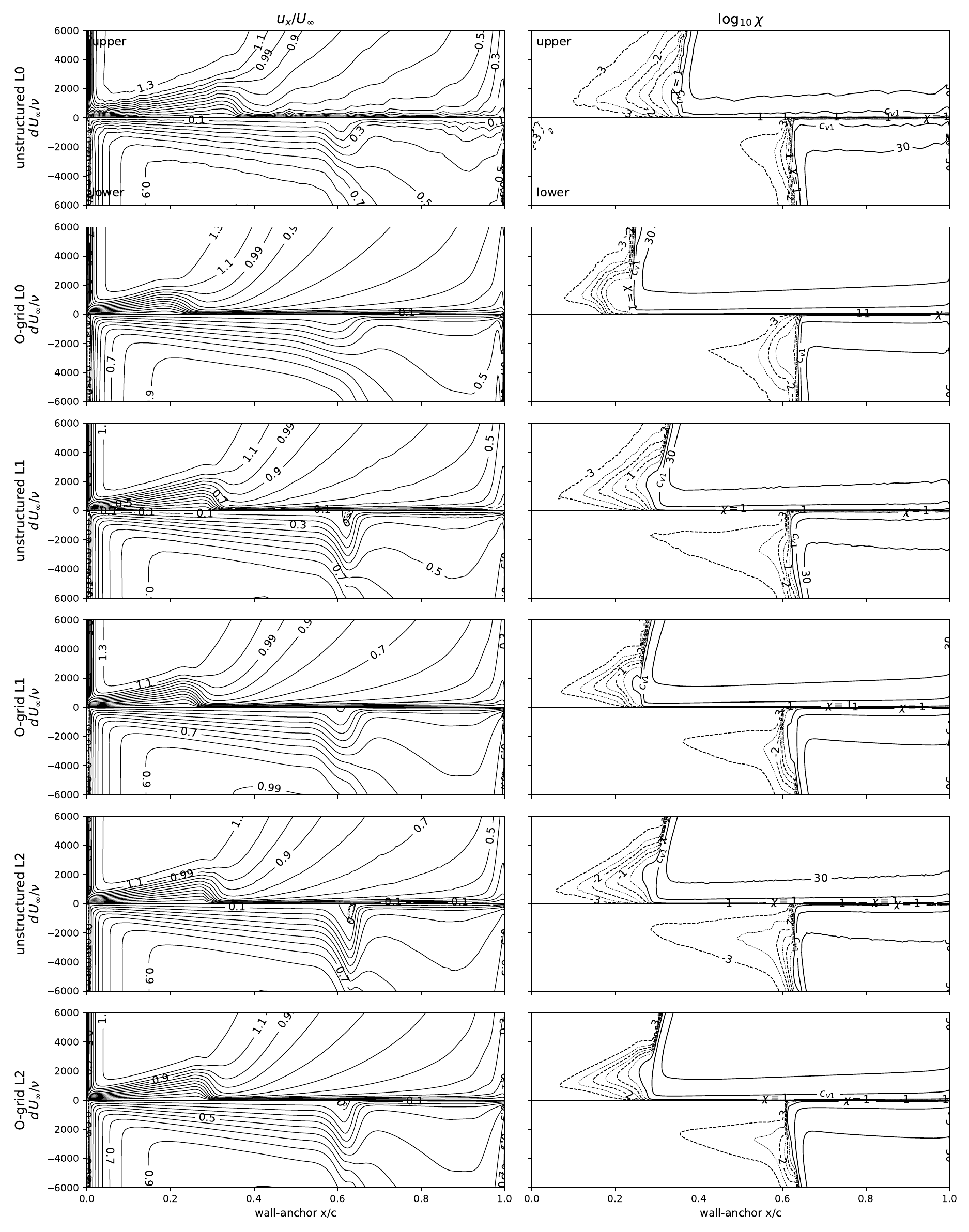}
\figcont{$\alpha=4^\circ$.}
\end{figure}

\begin{figure}[H]\centering
\includegraphics[width=\textwidth,height=0.94\textheight,keepaspectratio]{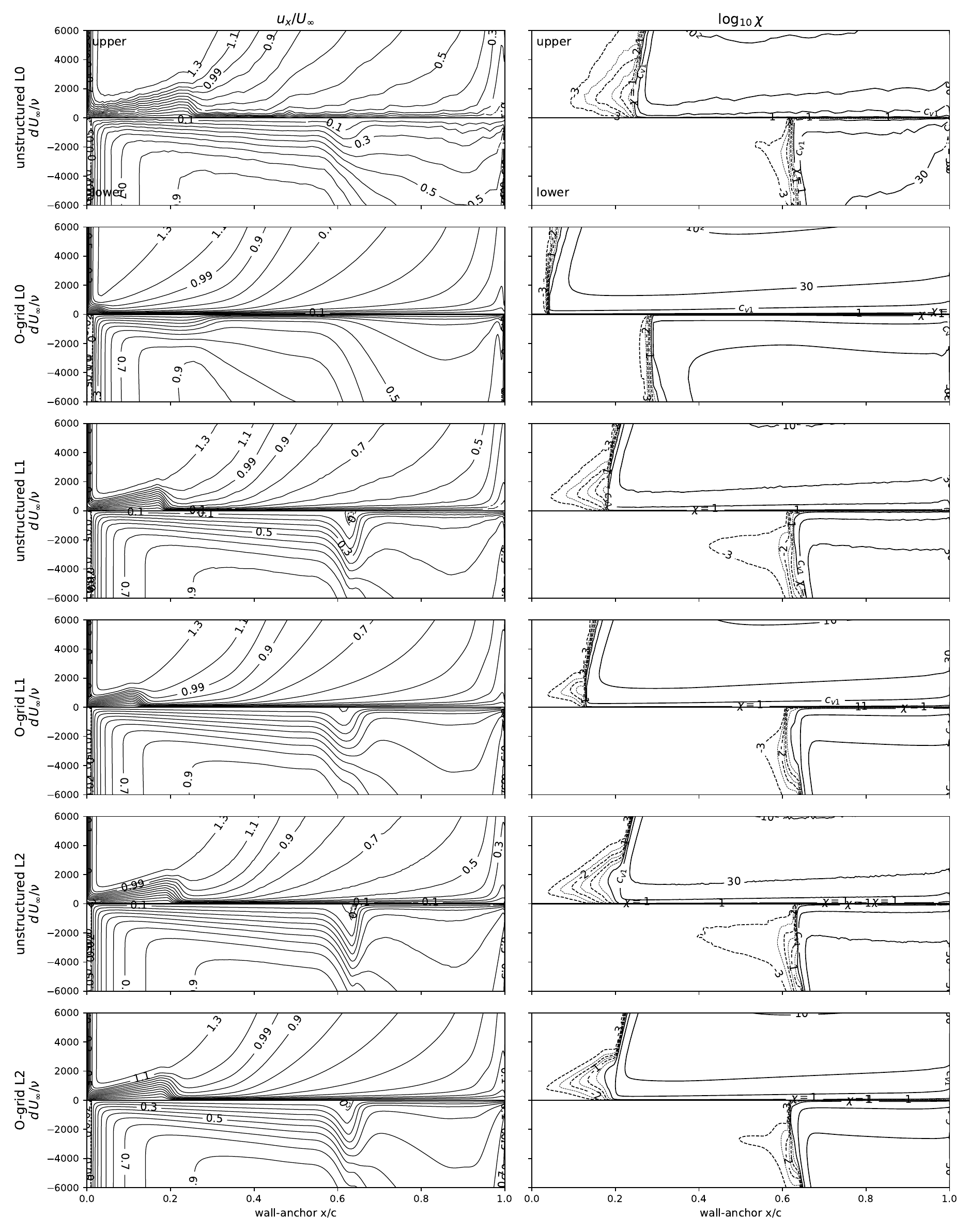}
\figcont{$\alpha=6^\circ$.}
\end{figure}

\begin{figure}[H]\centering
\includegraphics[width=\textwidth,height=0.94\textheight,keepaspectratio]{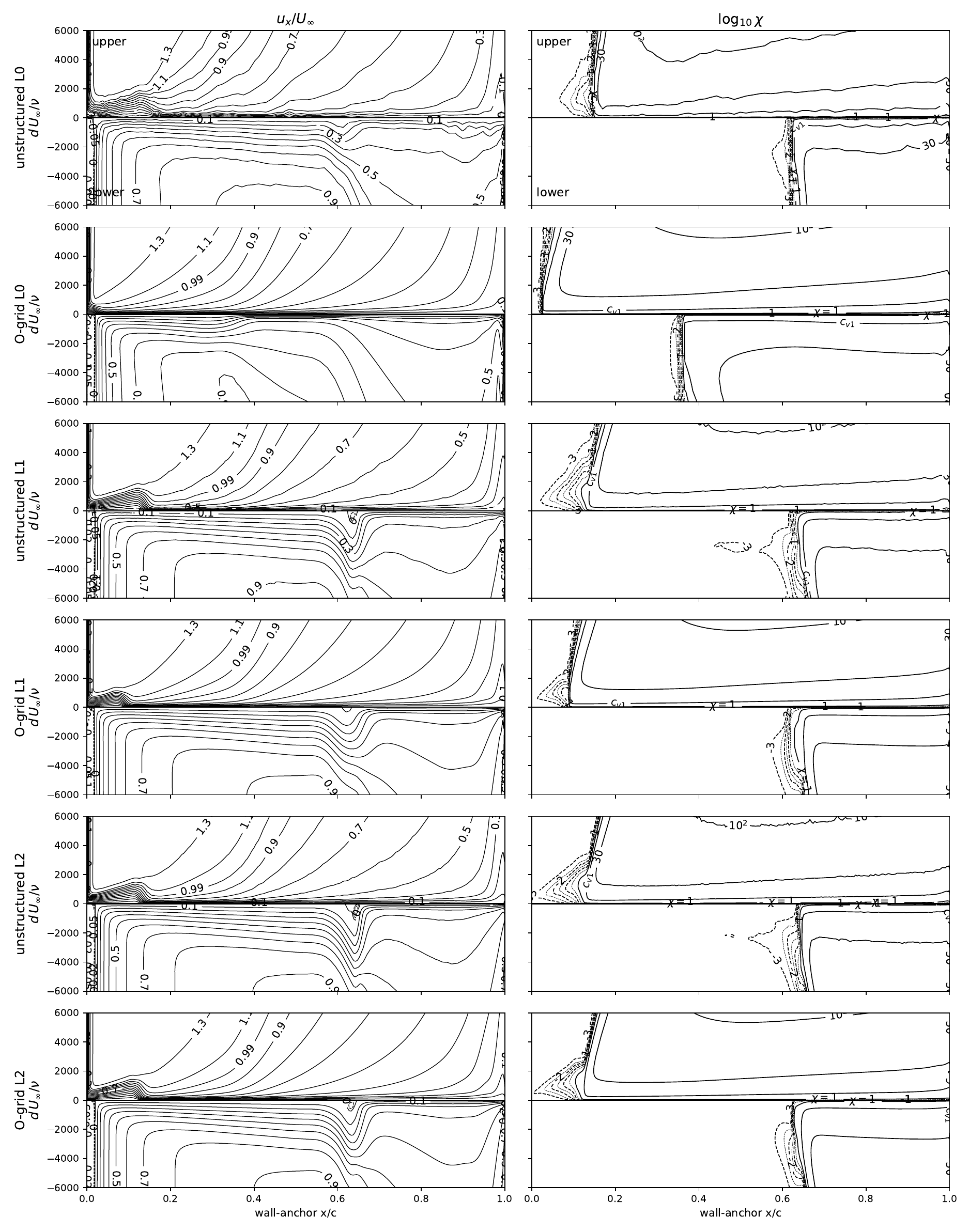}
\figcont{$\alpha=8^\circ$.}
\end{figure}

\begin{figure}[H]\centering
\includegraphics[width=\textwidth,height=0.94\textheight,keepaspectratio]{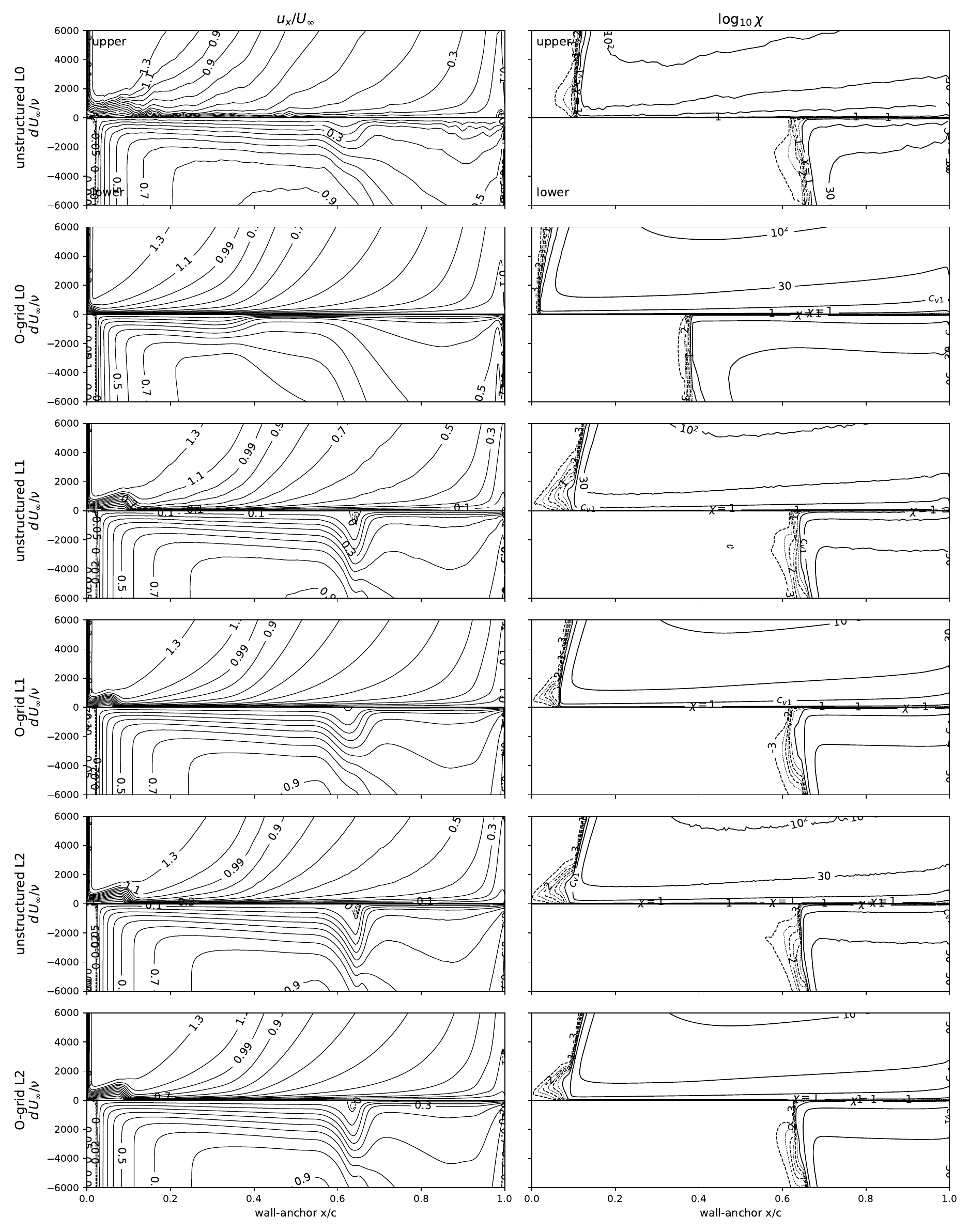}
\figcont{$\alpha=9^\circ$.}
\end{figure}

\begin{figure}[H]\centering
\includegraphics[width=\textwidth,height=0.94\textheight,keepaspectratio]{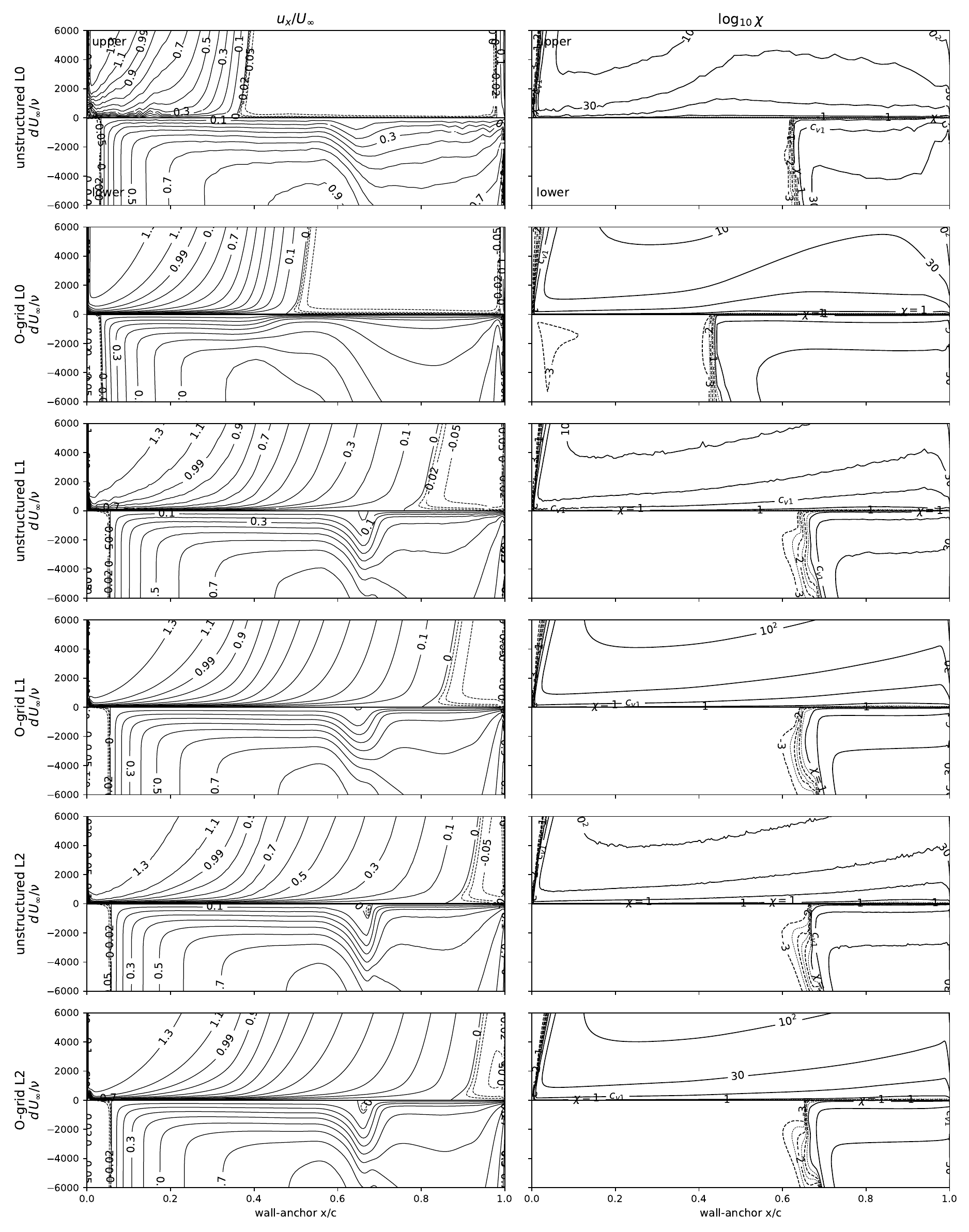}
\figcont{$\alpha=15^\circ$.}
\end{figure}

\subsection*{\tocentry{subsection}{The Eppler 387 Airfoil}The Eppler 387 Airfoil}

\begin{figure}[H]\centering
\includegraphics[width=0.99\textwidth]{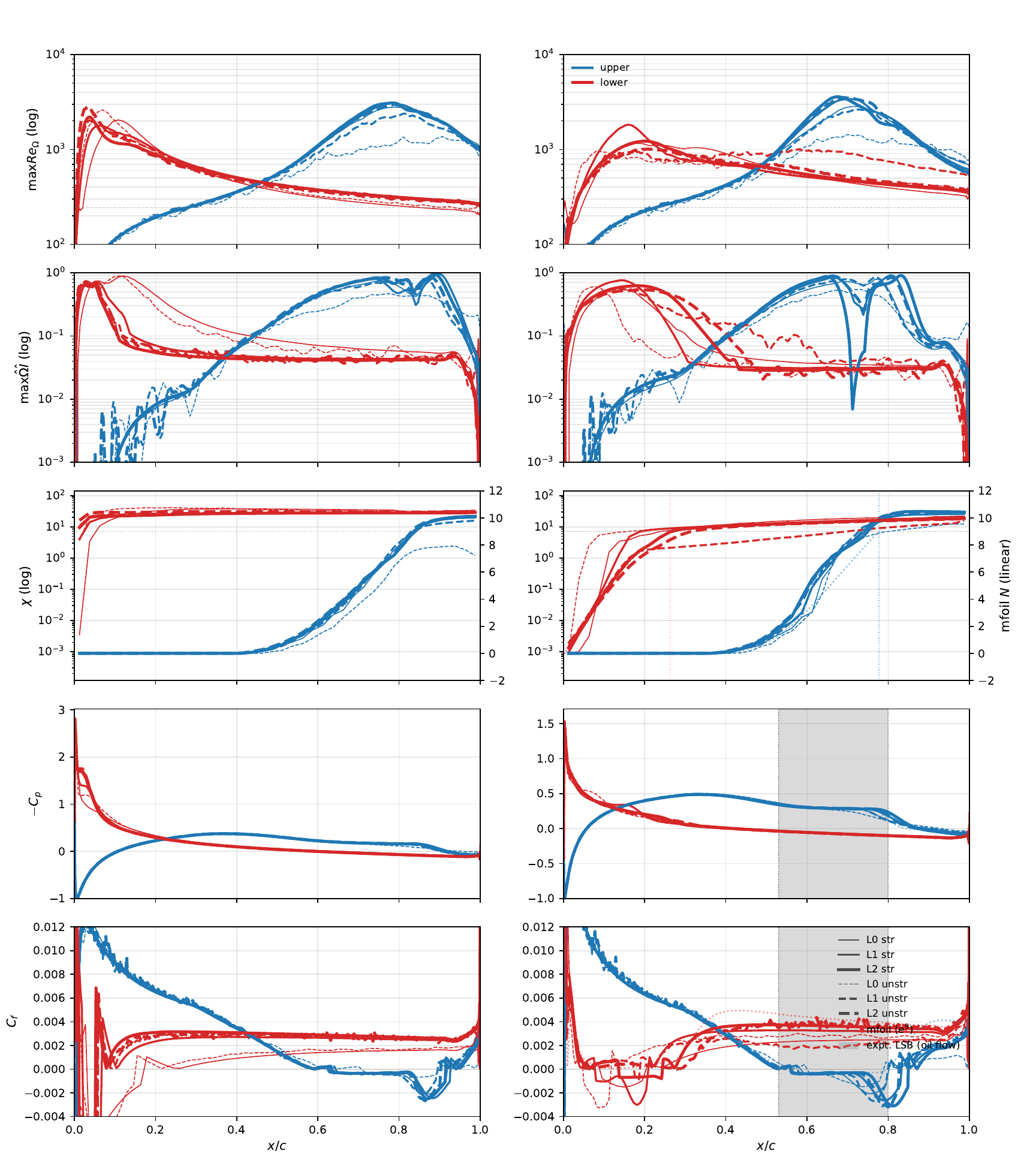}
\caption{Eppler~387 surface distributions, $Re=2\times10^5$, all six grids, in ascending incidence; rows as in Fig.~\ref{f:nlfcf0}. $\alpha=-4^\circ$, $\alpha=-2^\circ$.}\label{f:eppcflow}
\end{figure}

\begin{figure}[H]\centering
\includegraphics[width=0.99\textwidth]{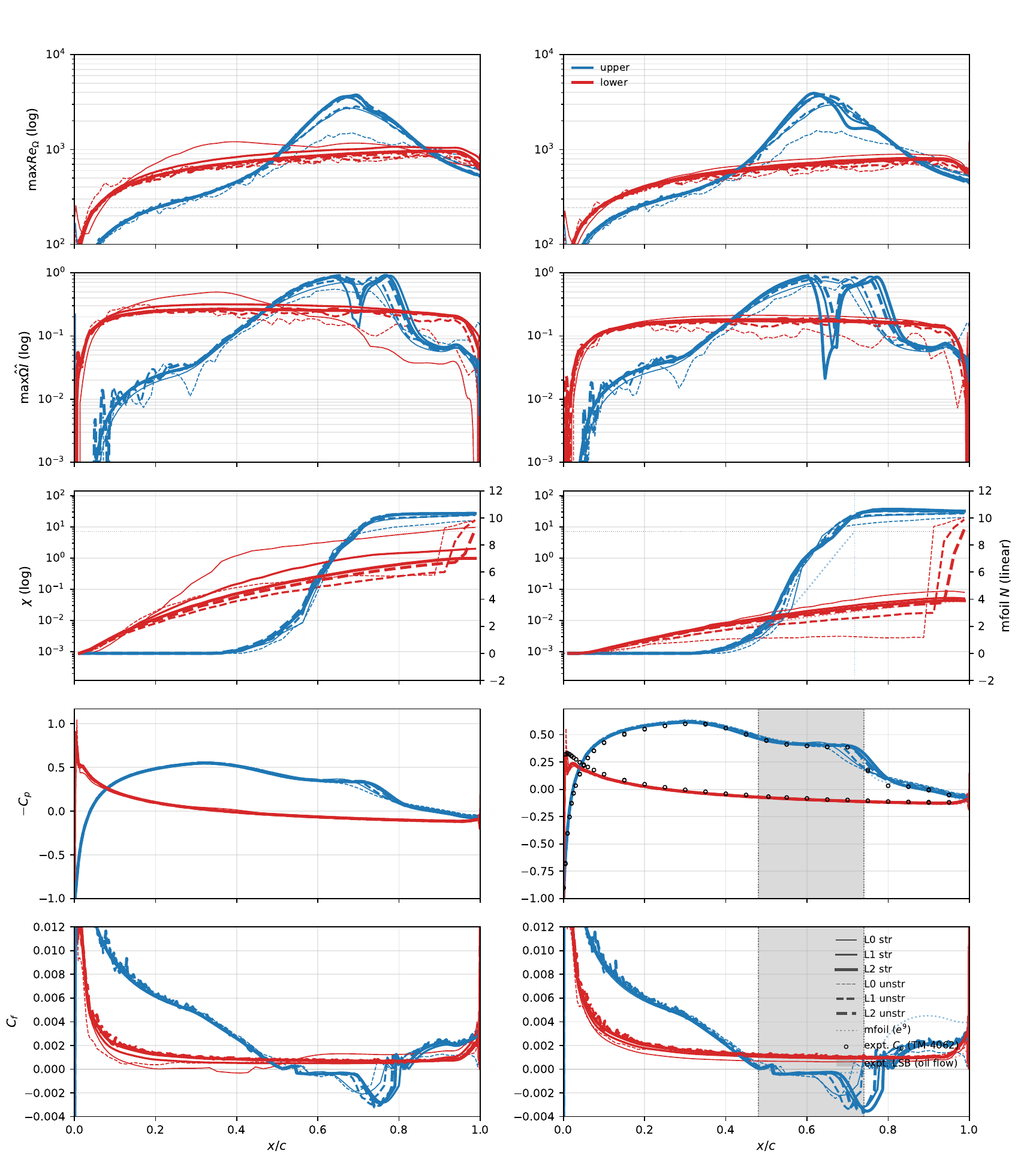}
\figcont{$\alpha=-1^\circ$, $\alpha=0^\circ$.}
\end{figure}

\begin{figure}[H]\centering
\includegraphics[width=0.99\textwidth]{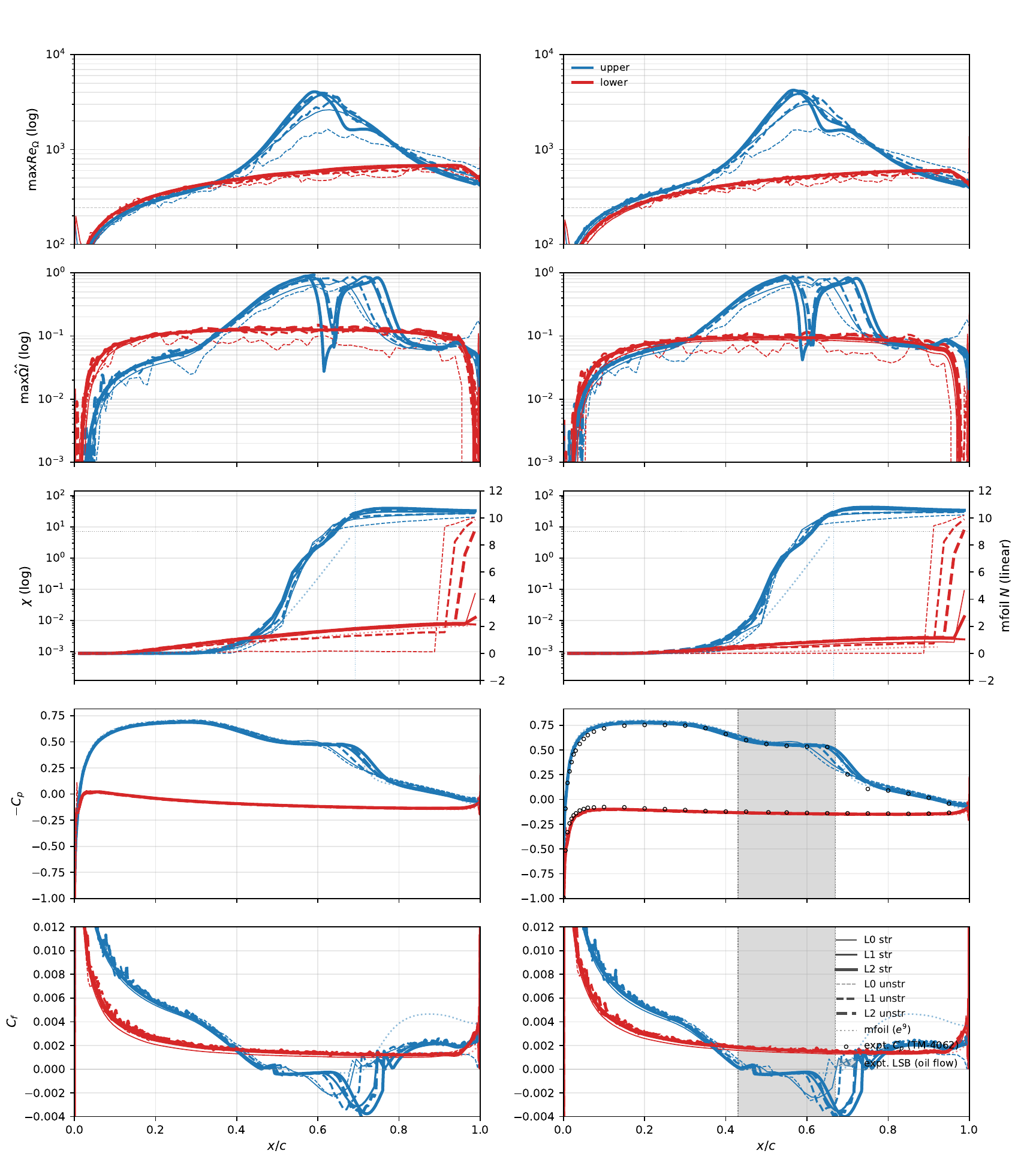}
\figcont{$\alpha=1^\circ$, $\alpha=2^\circ$.}
\end{figure}

\begin{figure}[H]\centering
\includegraphics[width=0.99\textwidth]{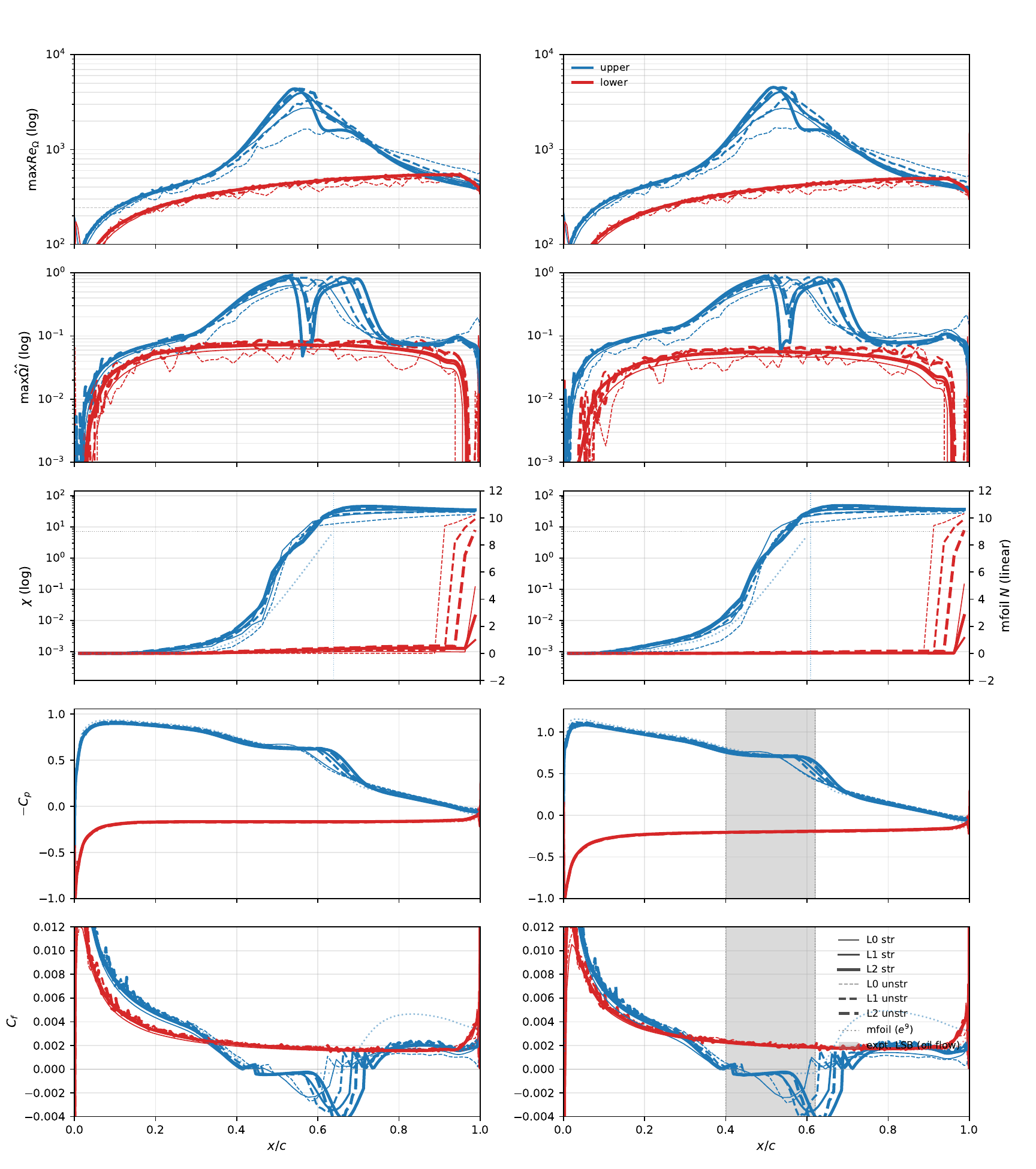}
\figcont{$\alpha=3^\circ$, $\alpha=4^\circ$.}
\end{figure}

\begin{figure}[H]\centering
\includegraphics[width=0.99\textwidth]{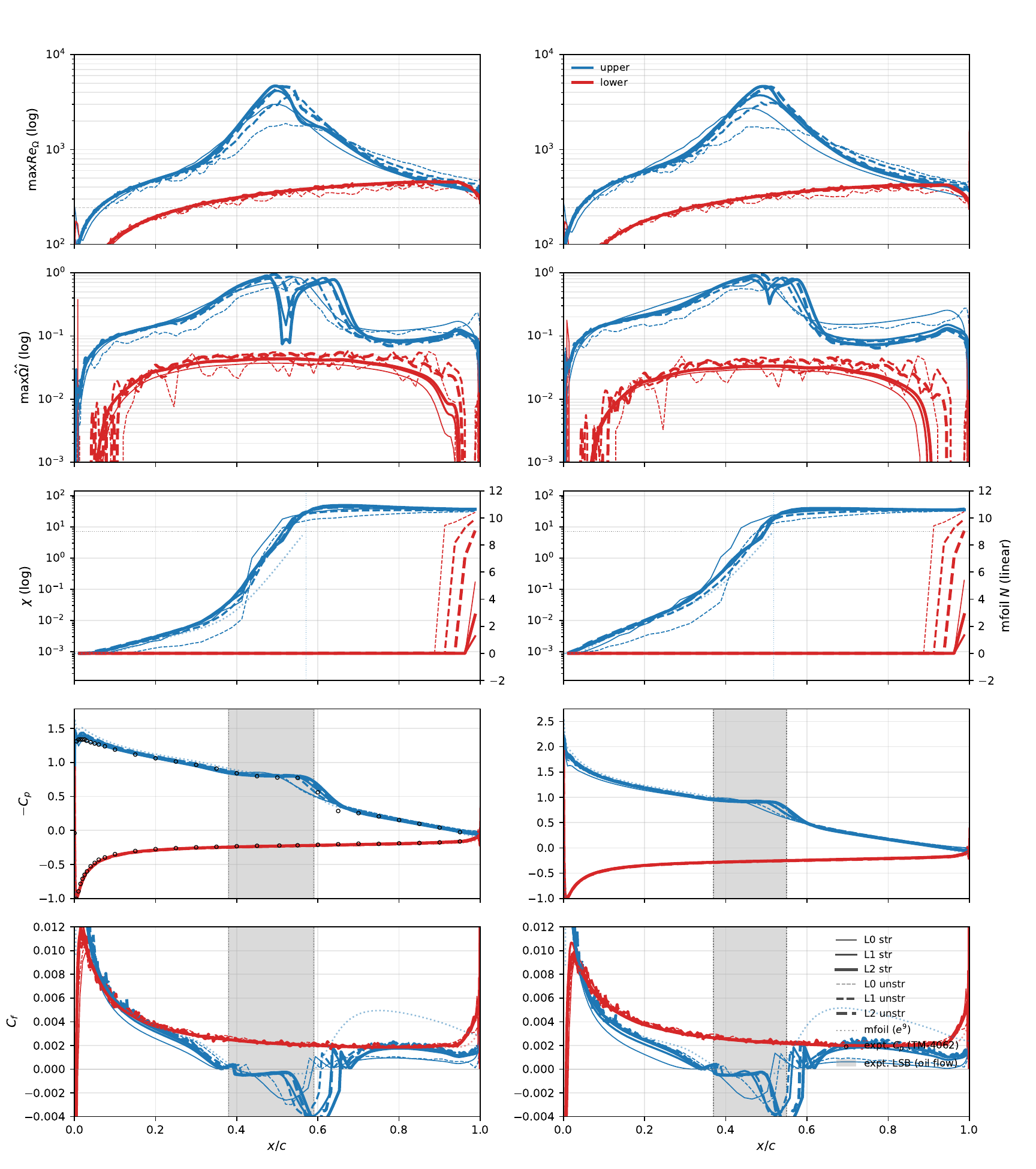}
\figcont{$\alpha=5^\circ$, $\alpha=6^\circ$.}
\end{figure}

\begin{figure}[H]\centering
\includegraphics[width=0.99\textwidth]{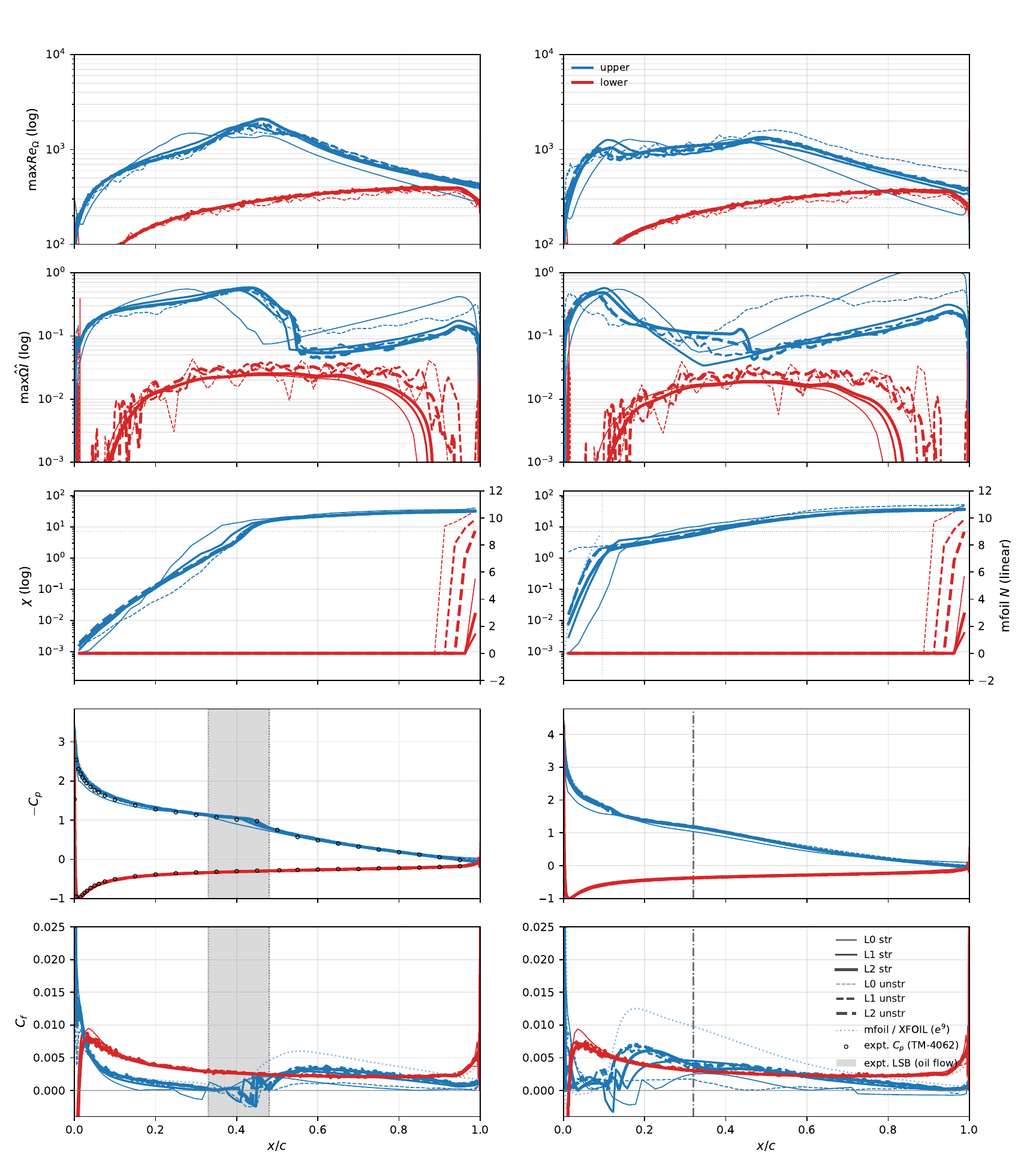}
\figcont{$\alpha=7^\circ$, $\alpha=8^\circ$.}
\end{figure}

\begin{figure}[H]\centering
\includegraphics[width=0.99\textwidth]{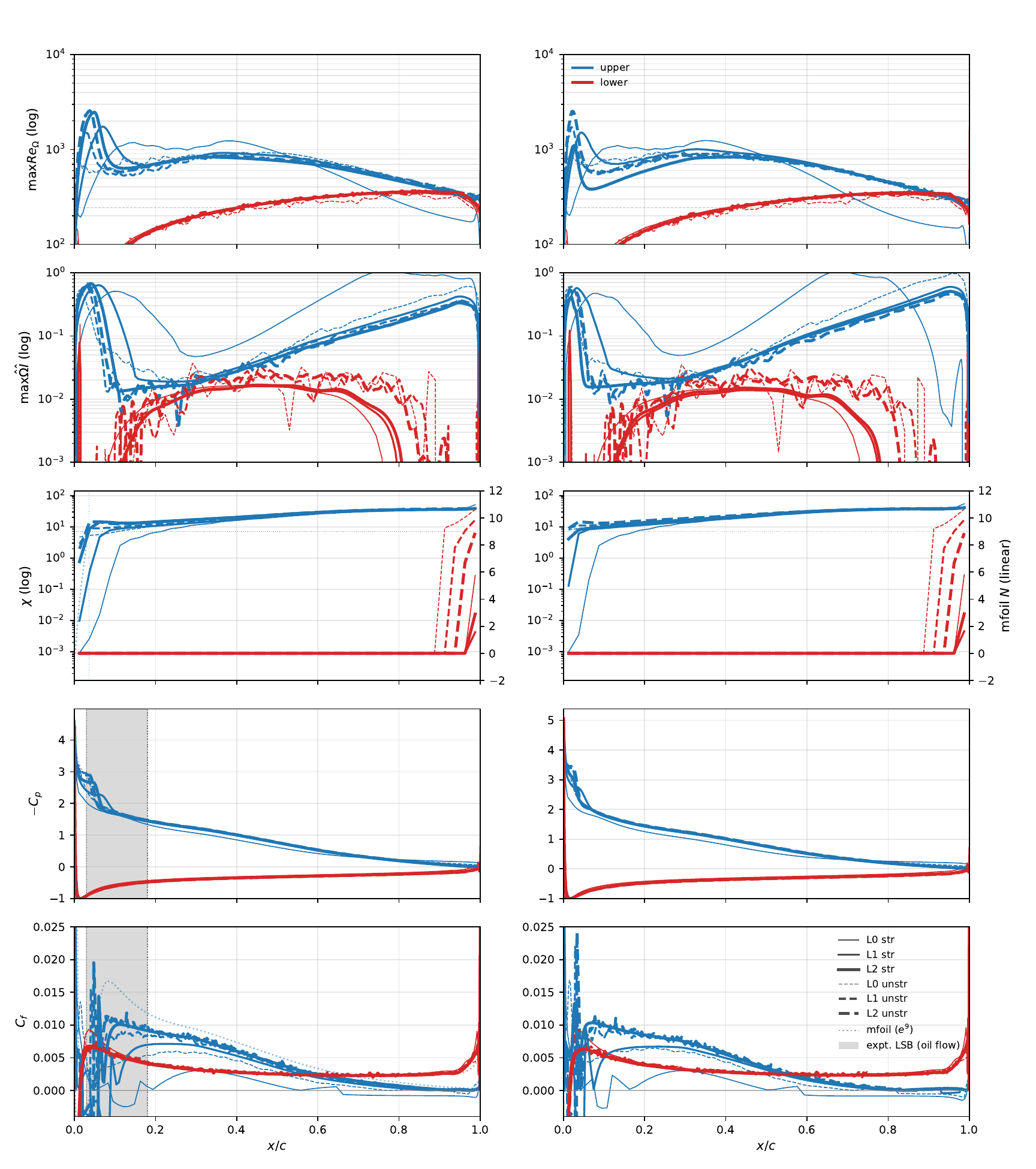}
\figcont{$\alpha=8.5^\circ$, $\alpha=9^\circ$.}
\end{figure}

\begin{figure}[H]\centering
\includegraphics[width=0.99\textwidth,height=0.94\textheight,keepaspectratio]{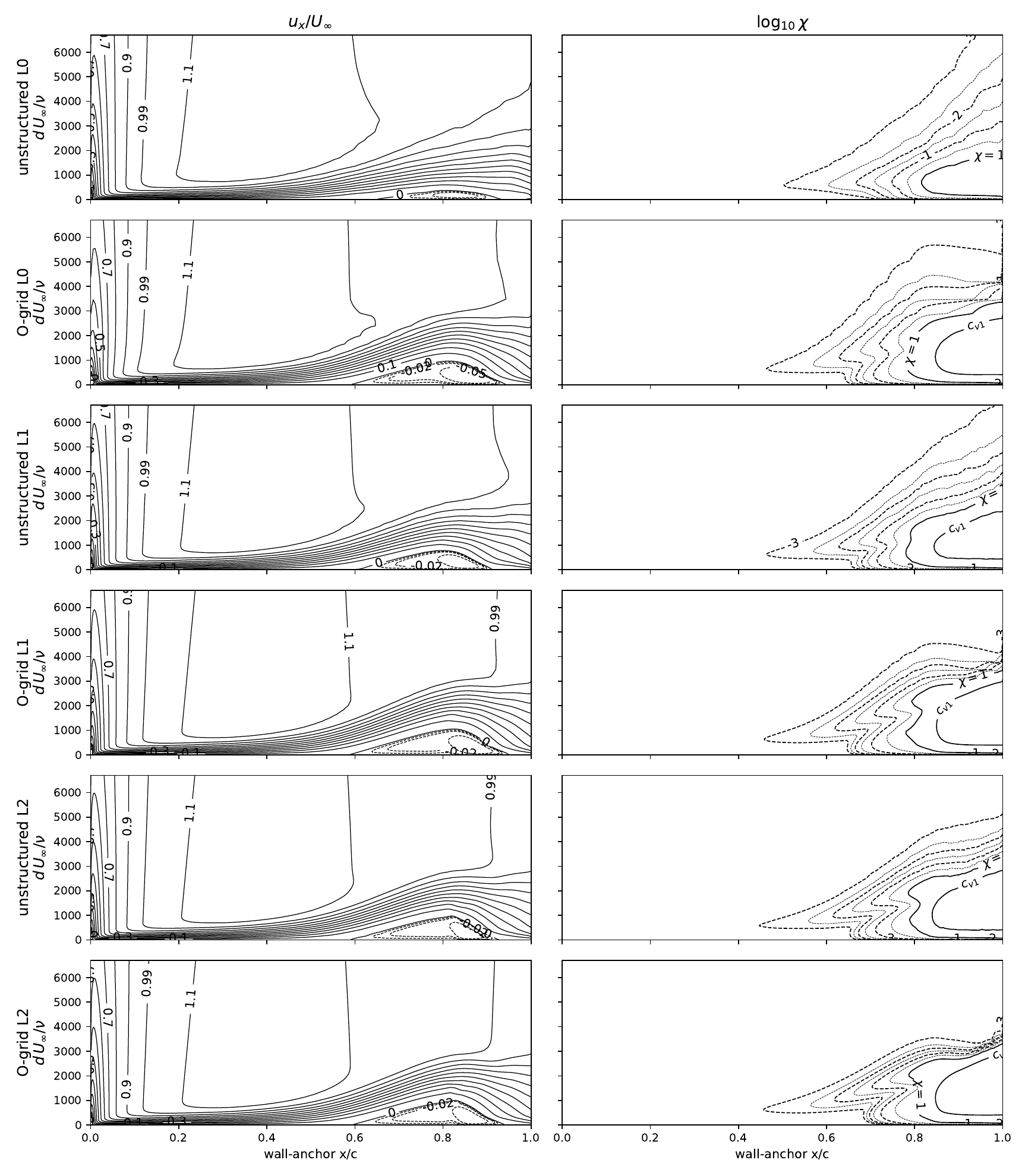}
\caption{Eppler~387 near the upper surface, $Re=2\times10^5$, all six grids: $u_x/U_\infty$ (left) and $\log_{10}\chi$ (right) against the chordwise station and the wall distance. $\alpha=-4^\circ$.}\label{f:eppsheets}
\end{figure}

\begin{figure}[H]\centering
\includegraphics[width=0.99\textwidth,height=0.94\textheight,keepaspectratio]{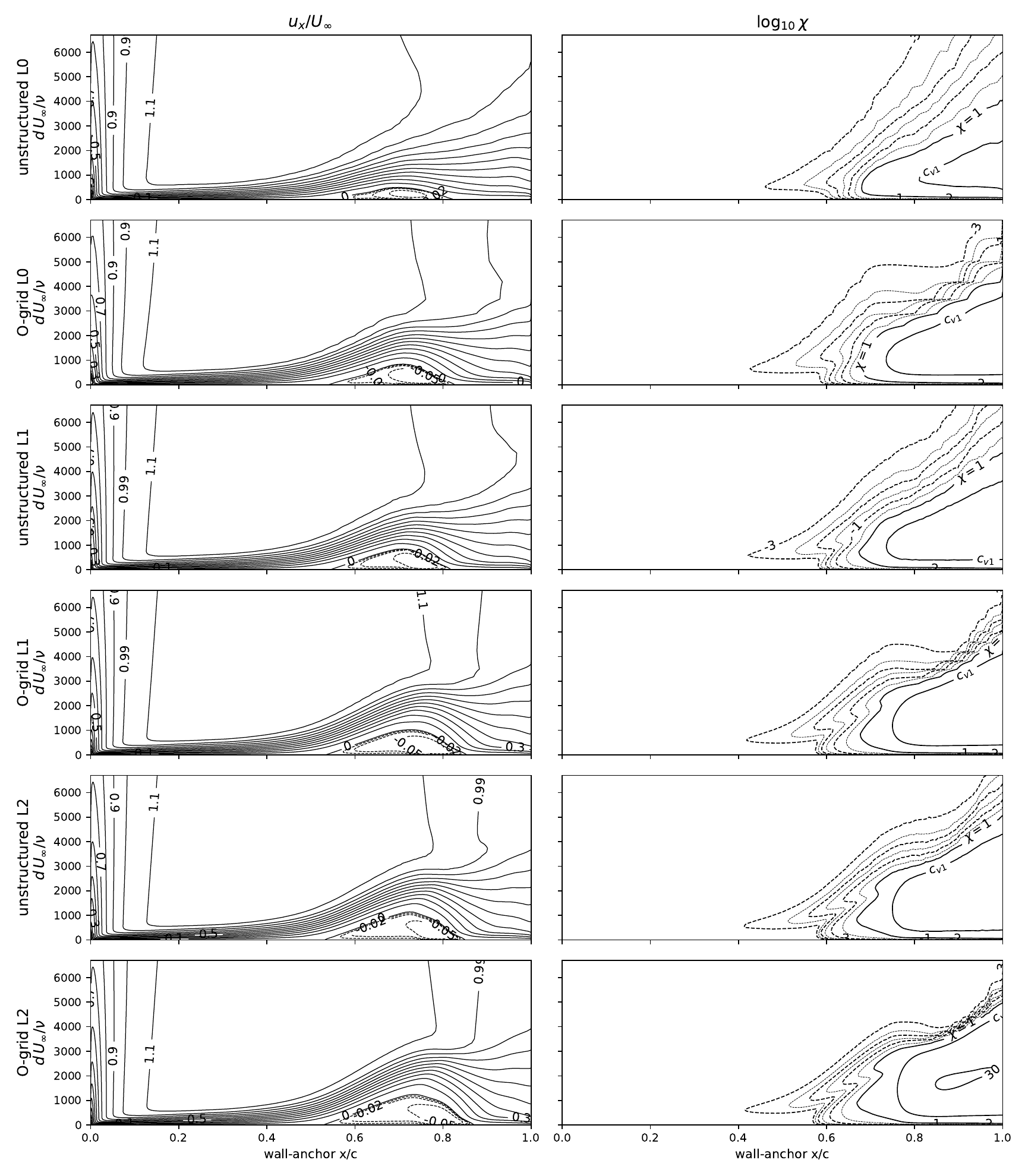}
\figcont{$\alpha=-2^\circ$.}
\end{figure}

\begin{figure}[H]\centering
\includegraphics[width=0.99\textwidth,height=0.94\textheight,keepaspectratio]{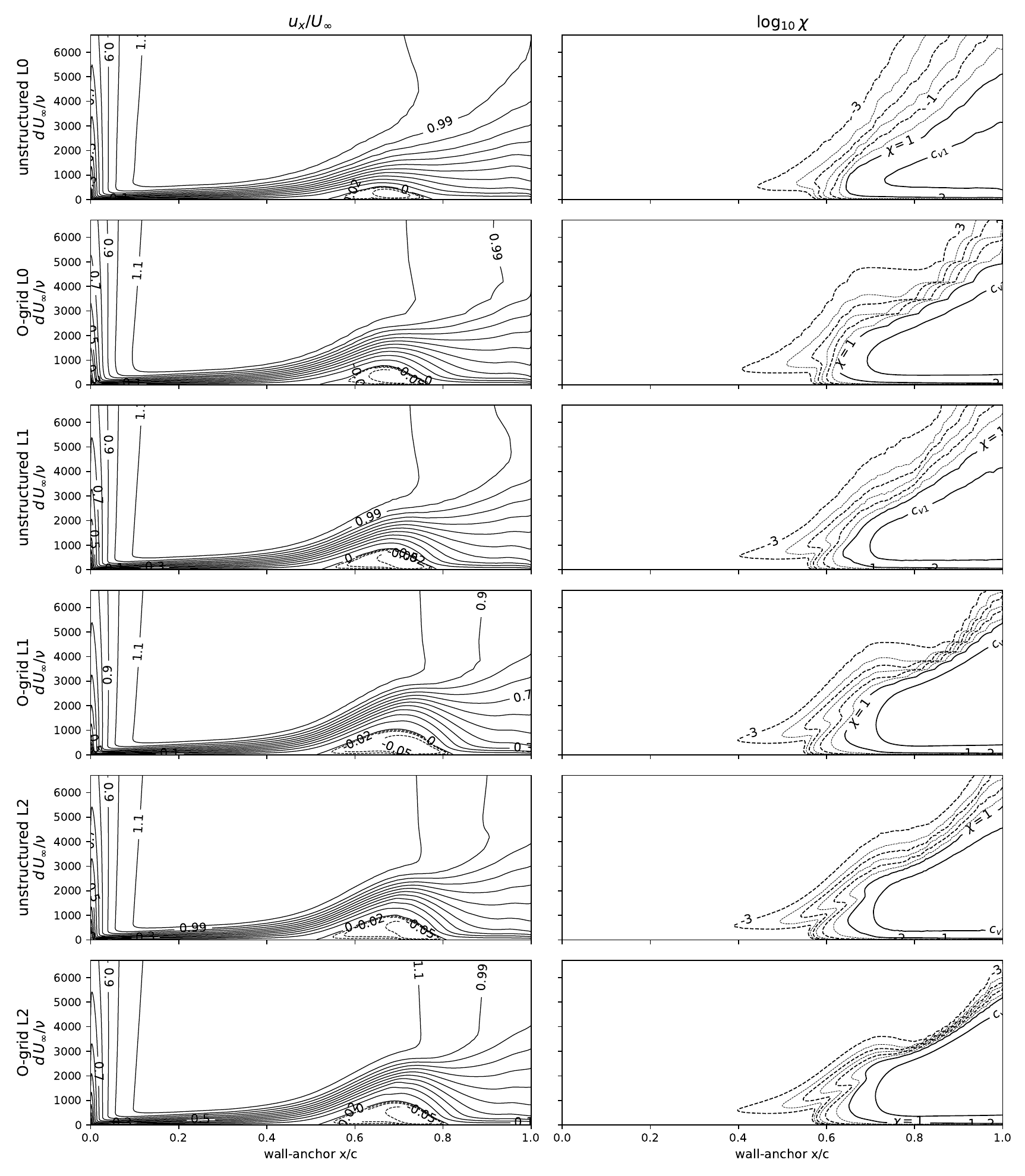}
\figcont{$\alpha=-1^\circ$.}
\end{figure}

\begin{figure}[H]\centering
\includegraphics[width=0.99\textwidth,height=0.94\textheight,keepaspectratio]{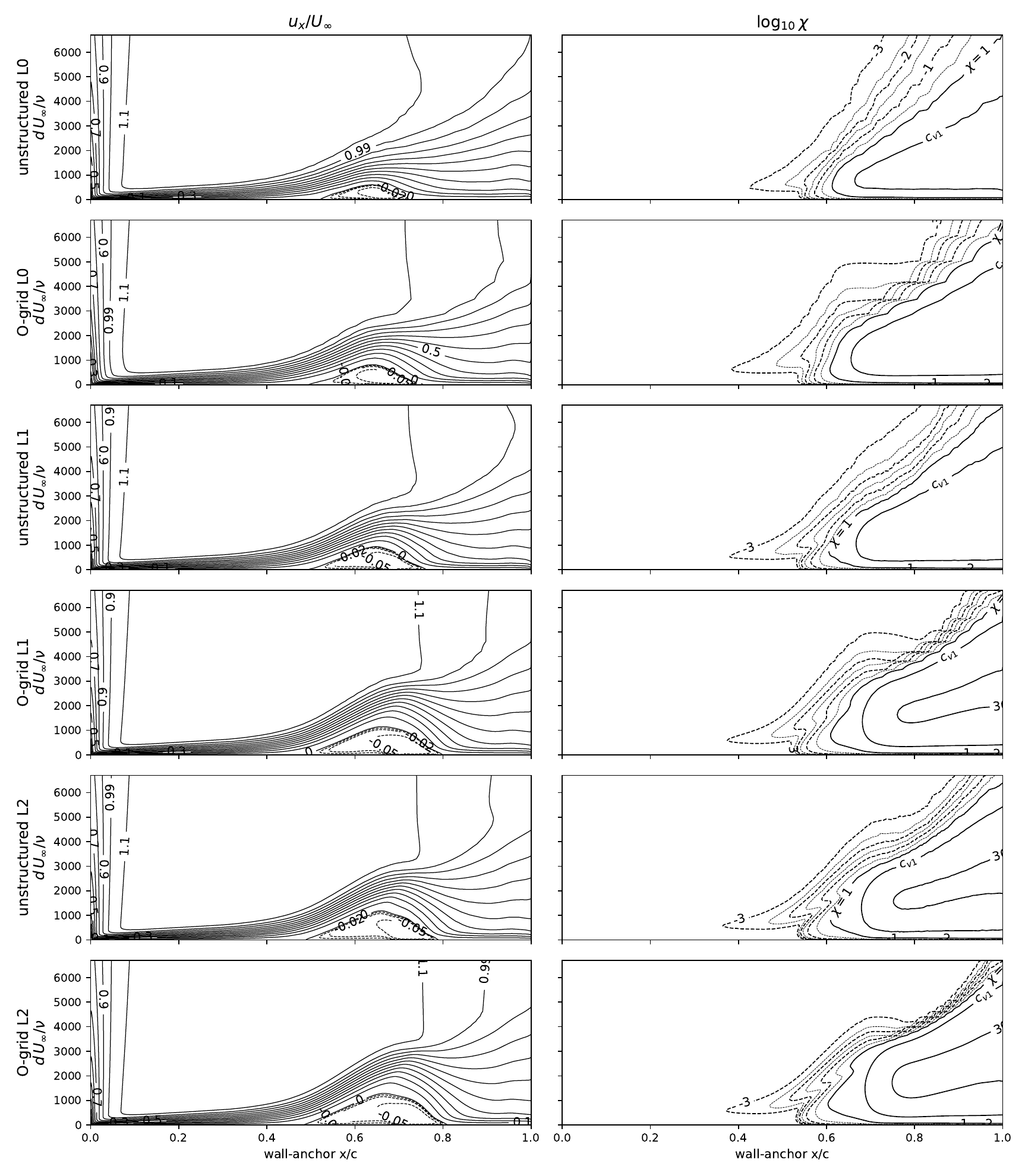}
\figcont{$\alpha=0^\circ$.}
\end{figure}

\begin{figure}[H]\centering
\includegraphics[width=0.99\textwidth,height=0.94\textheight,keepaspectratio]{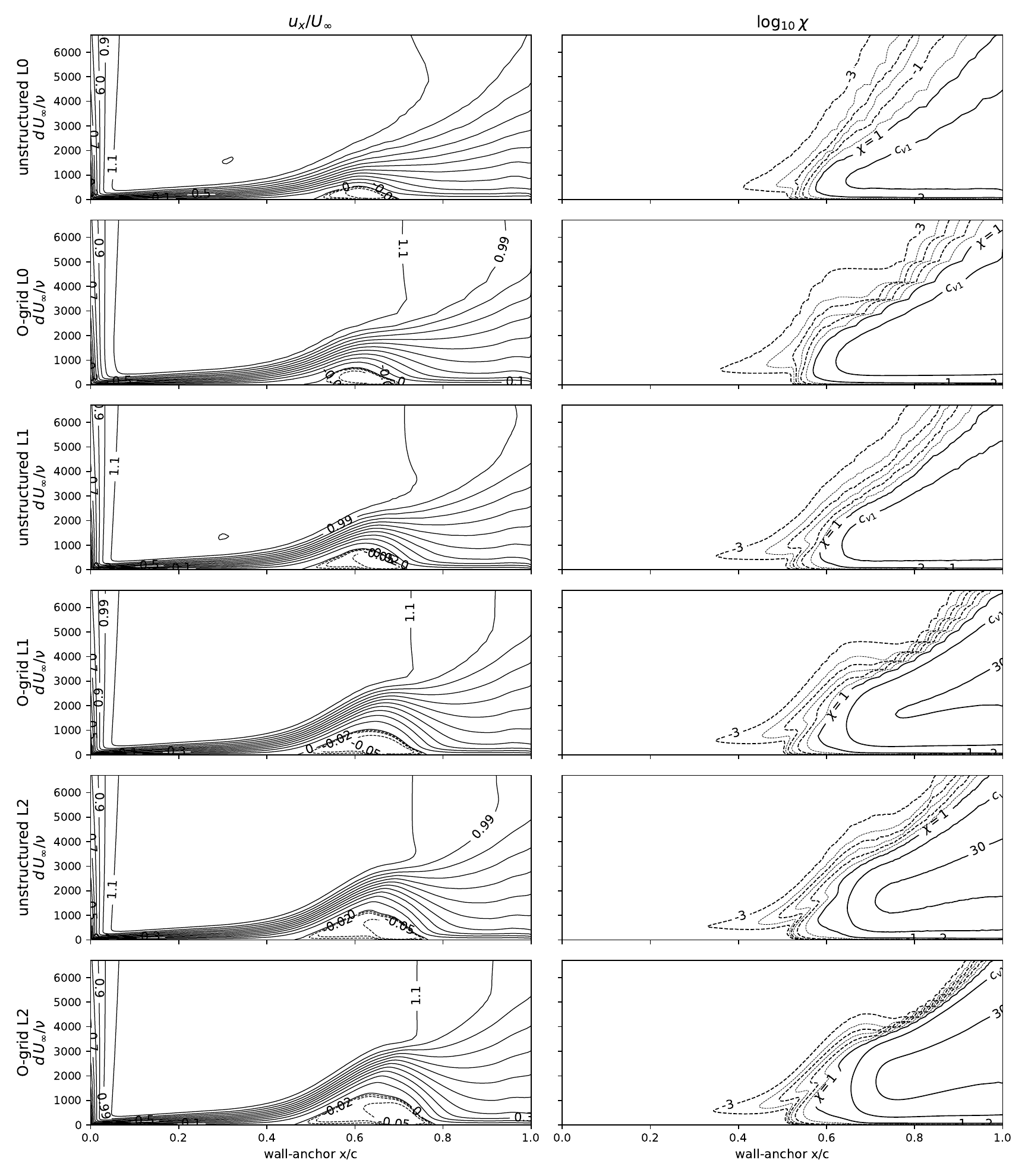}
\figcont{$\alpha=1^\circ$.}
\end{figure}

\begin{figure}[H]\centering
\includegraphics[width=0.99\textwidth,height=0.94\textheight,keepaspectratio]{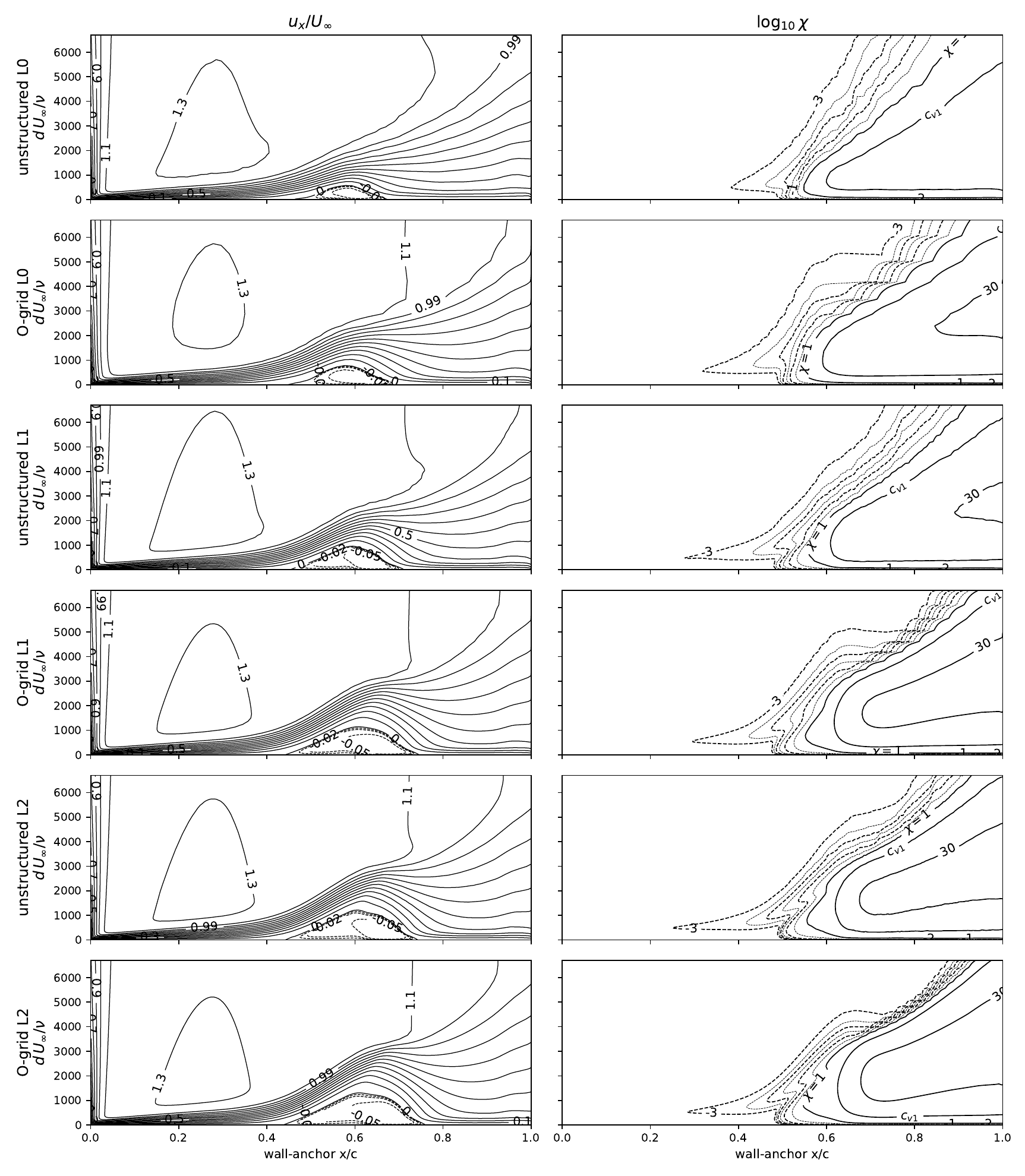}
\figcont{$\alpha=2^\circ$.}
\end{figure}

\begin{figure}[H]\centering
\includegraphics[width=0.99\textwidth,height=0.94\textheight,keepaspectratio]{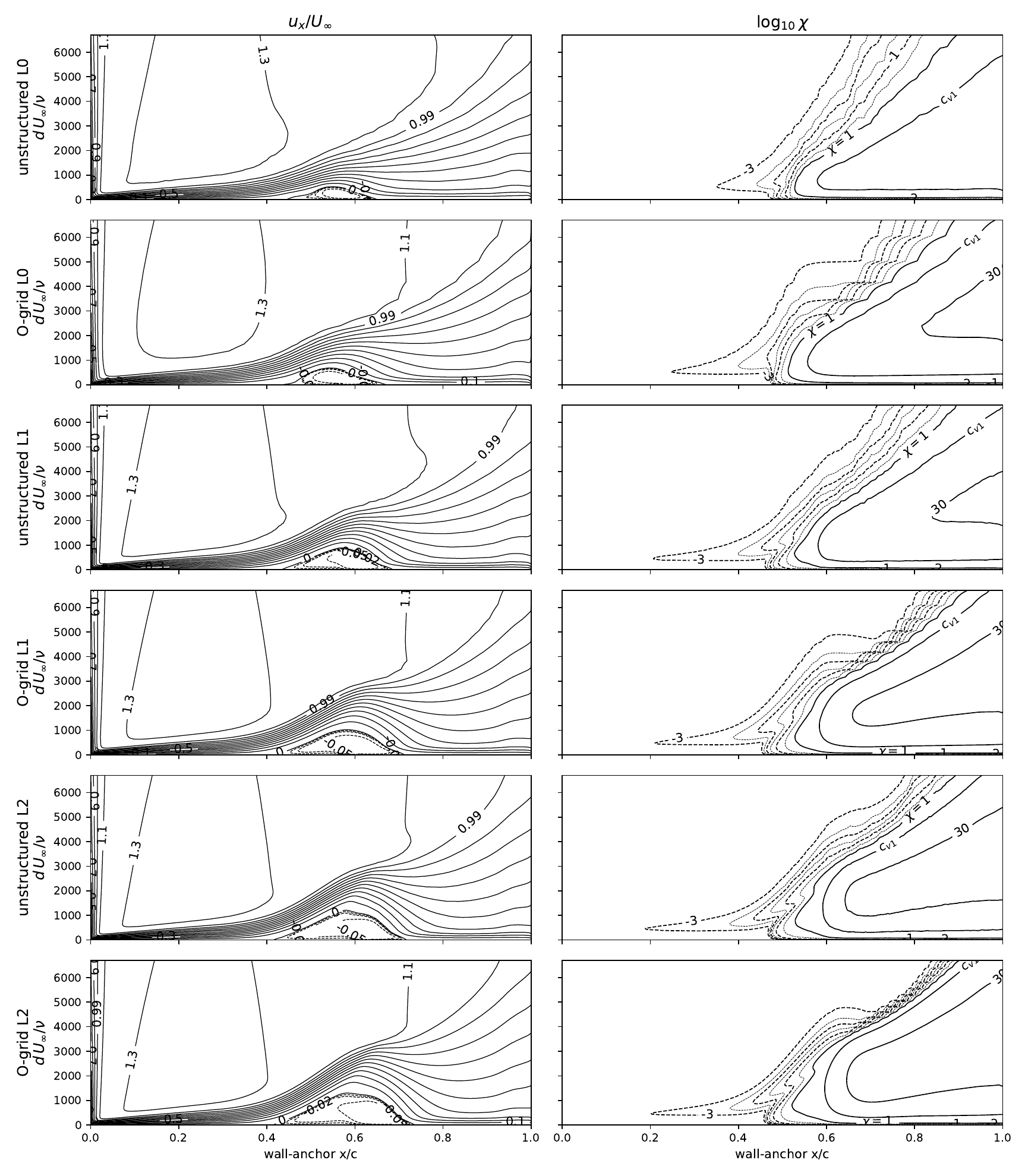}
\figcont{$\alpha=3^\circ$.}
\end{figure}

\begin{figure}[H]\centering
\includegraphics[width=0.99\textwidth,height=0.94\textheight,keepaspectratio]{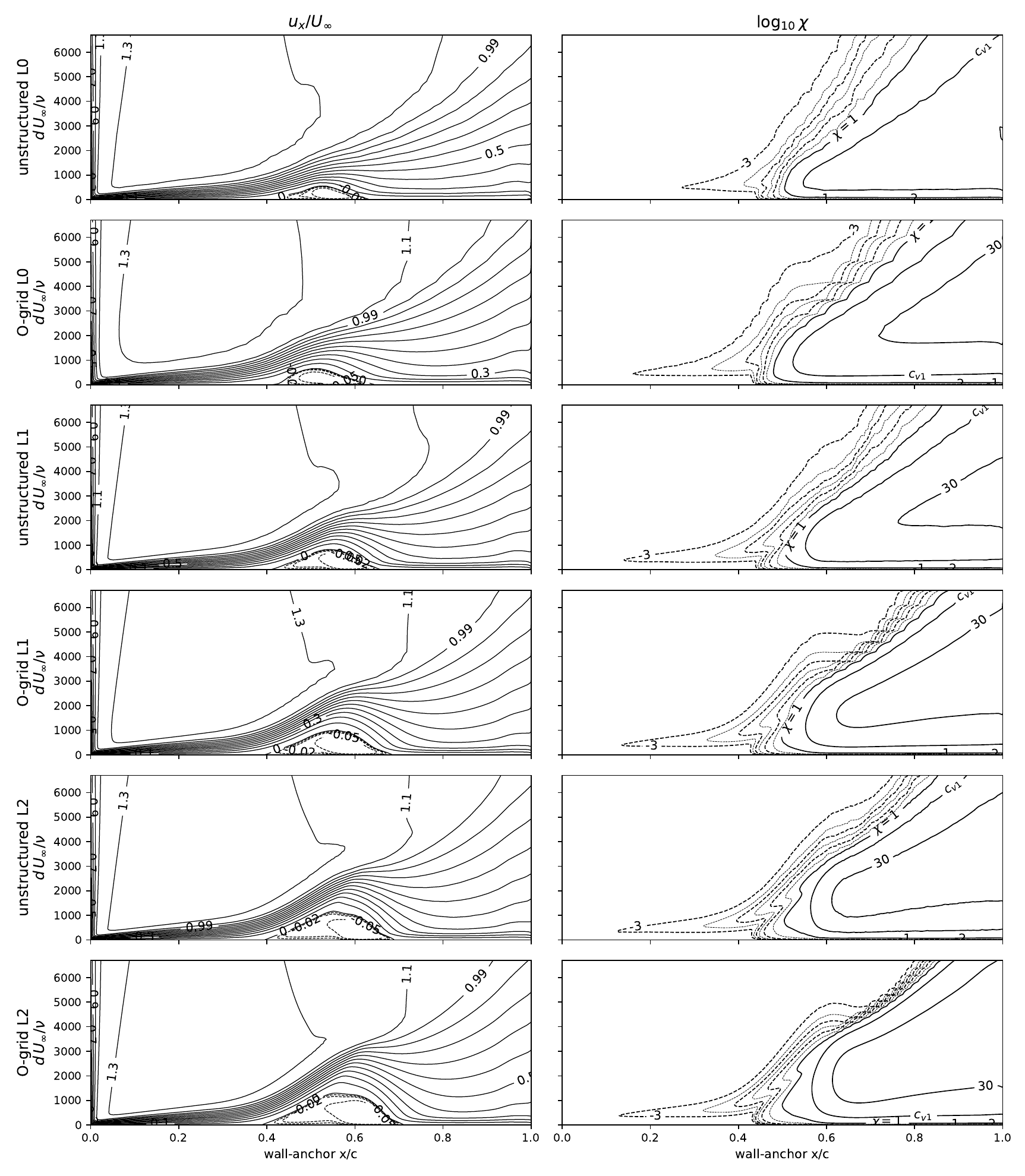}
\figcont{$\alpha=4^\circ$.}
\end{figure}

\begin{figure}[H]\centering
\includegraphics[width=0.99\textwidth,height=0.94\textheight,keepaspectratio]{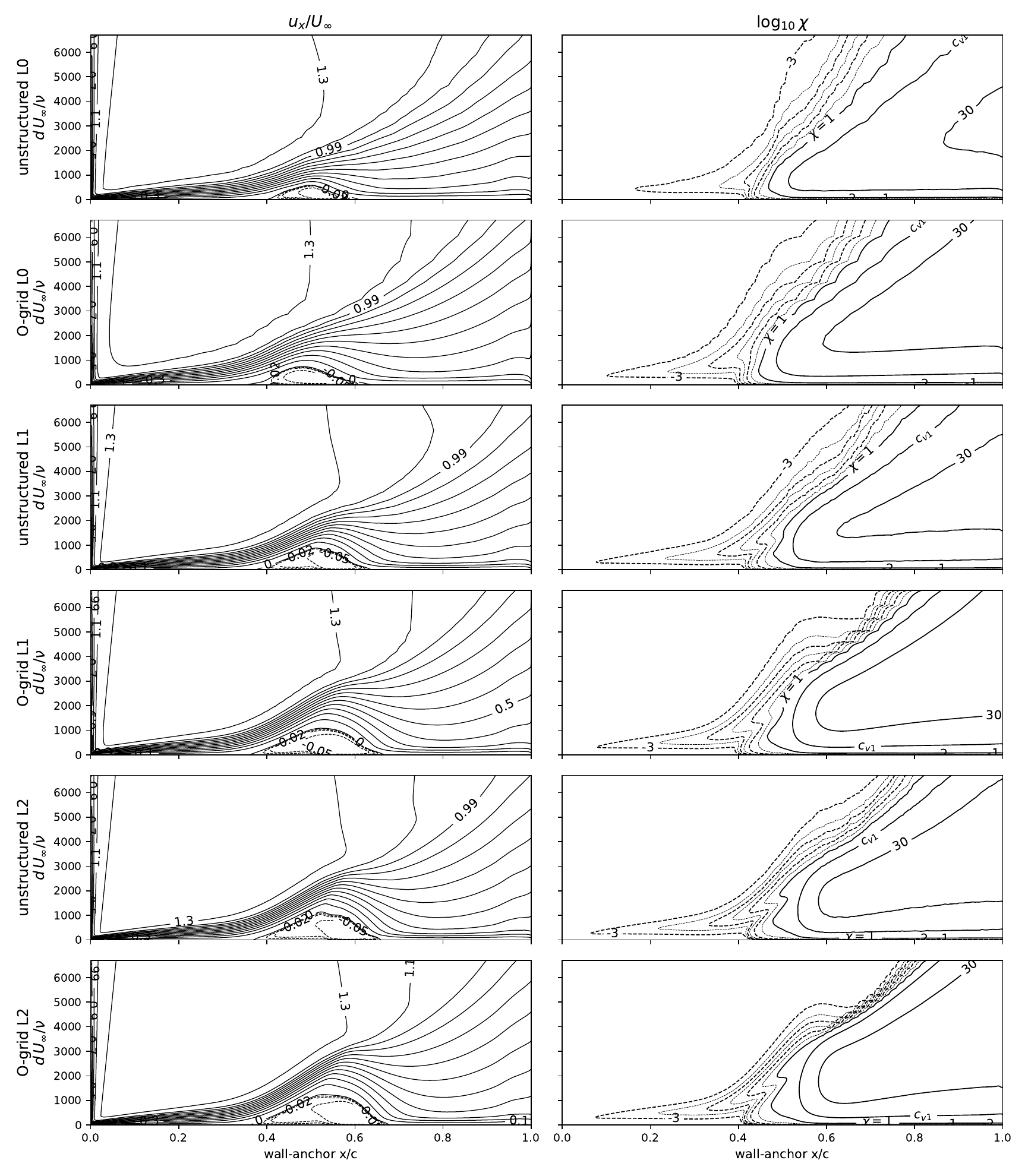}
\figcont{$\alpha=5^\circ$.}
\end{figure}

\begin{figure}[H]\centering
\includegraphics[width=0.99\textwidth,height=0.94\textheight,keepaspectratio]{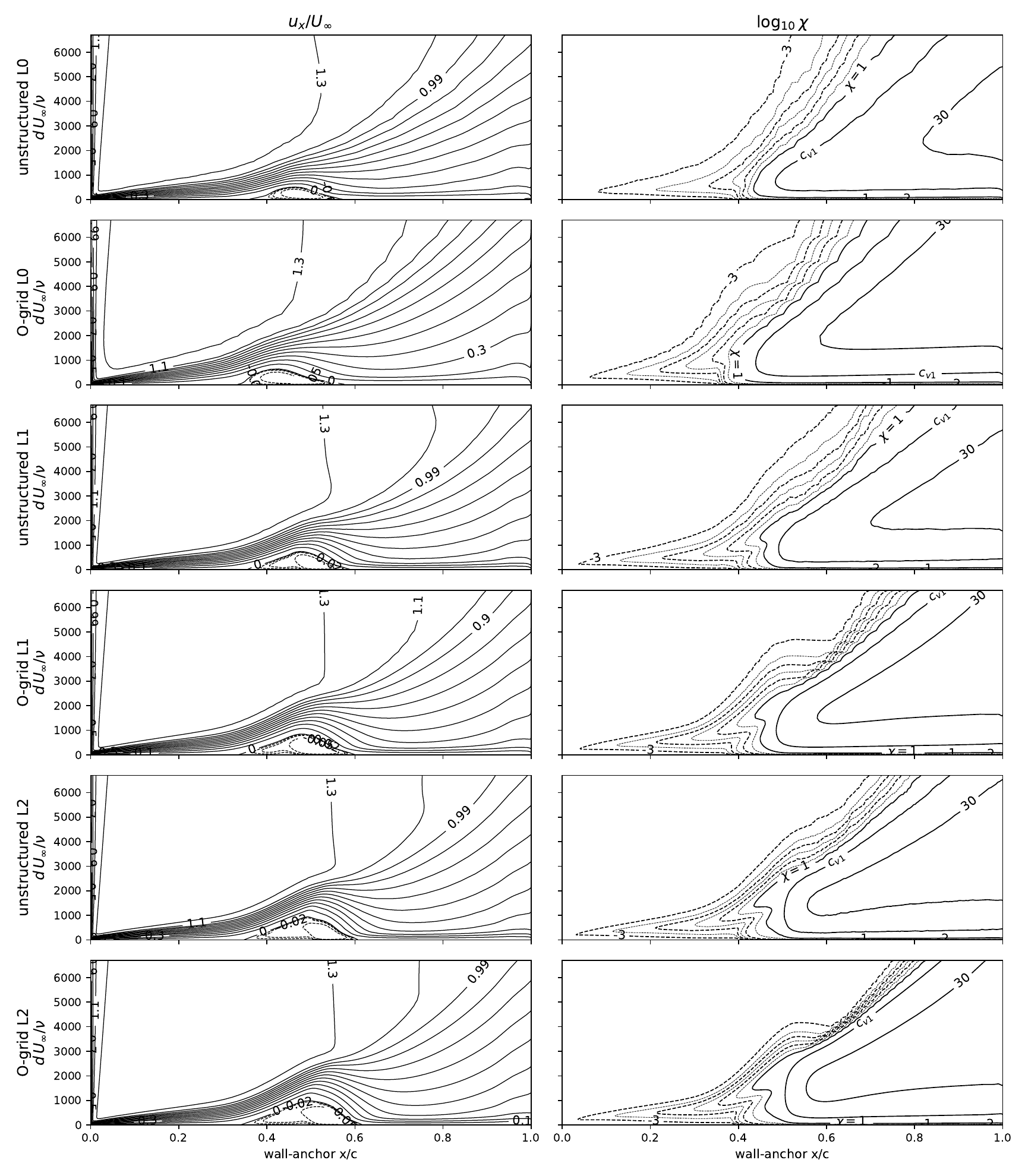}
\figcont{$\alpha=6^\circ$.}
\end{figure}

\begin{figure}[H]\centering
\includegraphics[width=0.99\textwidth,height=0.94\textheight,keepaspectratio]{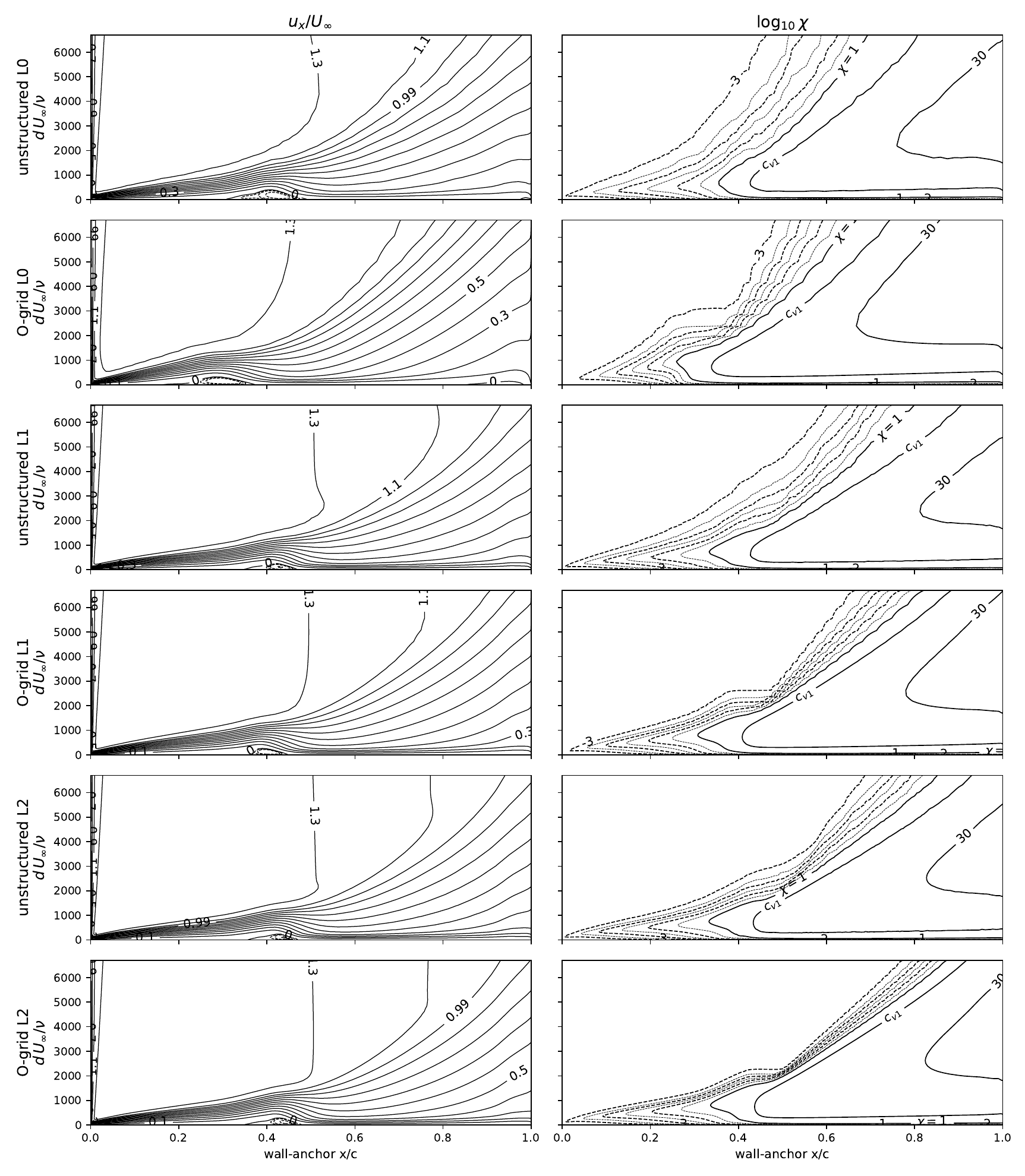}
\figcont{$\alpha=7^\circ$.}
\end{figure}

\begin{figure}[H]\centering
\includegraphics[width=0.99\textwidth,height=0.94\textheight,keepaspectratio]{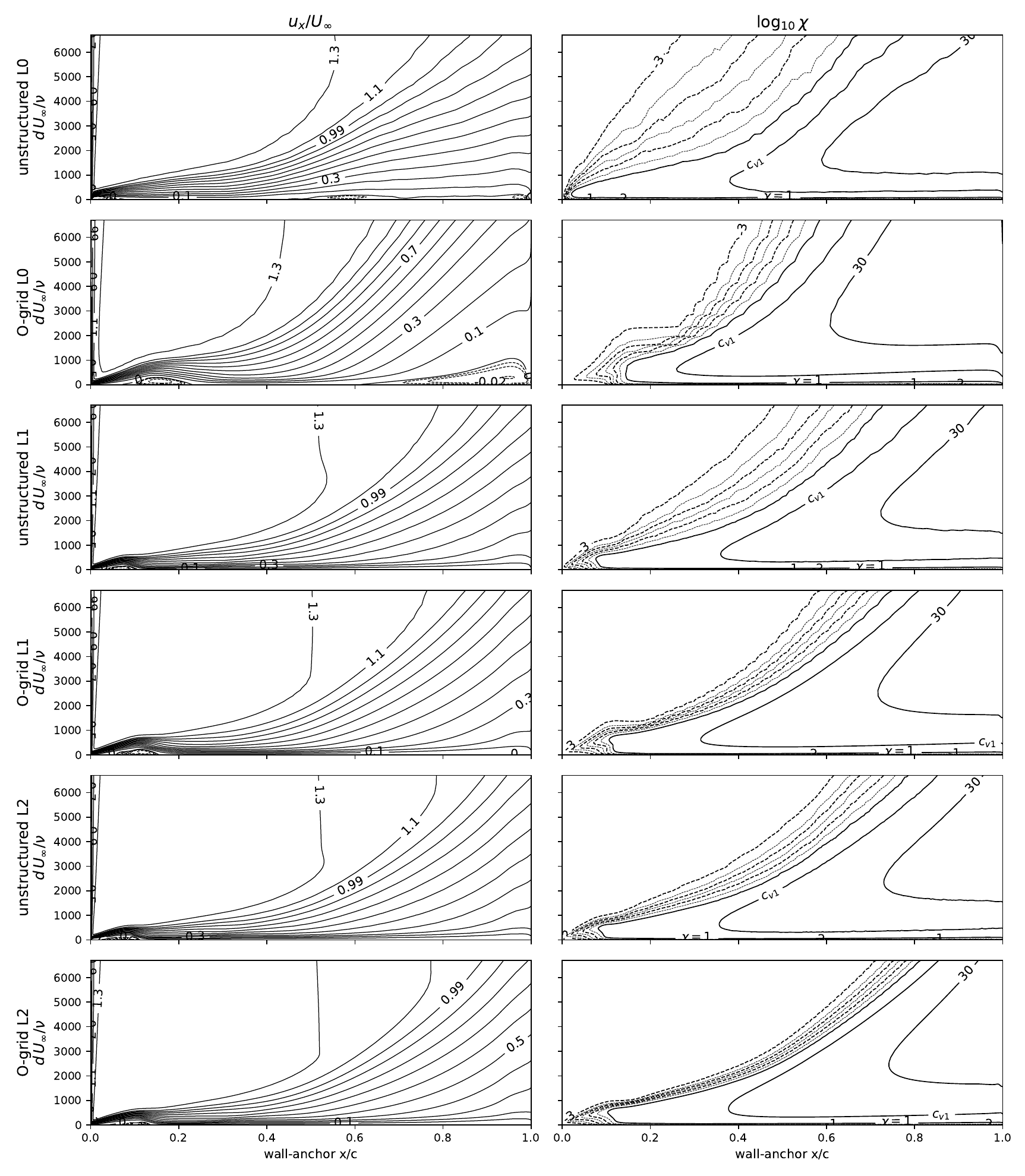}
\figcont{$\alpha=8^\circ$.}
\end{figure}

\begin{figure}[H]\centering
\includegraphics[width=0.99\textwidth,height=0.94\textheight,keepaspectratio]{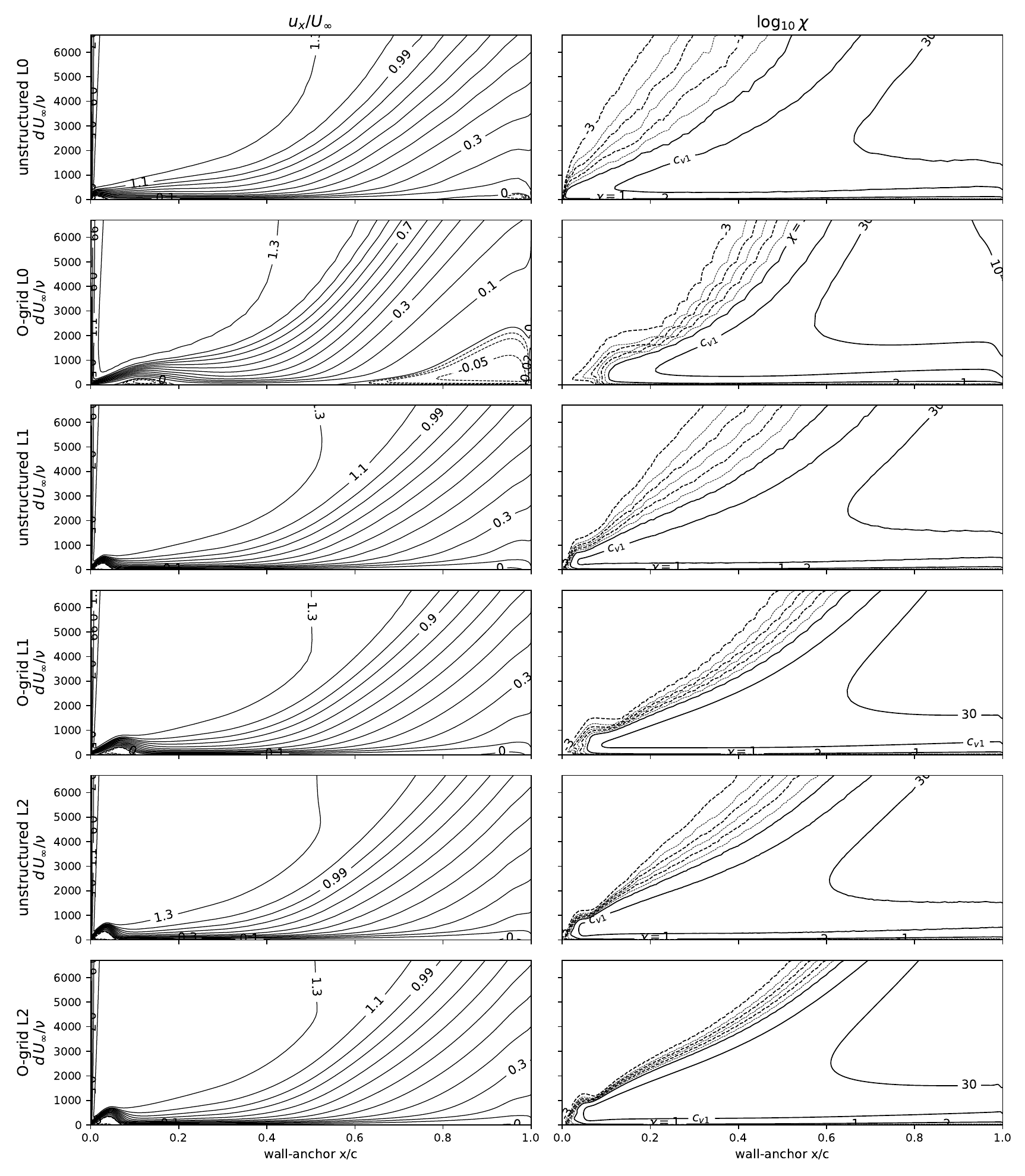}
\figcont{$\alpha=8.5^\circ$.}
\end{figure}

\begin{figure}[H]\centering
\includegraphics[width=0.99\textwidth,height=0.94\textheight,keepaspectratio]{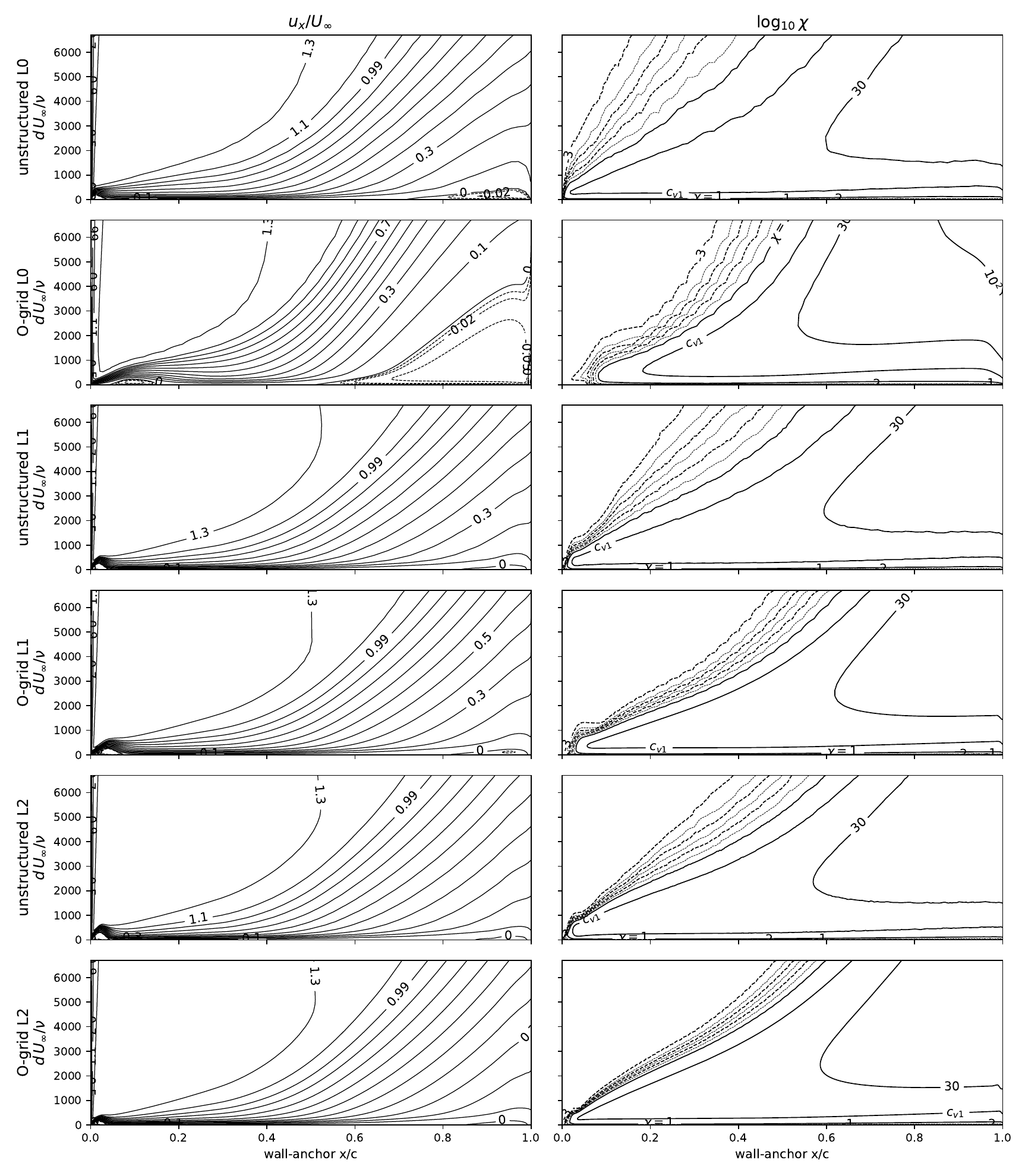}
\figcont{$\alpha=9^\circ$.}
\end{figure}

\subsection*{\tocentry{subsection}{Reynolds-Number Study of the Eppler 387 Airfoil}Reynolds-Number Study of the Eppler 387 Airfoil}

\begin{figure}[H]\centering
\includegraphics[width=0.99\textwidth,height=0.85\textheight,keepaspectratio]{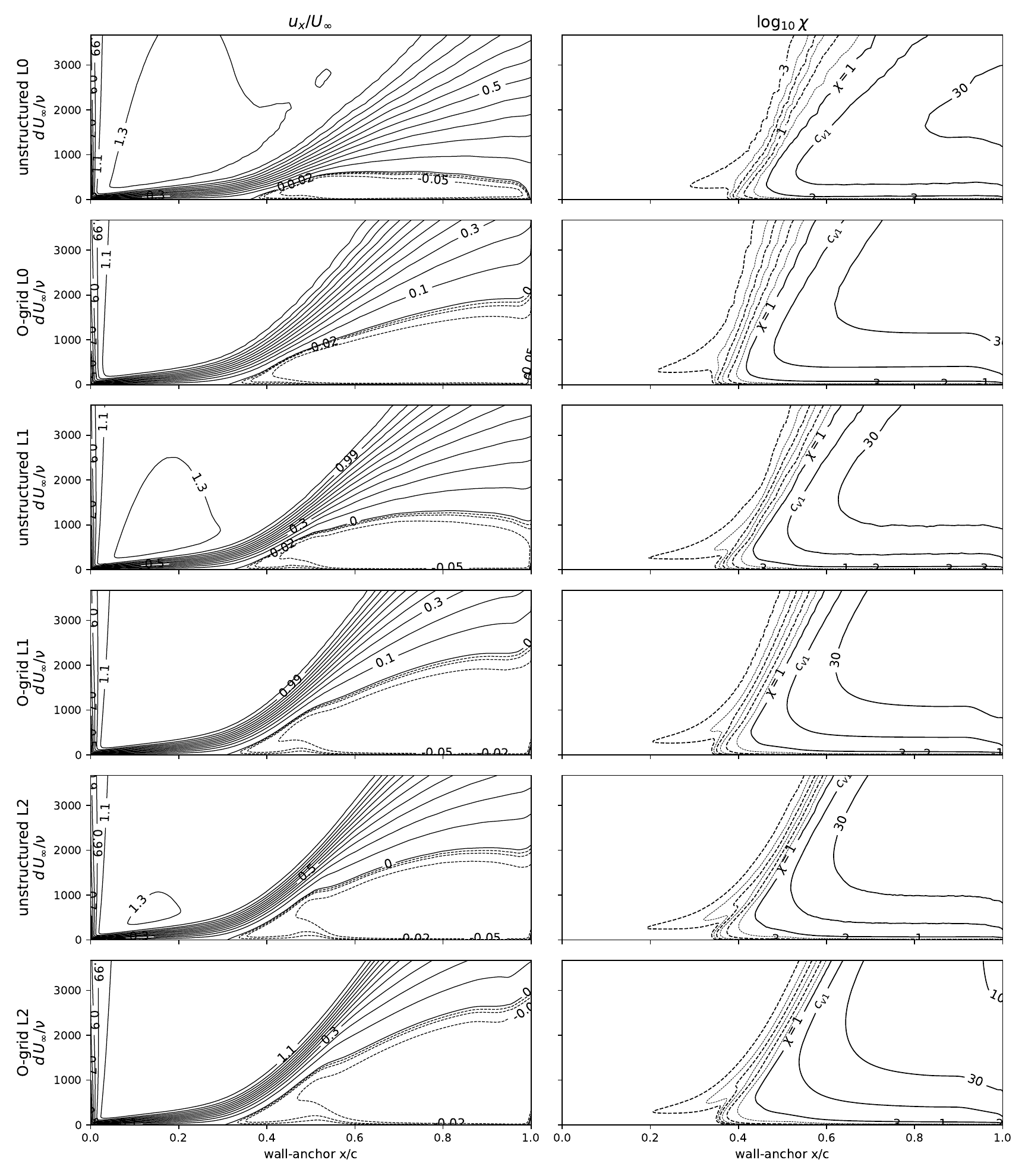}
\caption{Eppler~387 near the upper surface at $\alpha=5^\circ$, all six grids: $u_x/U_\infty$ (left) and $\log_{10}\chi$ (right) against the chordwise station and the wall distance. $Re=6\times10^4$.}\label{f:eppresweepsheets}
\end{figure}

\begin{figure}[H]\centering
\includegraphics[width=0.99\textwidth,height=0.94\textheight,keepaspectratio]{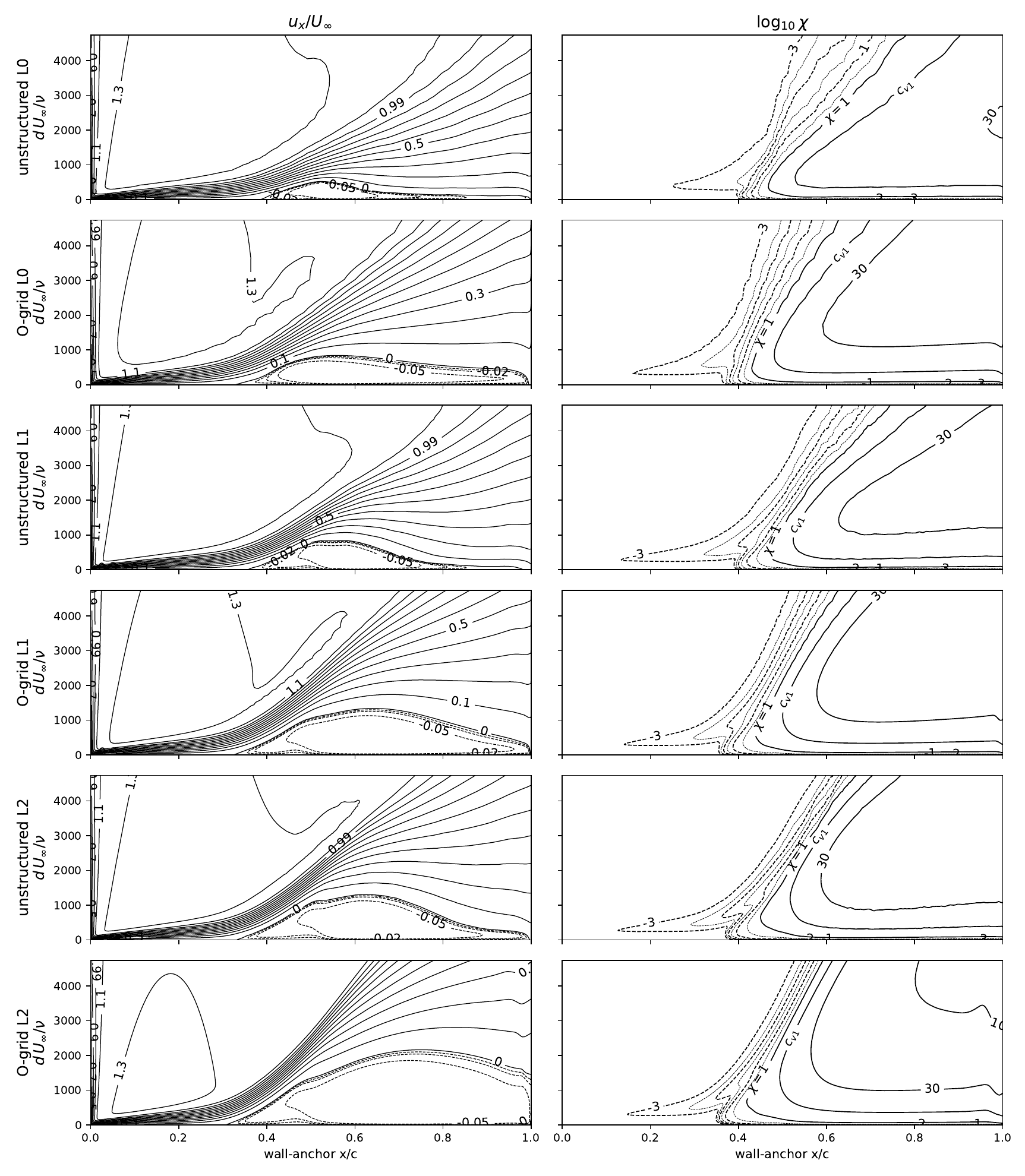}
\figcont{$Re=10^5$.}
\end{figure}

\begin{figure}[H]\centering
\includegraphics[width=0.99\textwidth,height=0.94\textheight,keepaspectratio]{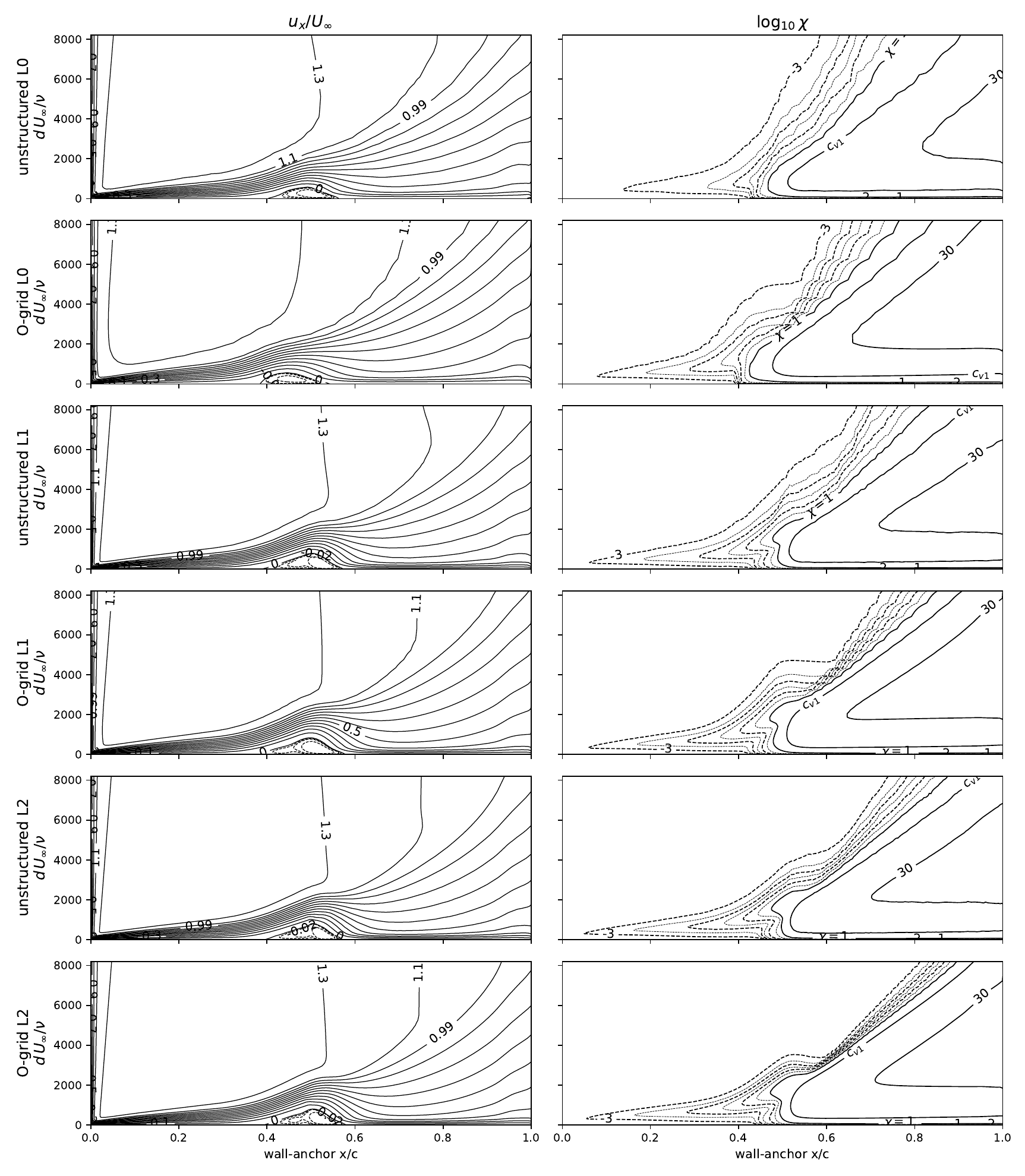}
\figcont{$Re=3\times10^5$.}
\end{figure}

\begin{figure}[H]\centering
\includegraphics[width=0.99\textwidth,height=0.94\textheight,keepaspectratio]{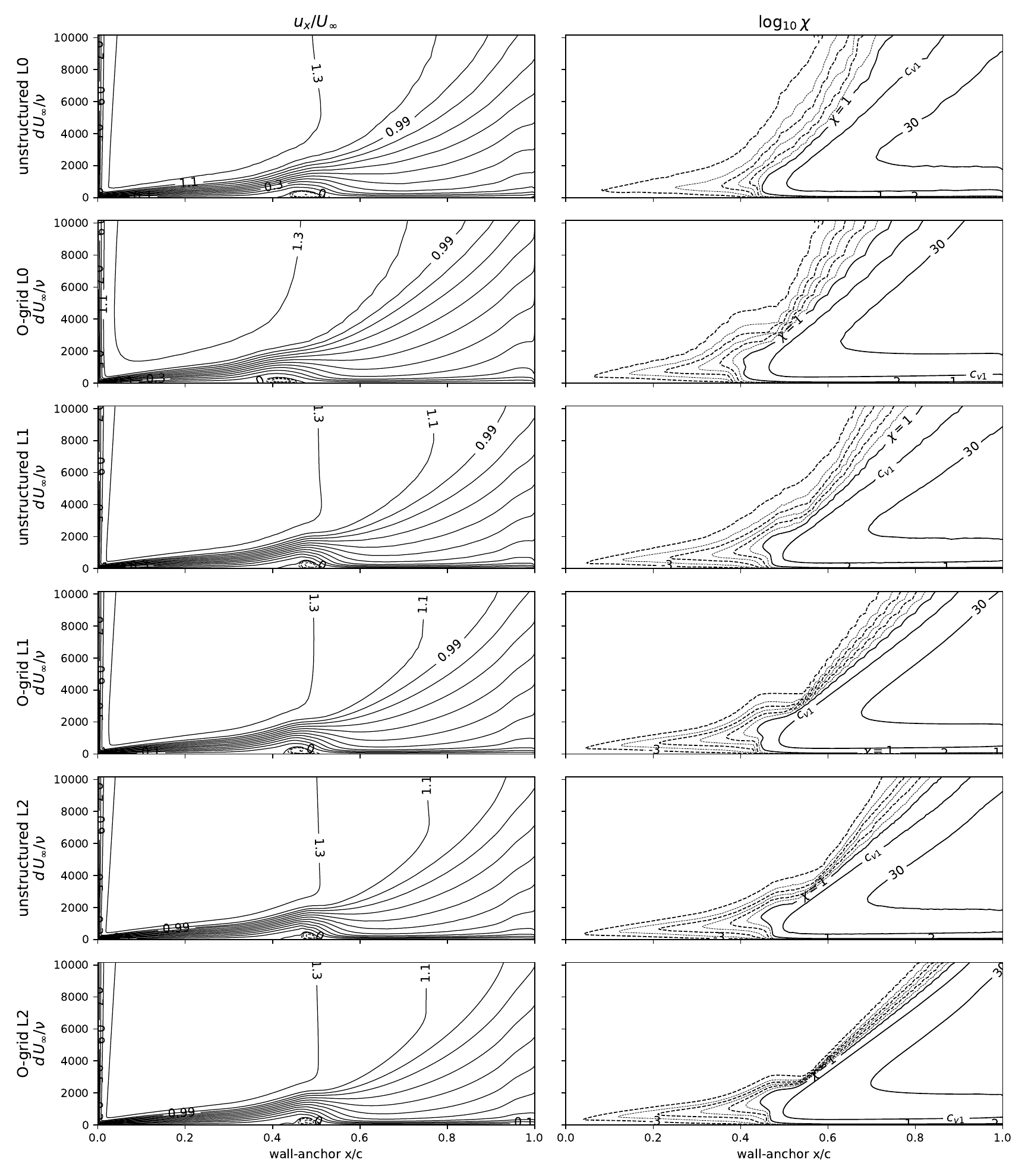}
\figcont{$Re=4.6\times10^5$.}
\end{figure}

\begin{figure}[H]\centering
\includegraphics[width=0.99\textwidth,height=0.94\textheight,keepaspectratio]{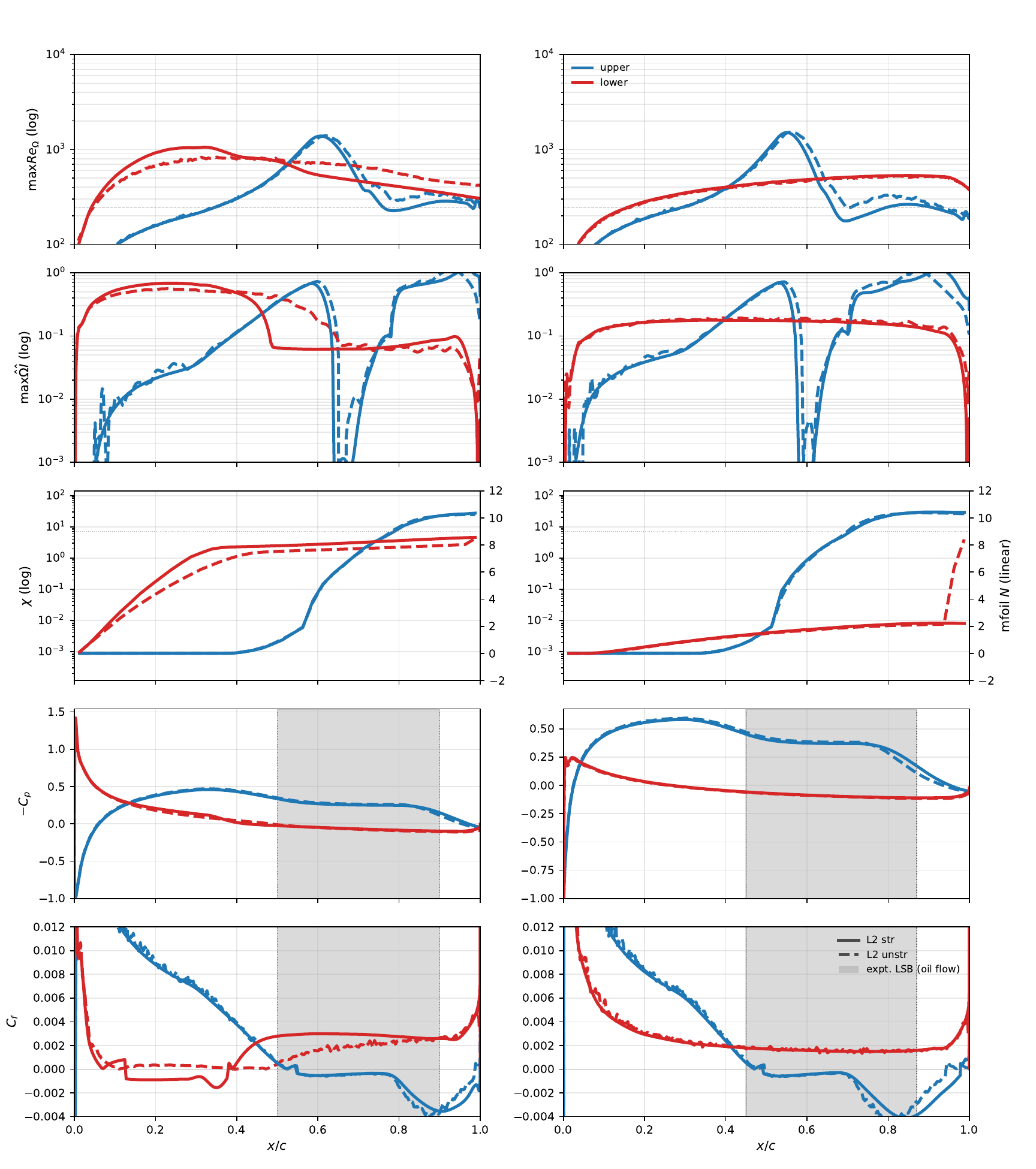}
\caption{Eppler~387 surface distributions, $Re=10^5$, L2 grids, in ascending incidence; rows as in Fig.~\ref{f:nlfcf0}. $\alpha=-2^\circ$, $\alpha=0^\circ$.}\label{f:eppcfRe100k}
\end{figure}

\begin{figure}[H]\centering
\includegraphics[width=0.99\textwidth,height=0.94\textheight,keepaspectratio]{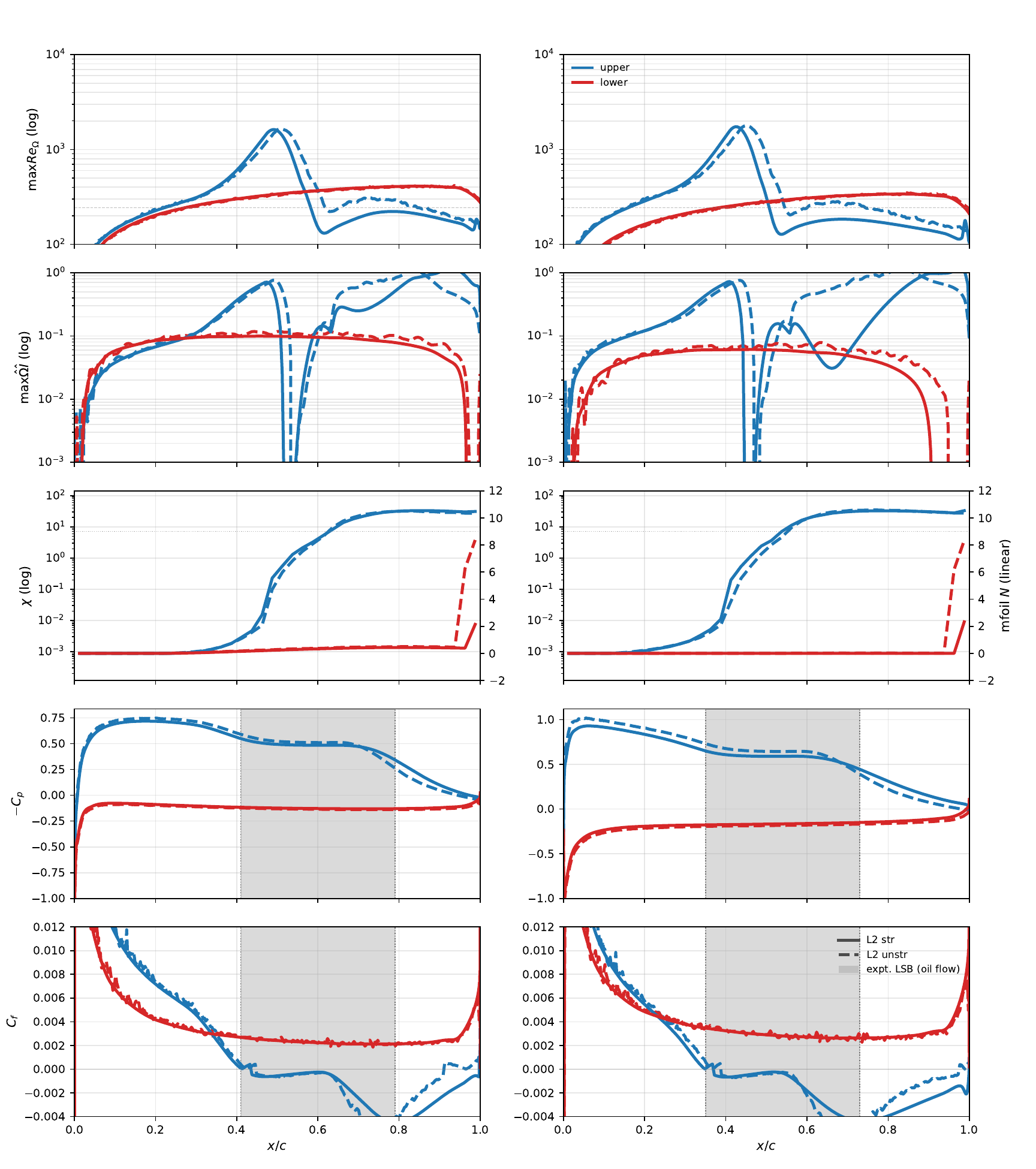}
\figcont{$\alpha=2^\circ$, $\alpha=4^\circ$.}
\end{figure}

\begin{figure}[H]\centering
\includegraphics[width=0.99\textwidth,height=0.94\textheight,keepaspectratio]{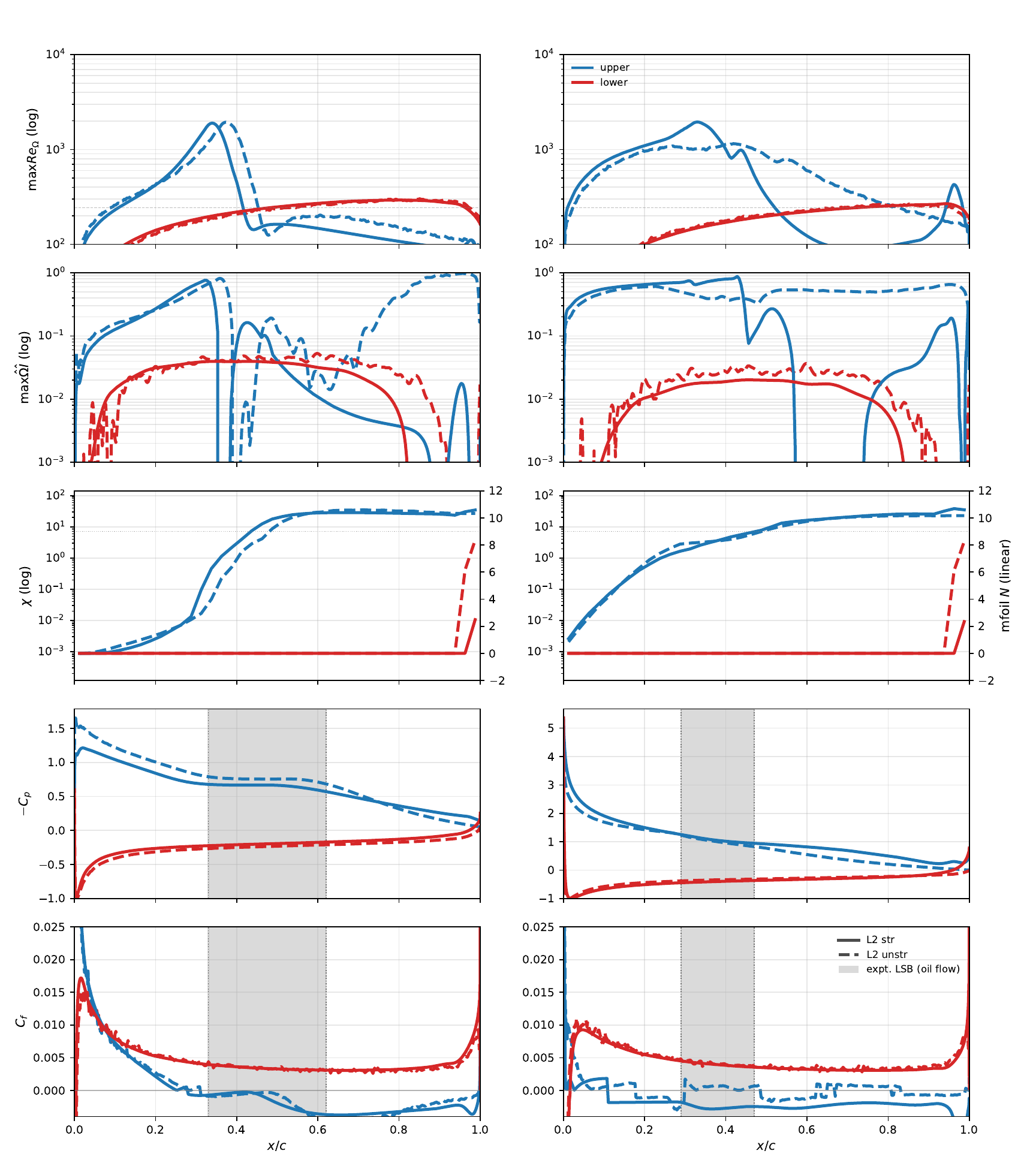}
\figcont{$\alpha=6^\circ$, $\alpha=8^\circ$.}
\end{figure}

\begin{figure}[H]\centering
\includegraphics[width=0.99\textwidth,height=0.94\textheight,keepaspectratio]{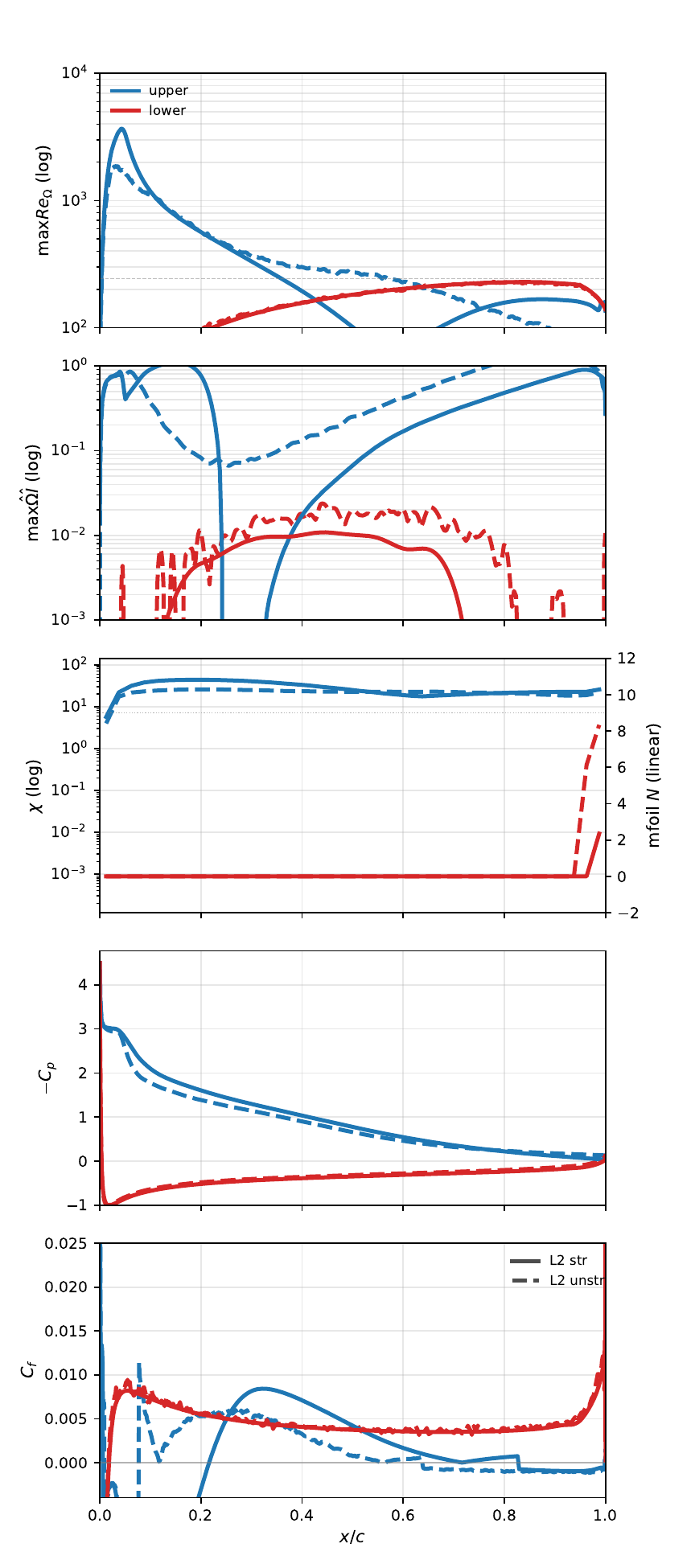}
\figcont{$\alpha=10^\circ$.}
\end{figure}

\begin{figure}[H]\centering
\includegraphics[width=0.99\textwidth,height=0.94\textheight,keepaspectratio]{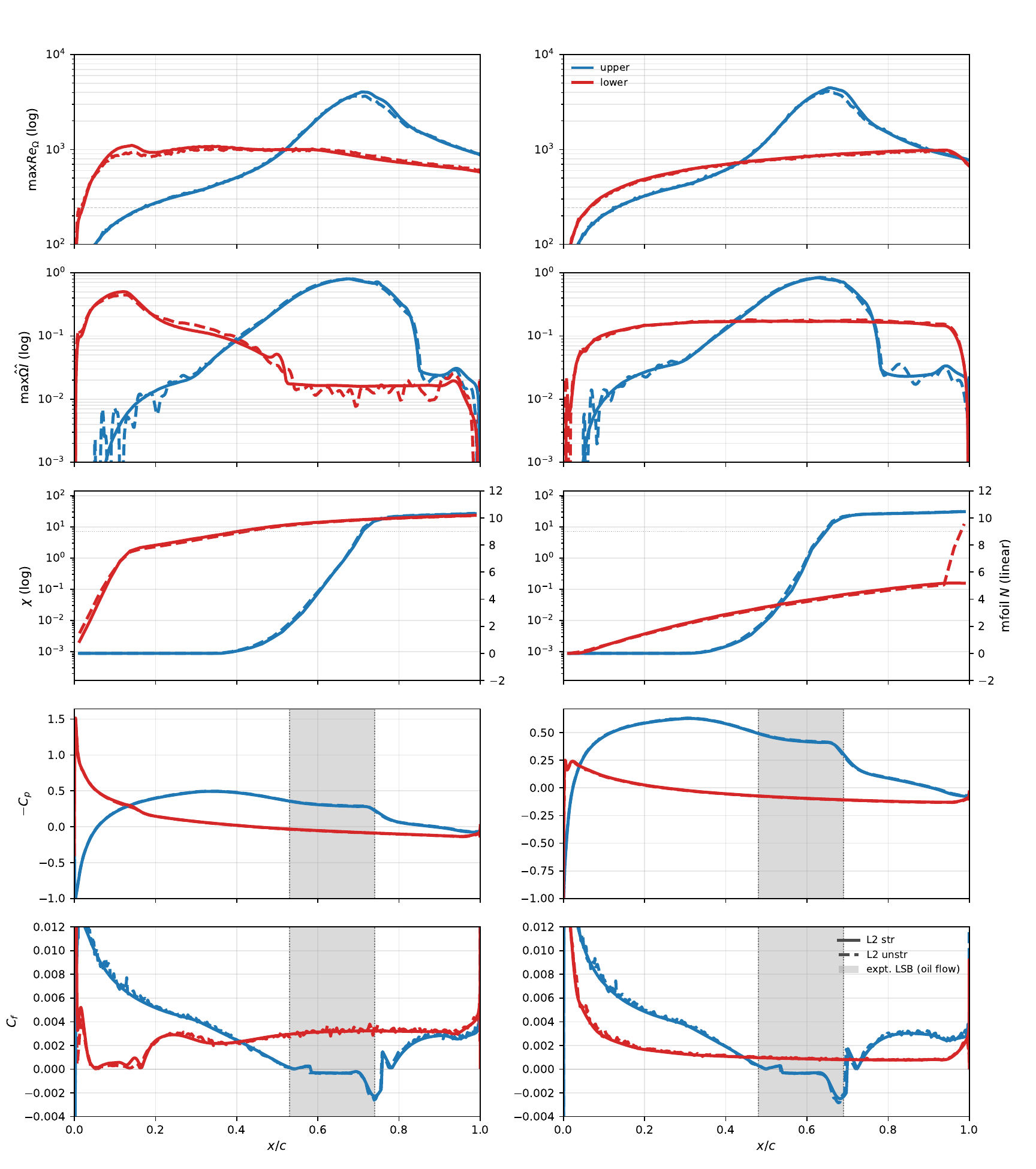}
\caption{Eppler~387 surface distributions, $Re=3\times10^5$, L2 grids, in ascending incidence; rows as in Fig.~\ref{f:nlfcf0}. $\alpha=-2^\circ$, $\alpha=0^\circ$.}\label{f:eppcfRe300k}
\end{figure}

\begin{figure}[H]\centering
\includegraphics[width=0.99\textwidth,height=0.94\textheight,keepaspectratio]{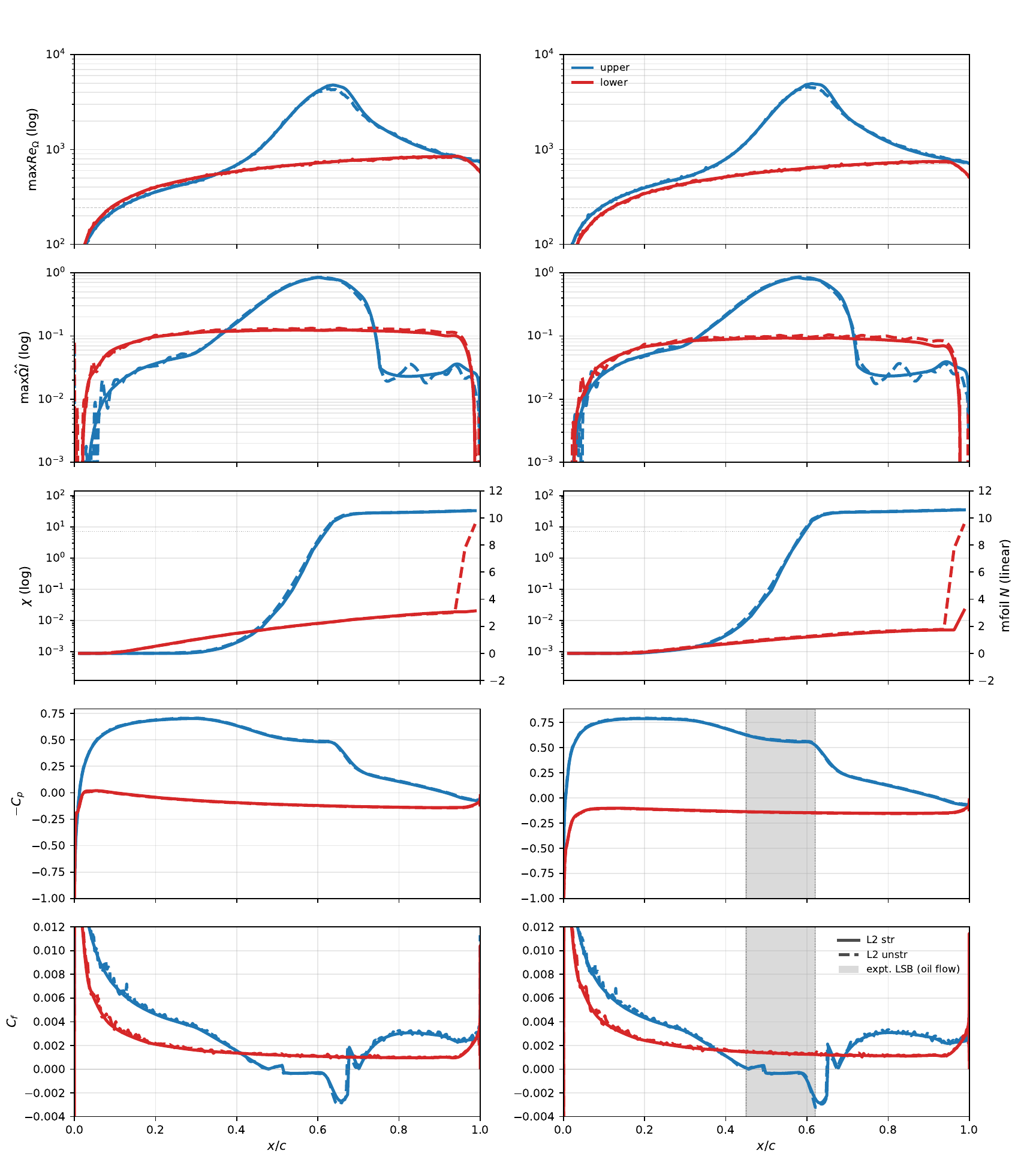}
\figcont{$\alpha=1^\circ$, $\alpha=2^\circ$.}
\end{figure}

\begin{figure}[H]\centering
\includegraphics[width=0.99\textwidth,height=0.94\textheight,keepaspectratio]{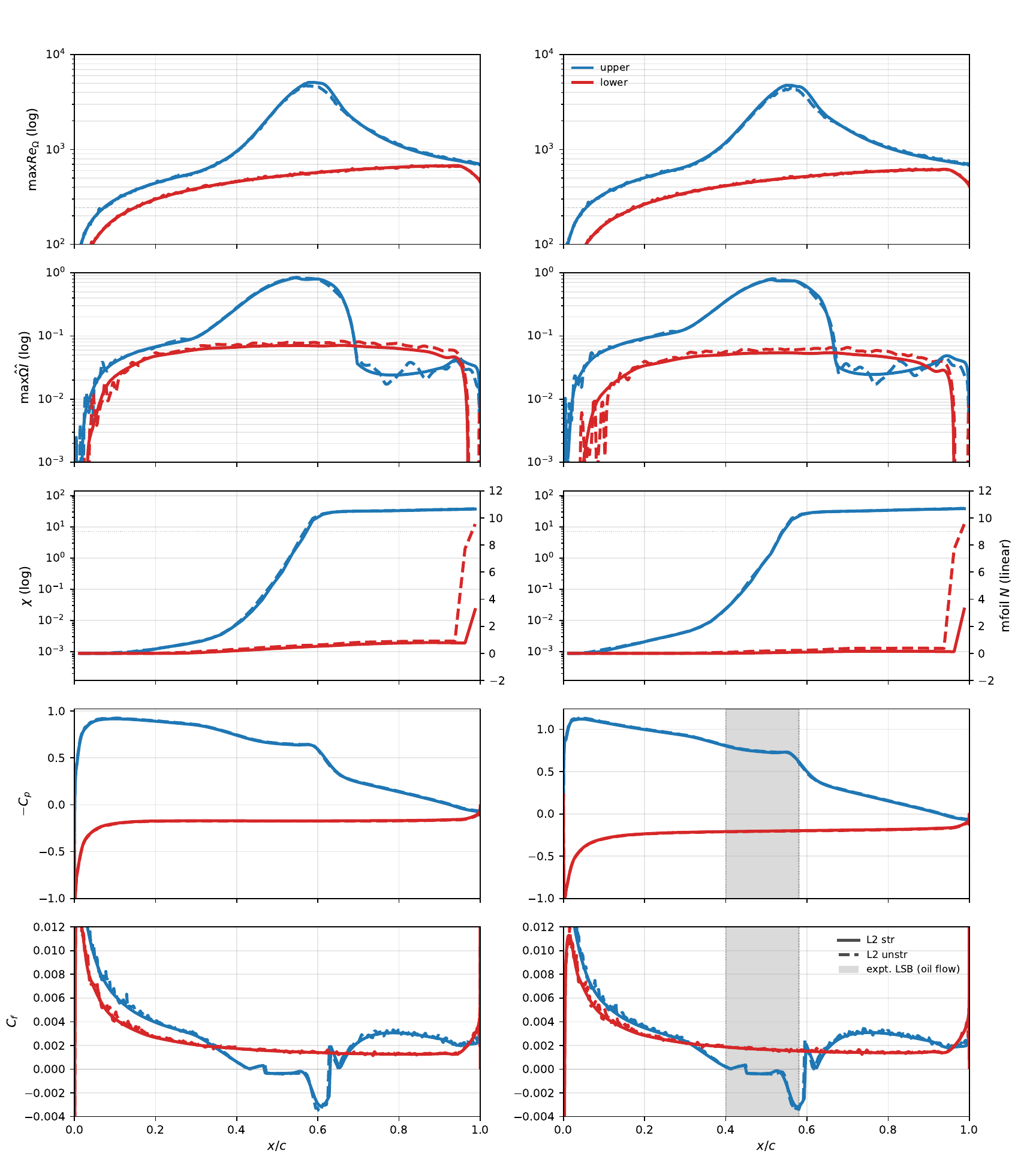}
\figcont{$\alpha=3^\circ$, $\alpha=4^\circ$.}
\end{figure}

\begin{figure}[H]\centering
\includegraphics[width=0.99\textwidth,height=0.94\textheight,keepaspectratio]{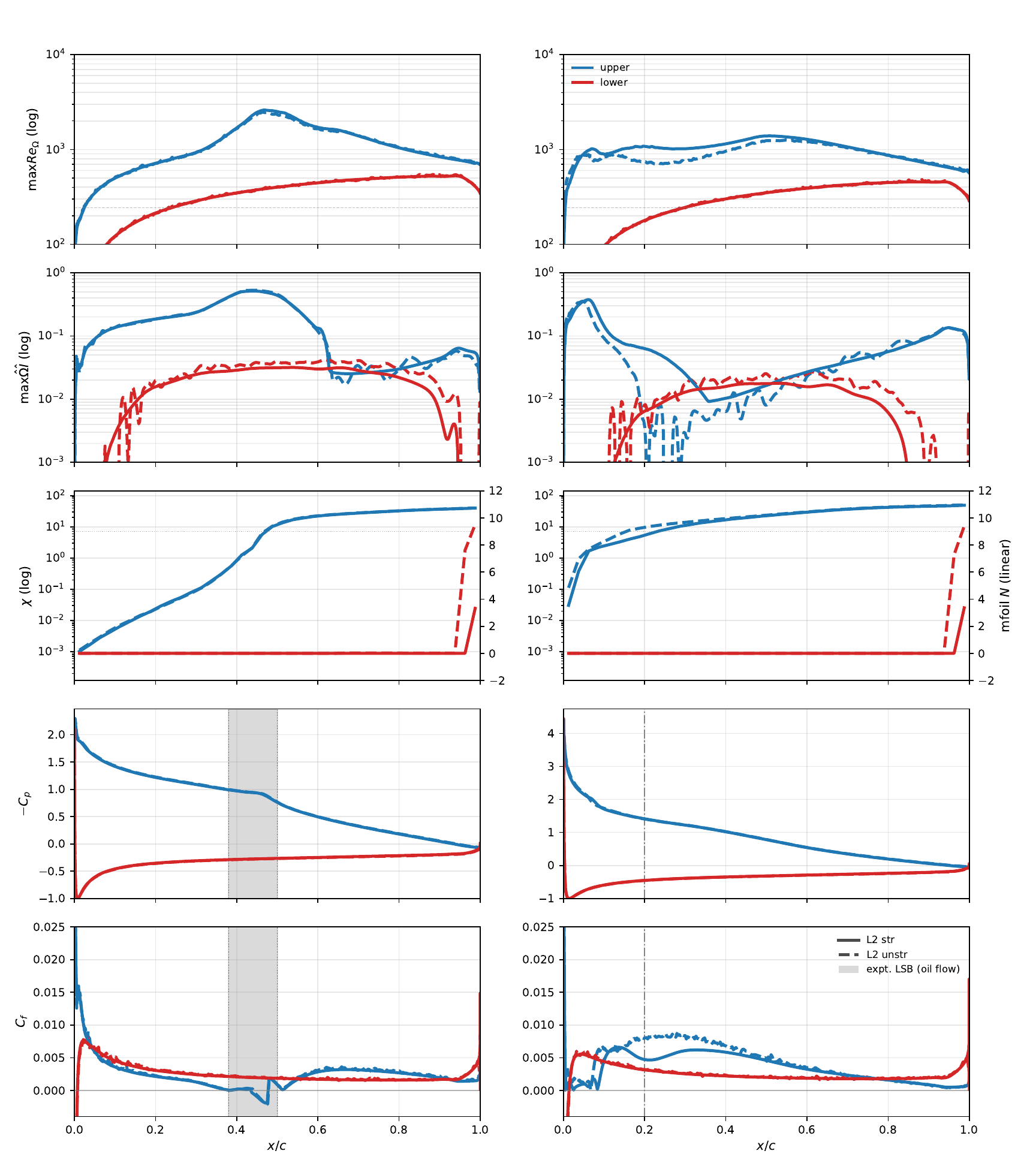}
\figcont{$\alpha=6^\circ$, $\alpha=8^\circ$.}
\end{figure}

\begin{figure}[H]\centering
\includegraphics[width=0.99\textwidth,height=0.94\textheight,keepaspectratio]{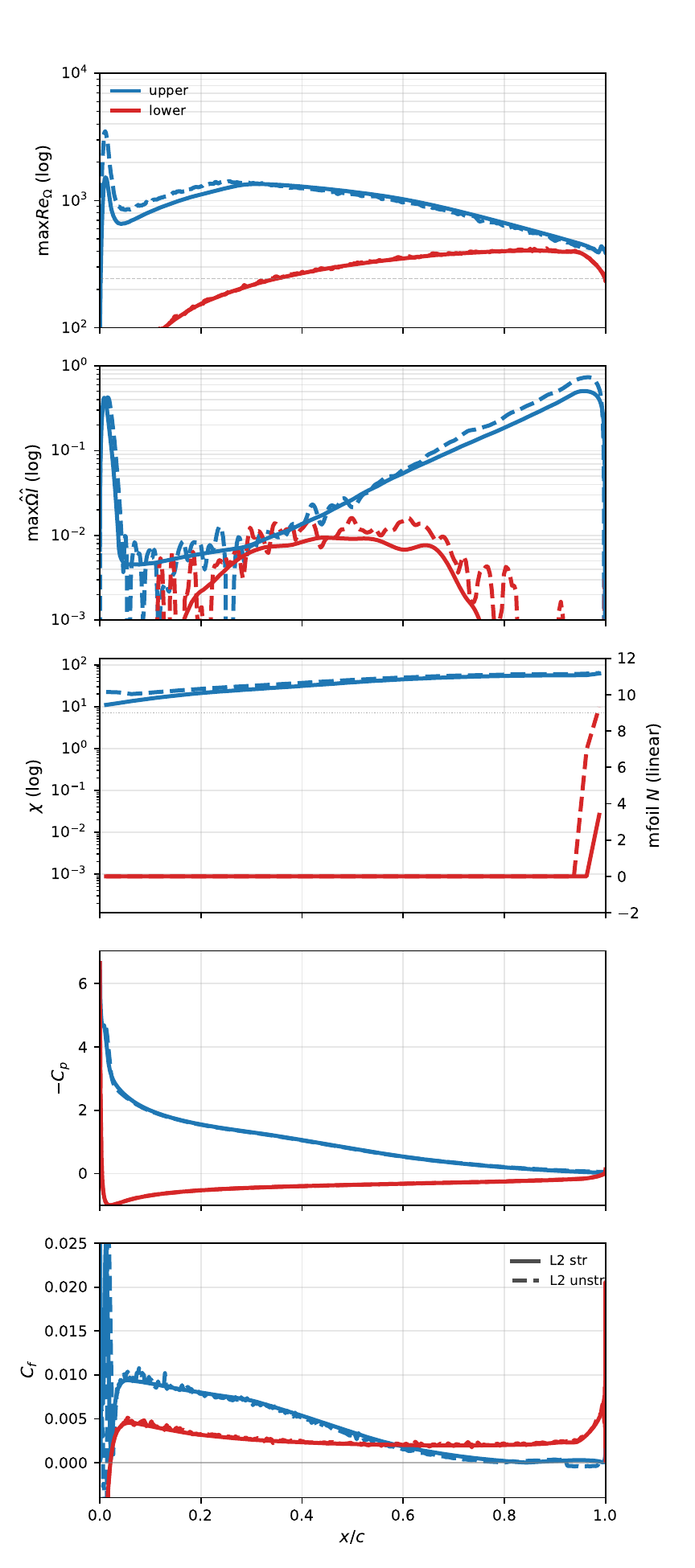}
\figcont{$\alpha=10^\circ$.}
\end{figure}

\begin{figure}[H]\centering
\includegraphics[width=0.99\textwidth,height=0.94\textheight,keepaspectratio]{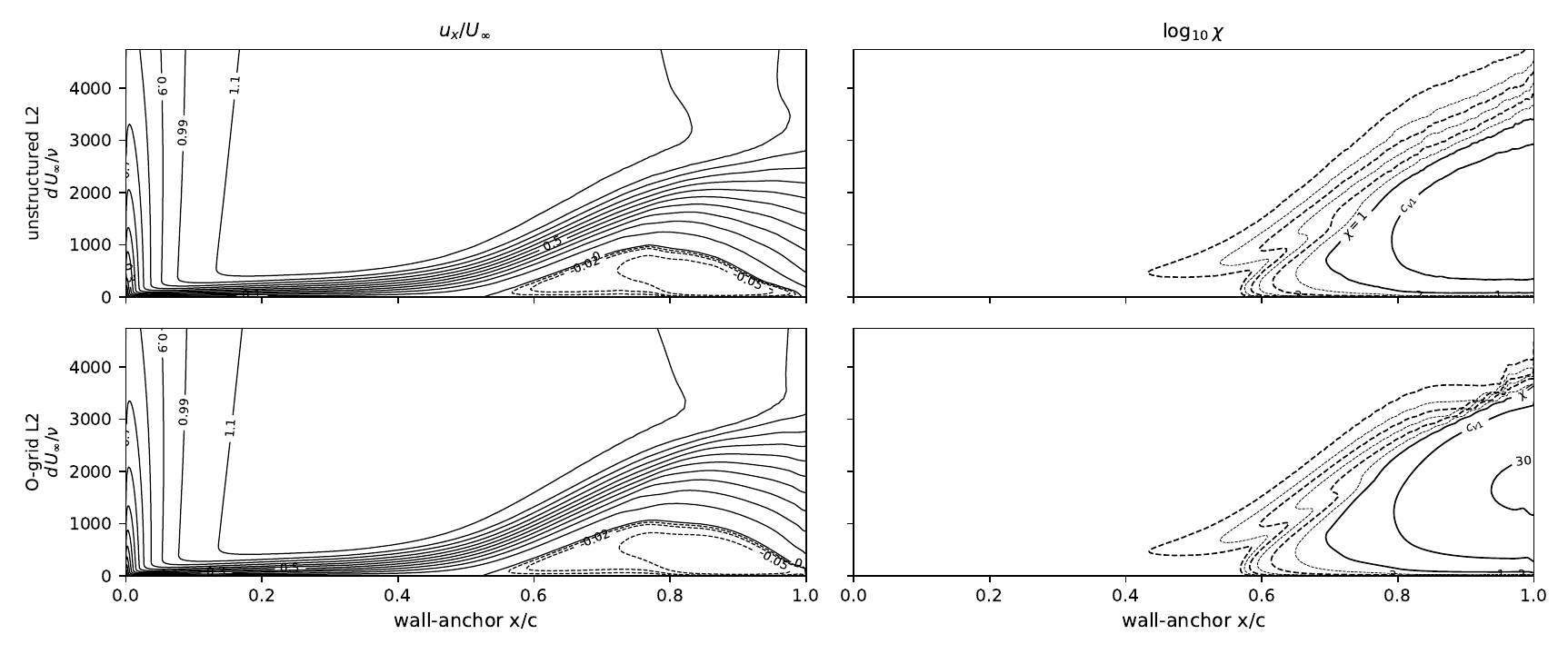}
\caption{Eppler~387 near the upper surface, $Re=10^5$, L2 grids: $u_x/U_\infty$ (left) and $\log_{10}\chi$ (right) against the chordwise station and the wall distance. $\alpha=-2^\circ$.}\label{f:eppsheetRe100k}
\end{figure}

\begin{figure}[H]\centering
\includegraphics[width=0.99\textwidth,height=0.94\textheight,keepaspectratio]{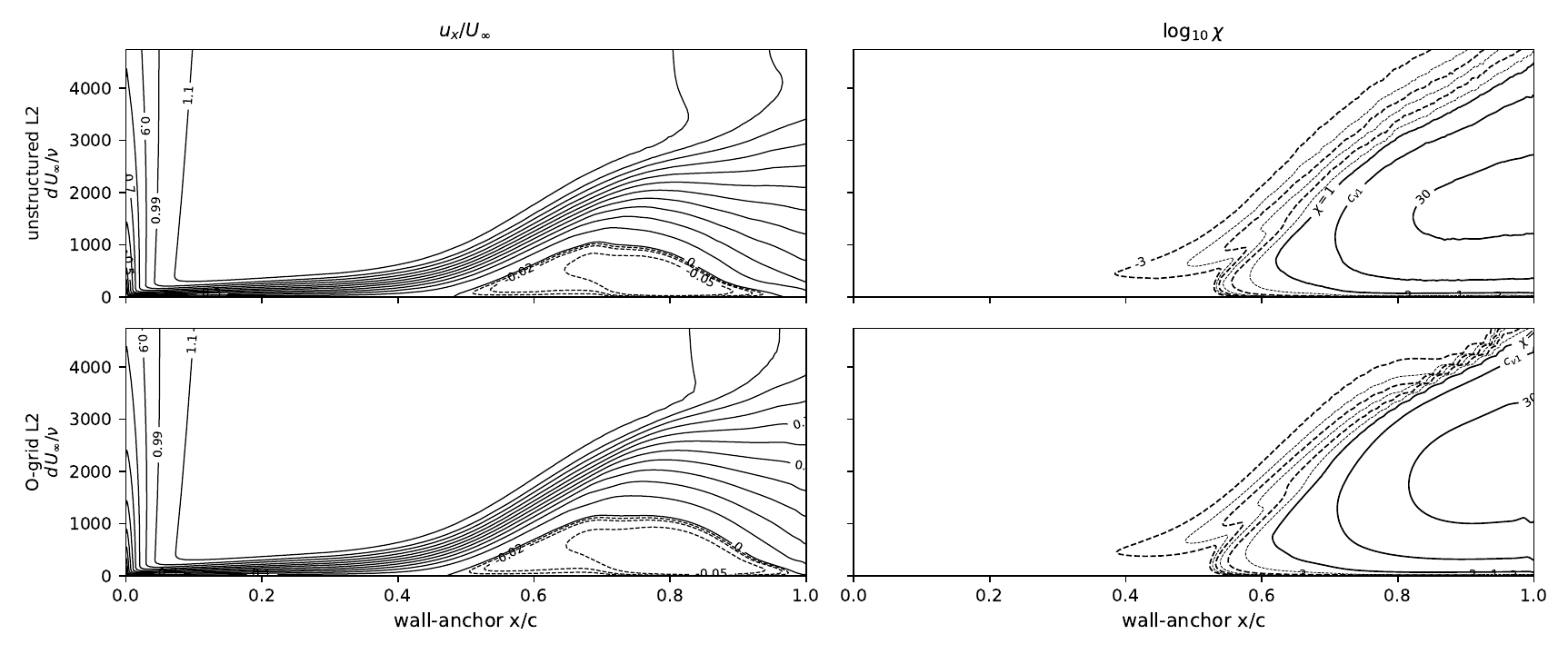}
\figcont{$\alpha=0^\circ$.}
\end{figure}

\begin{figure}[H]\centering
\includegraphics[width=0.99\textwidth,height=0.94\textheight,keepaspectratio]{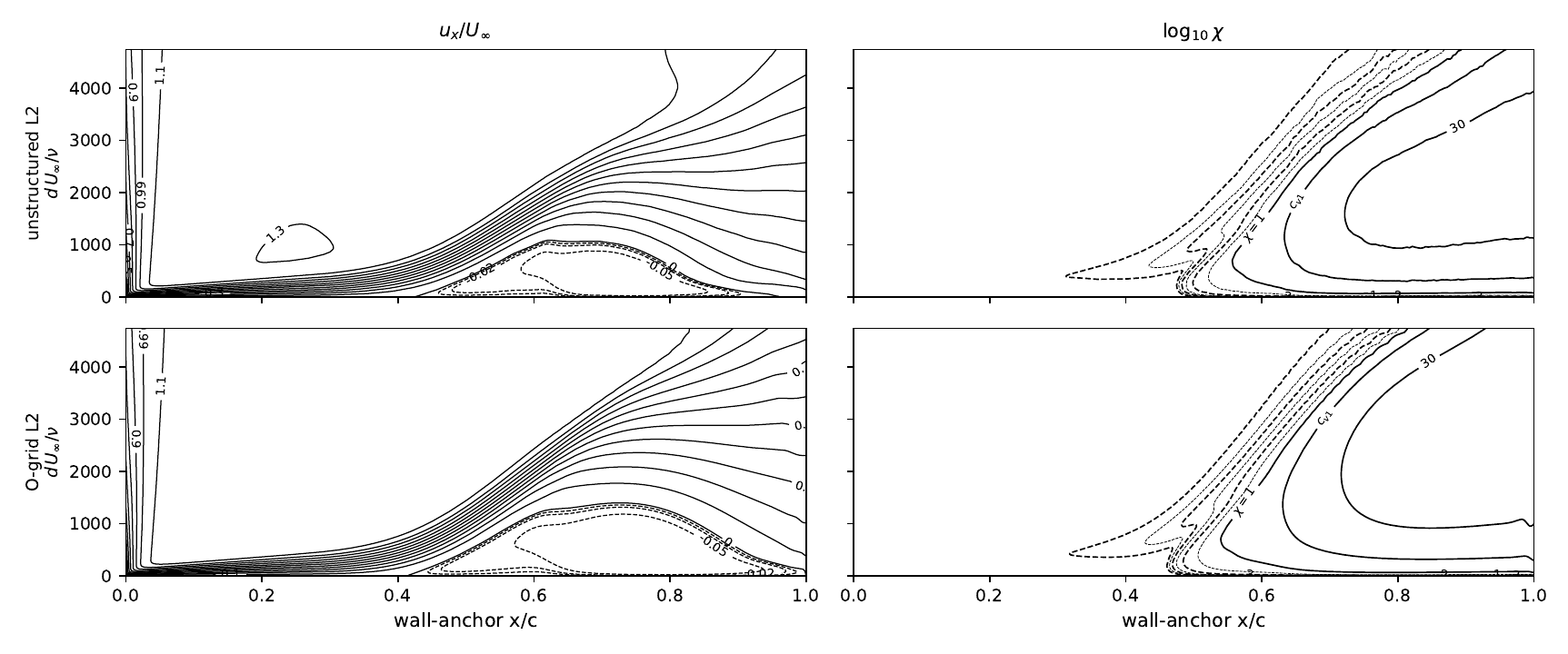}
\figcont{$\alpha=2^\circ$.}
\end{figure}

\begin{figure}[H]\centering
\includegraphics[width=0.99\textwidth,height=0.94\textheight,keepaspectratio]{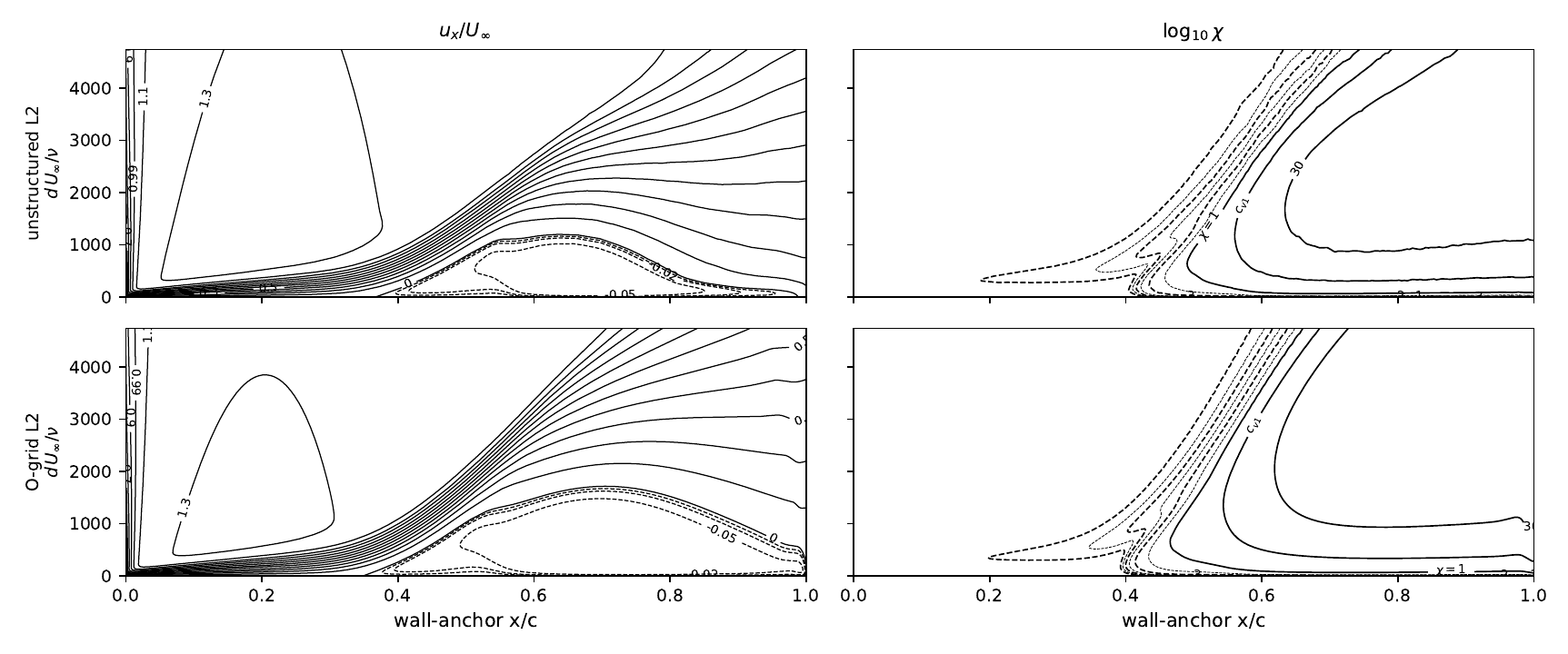}
\figcont{$\alpha=4^\circ$.}
\end{figure}

\begin{figure}[H]\centering
\includegraphics[width=0.99\textwidth,height=0.94\textheight,keepaspectratio]{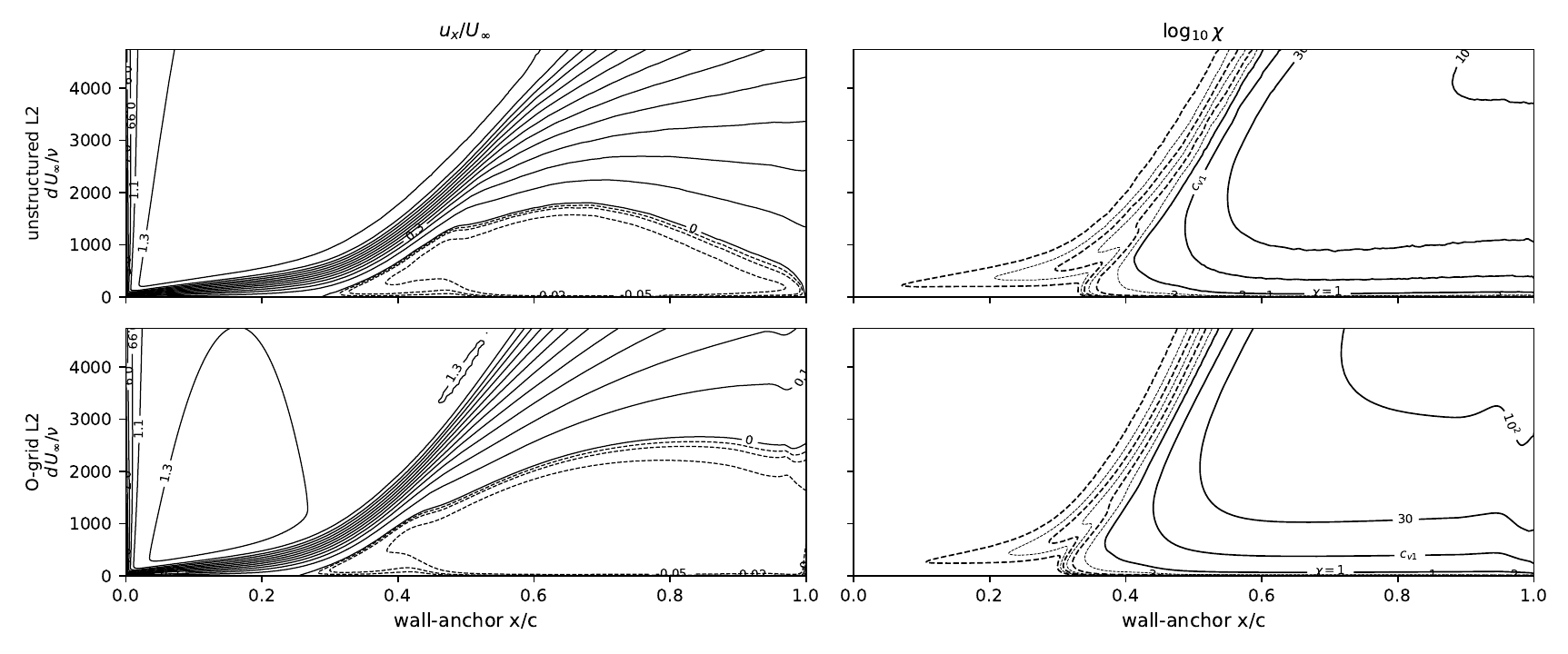}
\figcont{$\alpha=6^\circ$.}
\end{figure}

\begin{figure}[H]\centering
\includegraphics[width=0.99\textwidth,height=0.94\textheight,keepaspectratio]{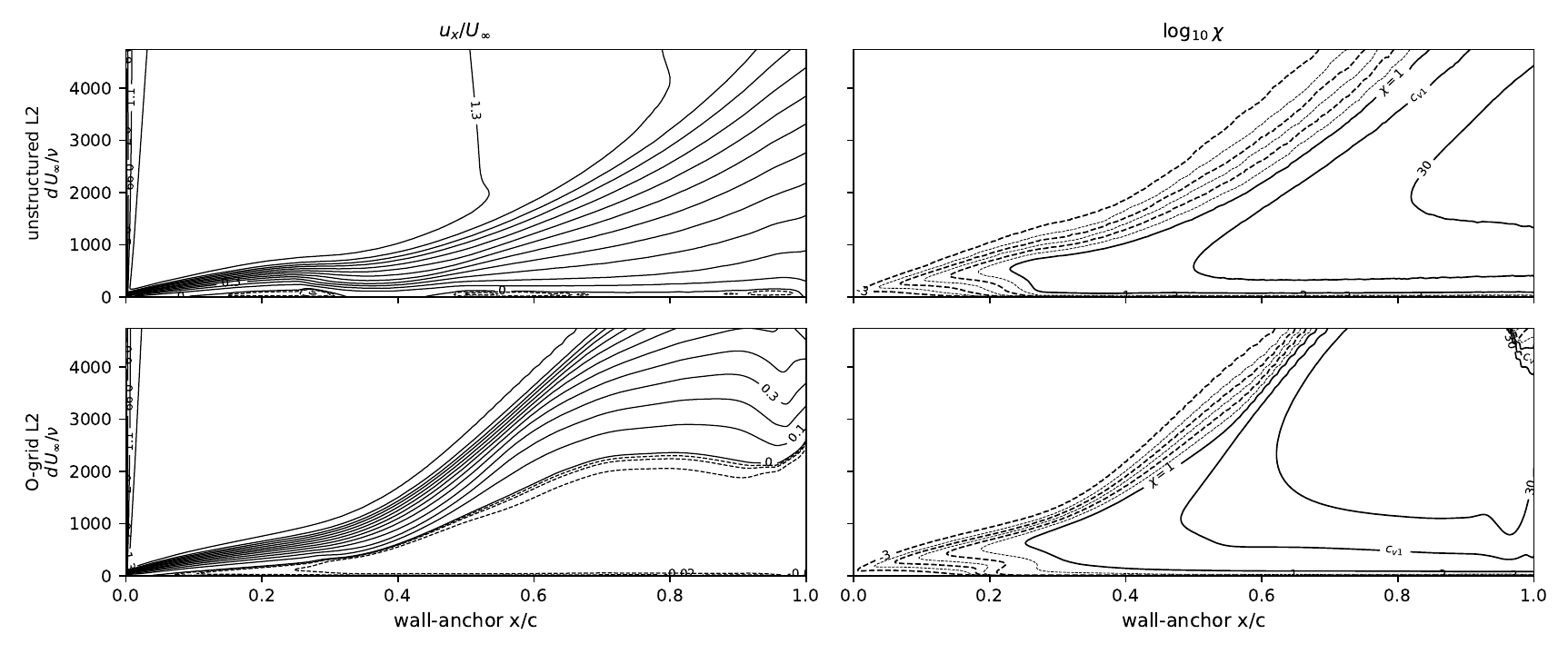}
\figcont{$\alpha=8^\circ$.}
\end{figure}

\begin{figure}[H]\centering
\includegraphics[width=0.99\textwidth,height=0.94\textheight,keepaspectratio]{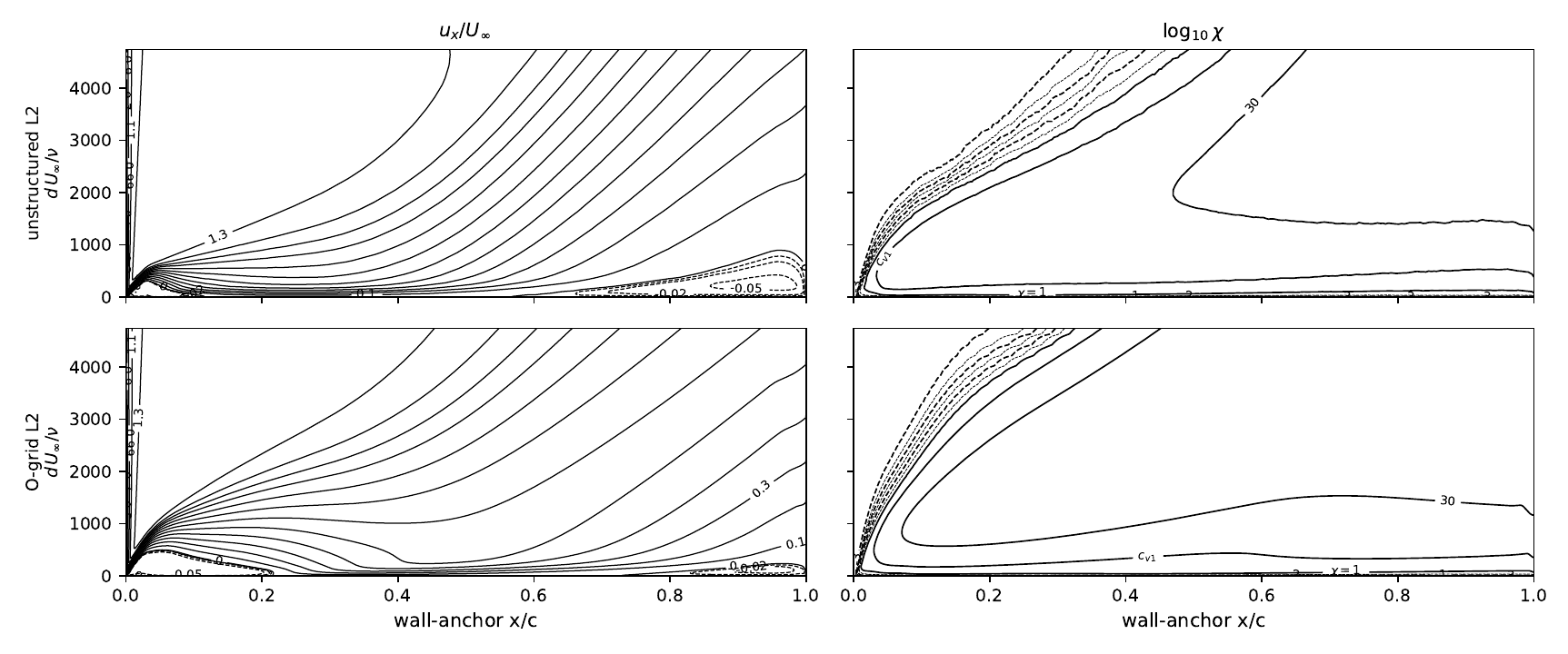}
\figcont{$\alpha=10^\circ$.}
\end{figure}

\begin{figure}[H]\centering
\includegraphics[width=0.99\textwidth,height=0.94\textheight,keepaspectratio]{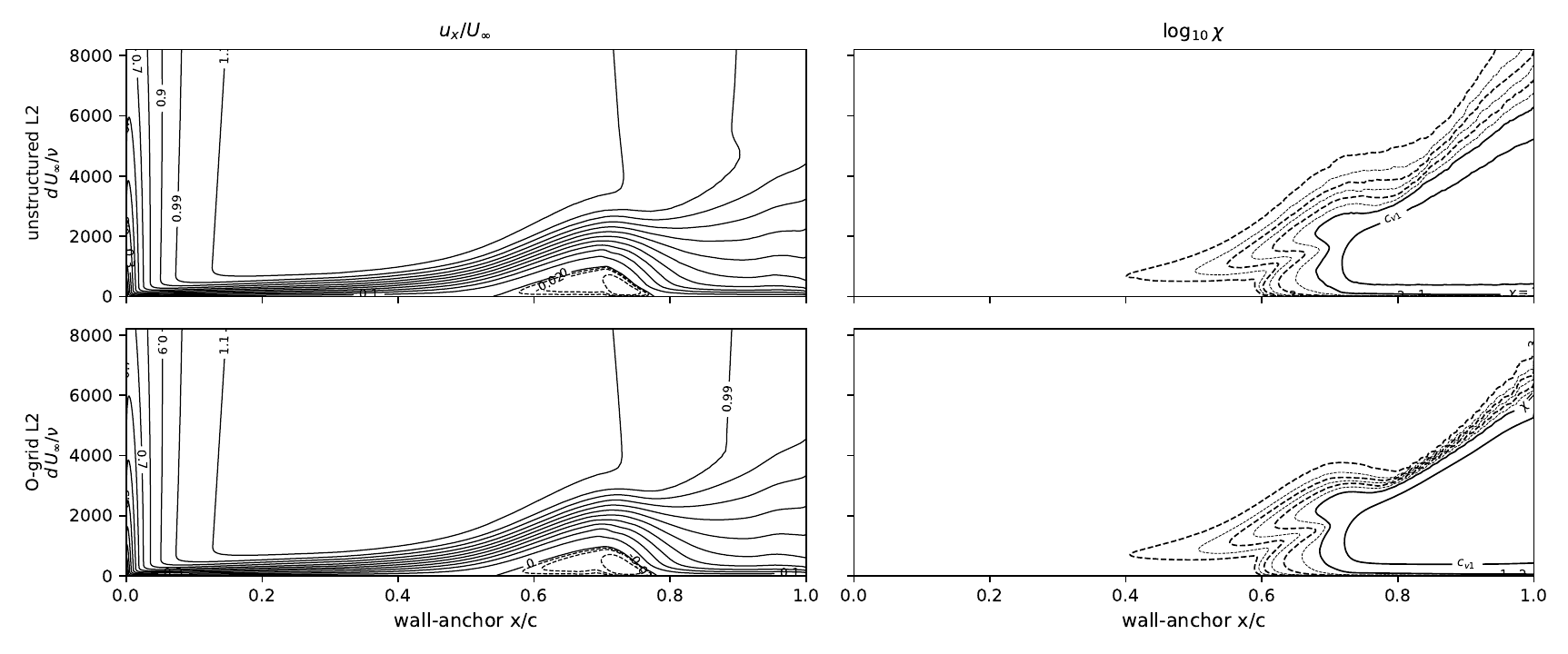}
\caption{Eppler~387 near the upper surface, $Re=3\times10^5$, L2 grids: $u_x/U_\infty$ (left) and $\log_{10}\chi$ (right) against the chordwise station and the wall distance. $\alpha=-2^\circ$.}\label{f:eppsheetRe300k}
\end{figure}

\begin{figure}[H]\centering
\includegraphics[width=0.99\textwidth,height=0.94\textheight,keepaspectratio]{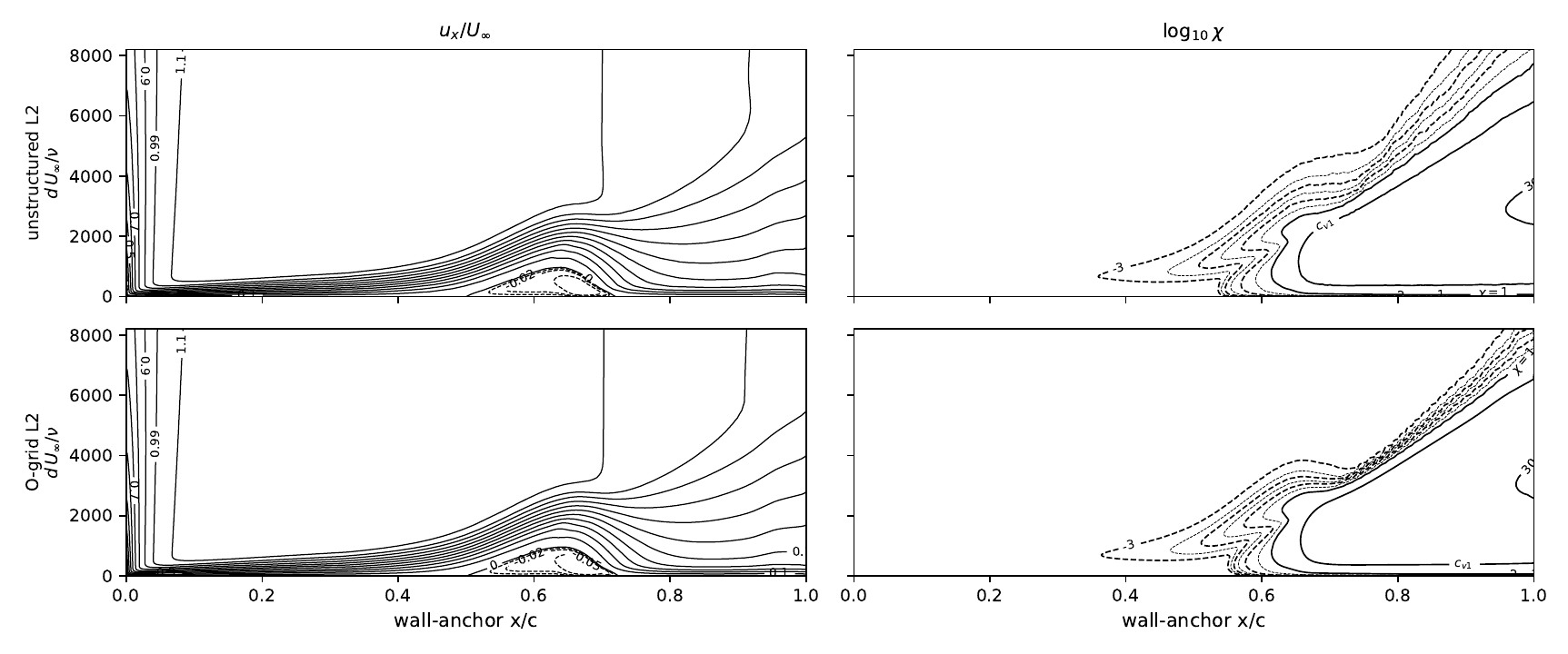}
\figcont{$\alpha=0^\circ$.}
\end{figure}

\begin{figure}[H]\centering
\includegraphics[width=0.99\textwidth,height=0.94\textheight,keepaspectratio]{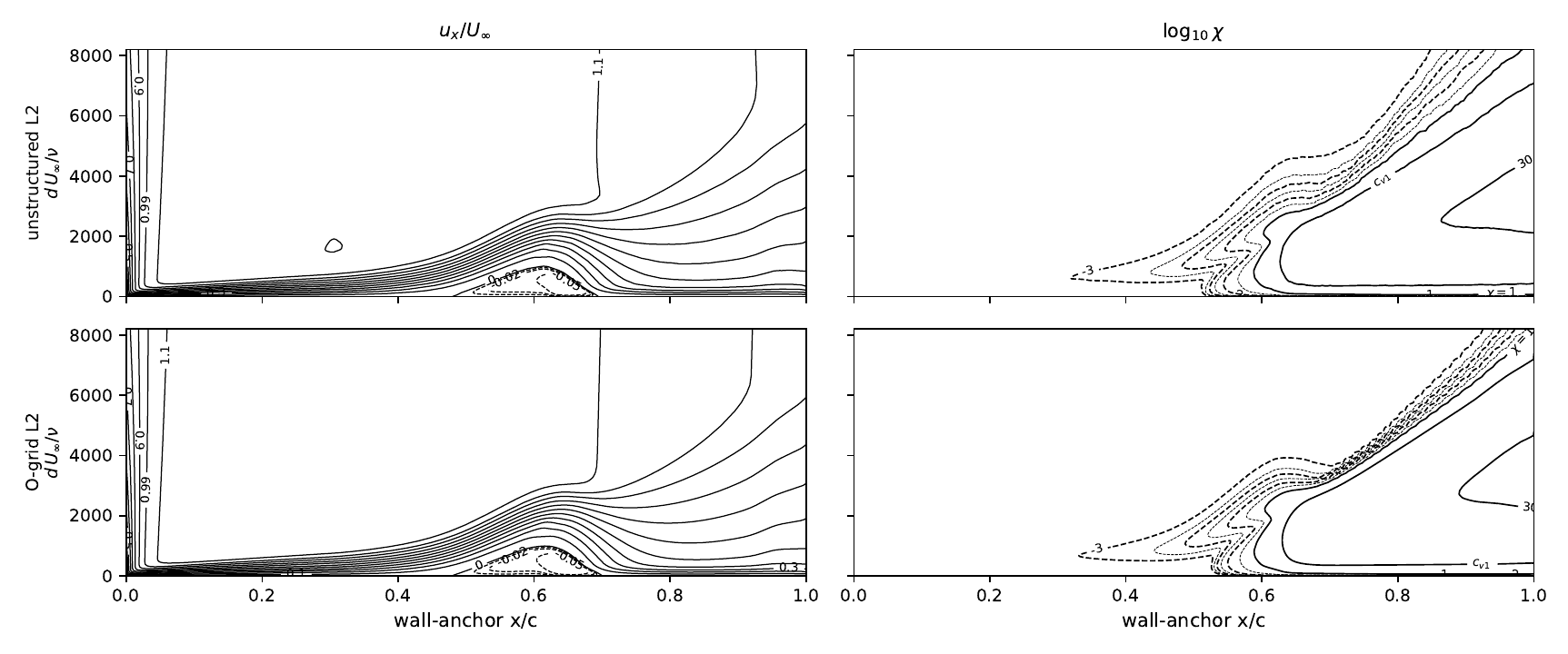}
\figcont{$\alpha=1^\circ$.}
\end{figure}

\begin{figure}[H]\centering
\includegraphics[width=0.99\textwidth,height=0.94\textheight,keepaspectratio]{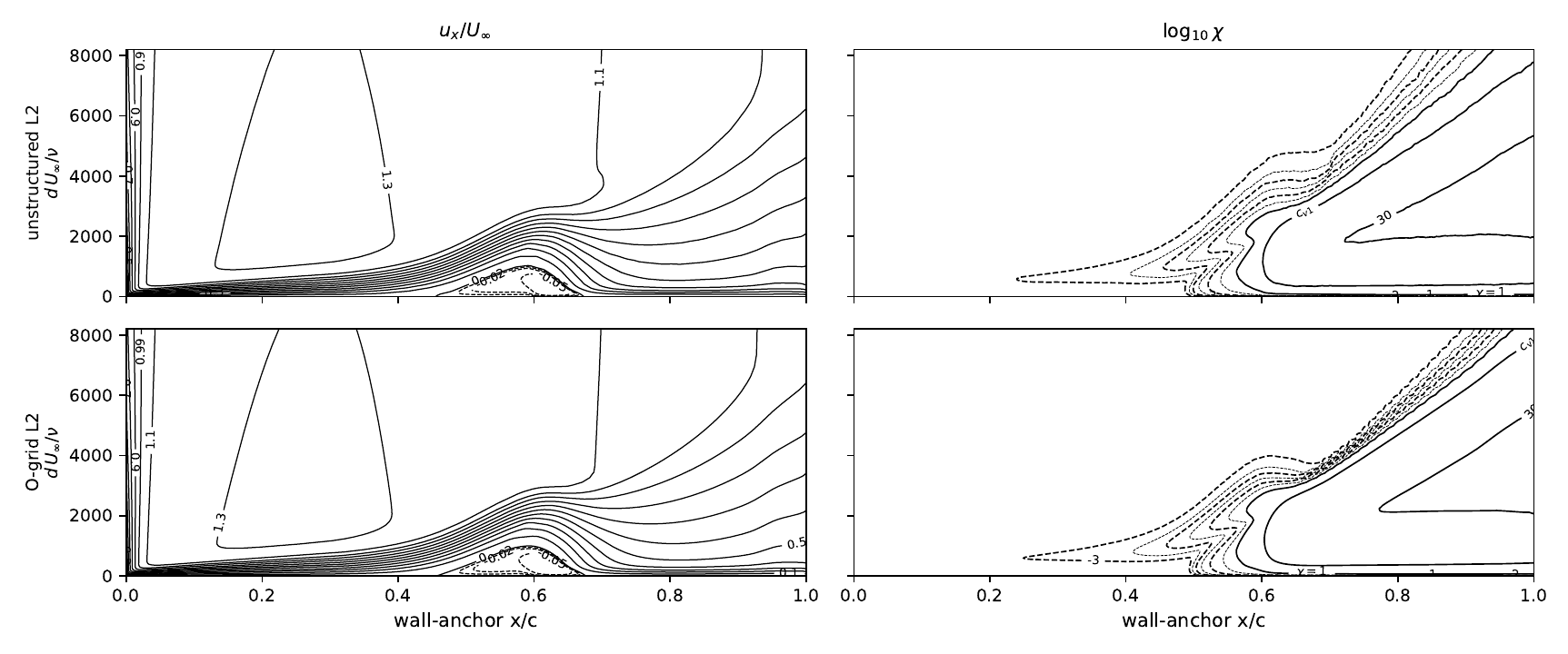}
\figcont{$\alpha=2^\circ$.}
\end{figure}

\begin{figure}[H]\centering
\includegraphics[width=0.99\textwidth,height=0.94\textheight,keepaspectratio]{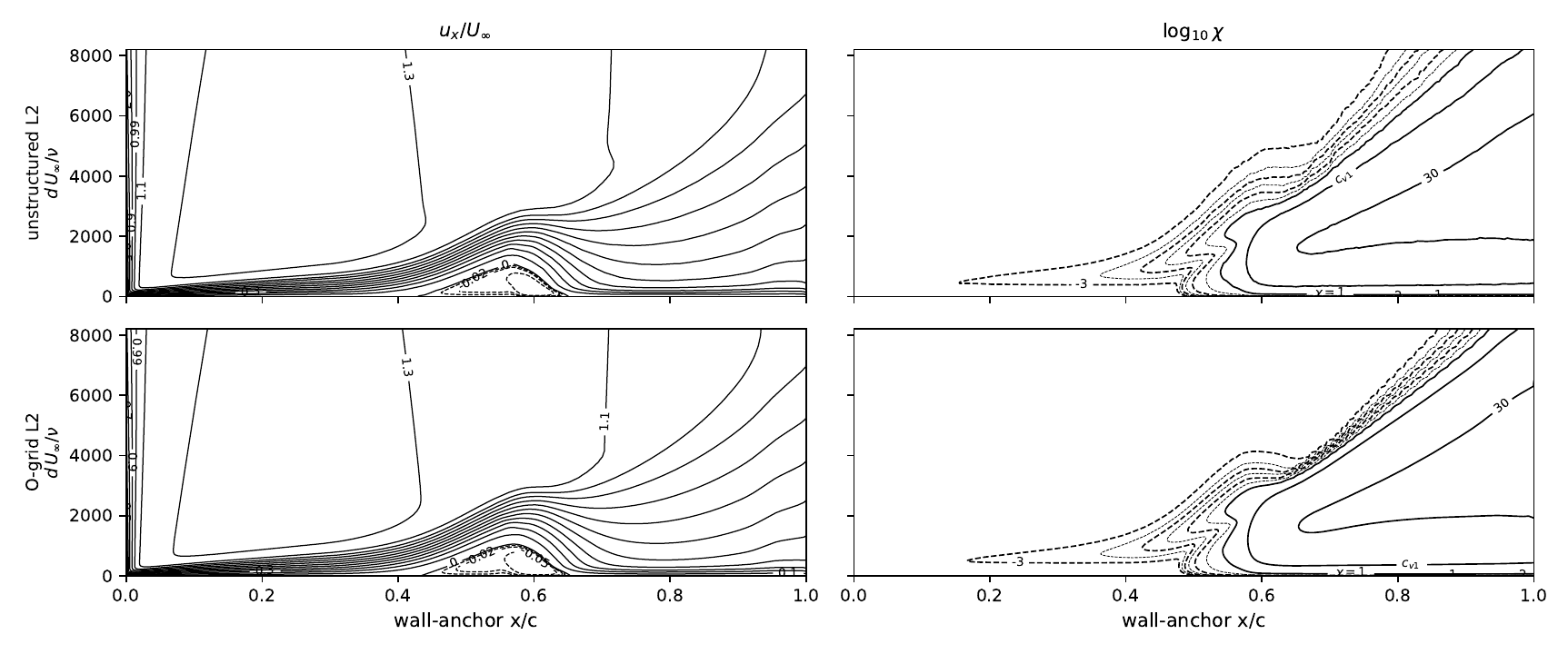}
\figcont{$\alpha=3^\circ$.}
\end{figure}

\begin{figure}[H]\centering
\includegraphics[width=0.99\textwidth,height=0.94\textheight,keepaspectratio]{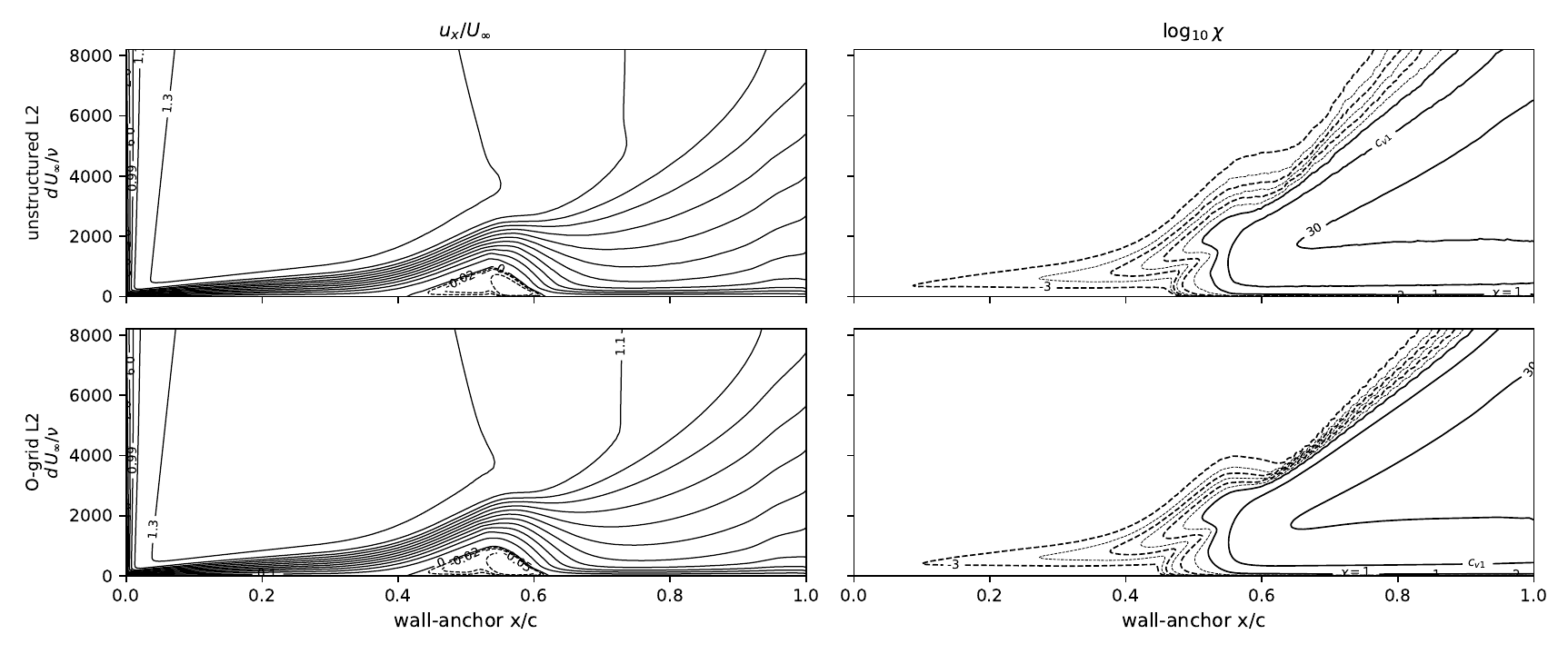}
\figcont{$\alpha=4^\circ$.}
\end{figure}

\begin{figure}[H]\centering
\includegraphics[width=0.99\textwidth,height=0.94\textheight,keepaspectratio]{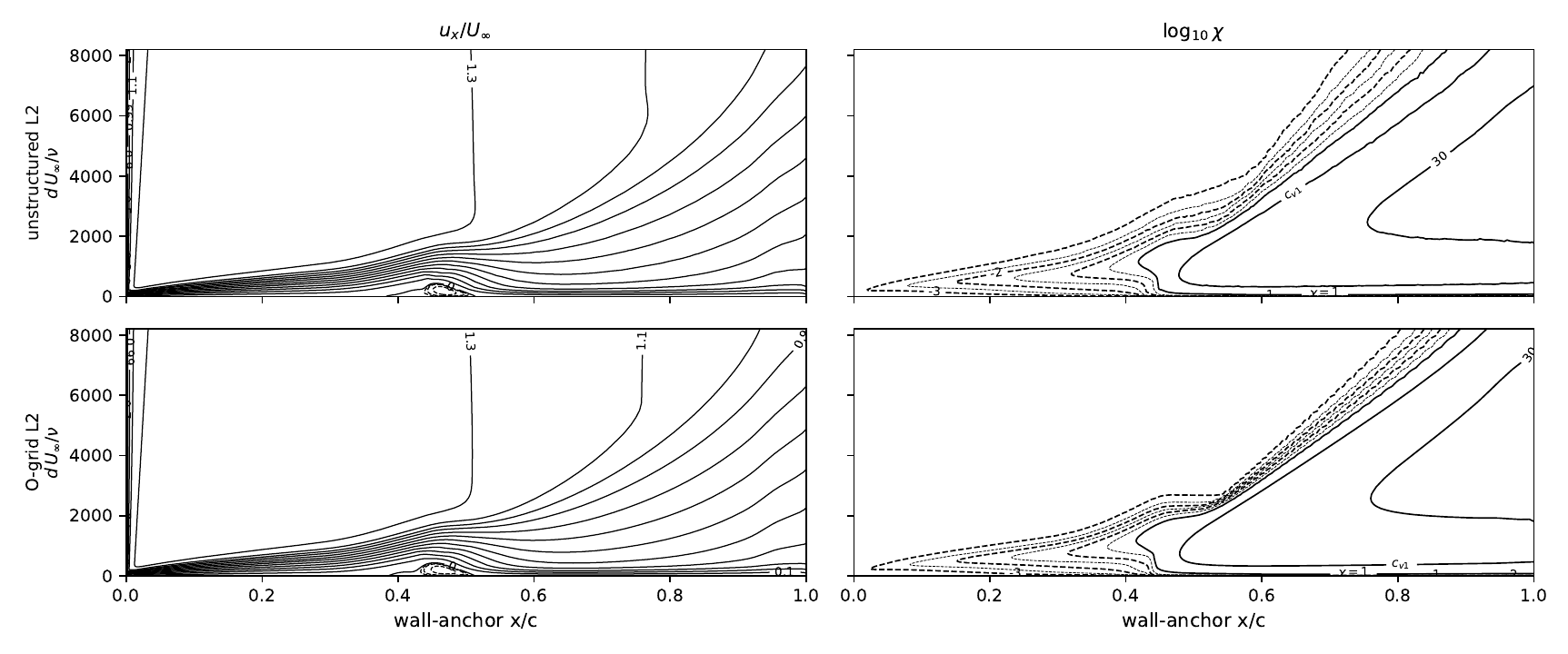}
\figcont{$\alpha=6^\circ$.}
\end{figure}

\begin{figure}[H]\centering
\includegraphics[width=0.99\textwidth,height=0.94\textheight,keepaspectratio]{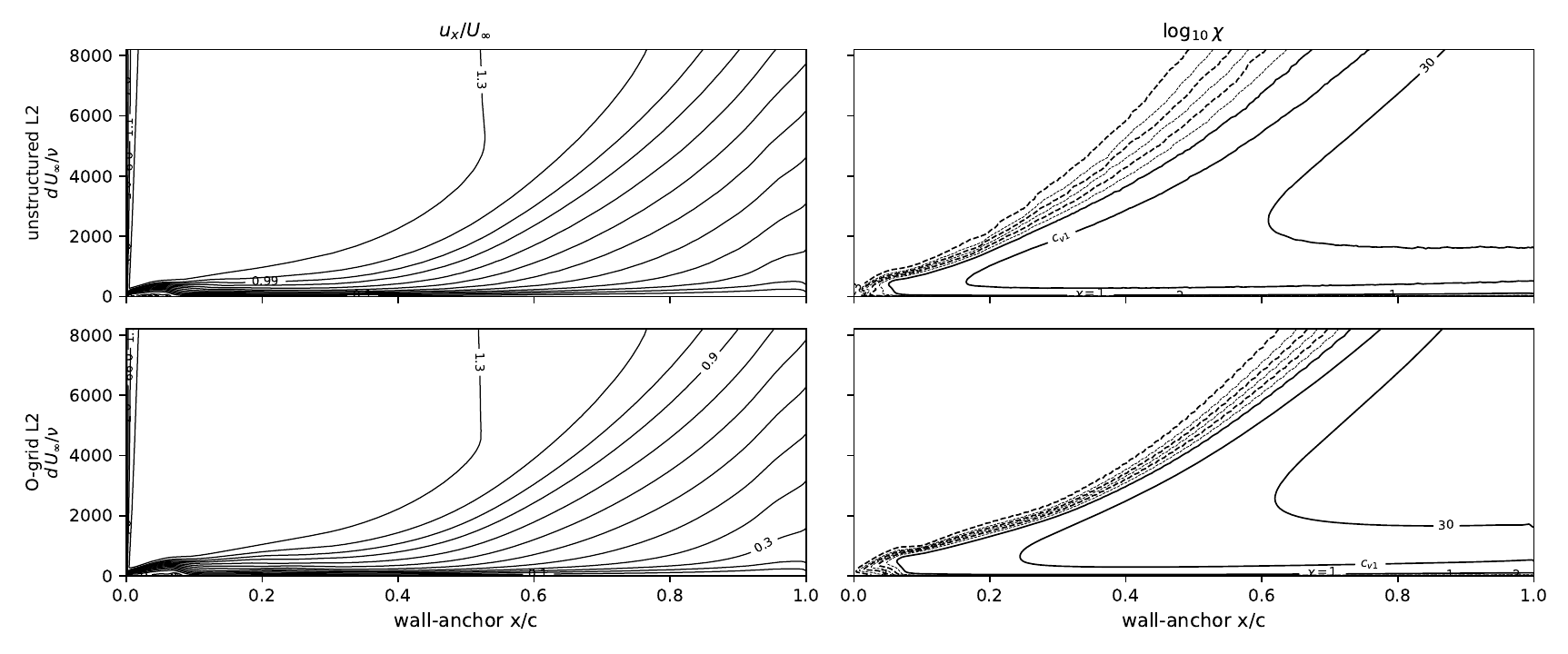}
\figcont{$\alpha=8^\circ$.}
\end{figure}

\begin{figure}[H]\centering
\includegraphics[width=0.99\textwidth,height=0.94\textheight,keepaspectratio]{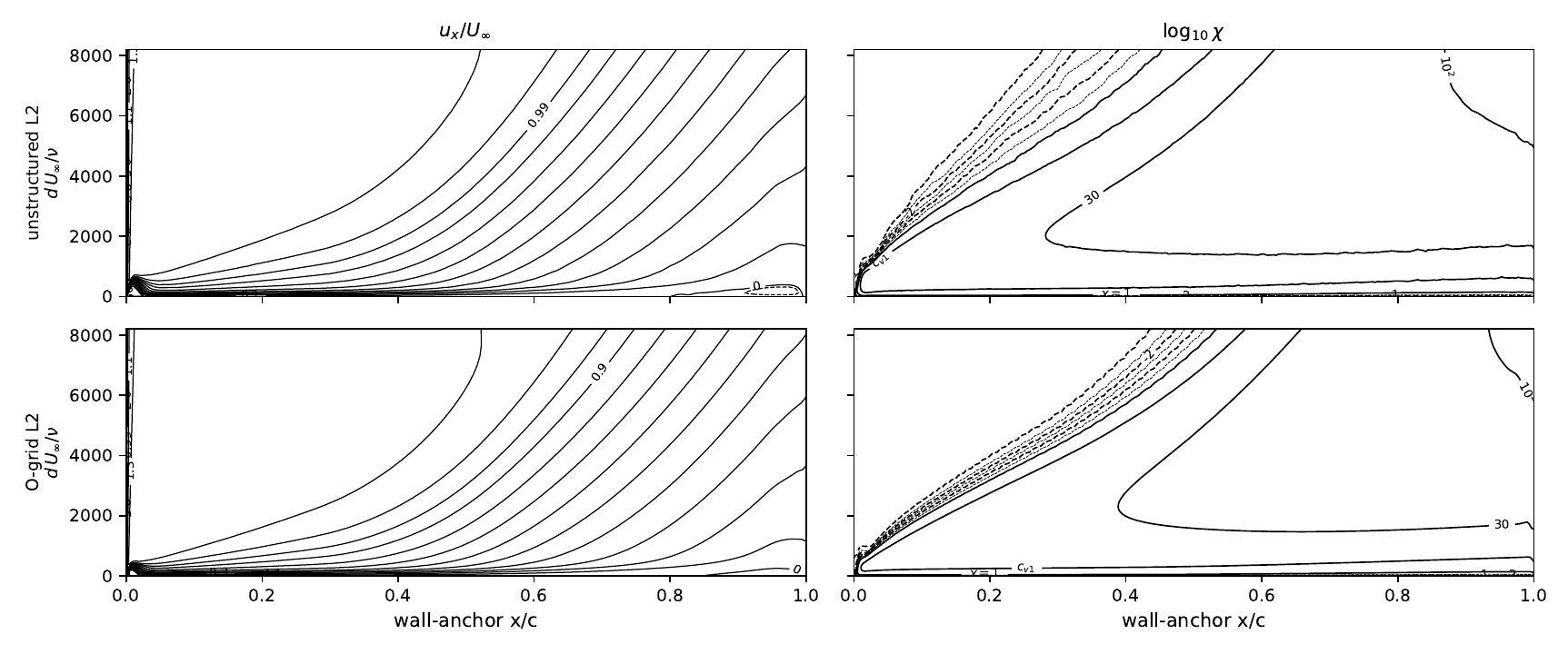}
\figcont{$\alpha=10^\circ$.}
\end{figure}

\subsection*{\tocentry{subsection}{A Two-Element Section}A Two-Element Section}

\begin{figure}[H]\centering
\includegraphics[width=\textwidth]{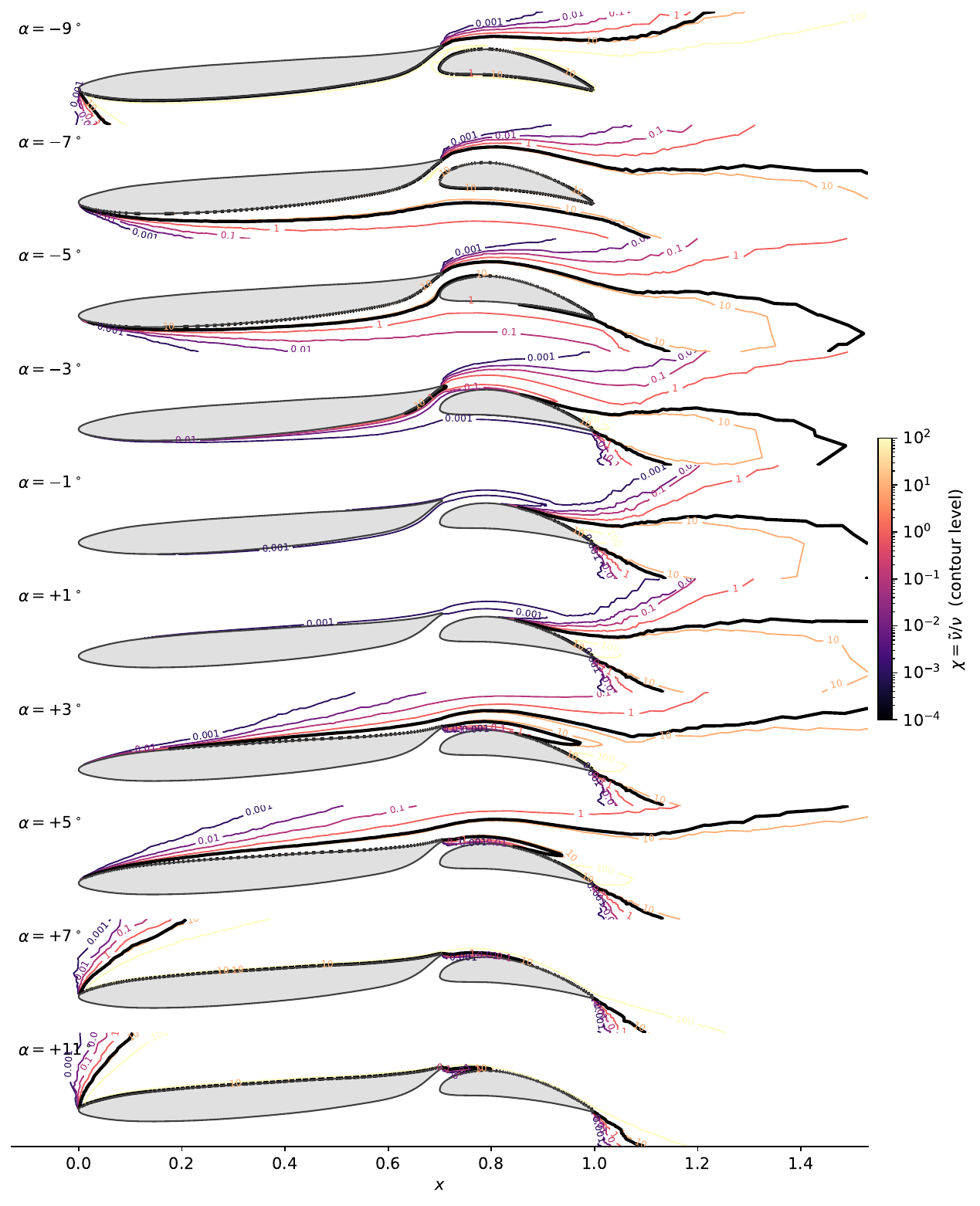}
\caption{Two-element section: contours of $\chi$ at each incidence on the three grid levels; levels as in Fig.~\ref{f:twoelemchiL2}. L0.}
\label{f:twoelemchiL1}
\end{figure}

\begin{figure}[H]\centering
\includegraphics[width=\textwidth]{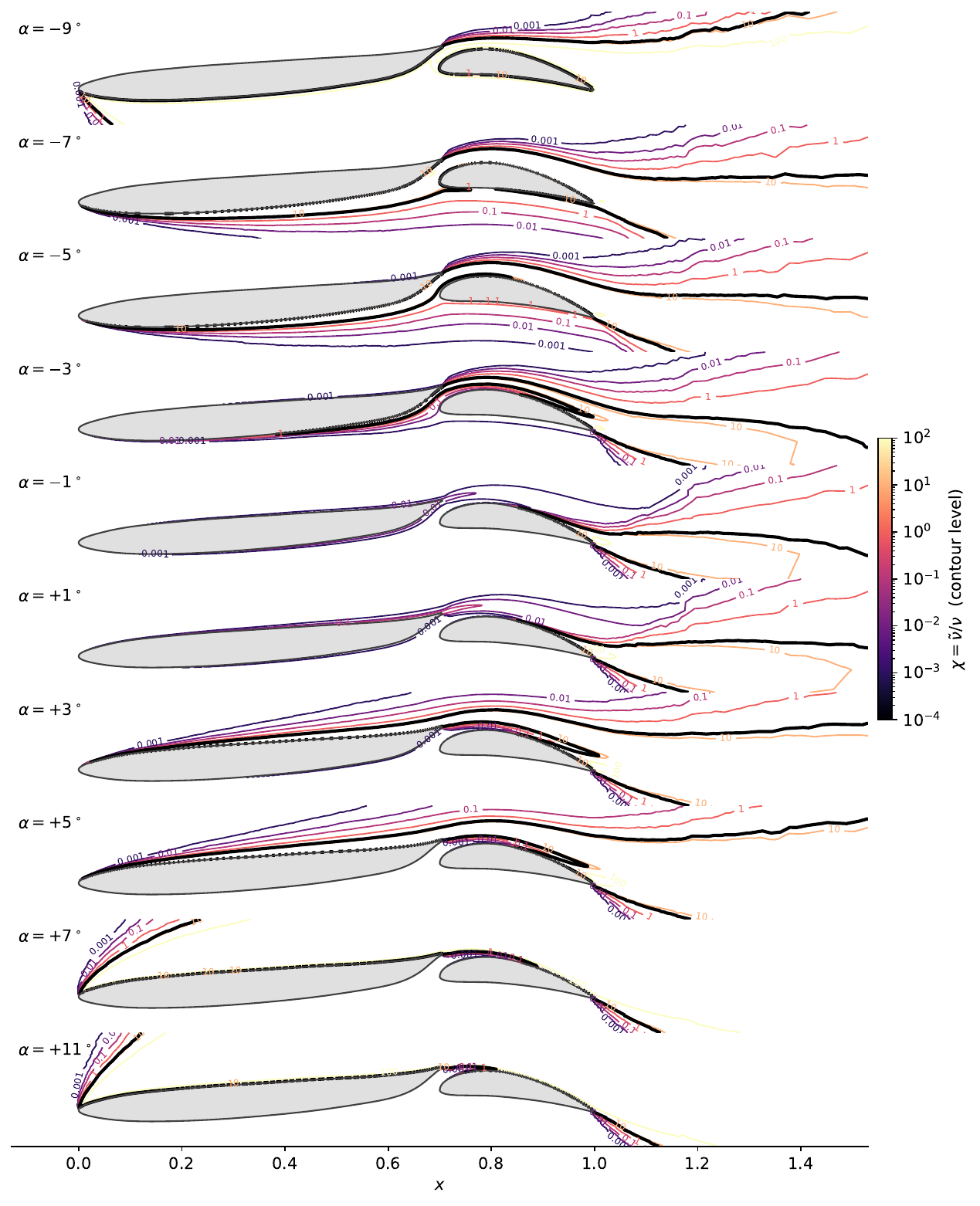}
\figcont{L1.}
\end{figure}

\begin{figure}[H]\centering
\includegraphics[width=\textwidth]{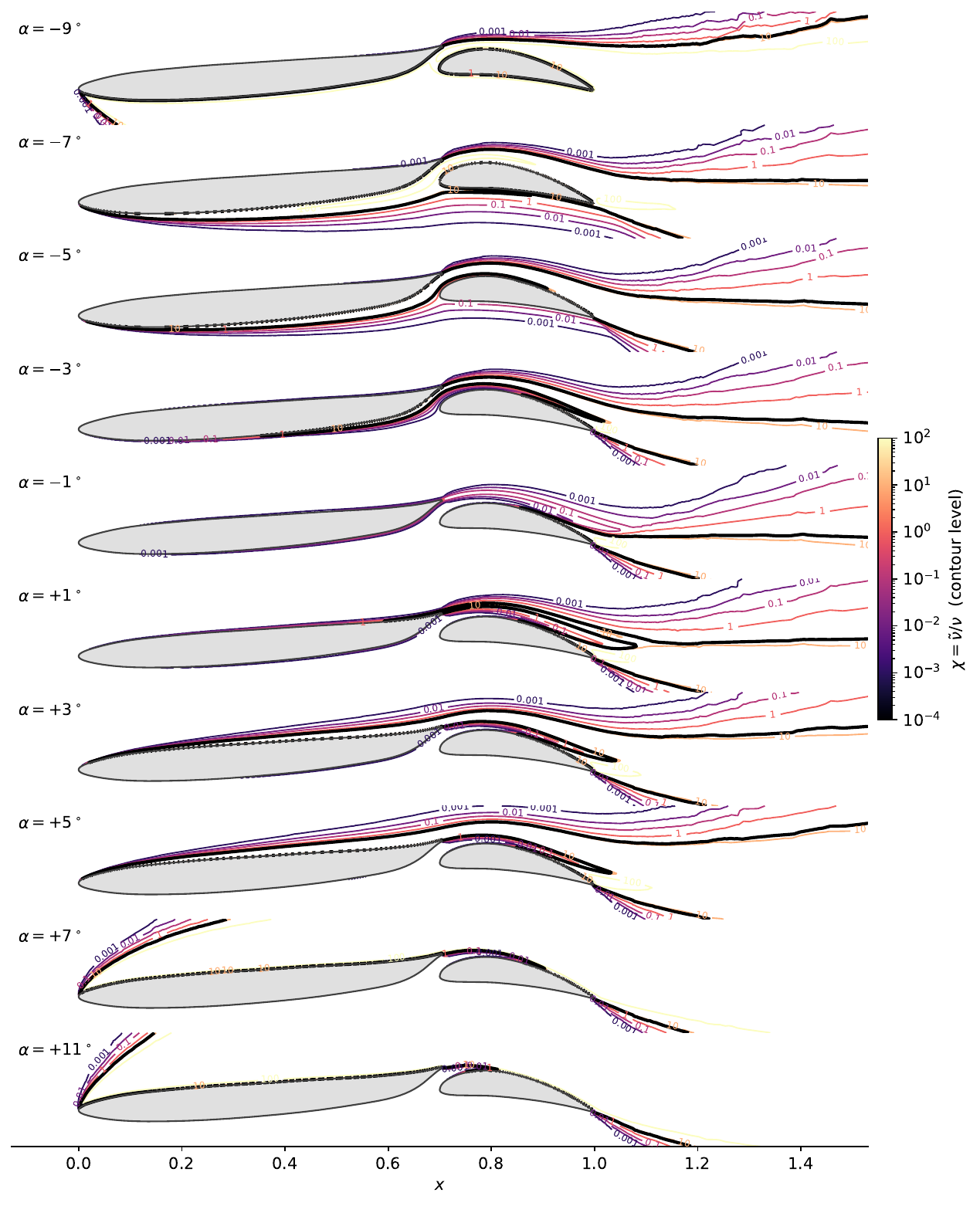}
\figcont{L2.}
\end{figure}

\begin{figure}[H]\centering
\includegraphics[width=\textwidth,page=1]{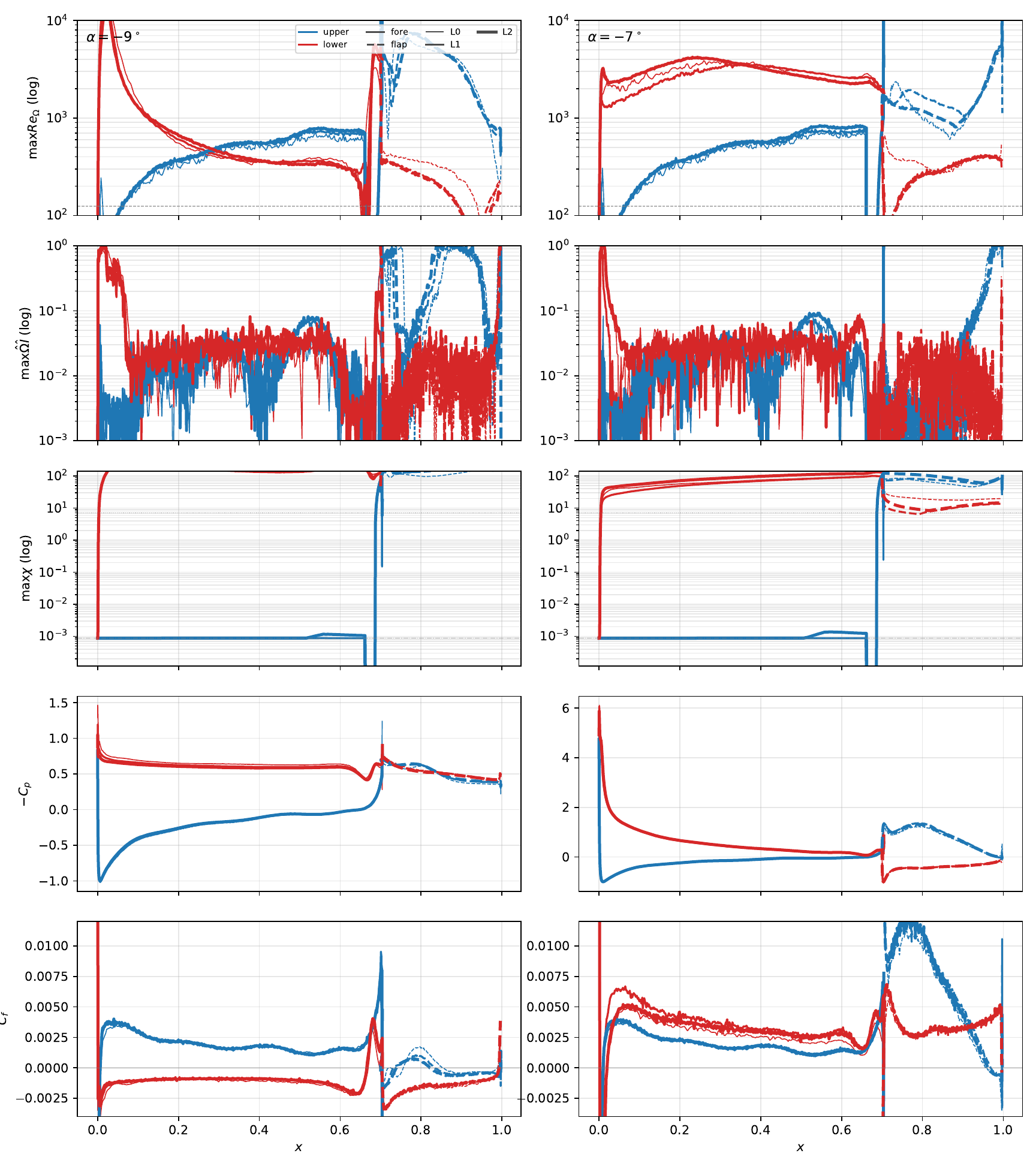}
\caption{Two-element section: surface distributions, two incidences per page. Rows: largest $Re_\Omega$ on the wall normal; largest $\hat\Omega\hat I$; largest $\chi$; $-C_p$; signed $C_f$. Blue: upper surface. Red: lower surface. Solid: main element. Dashed: flap. The line thickness increases from L0 to L2. $\alpha=-9^\circ$, $-7^\circ$.}\label{f:twoelemsuite1}
\end{figure}

\begin{figure}[H]\centering
\includegraphics[width=\textwidth,page=2]{twoelement_surface_suite.pdf}
\figcont{$\alpha=-5^\circ$, $-3^\circ$.}
\end{figure}

\begin{figure}[H]\centering
\includegraphics[width=\textwidth,page=3]{twoelement_surface_suite.pdf}
\figcont{$\alpha=-1^\circ$, $+1^\circ$.}
\end{figure}

\begin{figure}[H]\centering
\includegraphics[width=\textwidth,page=4]{twoelement_surface_suite.pdf}
\figcont{$\alpha=+3^\circ$, $+5^\circ$.}
\end{figure}

\begin{figure}[H]\centering
\includegraphics[width=\textwidth,page=5]{twoelement_surface_suite.pdf}
\figcont{$\alpha=+7^\circ$, $+11^\circ$.}
\end{figure}

\subsection*{\tocentry{subsection}{The Cylinder Drag Crisis}The Cylinder Drag Crisis}

\begin{figure}[H]\centering
\includegraphics[width=\textwidth,height=0.85\textheight,keepaspectratio,page=1]{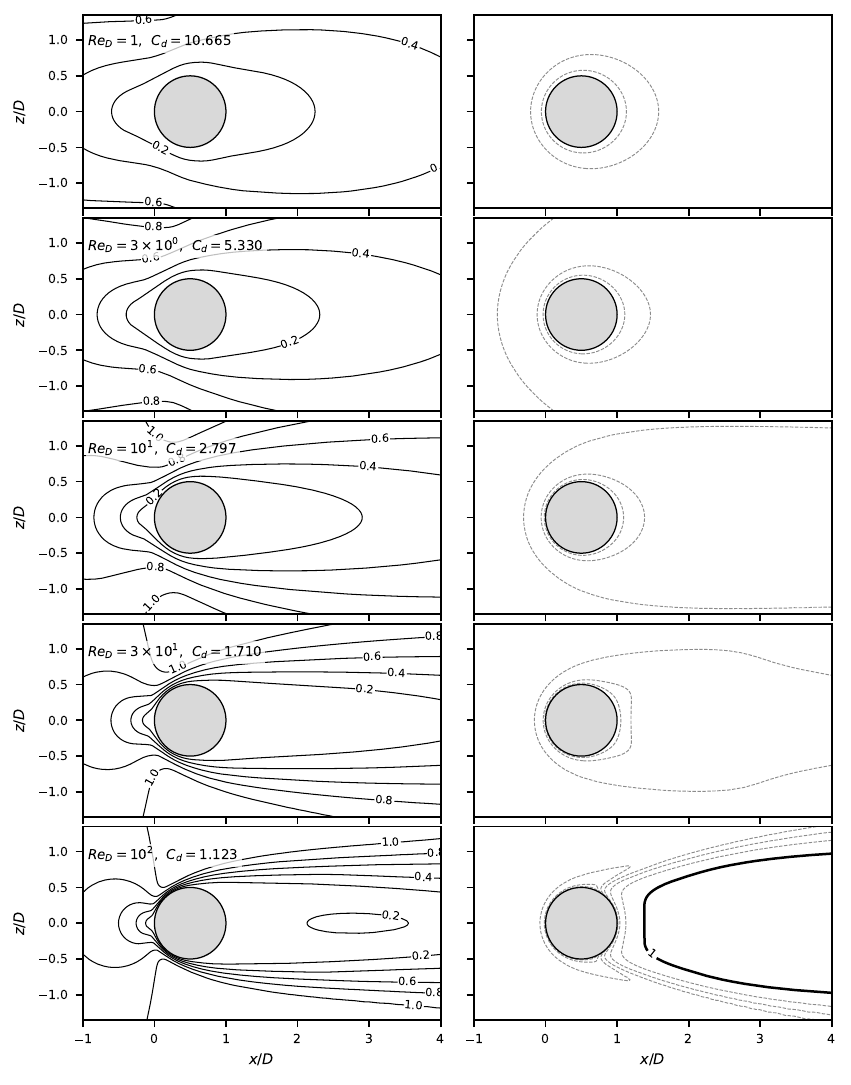}
\caption{Circular cylinder: $|\mathbf{u}|/U_\infty$ (left) and $\chi$ (right) at every Reynolds number of Fig.~\ref{f:dragcrisiscd}, in ascending order. Contours of $\chi$: dashed below one, solid at $1$, $c_{v1}$, $30$ and $10^2$ to $10^4$. $Re_D=1$ to $10^2$.}\label{f:dragcrisisfields}
\end{figure}

\begin{figure}[H]\centering
\includegraphics[width=\textwidth,height=0.90\textheight,keepaspectratio,page=2]{dragcrisis_fields.pdf}
\figcont{$Re_D=3\times10^2$--$3\times10^4$.}
\end{figure}

\begin{figure}[H]\centering
\includegraphics[width=\textwidth,height=0.90\textheight,keepaspectratio,page=3]{dragcrisis_fields.pdf}
\figcont{$Re_D=10^5$--$10^6$.}
\end{figure}

\begin{figure}[H]\centering
\includegraphics[width=\textwidth,height=0.90\textheight,keepaspectratio,page=4]{dragcrisis_fields.pdf}
\figcont{$Re_D=1.78\times10^6$--$3\times10^7$.}
\end{figure}

\begin{figure}[H]\centering
\includegraphics[width=\textwidth,height=0.90\textheight,keepaspectratio,page=5]{dragcrisis_fields.pdf}
\figcont{$Re_D=10^8$--$10^{10}$.}
\end{figure}

\subsection*{\tocentry{subsection}{The Daedalus Wing}The Daedalus Wing}

\begin{figure}[H]\centering
\includegraphics[width=0.99\textwidth,height=0.85\textheight,keepaspectratio]{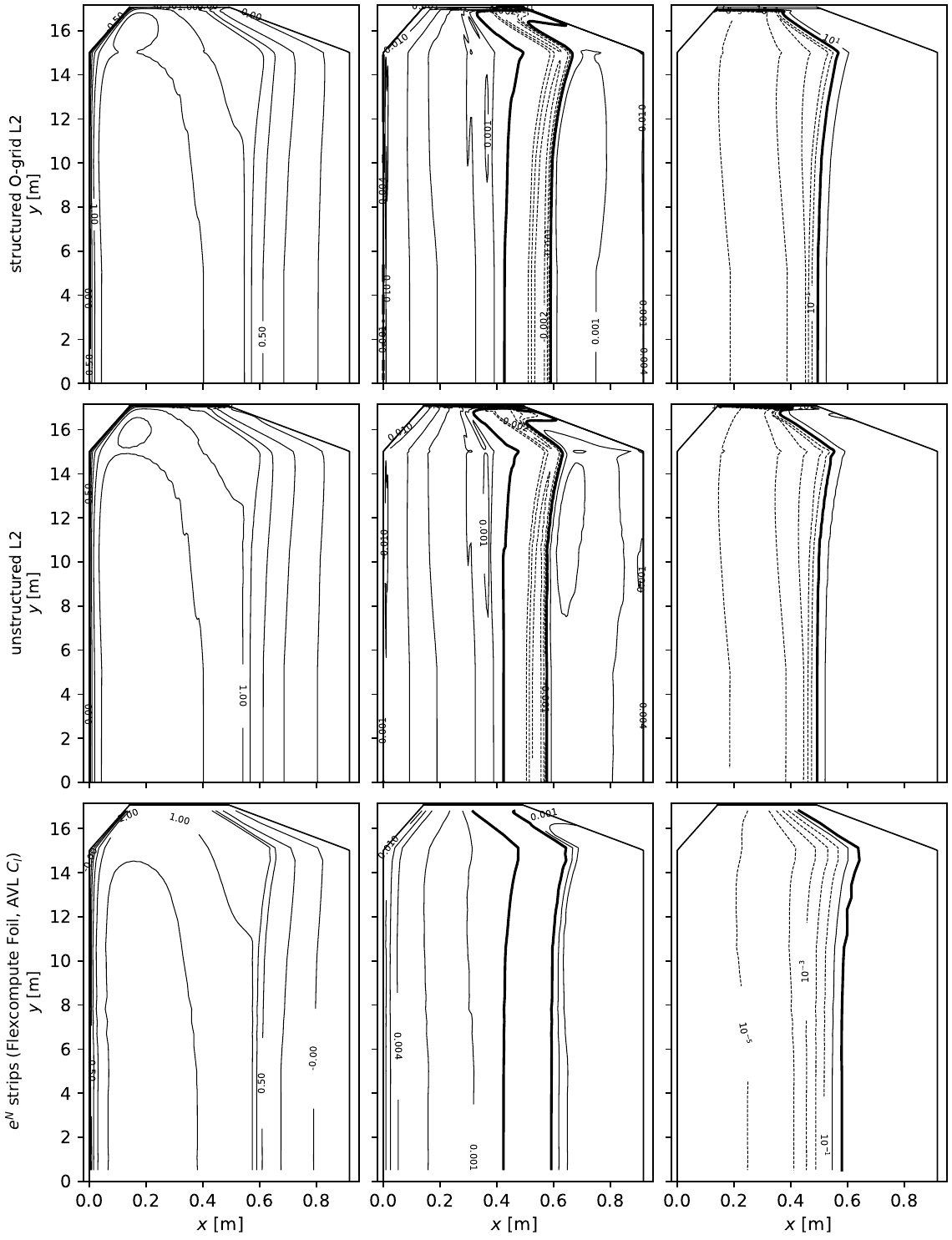}
\caption{Daedalus surface maps, $\alpha=5^\circ$; rows and columns as in Fig.~\ref{f:daesurf}.}\label{f:daesurfall}
\end{figure}

\begin{figure}[H]\centering
\includegraphics[width=0.99\textwidth,height=0.94\textheight,keepaspectratio]{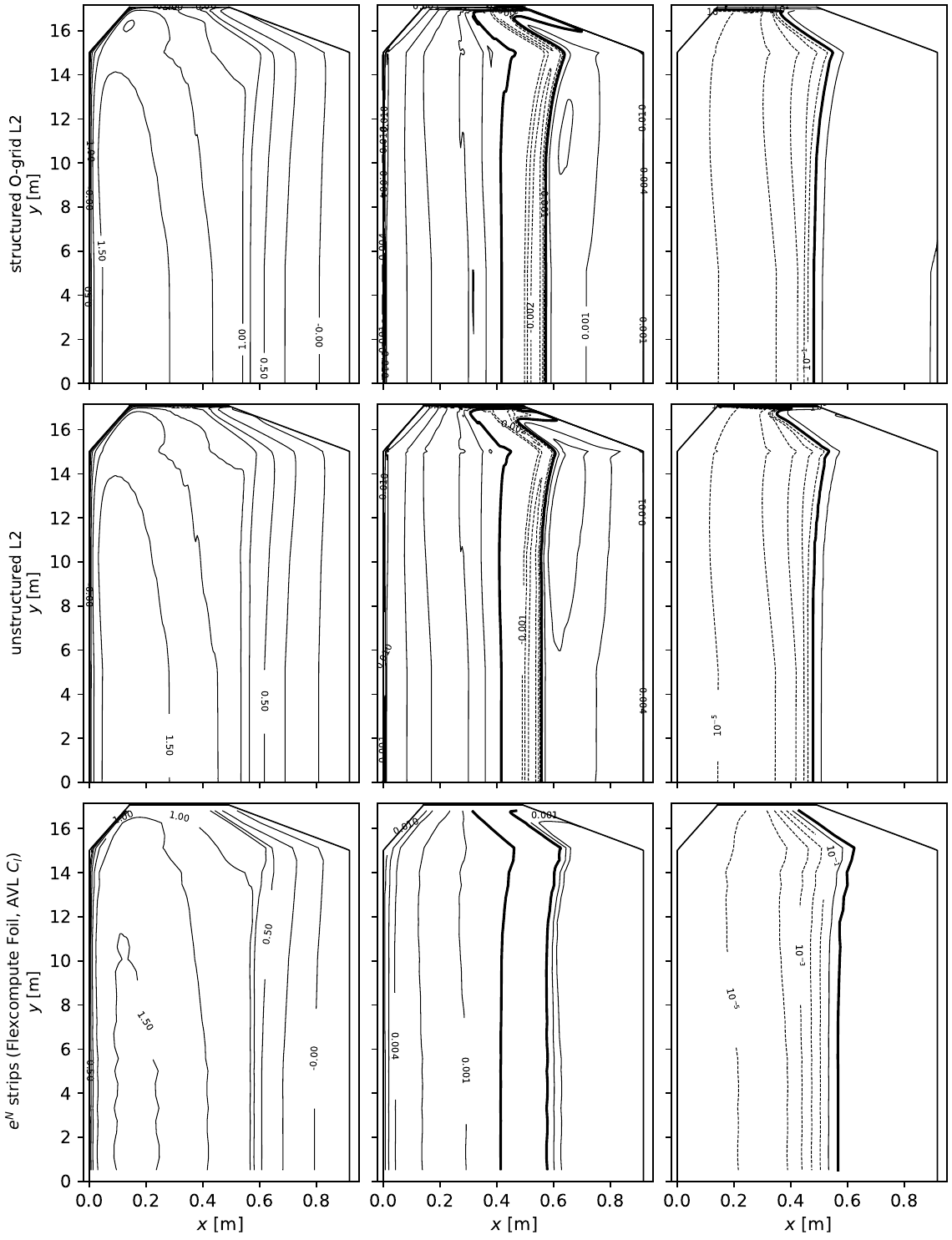}
\figcont{$\alpha=6^\circ$.}
\end{figure}

\begin{figure}[H]\centering
\includegraphics[width=0.99\textwidth,height=0.94\textheight,keepaspectratio]{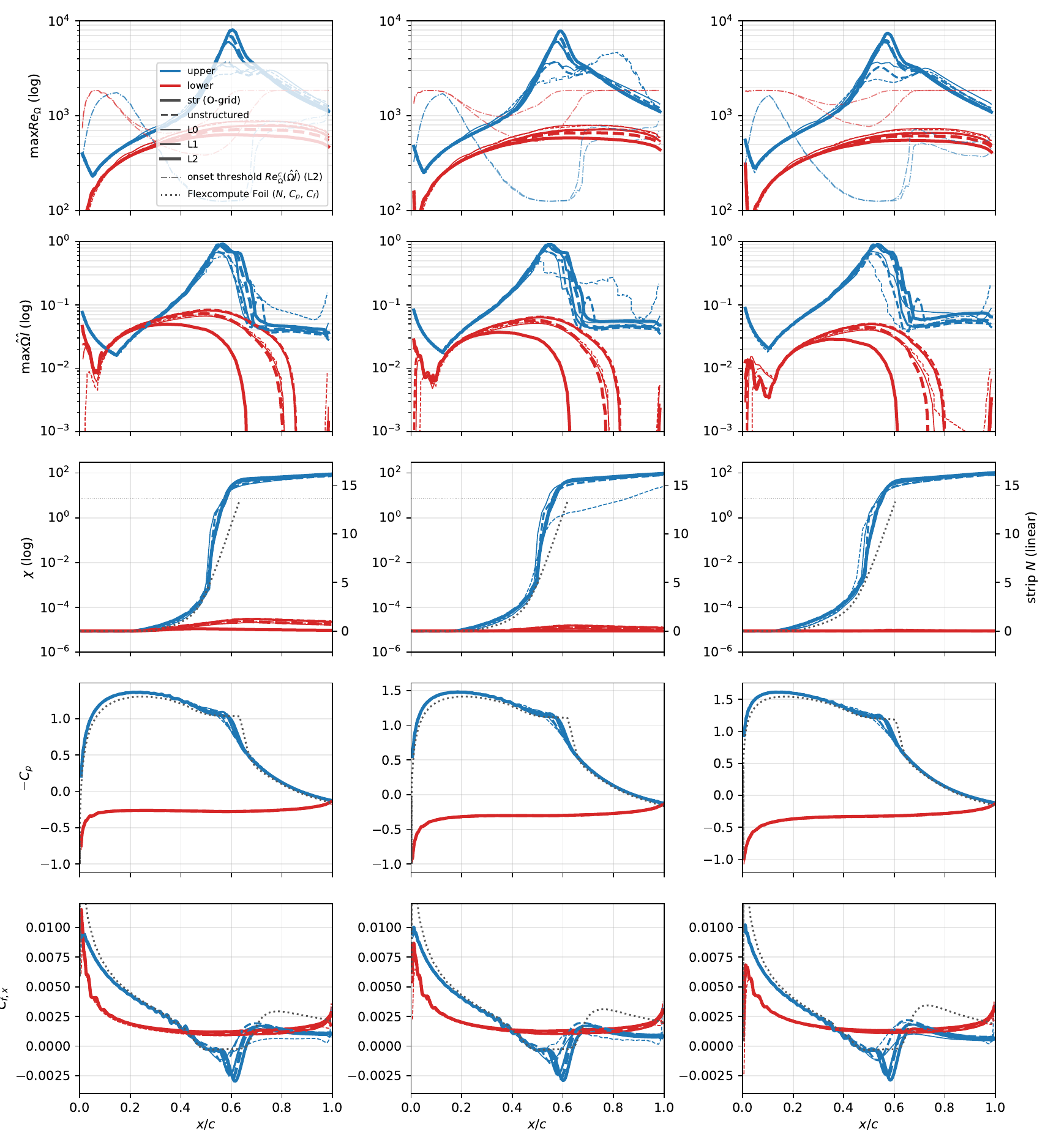}
\caption{Daedalus wing: section distributions on all six grids, with the reference, at $\alpha=4^\circ$, $5^\circ$ and $6^\circ$ (columns); rows as in Fig.~\ref{f:nlfcf0}. $\eta=0.10$.}\label{f:daesections}
\end{figure}

\begin{figure}[H]\centering
\includegraphics[width=0.99\textwidth,height=0.94\textheight,keepaspectratio]{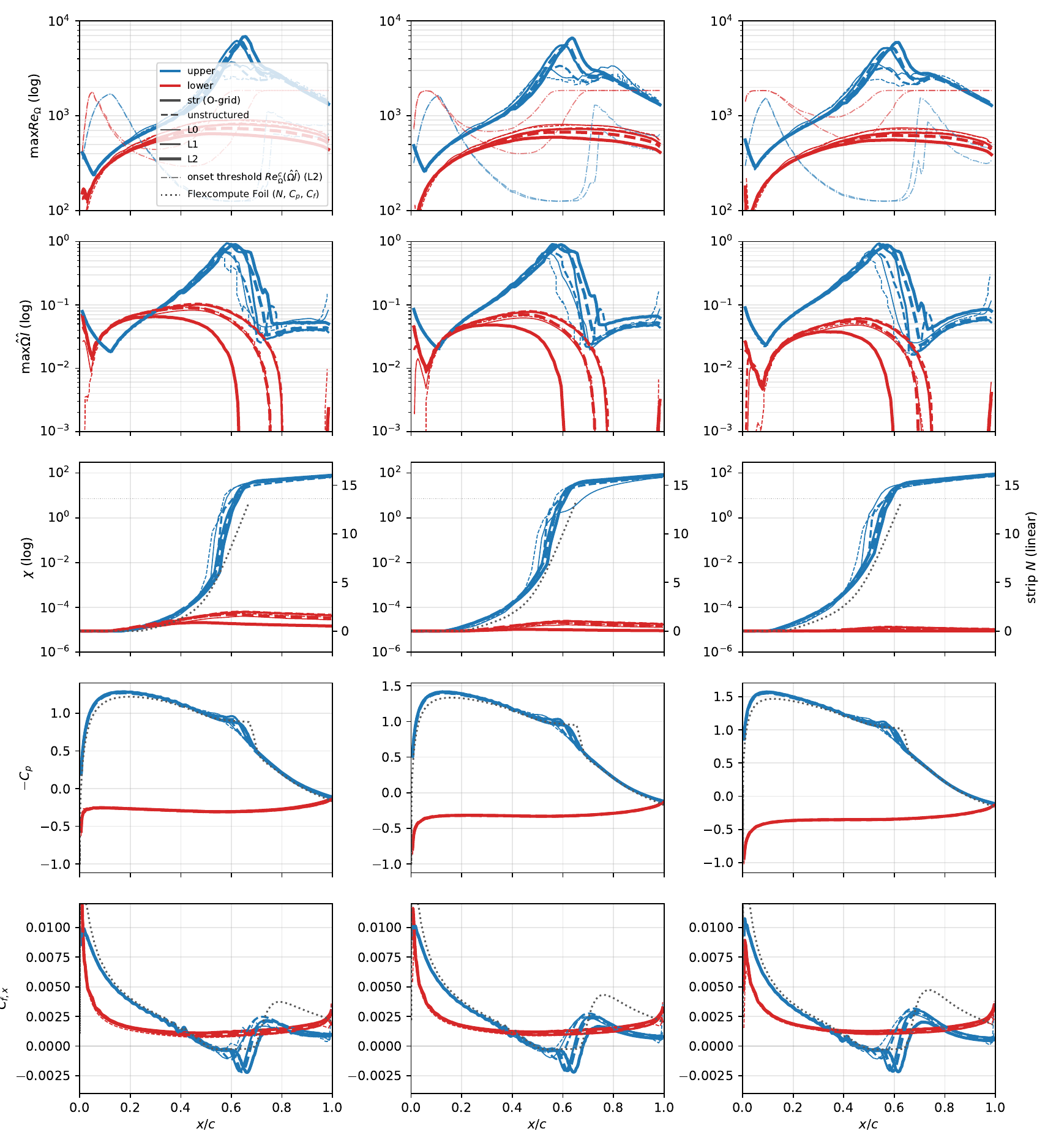}
\figcont{$\eta=0.75$.}
\end{figure}

\begin{figure}[H]\centering
\includegraphics[width=0.99\textwidth,height=0.94\textheight,keepaspectratio]{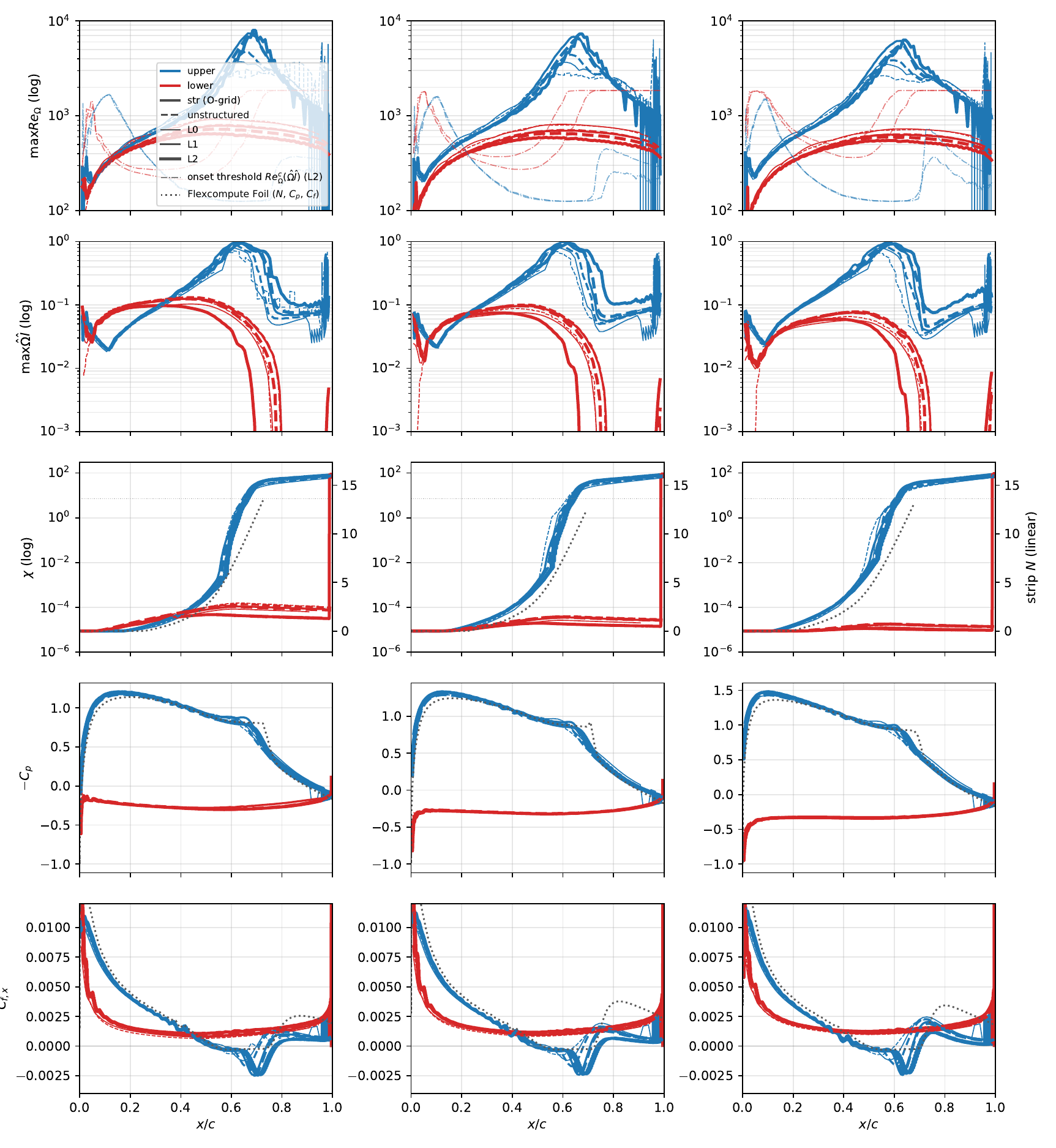}
\figcont{$\eta=0.92$.}
\end{figure}

\begin{figure}[H]\centering
\includegraphics[width=0.99\textwidth,height=0.94\textheight,keepaspectratio]{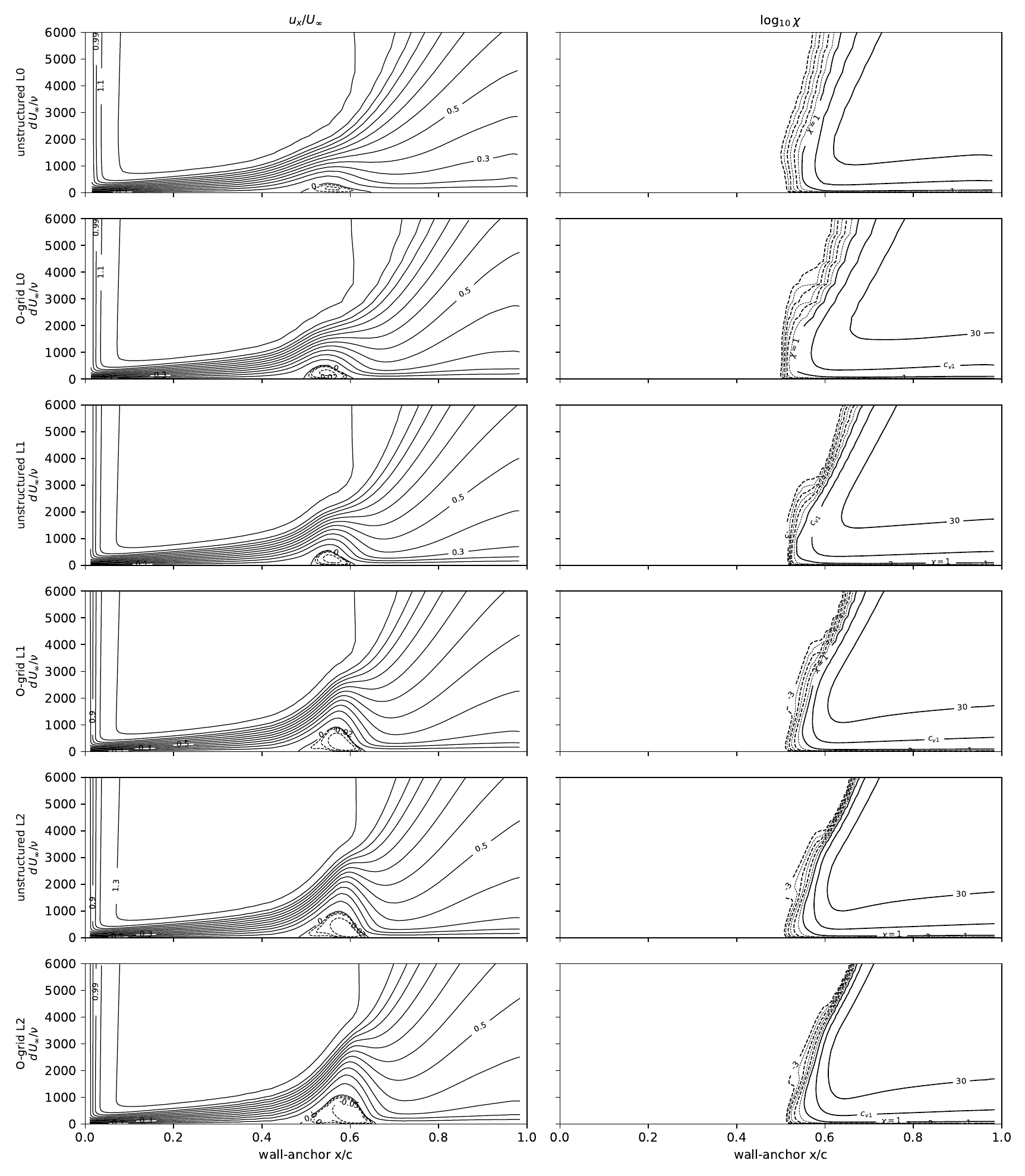}
\caption{Daedalus wing near the upper surface, on every grid: $u_x/U_\infty$ (left) and $\log_{10}\chi$ (right) against the chordwise station and the wall distance. $\alpha=4^\circ$, $\eta=0.10$.}\label{f:daesheets}
\end{figure}

\begin{figure}[H]\centering
\includegraphics[width=0.99\textwidth,height=0.94\textheight,keepaspectratio]{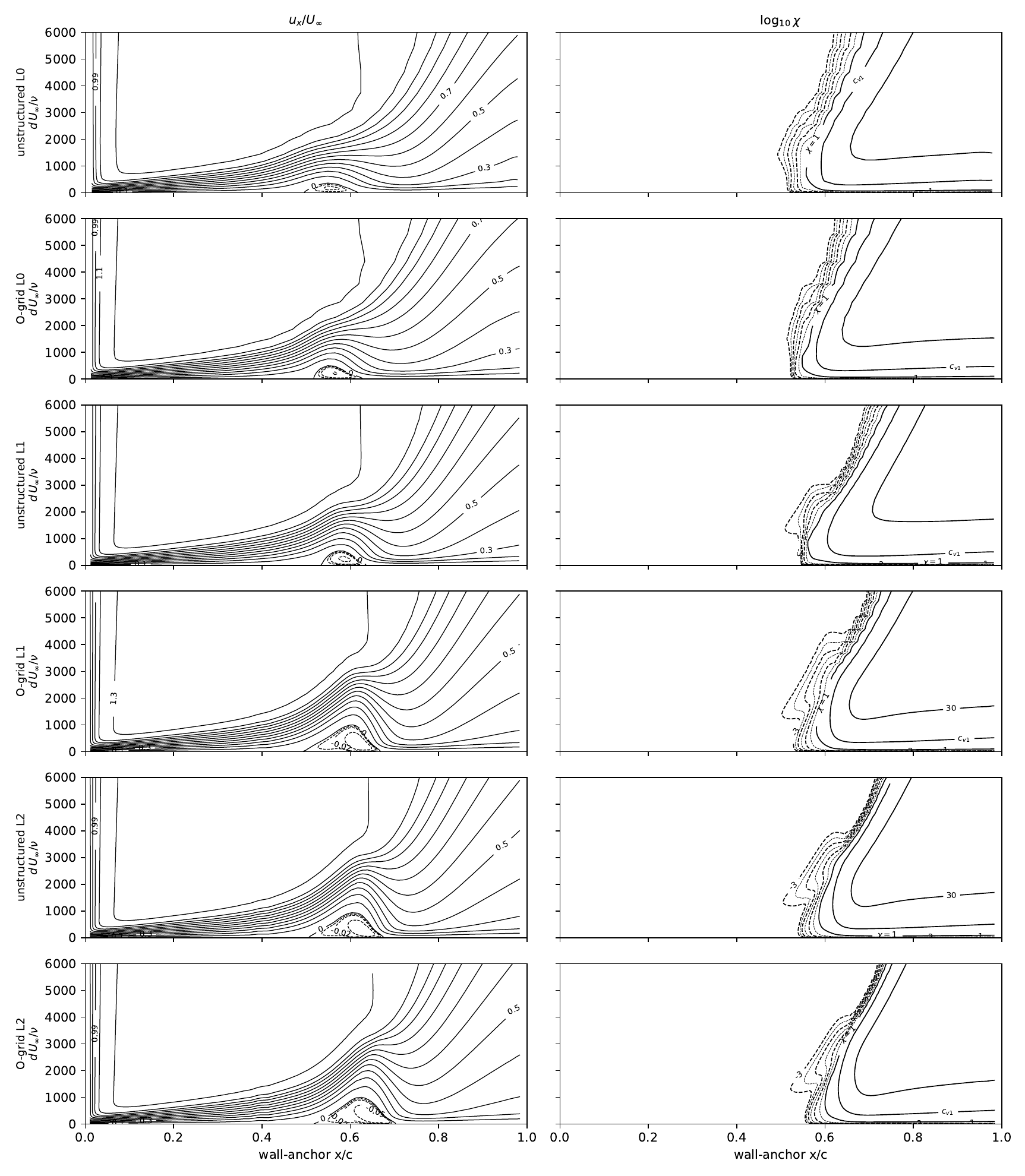}
\figcont{$\alpha=4^\circ$, $\eta=0.75$.}
\end{figure}

\begin{figure}[H]\centering
\includegraphics[width=0.99\textwidth,height=0.94\textheight,keepaspectratio]{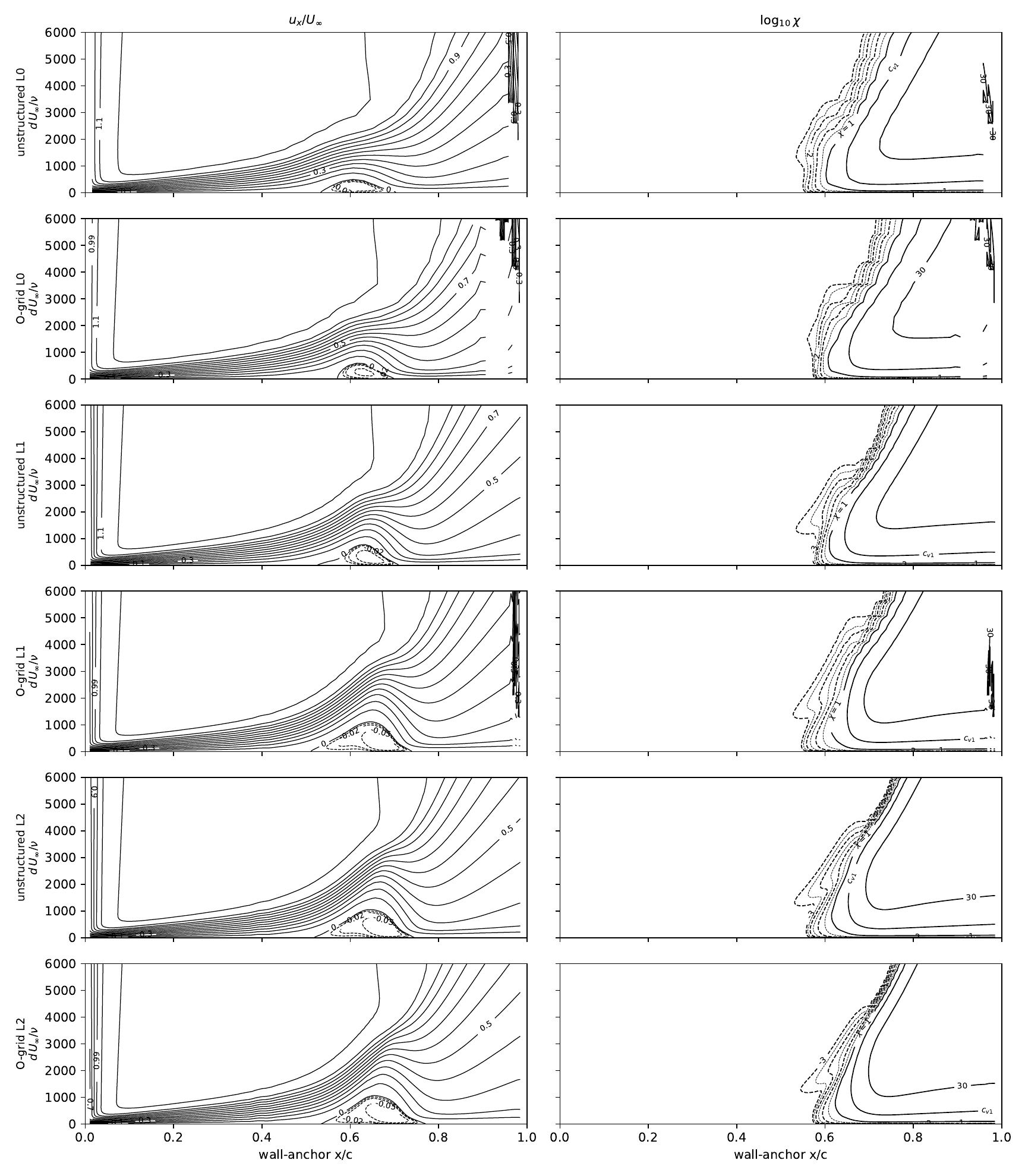}
\figcont{$\alpha=4^\circ$, $\eta=0.92$.}
\end{figure}

\begin{figure}[H]\centering
\includegraphics[width=0.99\textwidth,height=0.94\textheight,keepaspectratio]{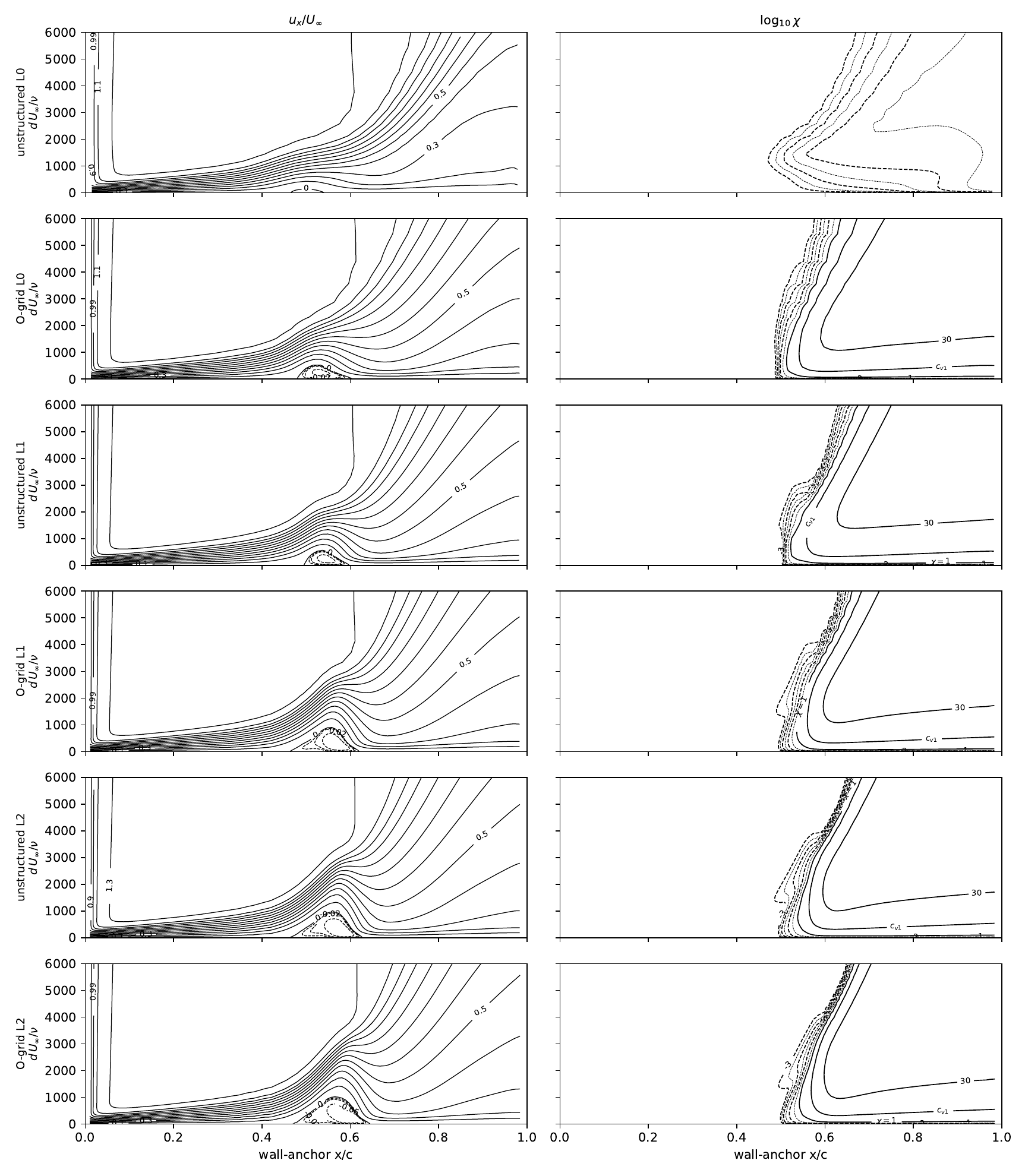}
\figcont{$\alpha=5^\circ$, $\eta=0.10$.}
\end{figure}

\begin{figure}[H]\centering
\includegraphics[width=0.99\textwidth,height=0.94\textheight,keepaspectratio]{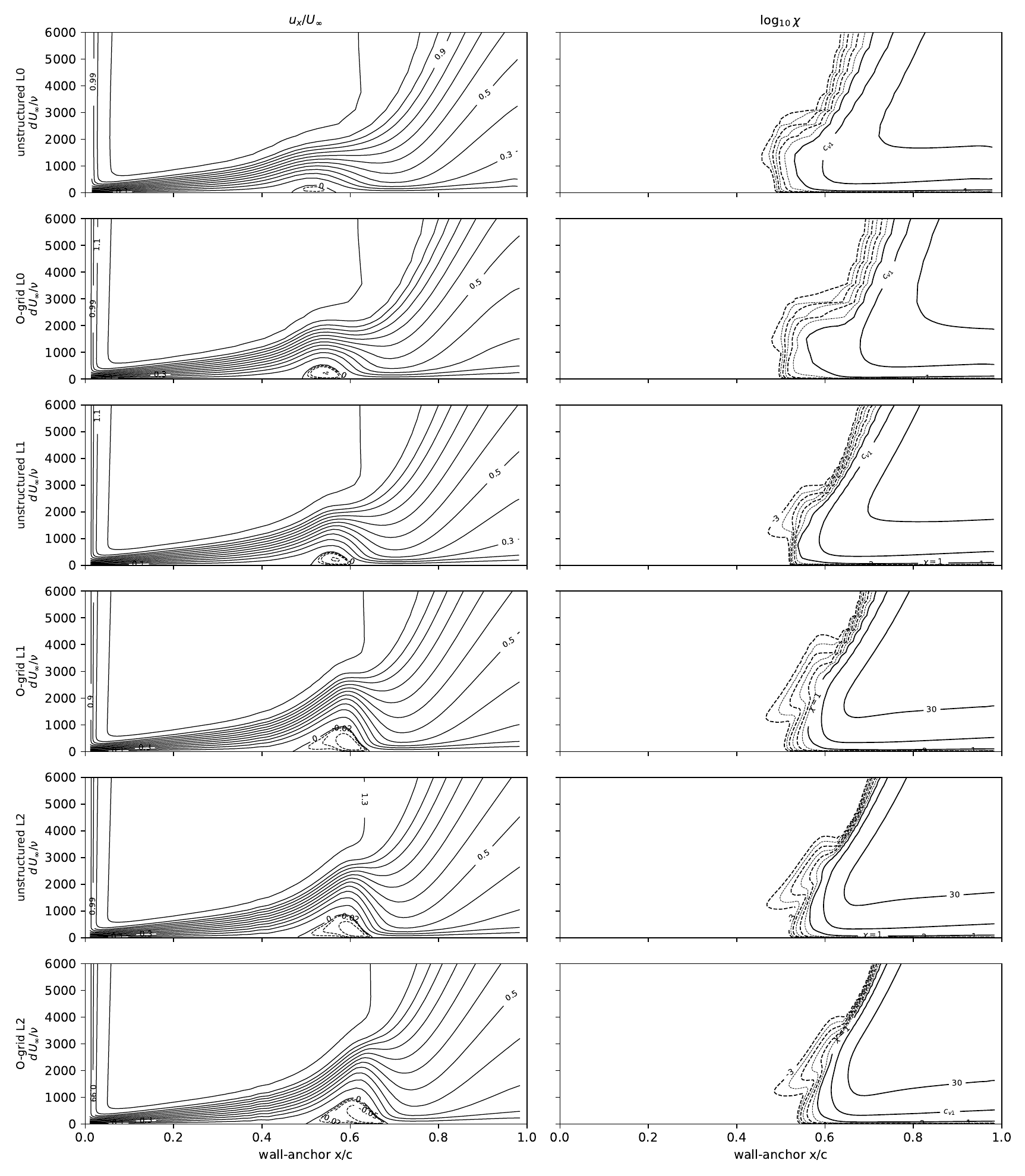}
\figcont{$\alpha=5^\circ$, $\eta=0.75$.}
\end{figure}

\begin{figure}[H]\centering
\includegraphics[width=0.99\textwidth,height=0.94\textheight,keepaspectratio]{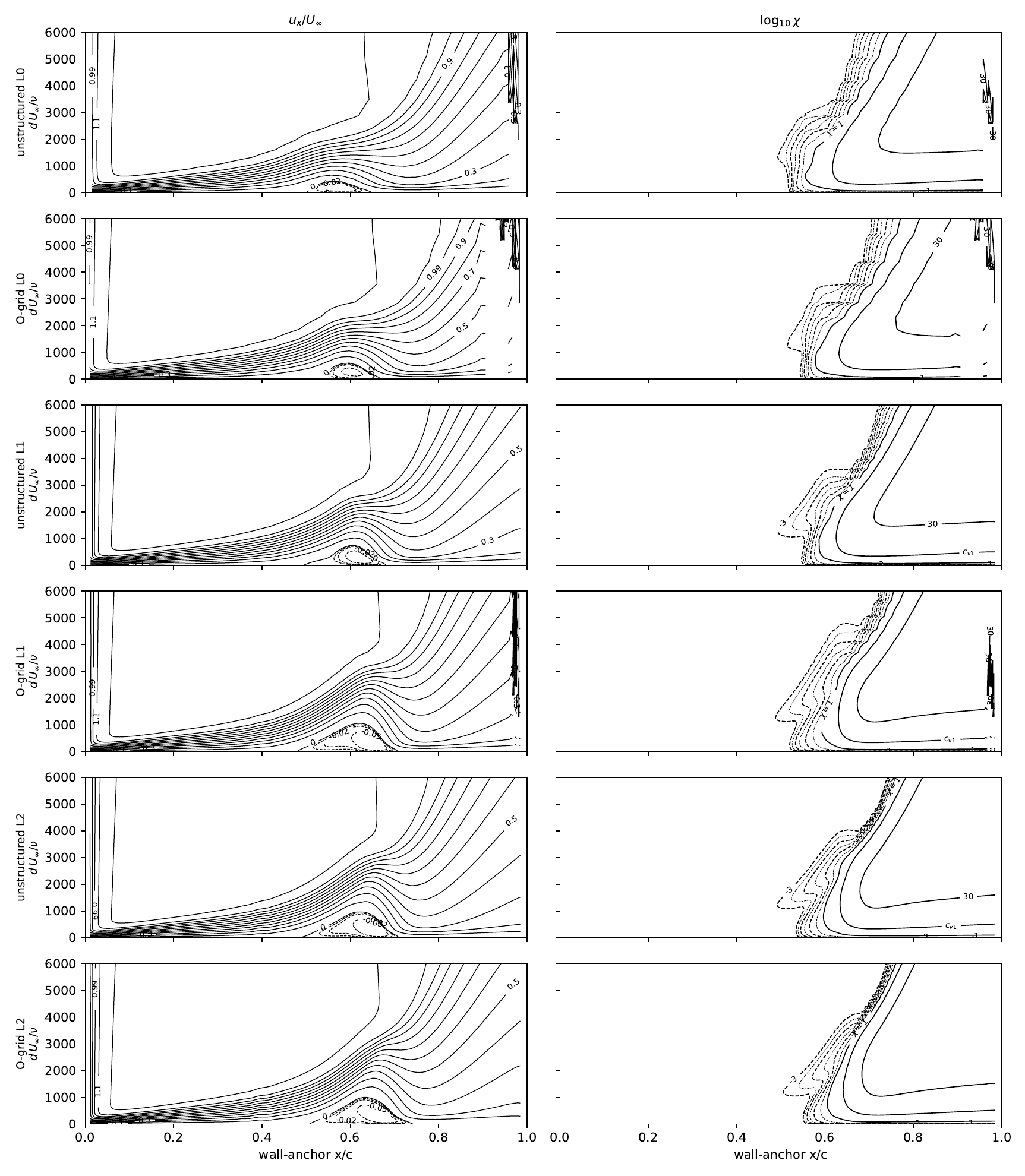}
\figcont{$\alpha=5^\circ$, $\eta=0.92$.}
\end{figure}

\begin{figure}[H]\centering
\includegraphics[width=0.99\textwidth,height=0.94\textheight,keepaspectratio]{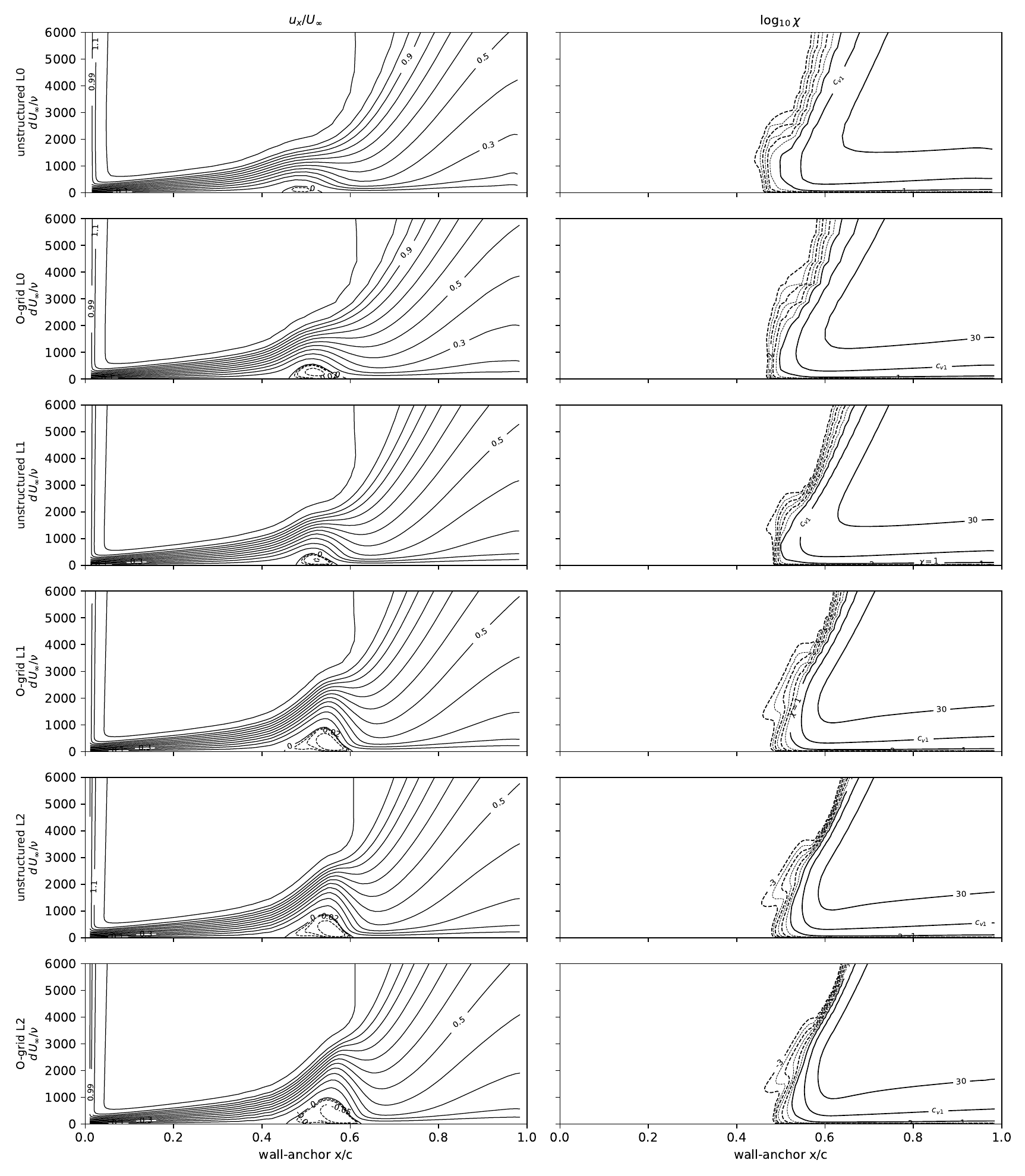}
\figcont{$\alpha=6^\circ$, $\eta=0.10$.}
\end{figure}

\begin{figure}[H]\centering
\includegraphics[width=0.99\textwidth,height=0.94\textheight,keepaspectratio]{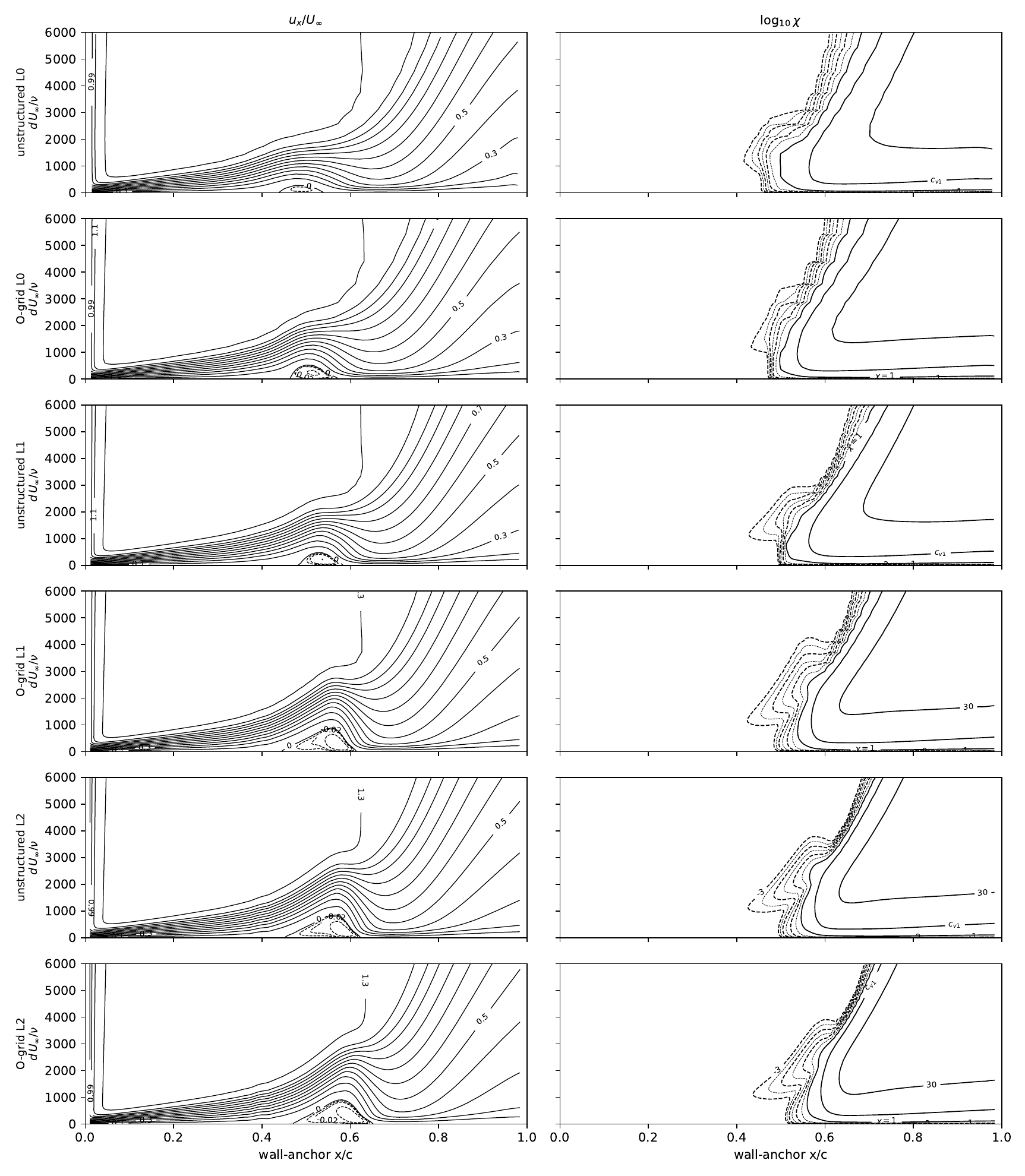}
\figcont{$\alpha=6^\circ$, $\eta=0.75$.}
\end{figure}

\begin{figure}[H]\centering
\includegraphics[width=0.99\textwidth,height=0.94\textheight,keepaspectratio]{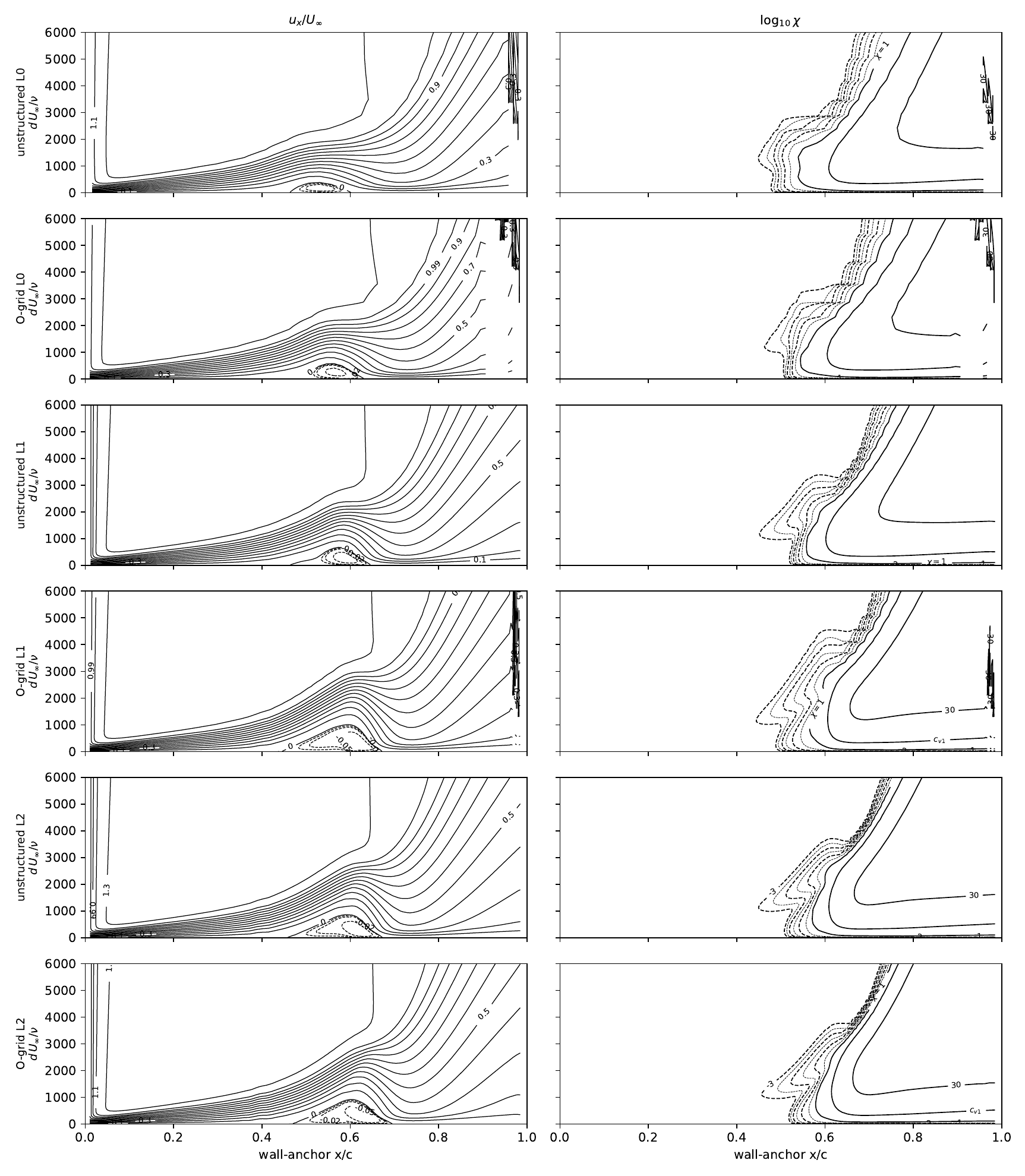}
\figcont{$\alpha=6^\circ$, $\eta=0.92$.}
\end{figure}

\subsection*{\tocentry{subsection}{The 6:1 Spheroid}The 6:1 Spheroid}

\begin{figure}[H]\centering
\includegraphics[width=0.86\textwidth]{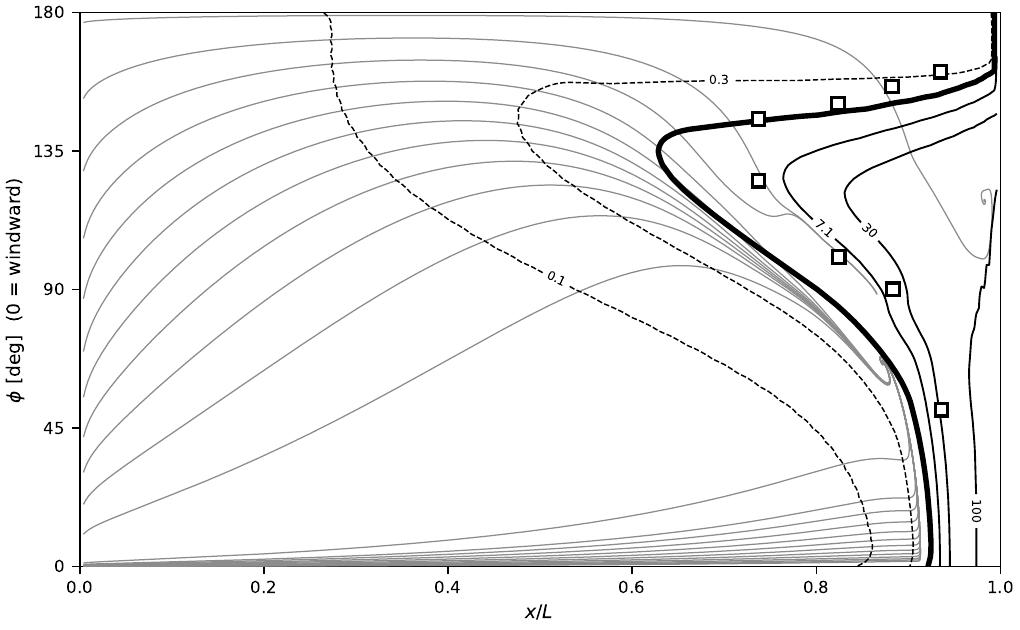}
\caption{6:1 spheroid with the measured turbulence intensity of the tunnel, one page per measured condition; layout as in Fig.~\ref{f:ov_a10hi}. $\alpha=5^\circ$, $Re_L=1.52\times10^6$.}\label{f:ov_a5lo_gmeas}
\end{figure}

\begin{figure}[H]\centering
\includegraphics[width=0.86\textwidth]{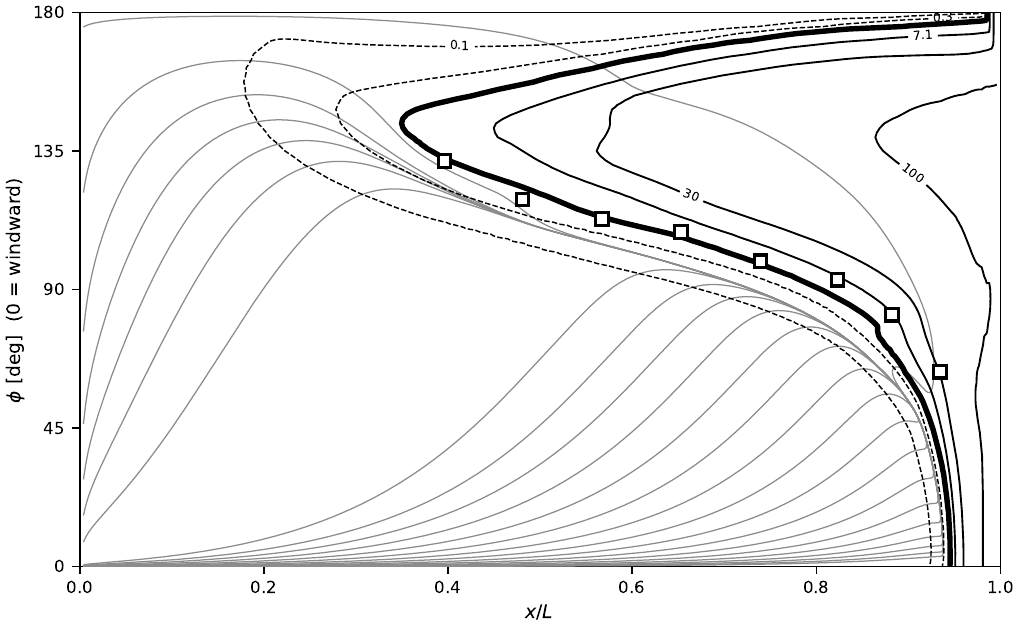}
\figcont{$\alpha=10^\circ$, $Re_L=1.52\times10^6$.}
\end{figure}

\begin{figure}[H]\centering
\includegraphics[width=0.86\textwidth]{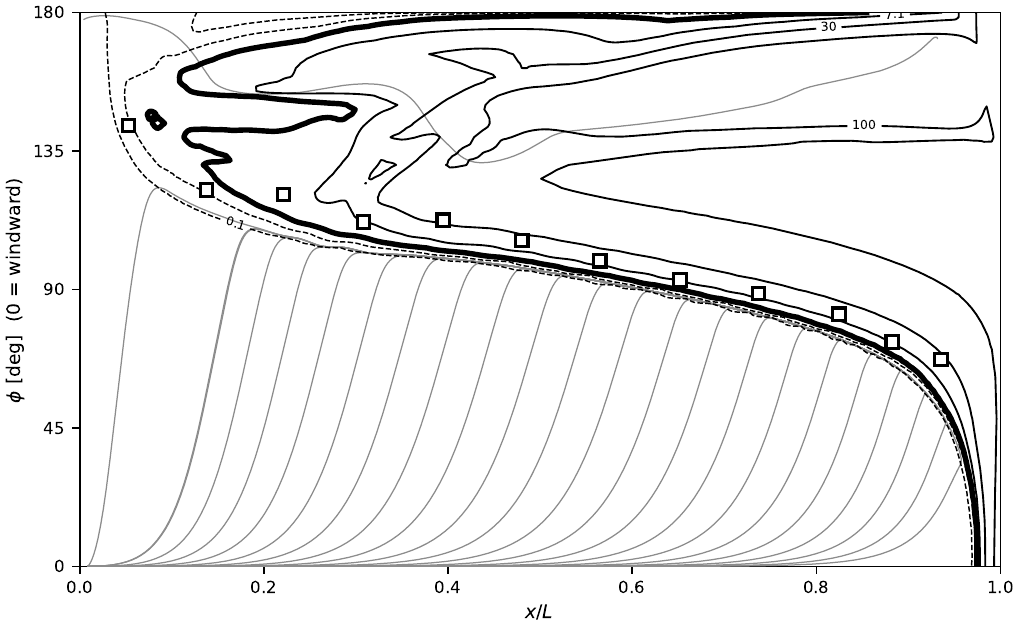}
\figcont{$\alpha=29.7^\circ$, $Re_L=1.53\times10^6$. The skin-friction lines converge on the separation line over most of the leeward half, and the twelve measured stations are on that line.}
\end{figure}

\begin{figure}[H]\centering
\includegraphics[width=0.86\textwidth]{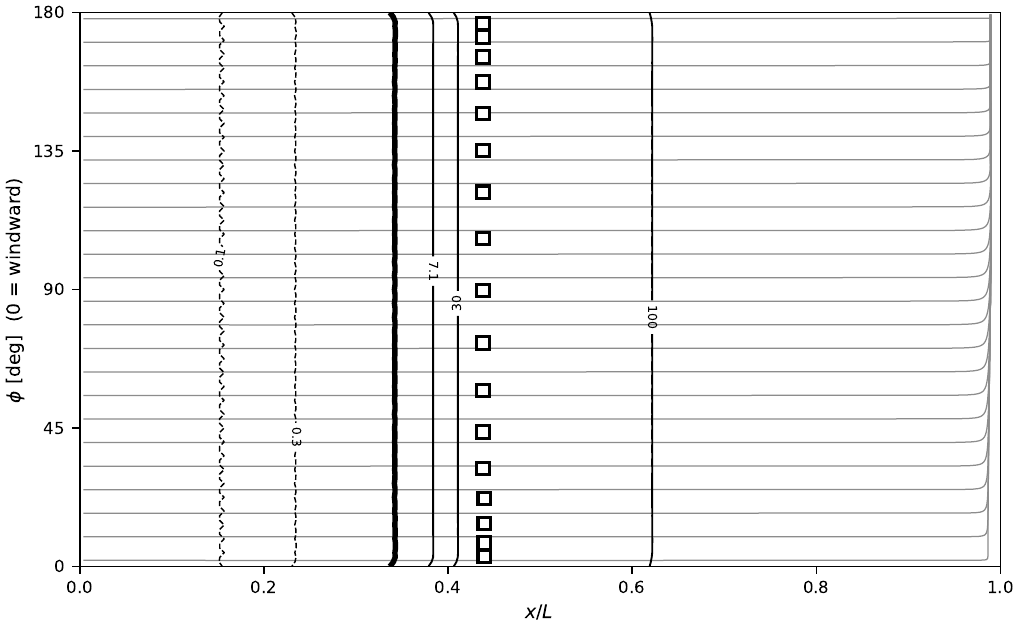}
\figcont{$\alpha=0^\circ$, $Re_L=7.2\times10^6$.}
\end{figure}

\begin{figure}[H]\centering
\includegraphics[width=0.86\textwidth]{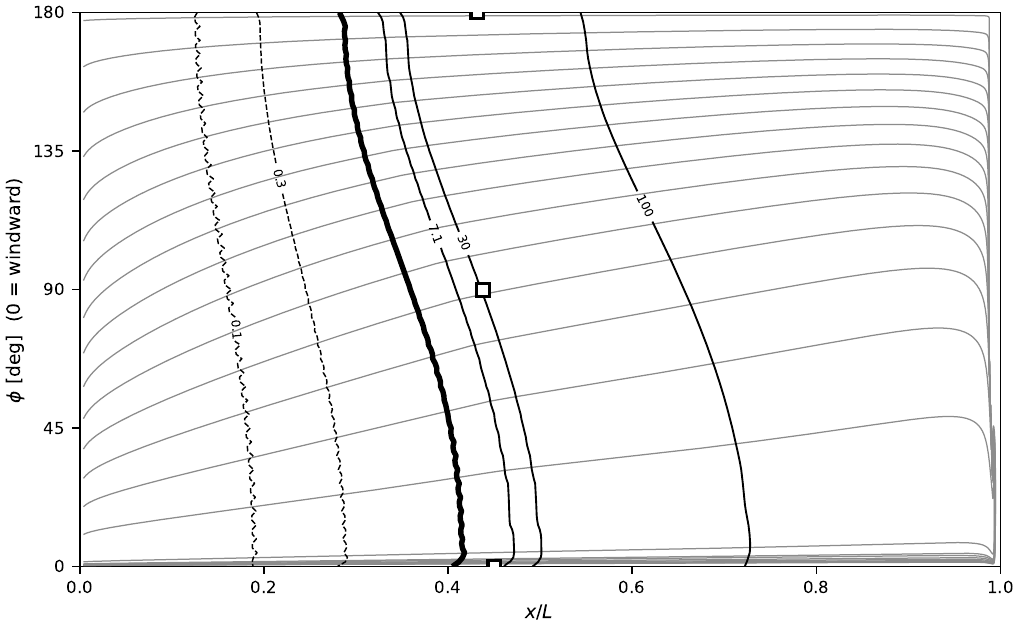}
\figcont{$\alpha=2.5^\circ$, $Re_L=7.2\times10^6$.}
\end{figure}

\begin{figure}[H]\centering
\includegraphics[width=0.86\textwidth]{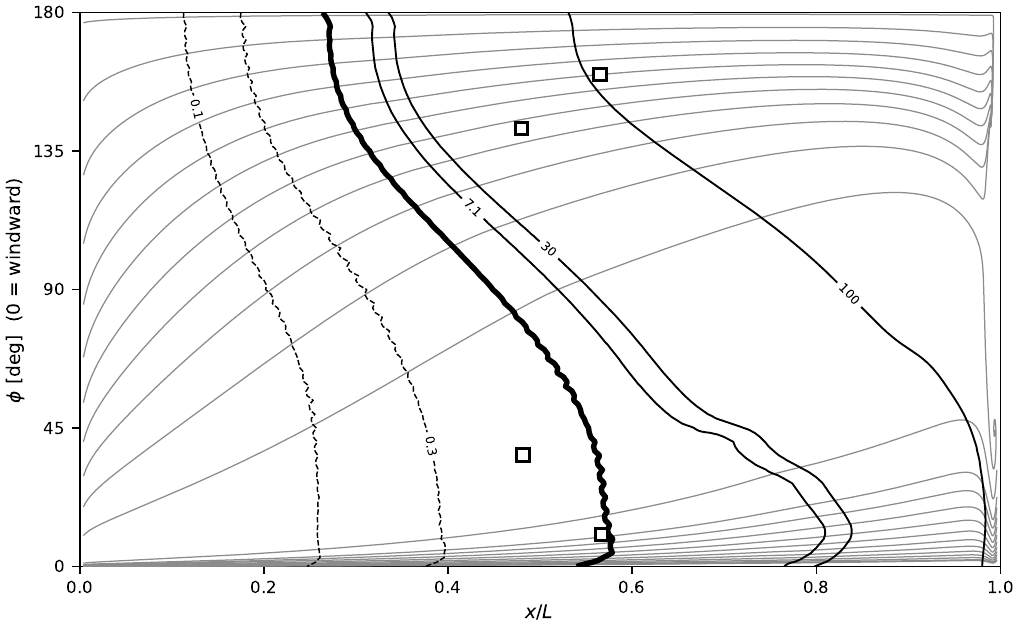}
\figcont{$\alpha=5^\circ$, $Re_L=6.49\times10^6$.}
\end{figure}

\begin{figure}[H]\centering
\includegraphics[width=0.86\textwidth]{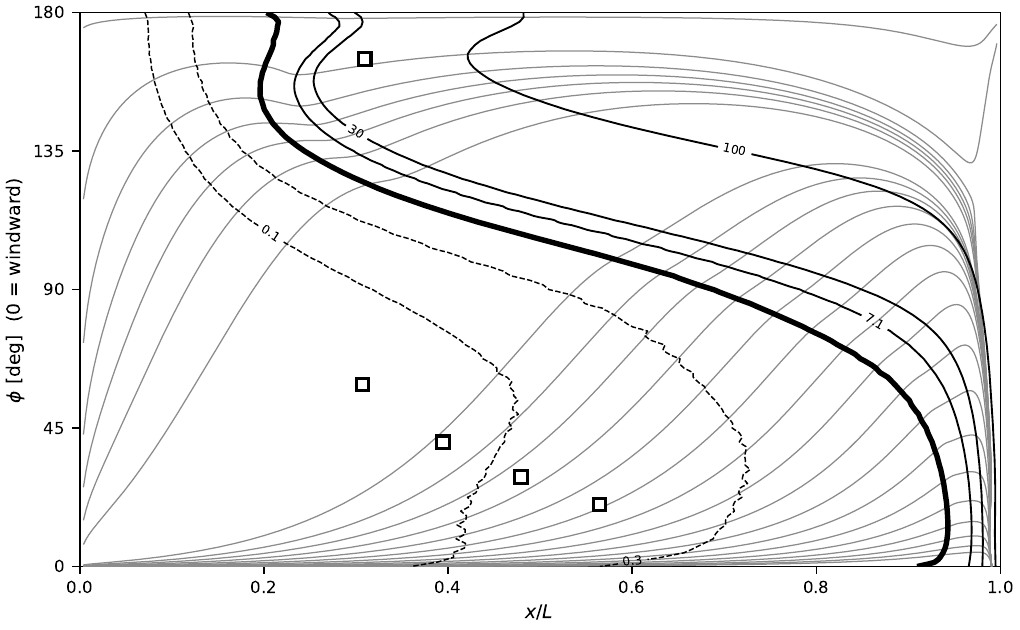}
\figcont{$\alpha=10^\circ$, $Re_L=6.56\times10^6$, G\"ottingen.}
\end{figure}

\begin{figure}[H]\centering
\includegraphics[width=0.86\textwidth]{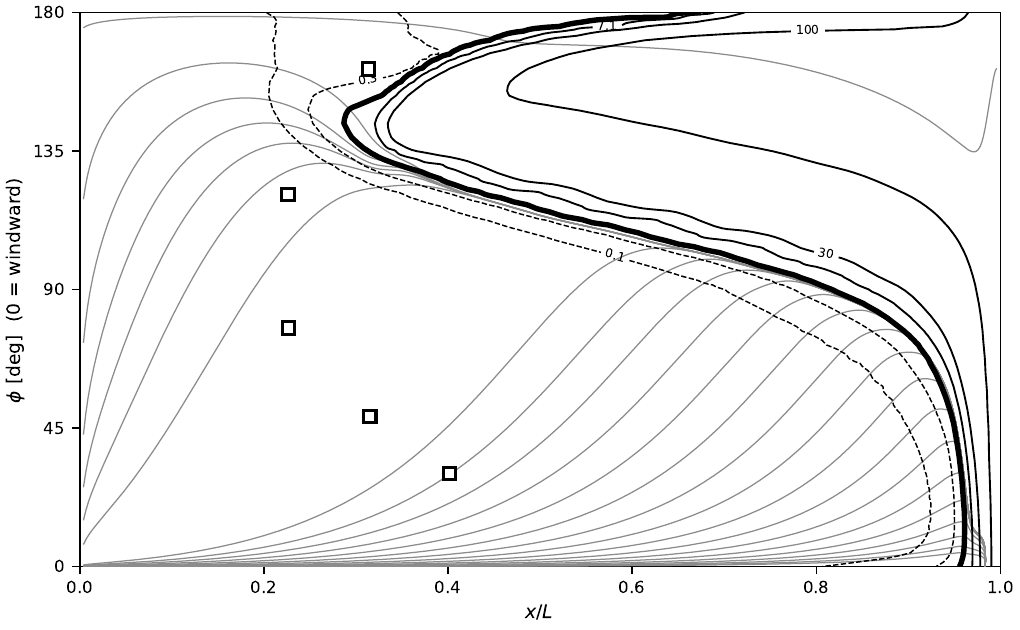}
\figcont{$\alpha=10^\circ$, $Re_L=6.56\times10^6$, ONERA F1.}
\end{figure}

\clearpage
\tocentry{section}{References}
\bibliography{references}

\begin{thebibliography}{61}
\newcommand{\enquote}[1]{``#1''}
\providecommand{\natexlab}[1]{#1}
\providecommand{\url}[1]{\texttt{#1}}
\providecommand{\urlprefix}{URL }
\expandafter\ifx\csname urlstyle\endcsname\relax
  \providecommand{\doi}[1]{doi:\discretionary{}{}{}#1}\else
  \providecommand{\doi}{doi:\discretionary{}{}{}\begingroup
  \urlstyle{rm}\Url}\fi

\bibitem[{Mack(1977)}]{mack_1977}
Mack, L.~M., \enquote{Transition and Laminar Instability,} JPL Publication
  77-15, Jet Propulsion Laboratory, 1977.

\bibitem[{Schmid and Henningson(2001)}]{schmid_henningson_2001}
Schmid, P.~J., and Henningson, D.~S., \emph{Stability and Transition in Shear
  Flows}, Applied Mathematical Sciences, Vol. 142, Springer, New York, 2001.
\newblock \doi{10.1007/978-1-4613-0185-1}.

\bibitem[{Schubauer and Klebanoff(1956)}]{schubauer_klebanoff_1956}
Schubauer, G.~B., and Klebanoff, P.~S., \enquote{Contributions on the Mechanics
  of Boundary-Layer Transition,} NACA Report 1289, National Advisory Committee
  for Aeronautics, 1956.
\newblock Supersedes NACA TN 3489 (1955).

\bibitem[{Smith and Gamberoni(1956)}]{smith_gamberoni_1956}
Smith, A. M.~O., and Gamberoni, N., \enquote{Transition, Pressure Gradient and
  Stability Theory,} Report ES~26388, Douglas Aircraft Company, 1956.

\bibitem[{van Ingen(2008)}]{vaningen_2008}
van Ingen, J.~L., \enquote{The $e^N$ Method for Transition Prediction.
  Historical Review of Work at {TU} Delft,} \emph{38th AIAA Fluid Dynamics
  Conference and Exhibit}, 2008.
\newblock \doi{10.2514/6.2008-3830}, {AIAA} Paper 2008-3830.

\bibitem[{Drela and Giles(1987)}]{drela_giles_1987}
Drela, M., and Giles, M.~B., \enquote{Viscous--Inviscid Analysis of Transonic
  and Low Reynolds Number Airfoils,} \emph{AIAA Journal}, Vol.~25, No.~10,
  1987, pp. 1347--1355.
\newblock \doi{10.2514/3.9789}.

\bibitem[{Spalart and Allmaras(1992)}]{spalart_allmaras_1992}
Spalart, P.~R., and Allmaras, S.~R., \enquote{A One-Equation Turbulence Model
  for Aerodynamic Flows,} \emph{30th Aerospace Sciences Meeting and Exhibit},
  1992.
\newblock \doi{10.2514/6.1992-439}, {AIAA} Paper 1992-0439.

\bibitem[{Allmaras et~al.(2012)Allmaras, Johnson, and Spalart}]{allmaras_2012}
Allmaras, S.~R., Johnson, F.~T., and Spalart, P.~R., \enquote{Modifications and
  Clarifications for the Implementation of the {Spalart--Allmaras} Turbulence
  Model,} \emph{Seventh International Conference on Computational Fluid
  Dynamics (ICCFD7)}, Big Island, Hawaii, 2012.
\newblock ICCFD7-1902.

\bibitem[{Thwaites(1949)}]{thwaites_1949}
Thwaites, B., \enquote{Approximate Calculation of the Laminar Boundary Layer,}
  \emph{Aeronautical Quarterly}, Vol.~1, 1949, pp. 245--280.

\bibitem[{Falkner and Skan(1931)}]{falkner_skan_1931}
Falkner, V.~M., and Skan, S.~W., \enquote{Some Approximate Solutions of the
  Boundary-Layer Equations,} \emph{Philosophical Magazine, Series 7}, Vol.~12,
  No.~80, 1931, pp. 865--896.

\bibitem[{Stewartson(1954)}]{stewartson_1954}
Stewartson, K., \enquote{Further Solutions of the {Falkner--Skan} Equation,}
  \emph{Mathematical Proceedings of the Cambridge Philosophical Society},
  Vol.~50, No.~3, 1954, pp. 454--465.

\bibitem[{Michalke(1964)}]{michalke_1964}
Michalke, A., \enquote{On the Inviscid Instability of the Hyperbolic-Tangent
  Velocity Profile,} \emph{Journal of Fluid Mechanics}, Vol.~19, No.~4, 1964,
  pp. 543--556.

\bibitem[{Spalart and Allmaras(1994)}]{spalart_allmaras_1994}
Spalart, P.~R., and Allmaras, S.~R., \enquote{A One-Equation Turbulence Model
  for Aerodynamic Flows,} \emph{La Recherche A{\'e}rospatiale}, , No.~1, 1994,
  pp. 5--21.

\bibitem[{Reichardt(1951)}]{reichardt_1951}
Reichardt, H., \enquote{Vollst{\"a}ndige Darstellung der turbulenten
  Geschwindigkeitsverteilung in glatten Leitungen,} \emph{Zeitschrift f{\"u}r
  Angewandte Mathematik und Mechanik}, Vol.~31, No.~7, 1951, pp. 208--219.

\bibitem[{Coles(1956)}]{coles_1956}
Coles, D., \enquote{The Law of the Wake in the Turbulent Boundary Layer,}
  \emph{Journal of Fluid Mechanics}, Vol.~1, No.~2, 1956, pp. 191--226.
\newblock \doi{10.1017/S0022112056000135}.

\bibitem[{Das(1987)}]{das_1987}
Das, D.~K., \enquote{A Numerical Study of Turbulent Separated Flows,}
  \emph{{ASME} Forum on Turbulent Flows}, FED, Vol.~51, 1987, pp. 85--90.

\bibitem[{Loseille et~al.(2017)Loseille, Alauzet, and
  Menier}]{loseille_alauzet_menier_2017}
Loseille, A., Alauzet, F., and Menier, V., \enquote{Unique Cavity-Based
  Operator and Hierarchical Domain Partitioning for Fast Parallel Generation of
  Anisotropic Meshes,} \emph{Computer-Aided Design}, Vol.~85, 2017, pp. 53--67.

\bibitem[{Fidkowski(2022)}]{fidkowski_mfoil_2022}
Fidkowski, K.~J., \enquote{A Coupled Inviscid--Viscous Airfoil Analysis Solver,
  Revisited,} \emph{AIAA Journal}, Vol.~60, No.~4, 2022.
\newblock \doi{10.2514/1.J061341}.

\bibitem[{Drela(1989)}]{drela_xfoil_1989}
Drela, M., \enquote{{XFOIL}: An Analysis and Design System for Low {Reynolds}
  Number Airfoils,} \emph{Low Reynolds Number Aerodynamics}, Lecture Notes in
  Engineering, Vol.~54, Springer, 1989, pp. 1--12.
\newblock \doi{10.1007/978-3-642-84010-4_1}.

\bibitem[{Smith(2026)}]{smith_flexfoil_2026}
Smith, H.~R., \enquote{Why {I} Built a Browser-Based {XFOIL} (and Why We're
  Investing in Open Source),} Flexcompute, Inc.,
  \url{https://flexcompute.com/blog/flexfoil}, Mar. 2026.

\bibitem[{Abu-Ghannam and Shaw(1980)}]{abu_ghannam_shaw_1980}
Abu-Ghannam, B.~J., and Shaw, R., \enquote{Natural Transition of Boundary
  Layers---The Effects of Turbulence, Pressure Gradient, and Flow History,}
  \emph{Journal of Mechanical Engineering Science}, Vol.~22, No.~5, 1980, pp.
  213--228.
\newblock \doi{10.1243/JMES_JOUR_1980_022_043_02}.

\bibitem[{Somers(1981)}]{somers_1981}
Somers, D.~M., \enquote{Design and Experimental Results for a
  Natural-Laminar-Flow Airfoil for General Aviation Applications,} NASA
  TP-1861, NASA Langley Research Center, 1981.

\bibitem[{Coder(2021)}]{coder_tmpw_summary}
Coder, J.~G., \enquote{Summary of the First {AIAA} {CFD} Transition Modeling
  and Prediction Workshop,} NASA Advanced Modeling and Simulation Seminar
  Series, NASA Ames Research Center, Mar. 2021.

\bibitem[{Coder(2014)}]{coder_thesis_2014}
Coder, J.~G., \enquote{Development of a {CFD}-Compatible Transition Model Based
  on Linear Stability Theory,} Ph.D. thesis, The Pennsylvania State University,
  2014.
\newblock ETDA catalog 22627.

\bibitem[{Piotrowski and Zingg(2021)}]{piotrowski_zingg_2021}
Piotrowski, M. G.~H., and Zingg, D.~W., \enquote{Smooth Local Correlation-based
  Transition Model for the {Spalart--Allmaras} Turbulence Model,} \emph{AIAA
  Journal}, Vol.~59, No.~2, 2021, pp. 474--492.

\bibitem[{Denison(2021)}]{denison_2021_overflow}
Denison, M., \enquote{1st {AIAA} {CFD} Transition Modeling and Prediction
  Workshop: {OVERFLOW} Results for the {SST} and {Langtry-Menter} Models,}
  \emph{AIAA CFD Transition Modeling and Prediction Workshop, AIAA SciTech
  Forum}, 2021.
\newblock NASA Ames Research Center; available from NASA NTRS.

\bibitem[{McGhee et~al.(1988)McGhee, Walker, and Millard}]{mcghee_1988}
McGhee, R.~J., Walker, B.~S., and Millard, B.~F., \enquote{Experimental Results
  for the {Eppler 387} Airfoil at Low {Reynolds} Numbers in the {Langley
  Low-Turbulence Pressure Tunnel},} NASA TM-4062, NASA Langley Research Center,
  1988.

\bibitem[{Selig(2003)}]{selig_2003}
Selig, M.~S., \enquote{Low {Reynolds} Number Airfoil Design Lecture Notes,}
  \emph{VKI Lecture Series: Low {Reynolds} Number Aerodynamics on Aircraft
  Including Applications in Emerging {UAV} Technology}, von Karman Institute
  for Fluid Dynamics, 2003.
\newblock NATO RTO/AVT-VKI-104, 24--28 November.

\bibitem[{Cole and Mueller(1990)}]{cole_mueller_1990}
Cole, G.~M., and Mueller, T.~J., \enquote{Experimental Measurements of the
  Laminar Separation Bubble on an {Eppler} 387 Airfoil at Low {Reynolds}
  Numbers,} NASA CR-186238, University of Notre Dame, 1990.
\newblock Also UNDAS-1419-FR; NTRS 19900006064.

\bibitem[{Shahjahan et~al.(2024)Shahjahan, Emmerson, and
  Verstraete}]{shahjahan_2024}
Shahjahan, S., Emmerson, B., and Verstraete, D., \enquote{On the Choice of
  Turbulence Model for the Simulation of Airfoils at {R}eynolds Numbers Below
  200,000,} \emph{Proceedings of the 34th Congress of the International Council
  of the Aeronautical Sciences}, Florence, Italy, 2024.
\newblock ICAS paper 2024-0327.

\bibitem[{Fr{\`e}re et~al.(2016)Fr{\`e}re, S{\o}rensen, Hillewaert, and
  Winckelmans}]{frere_2016}
Fr{\`e}re, A., S{\o}rensen, N.~N., Hillewaert, K., and Winckelmans, G.,
  \enquote{Discontinuous {Galerkin} Methodology for Large-Eddy Simulations of
  Wind Turbine Airfoils,} \emph{Journal of Physics: Conference Series}, Vol.
  753, 2016, p. 022037.
\newblock \doi{10.1088/1742-6596/753/2/022037}.

\bibitem[{Carre{\~n}o~Ruiz and D'Ambrosio(2022)}]{carreno_ruiz_2022}
Carre{\~n}o~Ruiz, M., and D'Ambrosio, D., \enquote{Validation of the
  {$\gamma$}-{$Re_\theta$} Transition Model for Airfoils Operating in the Very
  Low {Reynolds} Number Regime,} \emph{Flow, Turbulence and Combustion}, Vol.
  109, 2022, pp. 279--308.
\newblock \doi{10.1007/s10494-022-00331-z}.

\bibitem[{{Anushka R.} et~al.(2019){Anushka R.}, Chaudary, Singh, {Satish Wadel
  S.}, Narayana, and Gupta}]{anushka_2019}
{Anushka R.}, Chaudary, D., Singh, R.~R., {Satish Wadel S.}, Narayana, P.
  A.~A., and Gupta, A.~K., \enquote{Numerical Simulation of Flow over {Eppler}
  387 at Low {Reynolds} Number and Comparison with Experiment,}
  \emph{International Journal of Scientific and Research Publications}, Vol.~9,
  No.~1, 2019, pp. 204--219.
\newblock \doi{10.29322/IJSRP.9.01.2019.p8528}.

\bibitem[{Ghimire et~al.(2025)Ghimire, Ni, and Wang}]{ghimire_2025}
Ghimire, S., Ni, X., and Wang, Y., \enquote{Assessment of Transitional {RANS}
  Models and Implementation of Transitional {IDDES} Method for Boundary Layer
  Transition and Separated Flows in {OpenFOAM}-{V2312},} \emph{Fluids},
  Vol.~10, No.~9, 2025, p. 230.
\newblock \doi{10.3390/fluids10090230}.

\bibitem[{Dennis and Chang(1970)}]{dennis_chang_1970}
Dennis, S. C.~R., and Chang, G.-Z., \enquote{Numerical Solutions for Steady
  Flow Past a Circular Cylinder at {R}eynolds Numbers up to 100,} \emph{Journal
  of Fluid Mechanics}, Vol.~42, No.~3, 1970, pp. 471--489.
\newblock \doi{10.1017/S0022112070001428}.

\bibitem[{Wieselsberger(1922)}]{wieselsberger_1921}
Wieselsberger, C., \enquote{New Data on the Laws of Fluid Resistance,} Tech.
  Rep. TN-84, NACA, 1922.
\newblock Translation of Phys. Z. 22 (1921).

\bibitem[{Delany and Sorensen(1953)}]{delany_sorensen_1953}
Delany, N.~K., and Sorensen, N.~E., \enquote{Low-Speed Drag of Cylinders of
  Various Shapes,} Tech. Rep. TN 3038, NACA, 1953.

\bibitem[{Roshko(1961)}]{roshko_1961}
Roshko, A., \enquote{Experiments on the Flow Past a Circular Cylinder at Very
  High {R}eynolds Number,} \emph{Journal of Fluid Mechanics}, Vol.~10, No.~3,
  1961, pp. 345--356.

\bibitem[{Achenbach(1968)}]{achenbach_1968}
Achenbach, E., \enquote{Distribution of Local Pressure and Skin Friction around
  a Circular Cylinder in Cross-Flow up to {$Re = 5\times10^6$},} \emph{Journal
  of Fluid Mechanics}, Vol.~34, No.~4, 1968, pp. 625--639.
\newblock \doi{10.1017/S0022112068002120}.

\bibitem[{Achenbach and Heinecke(1981)}]{achenbach_heinecke_1981}
Achenbach, E., and Heinecke, E., \enquote{On Vortex Shedding from Smooth and
  Rough Cylinders in the Range of {R}eynolds Numbers {$6\times10^3$} to
  {$5\times10^6$},} \emph{Journal of Fluid Mechanics}, Vol. 109, 1981, pp.
  239--251.

\bibitem[{Schewe(1983)}]{schewe_1983}
Schewe, G., \enquote{On the Force Fluctuations Acting on a Circular Cylinder in
  Crossflow from Subcritical up to Transcritical {R}eynolds Numbers,}
  \emph{Journal of Fluid Mechanics}, Vol. 133, 1983, pp. 265--285.
\newblock \doi{10.1017/S0022112083001913}.

\bibitem[{Henderson(1995)}]{henderson_1995}
Henderson, R.~D., \enquote{Details of the Drag Curve Near the Onset of Vortex
  Shedding,} \emph{Physics of Fluids}, Vol.~7, No.~9, 1995, pp. 2102--2104.
\newblock \doi{10.1063/1.868459}.

\bibitem[{Qu et~al.(2013)Qu, Norberg, Davidson, Peng, and Wang}]{qu_2013}
Qu, L., Norberg, C., Davidson, L., Peng, S.-H., and Wang, F.,
  \enquote{Quantitative Numerical Analysis of Flow Past a Circular Cylinder at
  {Reynolds} Number Between 50 and 200,} \emph{Journal of Fluids and
  Structures}, Vol.~39, 2013, pp. 347--370.
\newblock \doi{10.1016/j.jfluidstructs.2013.02.007}.

\bibitem[{Dong and Karniadakis(2005)}]{dong_karniadakis_2005}
Dong, S., and Karniadakis, G.~E., \enquote{{DNS} of Flow Past a Stationary and
  Oscillating Cylinder at {$Re=10\,000$},} \emph{Journal of Fluids and
  Structures}, Vol.~20, No.~4, 2005, pp. 519--531.
\newblock \doi{10.1016/j.jfluidstructs.2005.02.004}.

\bibitem[{Rodr{\'i}guez et~al.(2015)Rodr{\'i}guez, Lehmkuhl, Chiva, Borrell,
  and Oliva}]{rodriguez_2015}
Rodr{\'i}guez, I., Lehmkuhl, O., Chiva, J., Borrell, R., and Oliva, A.,
  \enquote{On the Flow Past a Circular Cylinder from Critical to Super-Critical
  {R}eynolds Numbers: Wake Topology and Vortex Shedding,} \emph{International
  Journal of Heat and Fluid Flow}, Vol.~55, 2015, pp. 91--103.
\newblock \doi{10.1016/j.ijheatfluidflow.2015.05.009}.

\bibitem[{Catalano et~al.(2003)Catalano, Wang, Iaccarino, and
  Moin}]{catalano_2003}
Catalano, P., Wang, M., Iaccarino, G., and Moin, P., \enquote{Numerical
  Simulation of the Flow Around a Circular Cylinder at High {R}eynolds
  Numbers,} \emph{International Journal of Heat and Fluid Flow}, Vol.~24,
  No.~4, 2003, pp. 463--469.

\bibitem[{Stabnikov and Garbaruk(2020)}]{stabnikov_garbaruk_2020}
Stabnikov, A.~S., and Garbaruk, A.~V., \enquote{Prediction of Drag Crisis on a
  Circular Cylinder Using a New Algebraic Transition Model Coupled with {SST
  DDES},} \emph{Journal of Physics: Conference Series}, Vol. 1697, 2020, p.
  012224.

\bibitem[{Stringer et~al.(2014)Stringer, Zang, and Hillis}]{stringer_2014}
Stringer, R.~M., Zang, J., and Hillis, A.~J., \enquote{Unsteady {RANS}
  Computations of Flow Around a Circular Cylinder for a Wide Range of
  {R}eynolds Numbers,} \emph{Ocean Engineering}, Vol.~87, 2014, pp. 1--9.

\bibitem[{Drela(1988)}]{drela_daedalus_1988}
Drela, M., \enquote{Low-{R}eynolds-number airfoil design for the {MIT}
  {D}aedalus prototype: a case study,} \emph{Journal of Aircraft}, Vol.~25,
  No.~8, 1988, pp. 724--732.

\bibitem[{Drela and Youngren(2017)}]{drela_youngren_avl}
Drela, M., and Youngren, H., \enquote{{AVL}---{Athena Vortex Lattice},
  Ver.~3.36,} Software, Massachusetts Institute of Technology, 2017.
\newblock \url{https://web.mit.edu/drela/Public/web/avl/}.

\bibitem[{Kreplin et~al.(1985)Kreplin, Vollmers, and Meier}]{kreplin_1985}
Kreplin, H.-P., Vollmers, H., and Meier, H.~U., \enquote{Wall Shear Stress
  Measurements on an Inclined Prolate Spheroid in the {DFVLR} 3~m~$\times$~3~m
  Low-Speed Wind Tunnel, {G}\"ottingen,} Tech. Rep. IB 222-84 A 33, DFVLR-AVA,
  G\"ottingen, Germany, 1985.

\bibitem[{Stock(2006)}]{stock_2006}
Stock, H.~W., \enquote{$e^N$ Transition Prediction in Three-Dimensional
  Boundary Layers on Inclined Prolate Spheroids,} \emph{AIAA Journal}, Vol.~44,
  No.~1, 2006, pp. 108--118.
\newblock \doi{10.2514/1.16026}.

\bibitem[{Drazin and Reid(1981)}]{drazin_reid_1981}
Drazin, P.~G., and Reid, W.~H., \emph{Hydrodynamic Stability}, Cambridge
  University Press, 1981.

\bibitem[{Johnston et~al.(1972)Johnston, Halleen, and Lezius}]{johnston_1972}
Johnston, J.~P., Halleen, R.~M., and Lezius, D.~K., \enquote{Effects of
  Spanwise Rotation on the Structure of Two-Dimensional Fully Developed
  Turbulent Channel Flow,} \emph{Journal of Fluid Mechanics}, Vol.~56, No.~3,
  1972, pp. 533--557.

\bibitem[{Saric(1994)}]{saric_1994}
Saric, W.~S., \enquote{{G\"ortler} Vortices,} \emph{Annual Review of Fluid
  Mechanics}, Vol.~26, 1994, pp. 379--409.
\newblock \doi{10.1146/annurev.fl.26.010194.002115}.

\bibitem[{Tutty et~al.(2002)Tutty, Price, and Parsons}]{tutty_2002}
Tutty, O.~R., Price, W.~G., and Parsons, A.~T., \enquote{Boundary Layer Flow on
  a Long Thin Cylinder,} \emph{Physics of Fluids}, Vol.~14, No.~2, 2002, pp.
  628--637.

\bibitem[{Hall et~al.(1984)Hall, Malik, and Poll}]{hall_malik_poll_1984}
Hall, P., Malik, M.~R., and Poll, D. I.~A., \enquote{On the Stability of an
  Infinite Swept Attachment Line Boundary Layer,} \emph{Proceedings of the
  Royal Society of London A}, Vol. 395, No. 1809, 1984, pp. 229--245.
\newblock \doi{10.1098/rspa.1984.0099}.

\bibitem[{Poll(1979)}]{poll_1979}
Poll, D. I.~A., \enquote{Transition in the Infinite Swept Attachment Line
  Boundary Layer,} \emph{The Aeronautical Quarterly}, Vol.~30, No.~4, 1979, pp.
  607--629.

\bibitem[{Hocking(1975)}]{hocking_1975}
Hocking, L.~M., \enquote{Non-Linear Instability of the Asymptotic Suction
  Velocity Profile,} \emph{The Quarterly Journal of Mechanics and Applied
  Mathematics}, Vol.~28, No.~3, 1975, pp. 341--353.
\newblock \doi{10.1093/qjmam/28.3.341}.

\bibitem[{Mayle(1991)}]{mayle_1991}
Mayle, R.~E., \enquote{The Role of Laminar-Turbulent Transition in Gas Turbine
  Engines,} \emph{Journal of Turbomachinery}, Vol. 113, No.~4, 1991, pp.
  509--536.
\newblock \doi{10.1115/1.2929110}.

\bibitem[{Coder and Maughmer(2014)}]{coder_maughmer_2014}
Coder, J.~G., and Maughmer, M.~D., \enquote{Computational Fluid Dynamics
  Compatible Transition Modeling Using an Amplification Factor Transport
  Equation,} \emph{AIAA Journal}, Vol.~52, No.~11, 2014, pp. 2506--2512.
\newblock \doi{10.2514/1.J052905}.

\end{thebibliography}

\end{document}